\documentclass{aa}  

\usepackage[switch]{lineno}
\usepackage{graphicx}
\usepackage{txfonts}
\usepackage{scrextend}
\usepackage{hyperref}
\usepackage{orcidlink}
\usepackage{amsmath}
\hypersetup{colorlinks=true,linkcolor=blue,filecolor=blue,urlcolor=blue,citecolor=blue}
\usepackage{multirow}

\usepackage{subcaption}

\usepackage{xcolor}

\usepackage{tablefootnote}

\usepackage{threeparttable}
\newcommand{\fermilat}{\textit{Fermi}-LAT}
\newcommand{\gray}{$\gamma$-ray}
\newcommand{\grays}{$\gamma$-rays}
\renewcommand{\thesubfigure}{\Alph{subfigure}}

\newcommand{\RomanNumeralCaps}[1]
	{\MakeUppercase{\romannumeral #1}}

\begin{document}

   \title{Quasi-periodic oscillations in the $\gamma$-ray light curves\\ of bright active galactic nuclei}


   \author{Helena X. Ren \inst{1, 2\, \orcidlink{0000-0003-0221-2560}}
          \and Matteo Cerruti \inst{1, 3\,\orcidlink{0000-0001-7891-699X} }
          \and Narek Sahakyan\inst{4\, \orcidlink{0000-0003-2011-2731}}
          }

   \institute{Universitat de Barcelona, ICCUB, IEEC-UB, E-08028 Barcelona, Spain
              \and
             Max-Planck-Institut für Kernphysik, P.O. Box 103980, D 69029, Heidelberg, Germany\\
             \email{helena.ren@mpi-hd.mpg.de}
              \and
             Université Paris Cité, CNRS, Astroparticule et Cosmologie, F-75013 Paris, France\\
             \email{cerruti@apc.in2p3.fr}
             \and
             ICRANet-Armenia, at NAS RA, 0019 Yerevan, Armenia\\
             \email{narek@icra.it}
             }

   \date{\today}

 
  \abstract
   {The detection of quasi-periodic oscillations (QPOs) in the light curves of active galactic nuclei (AGNs) can provide insights on the physics of the super-massive black holes (SMBHs) powering these systems, and could represent a signature of the existence of SMBH binaries, setting fundamental constraints on SMBH evolution in the Universe.}
   {Identification of long term QPOs, with periods of the order of months to years, is particularly challenging and can only be achieved via all-sky monitoring instruments that can provide unbiased, continuous light curves of astrophysical objects. The \fermilat\ satellite, thanks to its monitoring observing strategy, is an ideal instrument to reach such a goal, and we aim to identify QPOs in the \gray\ light curves of the brightest AGNs within the \fermilat\ catalog.}
   {We analyze the light curves of the thirty-five brightest \fermilat\ AGNs, including data from the beginning of the Fermi mission (August 2008) to April 2021, and energies from 100 MeV to 300 GeV. Two time binnings are investigated, 7 and 30 days. The search for quasi-periodic features is then performed using the continuous wavelet transform. The significance of the result is tested via Monte Carlo simulations of artificial light curves with the same power spectral density and probability distribution function as the original light curves. The significances are then corrected for the look-elsewhere effect and provided as post-trials.}
   {We identify twenty-four quasars with candidate QPOs. Several of our candidates coincide with previous claims in the literature: PKS~0537-441, S5~0716+714, Mrk~421, B2~1520+31, and PKS 2247-131. All our candidates are transient. The most significant multi-year QPO, with a period of about $1100$ days, is observed in the quasar S5~1044+71, and is reported here for the first time. }
   {}


   \maketitle
%

\section{Introduction}
Super-massive black holes (SMBHs) dwell at the center of all galaxies, and when accreting material they are observed from Earth as active galactic nuclei (AGNs). In a minority of AGNs the accretion of matter onto the black hole is associated with the launching of a pair of relativistic jets of plasma along the polar axis \citep[see][for a recent review]{Blandfordreview}. Studying the inflows and outflows of matter on and from black holes provides an indirect access to the physics of these compact objects.\\

SMBHs have their low-mass analog in stellar-mass black holes produced in star collapses. These low-mass black holes, when in binary systems, have also been observed accreting surrounding matter and launching plasma jets. These systems are observationally classified as X-ray binaries and microquasars. A peculiar characteristic of X-ray binaries is the presence of quasi-periodic oscillations (QPOs) in their X-ray light curves, with periods of the order of 0.1-10 Hz \citep[see e.g. the review by][and references therein]{Ingramreview}. Such a period is consistent with the innermost stable orbit of the black hole and suggests that QPOs can be used to access the region closest to the horizon and to study the physics of black hole accretion. Given the analogy between stellar-mass black holes and SMBHs, QPOs are also expected in the latter, although with much longer periods due to the larger size of the Schwarzschild radius that scales linearly with the mass of the compact object. Several observational claims have been made for QPOs in the X-ray light curves of Seyfert galaxies (radio-quiet AGNs), with periods of the order of hours \citep{Gierlinski08, Alston15}. The much longer periods in AGNs make it difficult to unequivocally measure QPOs in their light curves: an uninterrupted sampling over several periods is needed in order to have a statistically significant measurement.\\

In the last decade the \fermilat\ instrument \citep{2009ApJ...697.1071A} has revolutionized our view of the $\gamma$-ray sky in the 100 MeV-100 GeV band thanks to its unprecedented sensitivity and all-sky monitoring observing strategy. Fourteen years after its launch we have now access to continuous light curves on hundreds of AGNs. In the GeV band the extra-galactic sky is dominated by a peculiar AGN class: blazars. Blazars are AGNs whose relativistic jet is closely aligned with our line of sight. The relativistic Doppler effect boosts the emission and makes these type of AGNs among the brightest sources in the Universe. From an observational point of view, blazars are divided into two sub-classes: BL Lacertae objects (BL Lacs for short), with weak or absent broad emission lines in their optical/UV spectrum; and Flat Spectrum Radio Quasars (FSRQs), with strong broad emission lines \citep{Urry95}.  \\

Searches for periodicities in the light curves of \fermilat\ blazars have been an active research topic since the beginning of the mission but the first claim had to wait six years of data taking: \citet{Ackermann15} presented the first evidence (at about $3\sigma$) for a periodic 2-years modulation in the light curve of the blazar PG~1553+113, coincident in period with a QPO seen at longer wavelengths. Interestingly, this first hint for a QPO in a blazar $\gamma$-ray light curve is at frequencies much lower than the ones detected in X-rays in Seyfert galaxies, suggesting that its origin is intrinsically different. Since 2015, several research groups have investigated QPOs in \fermilat\ light curves resulting in several positive claims. All $\gamma$-ray QPO candidates in blazars shown a period of the order of years. A notable exception is represented by the highly-significant ($5.2 \sigma$) QPO seen in PKS~2247-131 which has a period of about a month and, most importantly, seems to be a transient phenomenon \citep{Zhou2018}.   \\

The details of the radiative mechanism(s) at the origin of the $\gamma$-ray emission in blazars are a subject of investigation, but there is consensus on the fact that this high-energy emission is produced in the relativistic jet and far from the SMBH \citep[see][for a recent review on blazar emission models]{Cerrutireview}. The physical mechanisms for the production of QPOs in $\gamma$-ray blazars will thus be intrinsically different from the ones at work in radio-quiet AGNs, and linked to the physics of jets. Several models have been developed to explain these quasi-periodicities. They can be associated to the movement of plasmoids in the jet along helical paths \citep{Rieger04}, or related to precession of the jet itself. This precession could be driven by the gravitational perturbation of another SMBH \citep{Begelman80, Sobacchi17}, meaning that QPOs will provide key constraints on SMBH binaries in the Universe. SMBH binaries are expected to form during galaxy mergers, and it is thus clear how important their detection is to understand galaxy and SMBH evolution through the history of the Universe.  Alternatively, the periodicities could be related to a regular change in the accretion of matter onto the SMBH that is then translated into a QPO in the $\gamma$-ray emission in the jet. In this case as well, SMBH binaries can be the source of this periodicity via perturbation of the accretion flow \citep{Valtonen09}. The SMBH binary can also imprint a QPO in the light curve via gravitational stresses by one of the black holes on the jet of the other one \citep{Tavani18}. The most promising SMBH binary candidate identified thanks to a twelve-years QPO in its optical light curve is the blazar OJ~287 \citep{Silla88}. The source is a known \gray\ emitter, but its famous 12-years QPO has not been seen in its \fermilat\ light curve due to the long period, although claims of a \gray\ QPO with a period of about 300 days have been made \citep{Kushawa20}. \\

The goal of this work is to systematically study the light curves of bright \fermilat\ AGNs in order to identify QPO candidates. In order to get access to transient QPOs, and QPOs with varying periods that could be hidden in a time-integrated power spectral density, we make use of the continuous wavelet transform. The paper is organized as follows: in Section \ref{sec:sources} we discuss the selection of the targets; in Section \ref{sec:fermi} we provide the details of the \fermilat\ data analysis; in Section \ref{sec:wavelet} we describe the wavelet analysis, whose results are presented in Section \ref{sec:results}; the discussion and the conclusions are in Sections \ref{sec:discussion} and \ref{sec:conclusions}.\\


\section{Sources}
\label{sec:sources}

The only selection criterion applied is for the source to be bright enough to have continuous \fermilat\ light curves with time bins of seven days or one month. The main issue is to avoid the effect of large gaps in the continuous wavelet transform (see Section \ref{sec:wavelet}). This aspect, together with the need to limit the total computing time, drives the choice to limit the analysis to the thirty-four brightest sources in the 4LAC catalog \citep[data release 2,][sorting them by integral energy flux above 100 MeV and removing sources with a test significance lower than 100]{4LAC}. The least bright source has an integral flux above 100 MeV equal to $8.95\times 10^{-11}$ erg cm$^{-2}$ s$^{-1}$. The only exception in our source selection is represented by PKS 2247-131, which is manually added to our list due to the high significance QPO detected in this source by \citet{Zhou2018}. The source was not detected by \fermilat\ before entering in an active state in 2016, and thus its long-term average flux is biased towards lower values. If we consider only its emission after 2016, this source is definitely among the brightest \gray\ AGNs, and clearly fulfills our requirement to have an uninterrupted light curve. This manual addition brings the total of our sample to thirty-five AGNs. The sources we included in our analysis are listed in Table \ref{tab:sample} by their Right Ascension. The sample consists of 19 FSRQs, 15 BL Lacs, and 1 radio galaxy. For each source, together with the source name and 4FGL name, we provide the source class, coordinates in J2000 and the redshift (the three last quantities as provided in the 4LAC catalog). \\

\section{Fermi/LAT Data Analysis}
\label{sec:fermi}

The Large Area Telescope (LAT) on board the Fermi Gamma-ray Space Telescope is a pair conversion telescope sensitive to \grays\ in the energy band from $20$ MeV to  $500$ GeV. Since its launch on June 11, 2008, \fermilat\ scans the sky every $\sim$3 hours, regularly monitoring the \gray\ emission from different sources. More details on the \fermilat\ instrument are described in \citet{2009ApJ...697.1071A} and references therein.\\

In the current study, \gray\ data from the observation of the considered sources from August 8, 2008 to April 4, 2021 (from MJD 54686 to MJD 59308) were downloaded and analyzed using the standard analysis procedure provided by the \fermilat\ collaboration. The data for each of the considered sources are analyzed in an identical manner. The Pass 8 data in the energy range from 100 MeV to 500 GeV are analyzed using Fermi ScienceTools (1.2.1) and the P8R3\_SOURCE\_V3 instrument response functions. We selected only events within a maximum zenith angle of $90^{\circ}$, to reduce contamination by photons from Earth's atmosphere and used the filter expression (DATAQUAL $>$0) and (LAT CONFIG$==$1) to select good time intervals. The photons from a circular region with a radius of $12^{\circ}$ around each source under consideration were selected. Then, the events are binned within a $16.9^{\circ}\times16.9^{\circ}$ square region of interest (ROI) into $0.1^{\circ} \times 0.1^{\circ}$ pixels and into 37 equal logarithmically spaced energy bins. The model for which the likelihood is computed includes a combination of point-like and diffuse sources  and standard templates describing the diffuse emission from the Galaxy (gll\_iem\_v07) and the isotropic \gray\ background (iso\_P8R3\_SOURCE\_V3\_v1). The model file for each source is created using the \fermilat\ fourth source catalog \citep[4FGL;][]{2020ApJS..247...33A} where all sources falling within ROI+5$^\circ$ were all included with the same spectral models as in the catalog. The normalization parameters of the diffuse Galactic and the isotropic component, and both the normalization and spectral parameters of the point sources within the ROI were set as free parameters while that of the sources outside ROI were fixed to the catalog values. The binned likelihood analysis implemented in {\it gtlike} tool is used to optimize the parameters to best match the observations for the whole time period.\\

The model file obtained from the full time analysis is used to compute the light curves. We produced the light curves binned into weekly and monthly intervals by applying unbinned likelihood analysis in the 0.1–300 GeV energy range with the appropriate quality cuts mentioned above.\\

\begin{table*}[!hbtp]
\begin{center}
    \begin{tabular}{lccccc}
        \hline
        \hline
		Source name & 4FGL name &  Source class & R.A. (J2000) & Dec. (J2000) & Redshift \\
		\hline
		\object{4C~+01.02} & J0108.6+0134 & FSRQ & 01$^h$08$^m$40.7$^s$ & +01$^d$34$^m$54.8$^s$ & 2.099\\
        \object{3C~66A} & J0222.6+4302 & BL Lac & 02$^h$22$^m$40.7$^s$ & +43$^d$02$^m$08.5$^s$ & 0.444\\
		\object{4C~+28.07} & J0237.8+2848 & FSRQ & 02$^h$37$^m$53.7$^s$ & +28$^d$48$^m$15.8$^s$ & 1.206\\
        \object{PKS~0235+164} & J0238.6+1637 & BL Lac & 02$^h$38$^m$40.3$^s$ & +16$^d$37$^m$04.4$^s$ & 0.94\\
		\object{NGC 1275} & J0319.8+4130 & RDG FR-\RomanNumeralCaps{1} & 03$^h$19$^m$49.8$^s$ & +41$^d$30$^m$43.6$^s$ & 0.018\\
        \object{PKS~0402-362} & J0403.9-3605 & FSRQ & 04$^h$03$^m$54.0$^s$ & -36$^d$05$^m$13.2$^s$ & 1.417\\
		\object{PKS~0426-380} & J0428.6-3756 & BL Lac & 04$^h$28$^m$41.5$^s$ & -37$^d$56$^m$25.1$^s$ & 1.11\\
        \object{PKS~0447-439} & J0449.4-4350 & BL Lac & 04$^h$49$^m$26.0$^s$ & -43$^d$50$^m$06.0$^s$ & 0.205 \\
		\object{PKS~0454-234} & J0457.0-2324 & FSRQ & 04$^h$57$^m$02.6$^s$ & -23$^d$24$^m$53.6$^s$ & 1.003\\
		\object{PKS~0537-441} & J0538.8-4405 & BL Lac & 05$^h$38$^m$50.1$^s$ &  -44$^d$05$^m$10.3$^s$ & 0.892\\
		\object{S5~0716+714} & J0721.9+7120 & BL Lac & 07$^h$21$^m$57.2$^s$ & +71$^d$20$^m$25.8$^s$ & 0.300\\
	    \object{4C~+55.17} & J0957.6+5523 & FSRQ & 09$^h$57$^m$39.9$^s$ & +55$^d$23$^m$01.7$^s$ & 0.896 \\
	    \object{1H 1013+498} & J1015.0+4926 & BL Lac & 10$^h$15$^m$04.3$^s$ & +49$^d$26$^m$01.0$^s$ & 0.212 \\
		\object{S5~1044+71} & J1048.4+7143 & FSRQ & 10$^h$48$^m$25.6$^s$ & +71$^d$43$^m$46.9$^s$ & 1.15\\
		\object{Mrk~421} & J1104.4+3812 & BL Lac & 11$^h$04$^m$28.5$^s$  & +38$^d$12$^m$25.2$^s$ & 0.03\\
		\object{4C~+21.35} & J1224.9+2122 & FSRQ & 12$^h$24$^m$54.6$^s$ & +21$^d$22$^m$53.0$^s$ & 0.434\\
		\object{3C~273} & J1229.0+0202 & FSRQ & 12$^h$29$^m$04.2$^s$ & +02$^d$02$^m$43.4$^s$ & 0.158\\
		\object{3C~279} & J1256.1-0547 & FSRQ & 12$^h$56$^m$10.0$^s$ & -05$^d$47$^m$19.3$^s$ & 0.536\\
		\object{B3 1343+451} & J1345.5+4453 & FSRQ & 13$^h$45$^m$34.6$^s$ & +44$^d$53$^m$03.8$^s$ & 2.534\\
		\object{PKS~1424+240} & J1427.0+2348 & BL Lac & 14$^h$27$^m$0.4$^s$ & +23$^d$48$^m$0.6$^s$ & 0.601\\
		\object{PKS~1424-418} & J1427.9-4206 & FSRQ & 14$^h$27$^m$56.8$^s$ & -42$^d$06$^m$21.6$^s$ & 1.522\\
		\object{PKS~1502+106} & J1504.4+1029 & FSRQ & 15$^h$04$^m$24.8$^s$ & +10$^d$29$^m$52.1$^s$ & 1.839\\
		\object{PKS~1510-089} & J1512.8-0906 & FSRQ & 15$^h$12$^m$51.5$^s$ & -09$^d$06$^m$23.0$^s$ & 0.36\\
		\object{B2~1520+31} & J1522.1+3144 & FSRQ & 15$^h$22$^m$10.9$^s$ & +31$^d$44$^m$22.2$^s$ & 1.489\\
		\object{PG~1553+113} & J1555.7+1111 & BL Lac & 15$^h$55$^m$43.5$^s$ & +11$^d$11$^m$18.2$^s$ & 0.36\\
		\object{4C~+38.41} & J1635.2+3808 & FSRQ & 16$^h$35$^m$16.0$^s$ & +38$^d$08$^m$24.4$^s$ & 1.814\\
		\object{Mrk~501} & J1653.8+3945 & BL Lac & 16$^h$53$^m$53.7$^s$ & +39$^d$45$^m$34.2$^s$ & 0.033\\
		\object{1ES~1959+650} & J2000.0+6508 & BL Lac & 20$^h$00$^m$02.6$^s$ & +65$^d$08$^m$52.5$^s$ & 0.047 \\
		\object{PKS~2155-304} & J2158.8-3013 & BL Lac & 21$^h$58$^m$51.4$^s$ & -30$^d$13$^m$30.4$^s$ & 0.116\\
		\object{BL Lac} & J2202.7+4216 & BL Lac & 22$^h$02$^m$46.7$^s$ & +42$^d$16$^m$55.6$^s$ & 0.069\\
		\object{CTA 102} & J2232.6+1143 & FSRQ & 22$^h$32$^m$36.6$^s$ & +11$^d$43$^m$50.2$^s$ & 1.037\\
		\object{PKS~2247-131} & J2250.0-1250 & BL Lac & 22$^h$50$^m$01.2$^s$ & -12$^d$50$^m$54.6$^s$ & 0.22\\
		\object{3C~454.3} & J2253.9+1609 & FSRQ & 22$^h$53$^m$59.1$^s$ & +16$^d$09$^m$02.2$^s$ & 0.859\\
		\object{PKS~2326-502} & J2329.3-4955 & FSRQ & 23$^h$29$^m$19.1$^s$ & -49$^d$55$^m$56.6$^s$ & 0.518\\
		\object{PMN~J2345-1555} & J2345.2-1555 & FSRQ & 23$^h$45$^m$12.7$^s$ & -15$^d$55$^m$05.5$^s$ & 0.621 \\
		\hline
		\hline\\
	\end{tabular}
\caption{\label{tab:sample}The selected \textit{Fermi}-LAT sources with the catalog name, blazar type, coordinate in J2000 and redshift extracted from the \fermilat\ 4FGL catalog.}
\end{center}
\end{table*}



\section{Wavelet Analysis}
\label{sec:wavelet}

\subsection{Continuous Wavelet Transform}

In this work we make use of the continuous wavelet transform (CWT) technique, which is the convolution of a time series with a dilated and translated wavelet function, to analyse time-frequency properties of the light curves. Acting as a band-pass filter, the CWT maps the power of any particular periodic behaviour at different times in the time-frequency space. This is notably useful, since the CWT technique not only gives access to the frequencies of potential QPOs, but also when the periodicities appear and end, and how they evolve in time. \\

We use the Python implementation \texttt{PyCWT} provided by \citet{Torrence1998} and as mother wavelet the Morlet wavelet. This wavelet consists of a plane wave modulated by a Gaussian and is shown in \autoref{sec:AWavelet}. The \textit{wavelet power spectrum} is defined as the square of the amplitude of the wavelet coefficient. The \textit{global wavelet spectrum} can then be computed as the time-average of the wavelet spectrum. The global wavelet spectrum provides an unbiased and consistent estimation of the true power spectrum of a time series \citep{Percival1995}. Because of the finite length of the time series, border effects occur at the edges of the wavelet power spectrum. The cone of influence (COI) is defined  as the region of the wavelet power spectrum in which edge effects become important, and is calculated as suggested by \cite{Torrence1998}. \\

In \autoref{sec:AWavelet} we show the CWT map and global wavelet spectrum of a periodic signal with two periods at 0.1 and 0.02 s (Fig.~\ref{fig:Sinusoid}) and a Dirac delta signal (Fig.~\ref{fig:Dirac}). The CWT has an intrinsic resolution in reconstructing a periodic signal, that can be visualized as broad bands in the CWT map. We estimate this uncertainty as half of the full-width at half-maximum (FWHM) of the global wavelet spectrum. For the Dirac delta function, a vertical feature appears in the wavelet power spectrum which reveals the response of this technique when a flare-like signal is present in the light curve. This response of the CWT to flares is of particular relevance for our study. A high power in the CWT by itself does not mean that a periodicity is present in the light curve: all CWT maps have to be inspected to visually confirm that the power is distributed horizontally, and is not related to a flaring behaviour.\\

\subsection{Significance Estimation}

To determine the significance and confidence levels of the analysis, we simulated artificial light curves following the work by \citet{Emmanoulopoulos13}, using the Python version provided by \citet{Connolly15}. This algorithm generates artificial light curves having the same power spectral density (PSD) and probability distribution function (PDF) as the original light curve and represents an improvement of the procedure of \cite{Timmer1995}, which produces normally distributed time series from a given PSD. The algorithm of \cite{Emmanoulopoulos13} is able to precisely reproduce light curves which match both the PSD and the PDF of a given observed light curve or a theoretical model, where the PSD estimate is performed using a maximum likelihood methodology, and assuming a smoothly bending power-law model plus a constant. \\

For each \fermilat\ light curve, we produced 10000 artificial light curves. In \autoref{sec:AStatistical} we show a comparison of the PDF and PSD of one of the simulated light curve to the original one. The global wavelet spectrum is computed for each simulated light curve, such that a histogram of the power spectrum can be produced at every period, or scale. We then fit the histogram with a $\chi^2$ distribution with $k$ degrees of freedom 
\begin{equation}
    \chi(x, k)^2 =\frac{1}{2^{k / 2} \Gamma(k / 2)} x^{k / 2-1} \exp (-x / 2).
    \label{eq:Chi2}
\end{equation}
The confidence levels are obtained by using the percentiles of the power for each scale, which define the global significance of the results. An example of the $\chi^2$ fitting to the histogram is included in \autoref{sec:ASignificance}, as well as the resulting confidence levels for the AGN S5~0716+714. \\


When searching for significant results in physics, one needs to take into account the "look-elsewhere effect", which can be quantified in term of number of trials. We estimate the post-trial confidence levels of the global and local wavelet spectrum following the work presented by \cite{Auchere2016}. Taking into account the fact that the bins in the wavelet spectrum are not statistically independent, they showed that the post-trial probability $P_G$ can be computed from the pre-trial one $P_L$ as: 

\begin{equation}
	P_G = 1-(1-P_L^{\ a})^n, 
\end{equation}
where $a$ and $n$ are empirically derived coefficients, which are relate to the number of bins in the spectrum and the resolution of the CWT that we are using $\delta j$, and are specific for each mother wavelet. The resolution $\delta j$ in our analysis is $1/12$. Different parameterization for the global and local wavelet spectrum are needed, since in the local one, trials are made in the time-period bins of the power spectrum map, whereas after the time-average, only the period bins accounts for the global wavelet spectrum. \cite{Auchere2016} parameterized $a = 0.810 ~(N_{\rm{out}}~ \delta j)^{0.011}$ and $n = 0.491 ~(N_{\rm{out}}~ \delta j)^{0.926}$ for the local wavelet spectrum, and  $a = 0.805 + 0.45 \times 2^{-S_{\rm{out}} \delta j }$ and $n = 1.136 ~(S_{\rm{out}} ~\delta j)^{1.2}$ for the global one, where $N_{\rm{out}}$ and $S_{\rm{out}}$ are the time-period bins and the period bins of the wavelet spectrum outside the COI respectively. These coefficients are valid for time series affected by power-law noise, but adequate only for wavelet analysis using the Morlet wavelet \citep[for more details see][]{Auchere2016}. \\

Additionally, we are considering 35 sources in this work and 2 time-binnings for each of them. Therefore, for each time-binning, we compute the number of bins $N_{\rm{out}}$ and $S_{\rm{out}}$, and we multiply them to the number of sources in order to account the total number of trials made. One special case is the AGN PKS~2247-131, whose light curve is cut at MJD~57427, since there was no detection by \fermilat~ before that date. Hence, the number of time-period bins $N_{out}$ for this source is lower. \\

The total number of bins $N_{out, T}$ and $S_{out, T}$ are, thus, computed as:

\begin{equation}
    \begin{split}
        N_{\rm{out}, T} =&~(34 \times N_{\rm{out},1month} + N^{PKS~2247-131}_{\rm{out},1month}) + \\ 
        &+ (34 \times N_{\rm{out},7days} + N^{PKS~2247-131}_{\rm{out},7days})~ = 1698893, \\
        \\
    	S_{\rm{out}, T} =&~(34 \times S_{\rm{out},1month} + S^{PKS~2247-131}_{\rm{out},1month}) + \\ 
        &+ (34 \times S_{\rm{out},7days} + S^{PKS~2247-131}_{\rm{out},7days})~ = 4833. \\
    \end{split}
\end{equation}

\begin{table*}[!hbtp]
\centering
\begin{threeparttable}
\begin{tabular}{lccccc}
\hline \hline Source  & Period(d) 30~d LC& Significance ($\sigma$) & Period(d) 7~d LC & Significance ($\sigma$) & Nr. fitted cycles\\
\hline \multirow{2}{*}{4C~+01.02}     & 268$\pm$55       & >5            & 268$\pm$54        & >5      & 4 \\
                                      & 123$\pm$26       & 4.7           & 122$\pm$26        & >5      & 5 \\
\hline PKS~0537-441                   & 285$\pm$67       & >5            & 286$\pm$73        & >5      & 4 \\
\hline \multirow{2}{*}{S5~1044+71}    & 1133$\pm$229     & 4.9           & 1127$\pm$226      & 4.6     & 3 \\
                                      & 116$\pm$33       & >5            & 117$\pm$38        & >5      & 4 \\
\hline \multirow{3}{*}{B2~1520+31}    & 176$\pm$48       & >5            & 179$\pm$42        & >5      & 6 \\
                                      &                  &               & 71$\pm$15         & >5      & 14 \\
                                      &                  &               & 39$\pm$11         & >5      & 17 \\
\hline \multirow{3}{*}{PKS~2247-131}  & 217$\pm$38       & >5            & 214$\pm$43        & >5      & 6 \\
                                      &                  &               & 34$\pm$13         & >5      & 10 \\

\hline 
\hline 4C~+28.07                      
                                      & 230$\pm$90       & >5            & 244$\pm$88        & >5      & 3 \\
\hline \multirow{2}{*}{NGC 1275}      & 282$\pm$84       & 3.3           & 247$\pm$63        & >5      & 3 \\
                                      &                  &               & 92$\pm$33         & >5      & 4 \\
\hline \multirow{2}{*}{PKS~0402-362}  & 221$\pm$56       & >5            & 221$\pm$60        & >5      & 3 \\
                                      &                  &               & 122$\pm$42        & >5      & 5 \\
\hline PKS~0426-380                   &                  &               & 85$\pm$26         & >5      & 8 \\
\hline PKS~0447-439                   & 120$\pm$37       & >5            & 111$\pm$42        & >5      & 7 \\
\hline PKS~0454-234                   
                                      &                  &               & 69$\pm$21         & >5      & 4 \\
\hline S5~0716+714                    & 325$\pm$75       & 2.4           & 324$\pm$77        & 3.2     & 5 \\
\hline \multirow{3}{*}{1H~1013+498}   
                                      & 263$\pm$52       & >5            & 264$\pm$59        & 4.9     & 4 \\
                                      &                  &               & 100$\pm$25        & >5      & 4 \\
                                      &                  &               & 52$\pm$15         & >5      & 12 \\
\hline Mrk~421                        & 300$\pm$64       & >5            & 300$\pm$65        & >5      & 3 \\
\hline 4C~+21.35                      &                  &               & 66$\pm$17         & >5      & 6 \\
\hline \multirow{2}{*}{3C~273}        & 177$\pm$36       & >5            & 177$\pm$38        & >5      & 4 \\
                                      & 99$\pm$26        & >5            & 97$\pm$25         & >5      & 3 \\
\hline  \multirow{2}{*}{3C~279}       
                                      & 102$\pm$26       & >5            & 101$\pm$27        & >5      & 6 \\
                                      &                  &               & 40$\pm$8          & >5      & 4 \\
\hline PKS~1424-418                   & 94$\pm$25        & >5            & 90$\pm$22         & >5      & 5 \\
\hline \multirow{1}{*}{PKS~1510-089}  
                                      & 119$\pm$31       & >5            & 120$\pm$36        & >5      & 3 \\
 
\hline Mrk~501                        & 315$\pm$98       & 2.9           & 326$\pm$76        & >5      & 7\tnote{a} \\
\hline PKS~2155-304                   & 334$\pm$107      & 2.2           & 341$\pm$106       & 3.5     & 4 \\
\hline \multirow{2}{*}{CTA 102}       & 370$\pm$85       & >5            & 366$\pm$81        & >5      & 3 \\
                                      & 179$\pm$40       & >5            & 178$\pm$40        & >5      & 5 \\
\hline 3C~454.3                       & 117$\pm$23       & >5            & 120$\pm$27        & >5      & 4\\ 
\hline PMN~J2345-1555                 & 197$\pm$50       & >5            & 191$\pm$44        & >5      & 4 \\
                                                       
\hline \hline
\end{tabular}
\caption{\label{tab:QPO} QPOs candidates identified by the CWT of the \fermilat\ light curves in time bins of 30 days and 7 days. }

\begin{tablenotes}

\item [a] \footnotesize{Number fitted cycles observed only in the weekly binned light curve.}

\end{tablenotes}

\end{threeparttable}
\end{table*}


\section{Results}
\label{sec:results}

In this section we present the results for those light curves showing possible QPOs and we discuss in more details the five sources exhibiting the most significant features: 4C~+01.02, PKS~0537-441, S5~1044+71, B2~1520+31 and PKS~2247-131. Furthermore, results consistent with previous QPO searches are also found in several sources, although at lower significance: PKS~0426-380, S5~0716+714, Mrk~421, PKS~1424-418, Mrk~501 and PKS~2155-304. These results are also commented briefly in this Section, although their CWT maps will be provided together with all the remaining sources in \autoref{sec:AAll}. Table~\ref{tab:QPO} lists the candidate QPOs found by CWT.\\

\begin{figure*}[!htbp]
	\centering
	\begin{subfigure}[b]{0.48\textwidth}
		\centering
		\includegraphics[width=\textwidth]{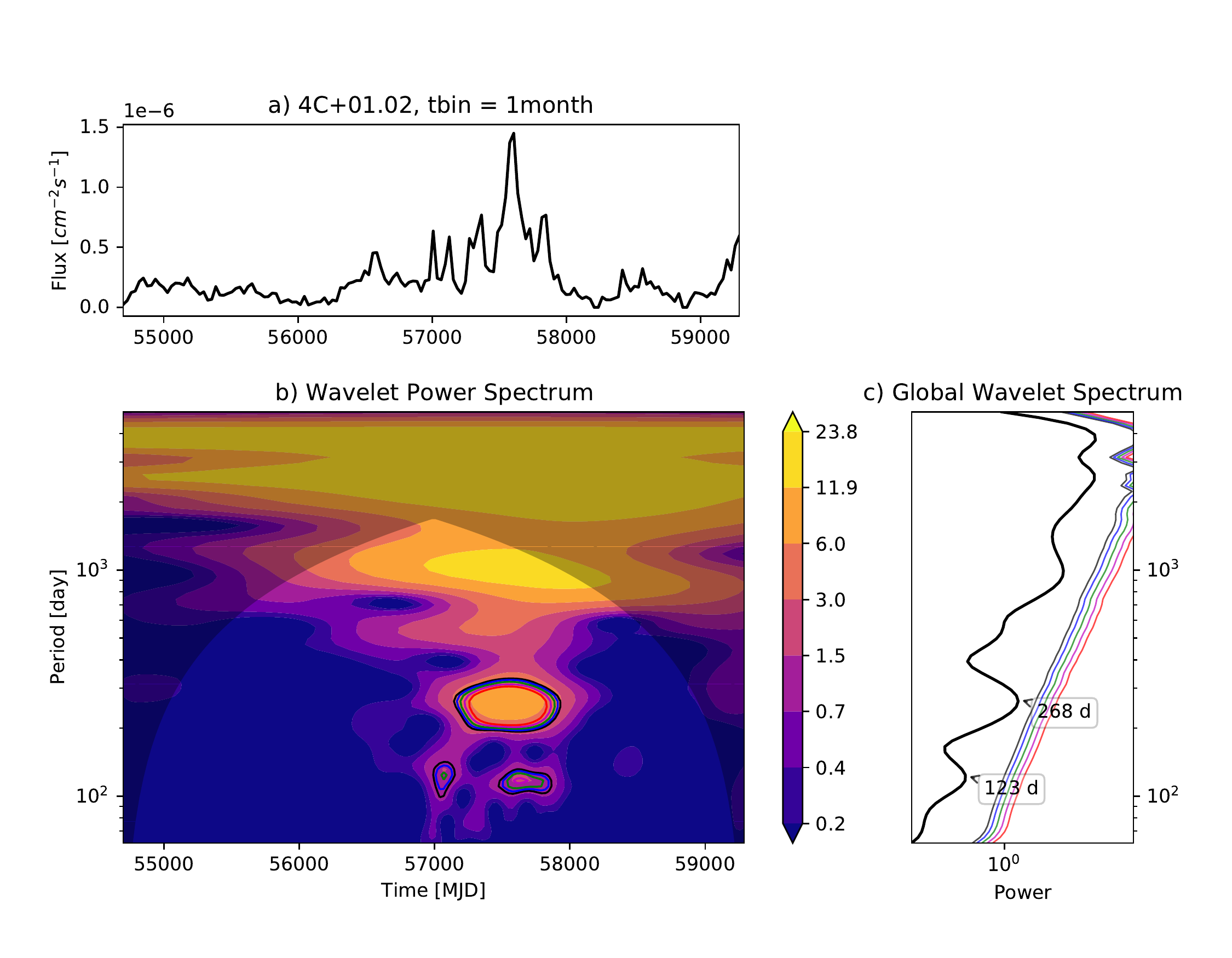}
		\caption{}
	\end{subfigure}
	\hfill
	\begin{subfigure}[b]{0.48\textwidth}
		\centering
		\includegraphics[width=\textwidth]{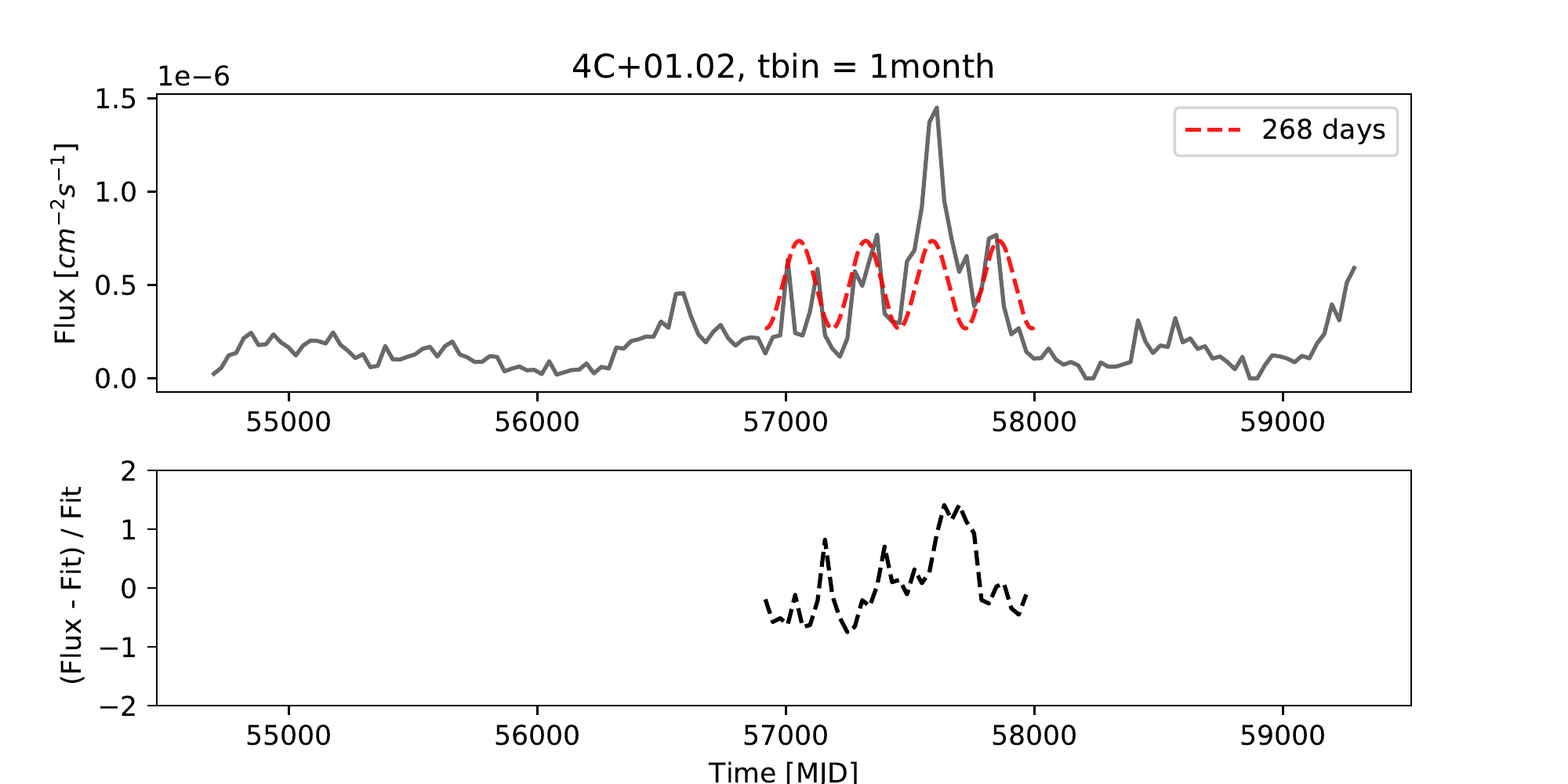}
		\caption{}
	\end{subfigure}
	\vskip\baselineskip
	
	\begin{subfigure}[b]{0.48\textwidth}  
		\centering 
		\includegraphics[width=\textwidth]{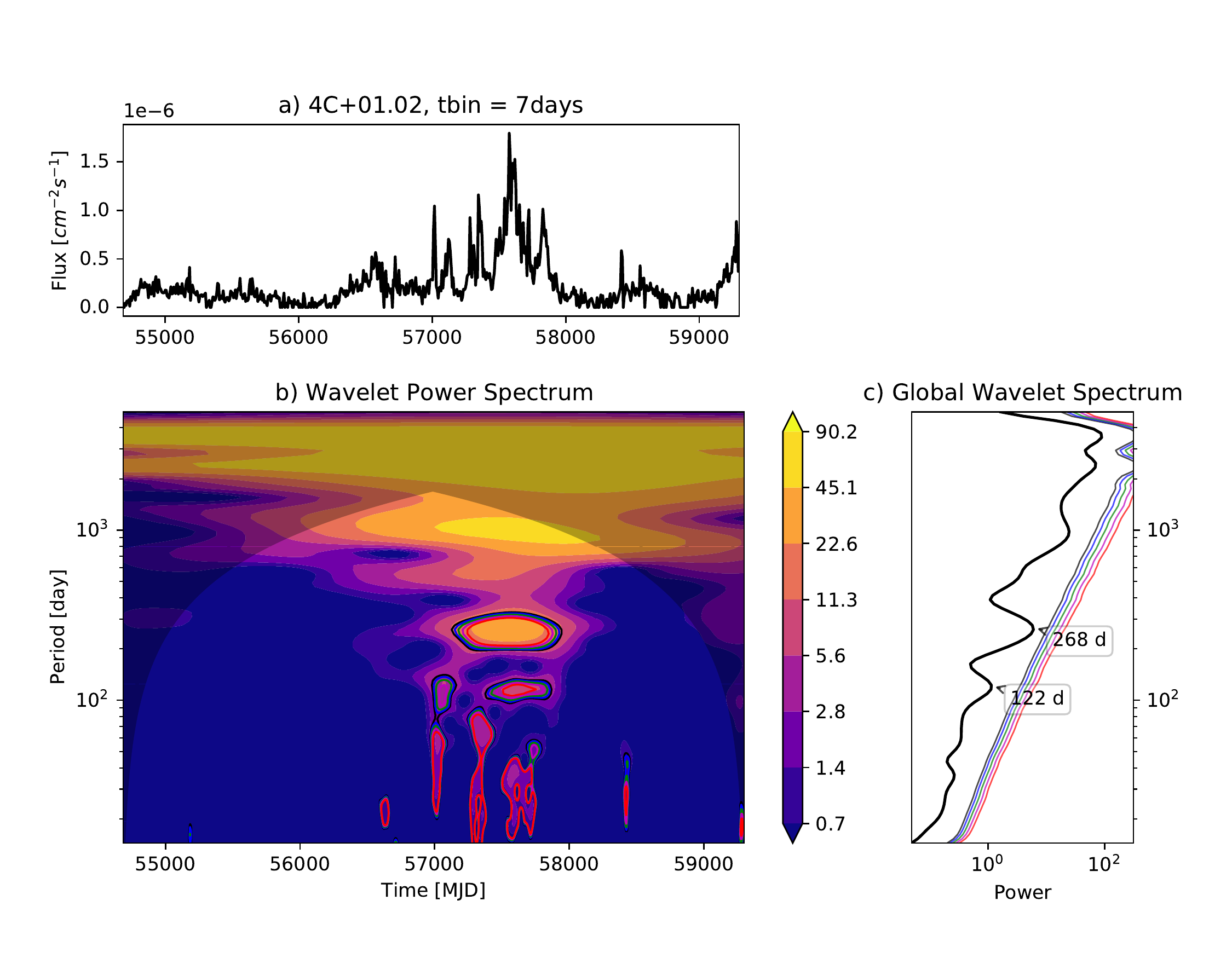}
		\caption{}
	\end{subfigure}
	\hfill
	\begin{subfigure}[b]{0.48\textwidth}  
		\centering 
		\includegraphics[width=\textwidth]{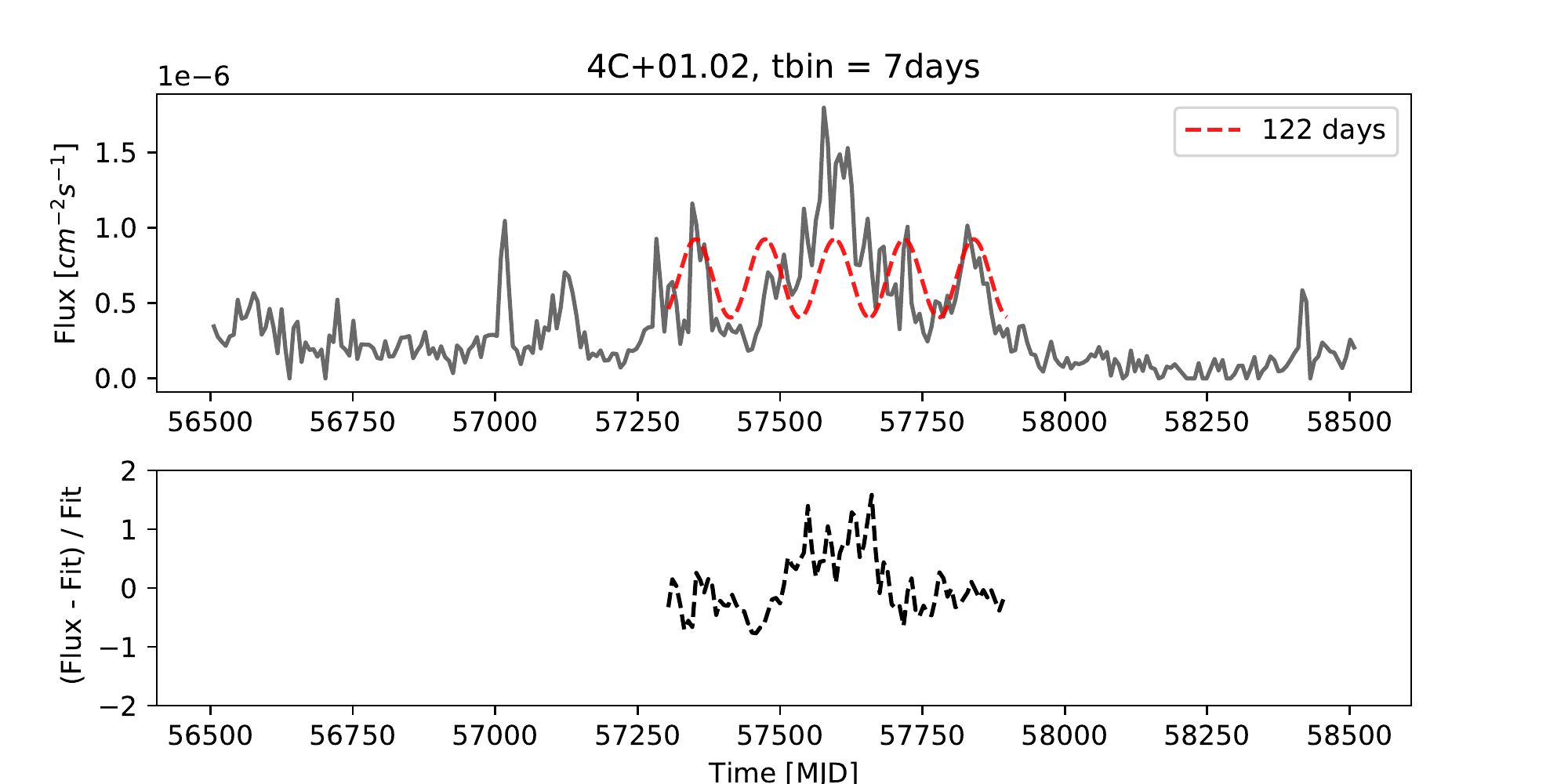}
		\caption{}
	\end{subfigure}
	
	\caption{Left: CWT map for the monthly binned (upper) and the weekly binned light curve (bottom) for 4C~+01.02. In each subplot, the panels represent a) Fermi/LAT light curve, b) wavelet power spectrum and c) global wavelet power spectrum. The solid coloured contours in b) and the dashed coloured lines in c) are the confidence levels (1 to 5 $\sigma$ in black, blue, green, violet, and red). Right: monthly binned and weekly binned light curves with the fitted periodic signal in red dashed line in the upper panels and the relative error in the bottom panels. }
	\label{fig:CWT0102}
\end{figure*}

\begin{figure*}[!htbp]
    \begin{subfigure}[b]{0.48\textwidth}   
		\centering 
		\includegraphics[width=\textwidth]{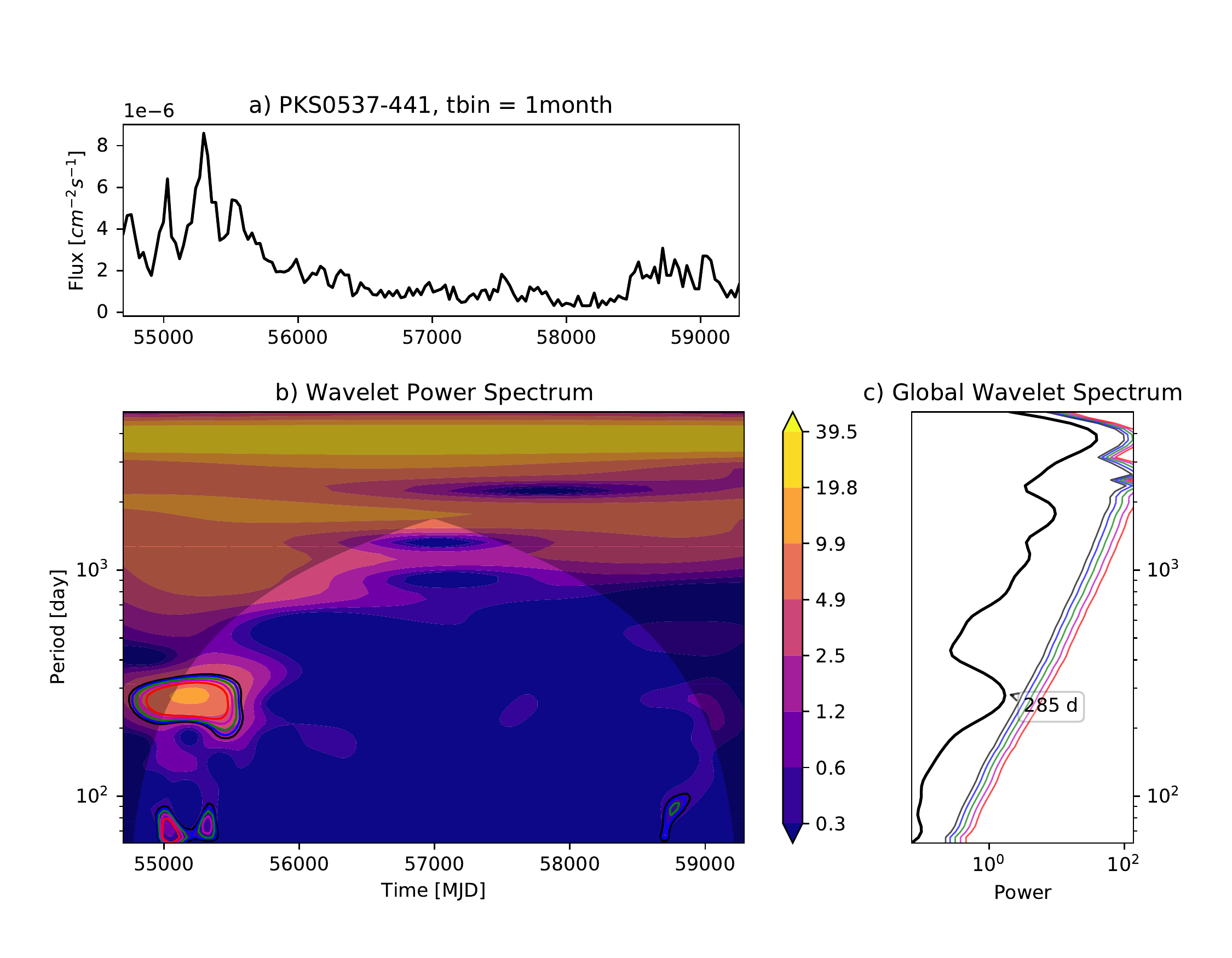}
		\caption{}
	\end{subfigure}
	\hfill
	\begin{subfigure}[b]{0.48\textwidth}   
		\centering 
		\includegraphics[width=\textwidth]{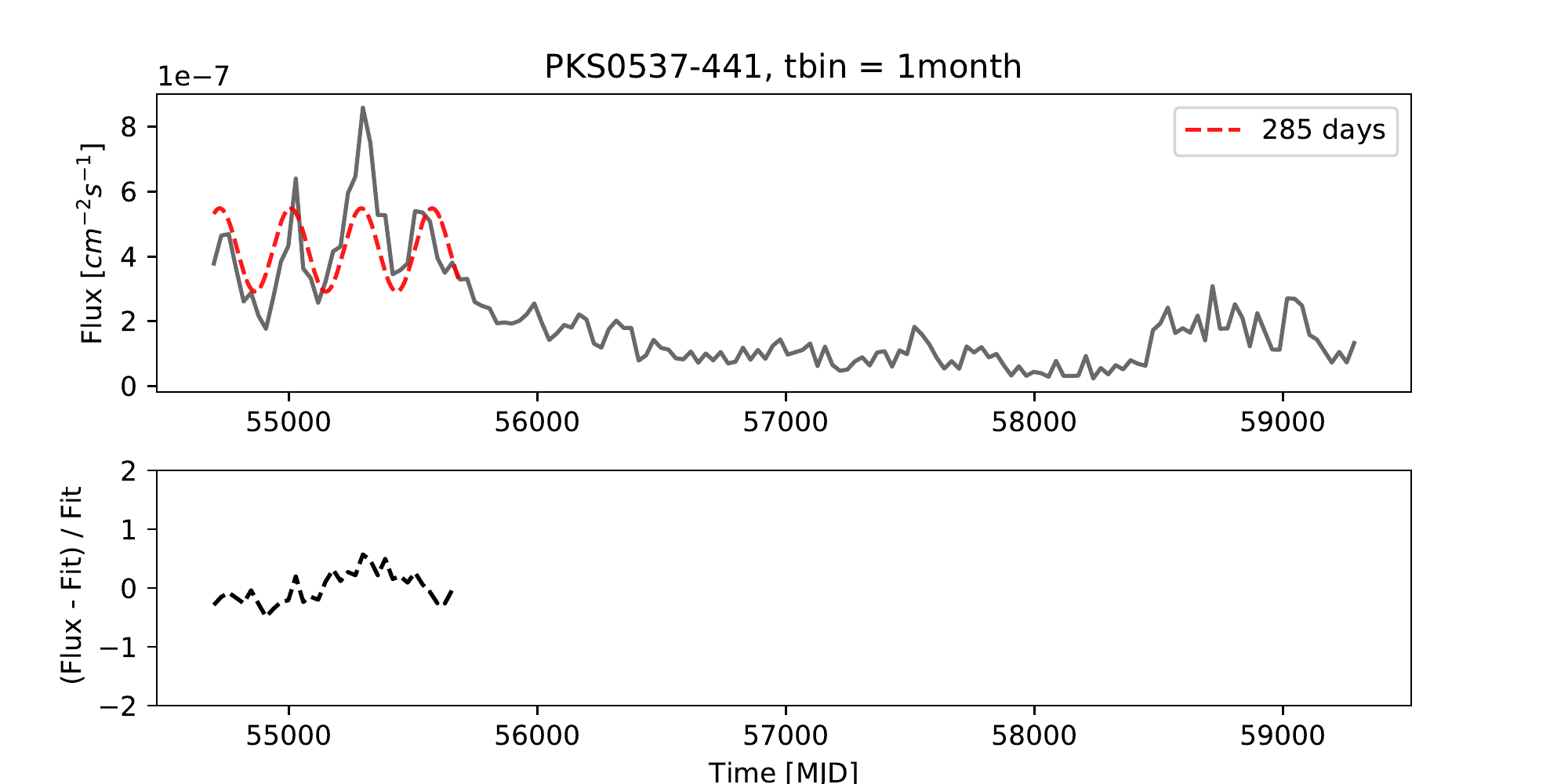}
		\caption{}
	\end{subfigure}
	\vskip\baselineskip
	
	\begin{subfigure}[b]{0.48\textwidth}  
		\centering 
		\includegraphics[width=\textwidth]{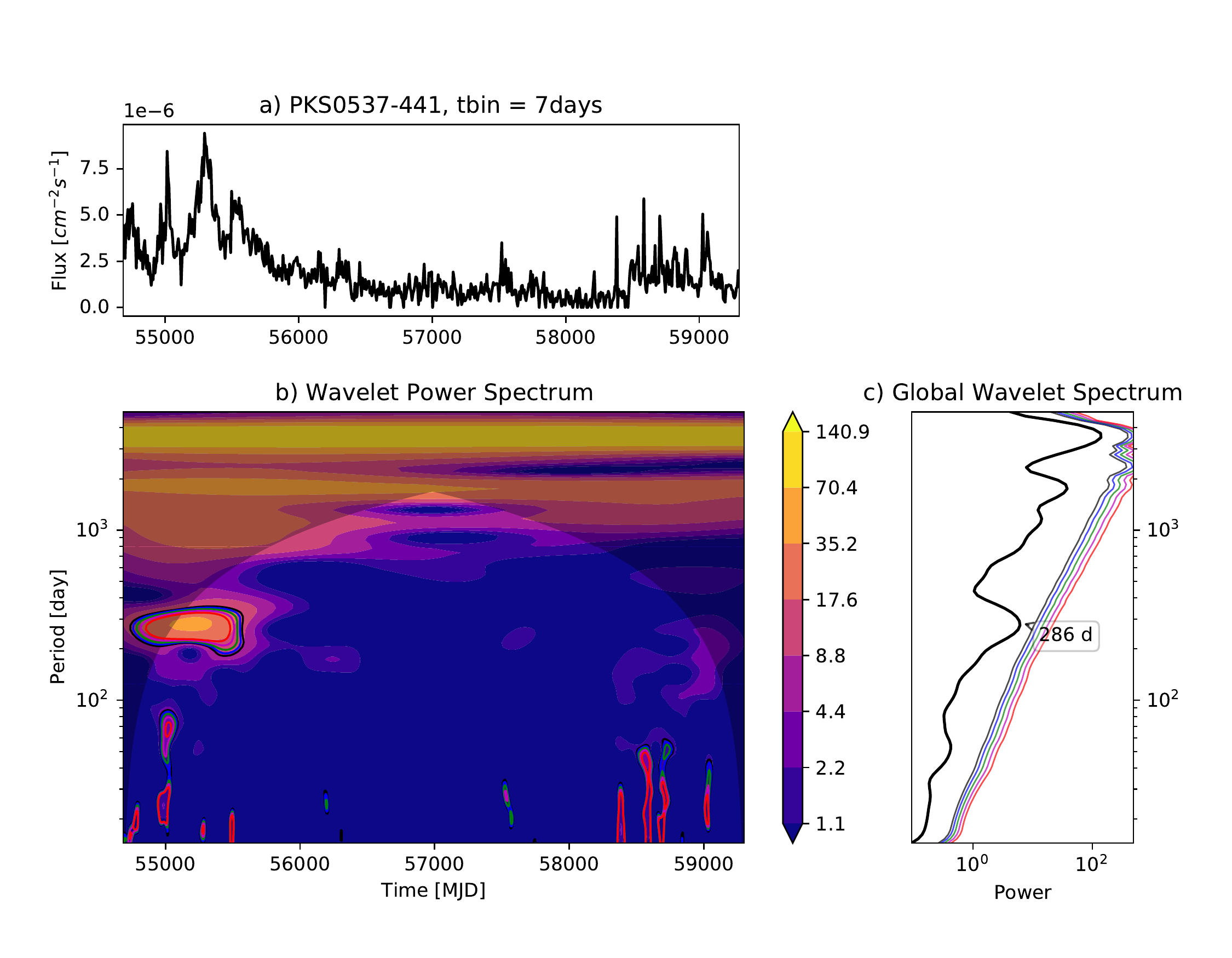}
		\caption{}
	\end{subfigure}
	\hfill
	\begin{subfigure}[b]{0.48\textwidth}  
		\centering 
		\includegraphics[width=\textwidth]{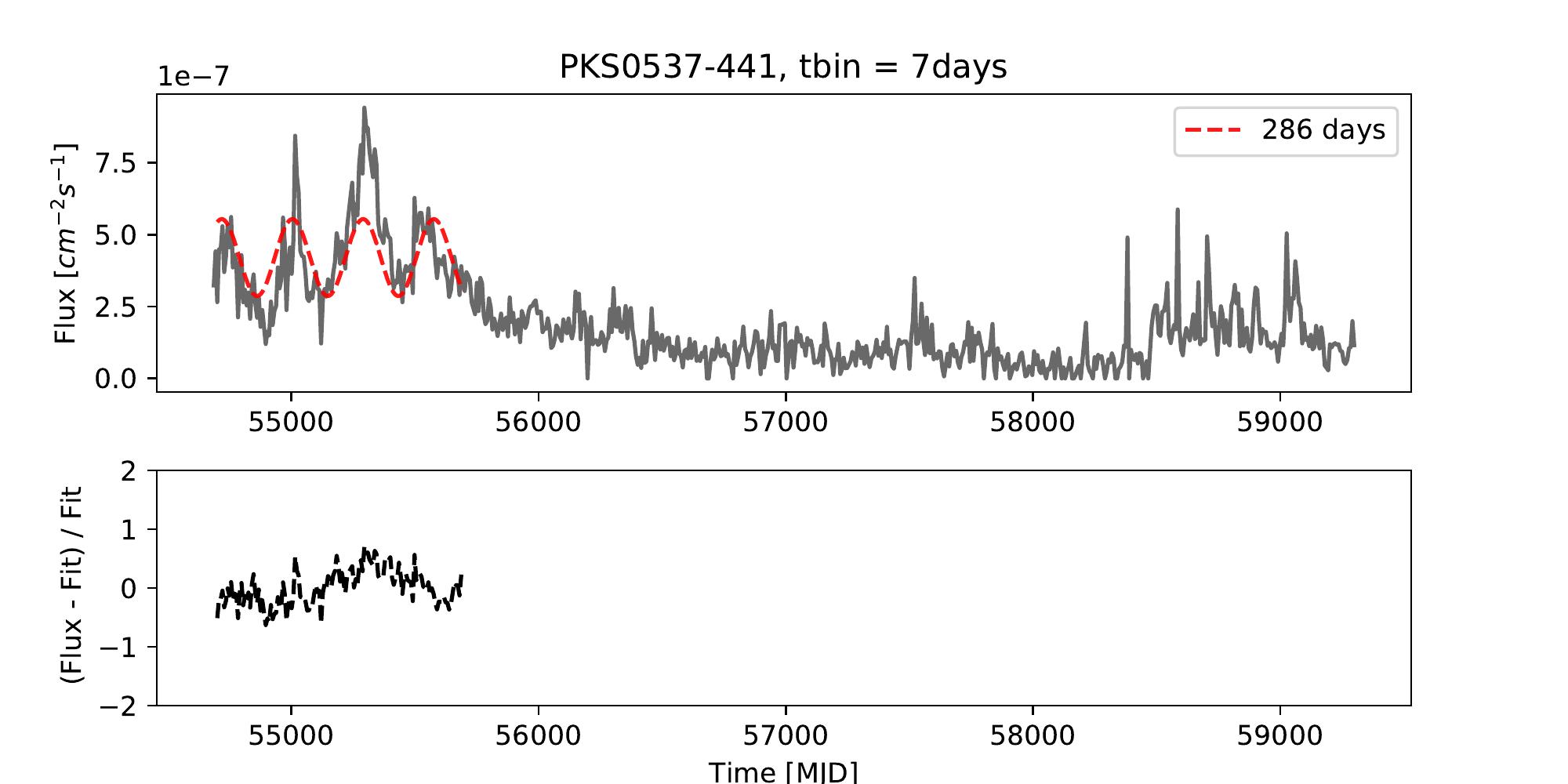}
		\caption{}
	\end{subfigure}
	
	\caption{Same description as Fig.~\ref{fig:CWT0102} for PKS~0537-441.}
	\label{fig:CWT0537}
\end{figure*}

\begin{figure*}[!htbp]
	\centering
	\begin{subfigure}[b]{0.48\textwidth}
		\centering
		\includegraphics[width=\textwidth]{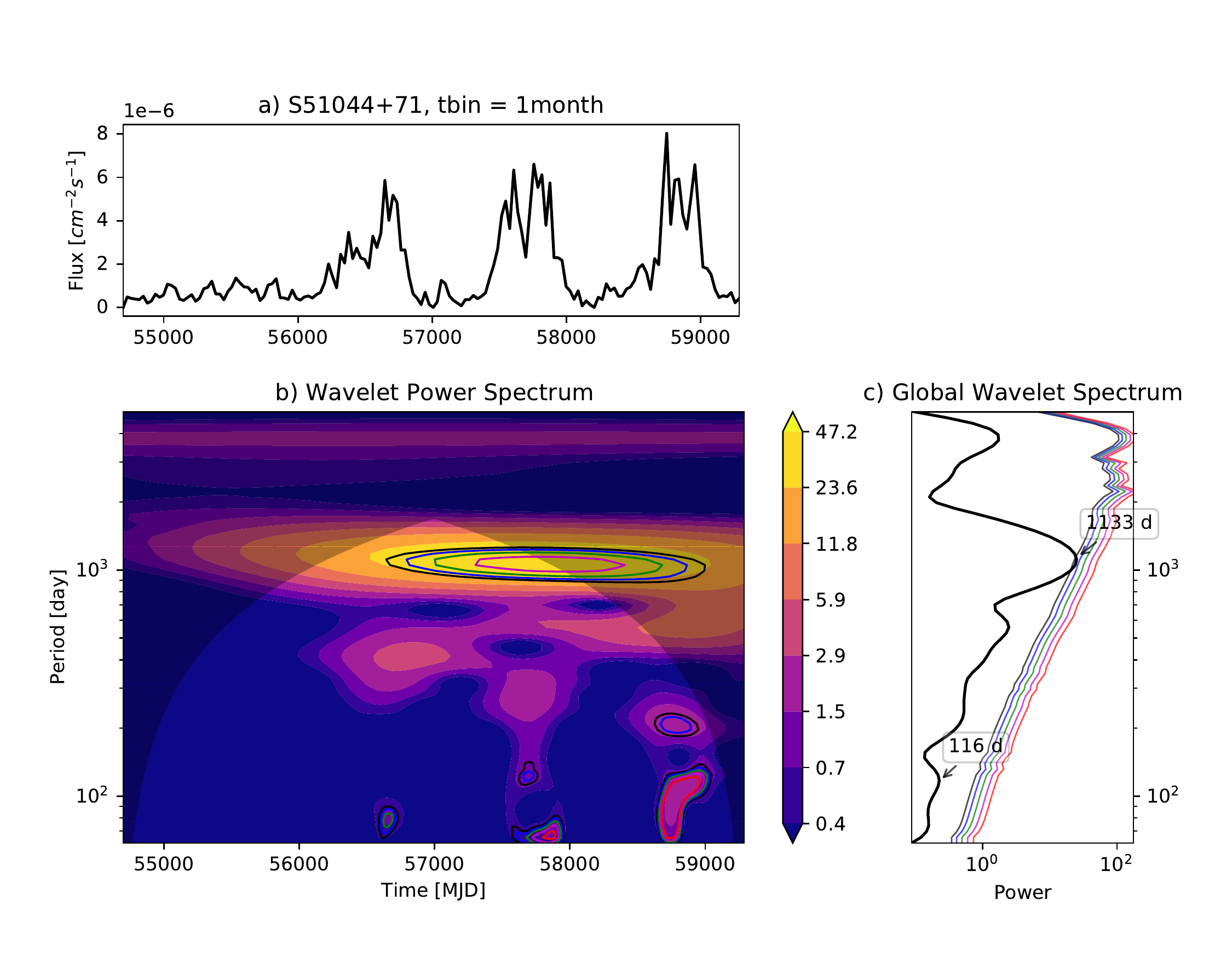}
		\caption{}
	\end{subfigure}
	\hfill
	\begin{subfigure}[b]{0.48\textwidth}
		\centering
		\includegraphics[width=\textwidth]{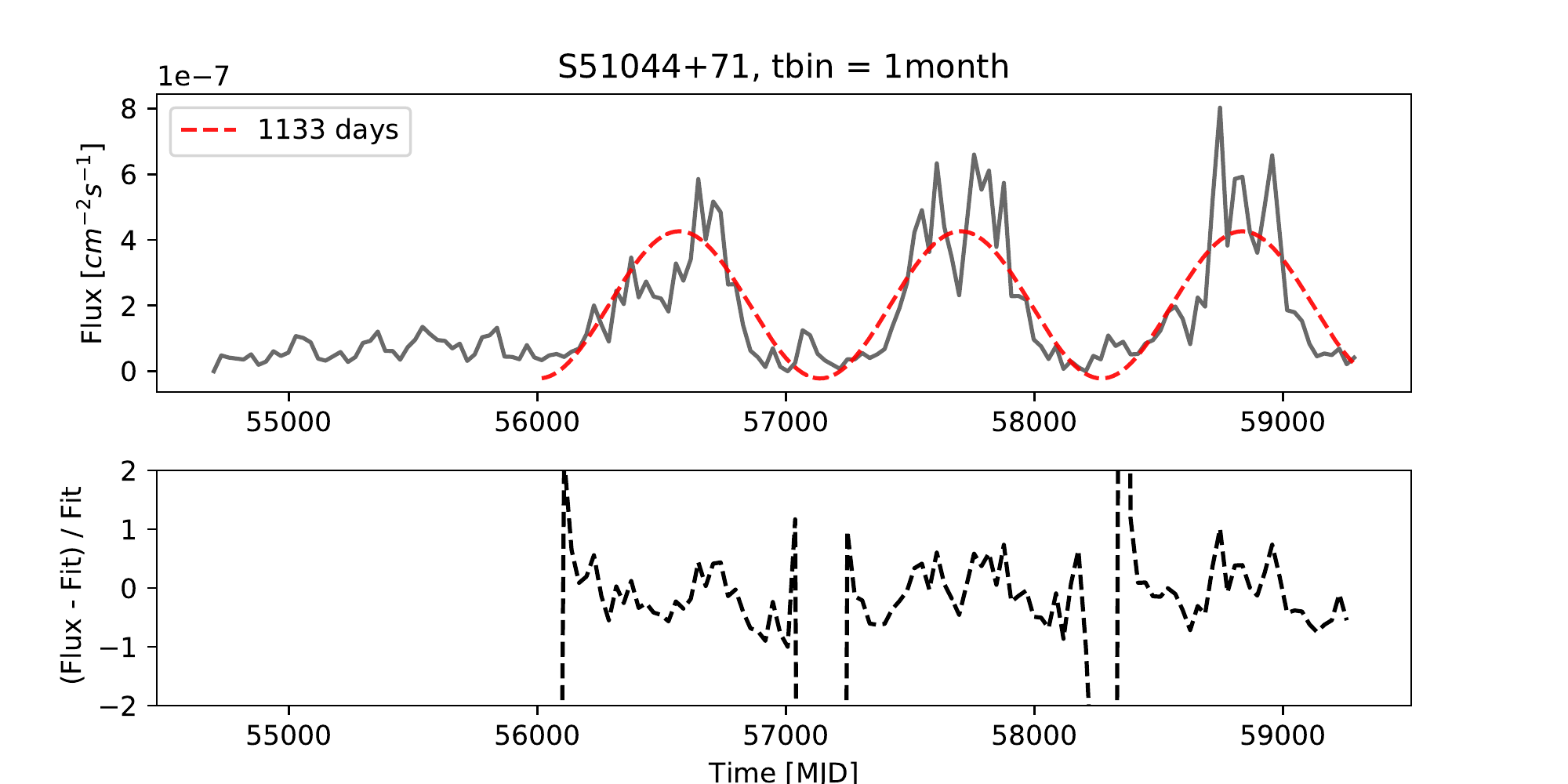}
		\caption{}
	\end{subfigure}
	\vskip\baselineskip
	
	\begin{subfigure}[b]{0.48\textwidth}  
		\centering 
		\includegraphics[width=\textwidth]{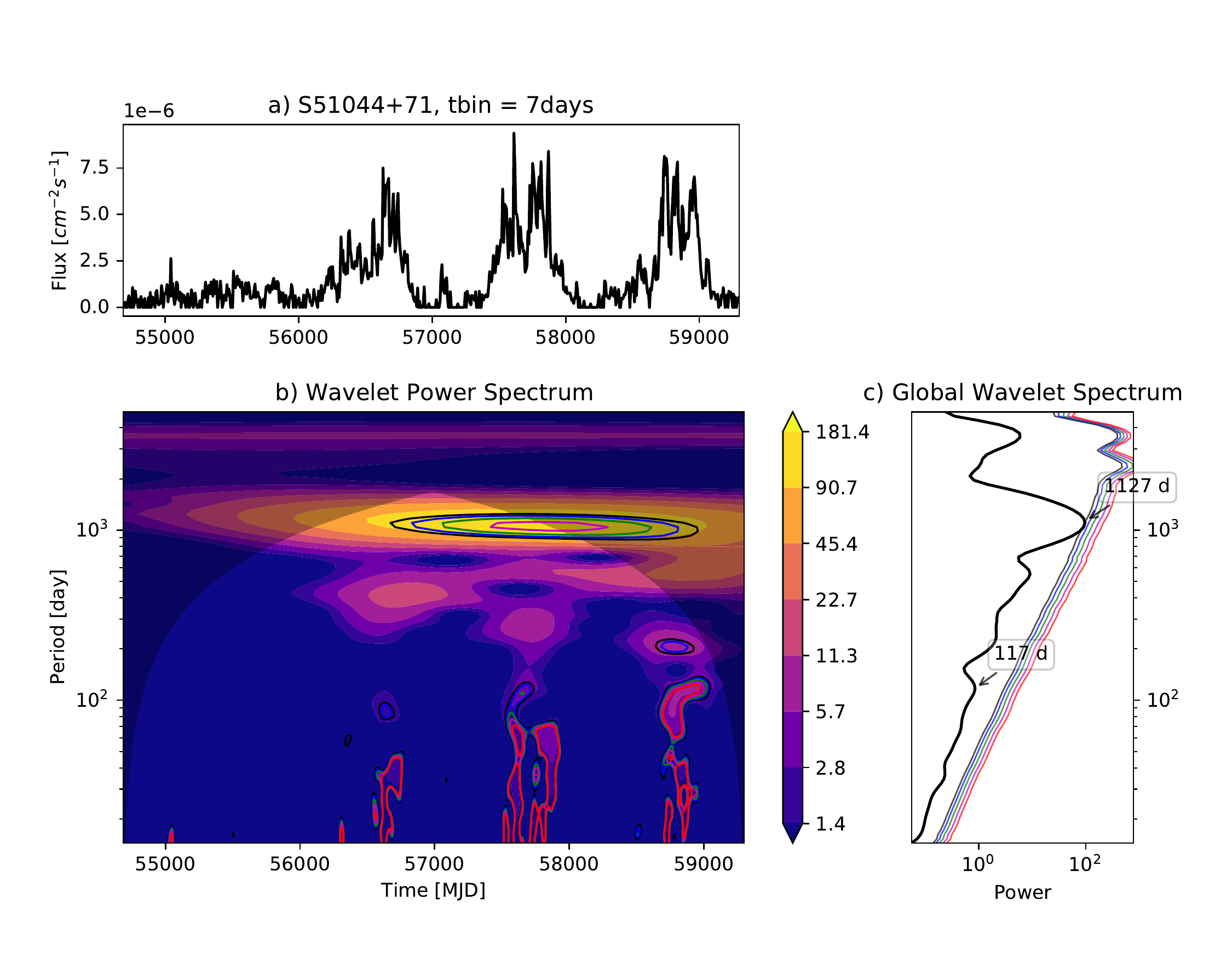}
		\caption{}
	\end{subfigure}
	\hfill
	\begin{subfigure}[b]{0.48\textwidth}  
		\centering 
		\includegraphics[width=\textwidth]{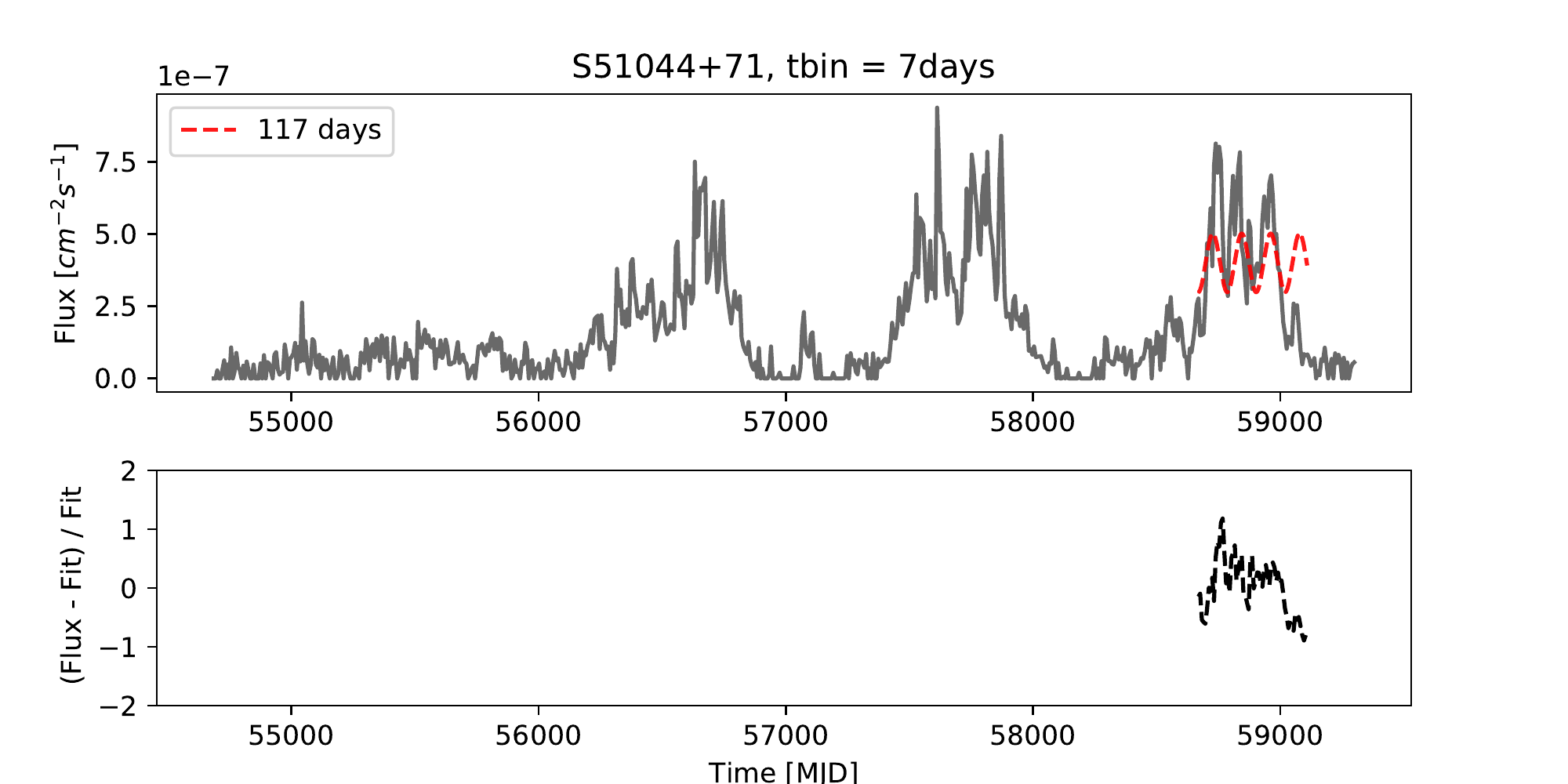}
		\caption{}
	\end{subfigure}
	\vskip\baselineskip
	\caption{Same description as Fig.~\ref{fig:CWT0102} for S5~1044+71.}
	\label{fig:CWT1044}
\end{figure*}

\begin{figure*}[!htbp]
	\centering
	\begin{subfigure}[b]{0.48\textwidth}
		\centering
		\includegraphics[width=\textwidth]{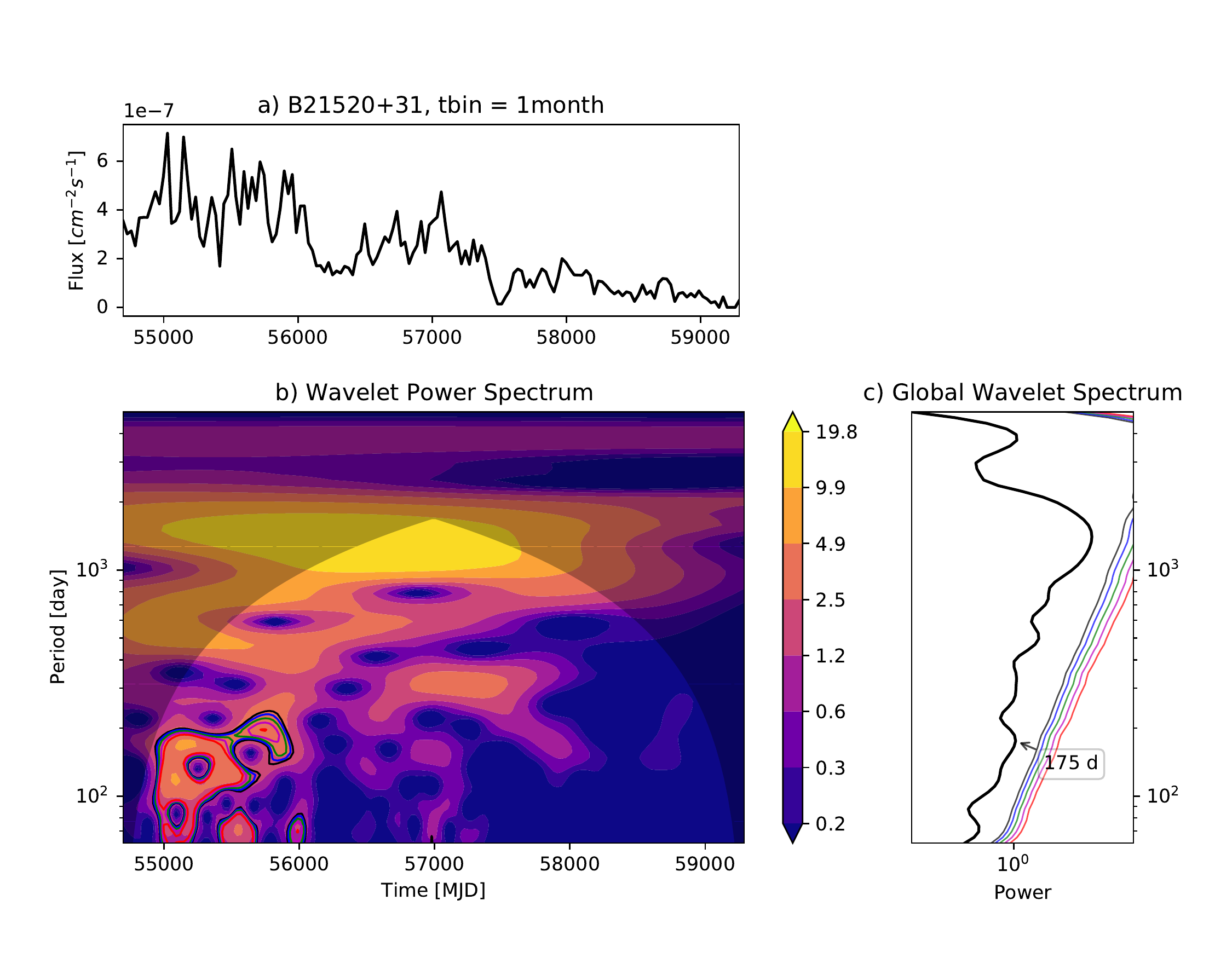}
		\caption{}
	\end{subfigure}
	\hfill
	\begin{subfigure}[b]{0.48\textwidth}
		\centering
		\includegraphics[width=\textwidth]{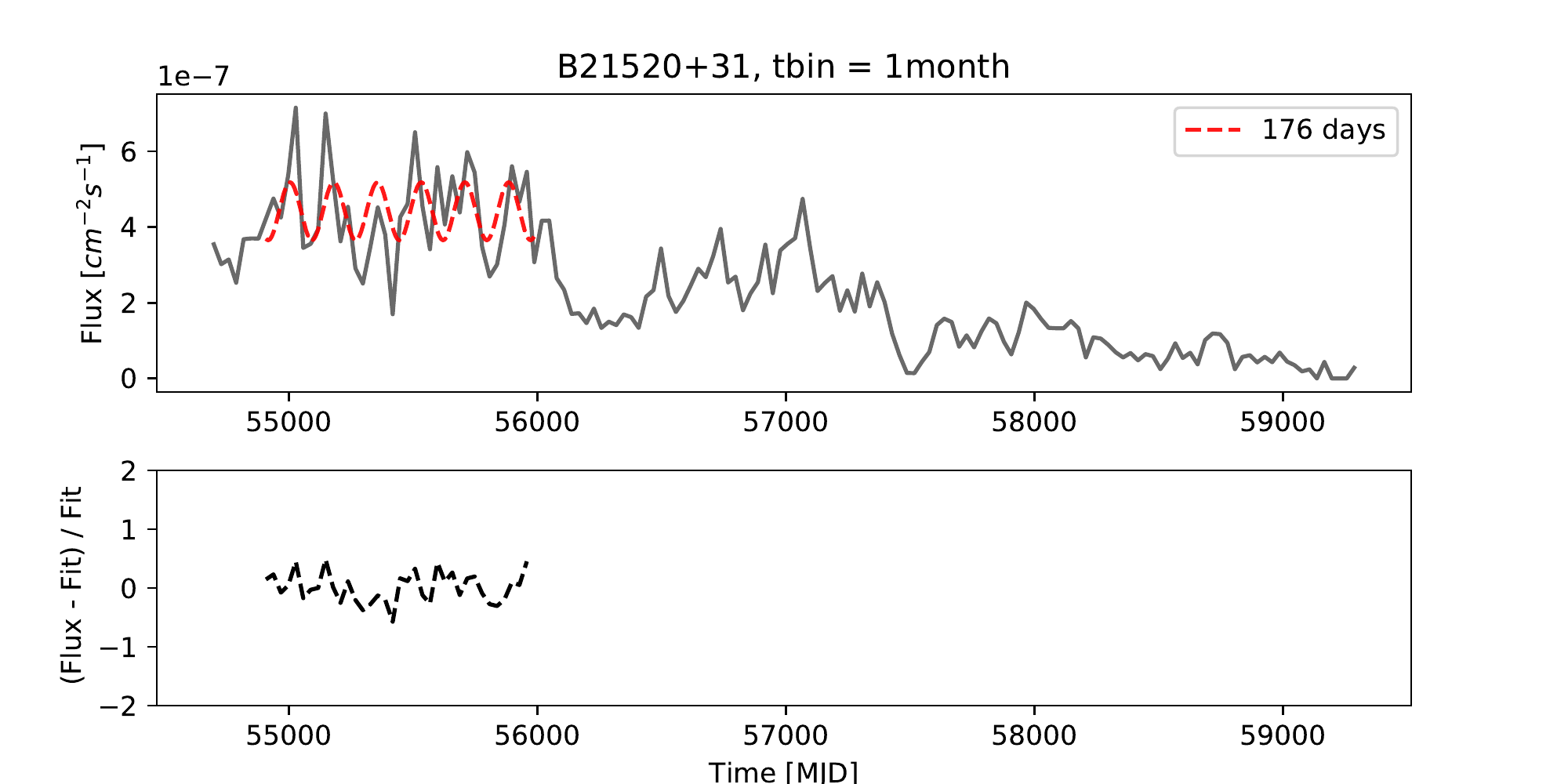}
		\caption{}
	\end{subfigure}
	\vskip\baselineskip
	
	\begin{subfigure}[b]{0.48\textwidth}  
		\centering 
		\includegraphics[width=\textwidth]{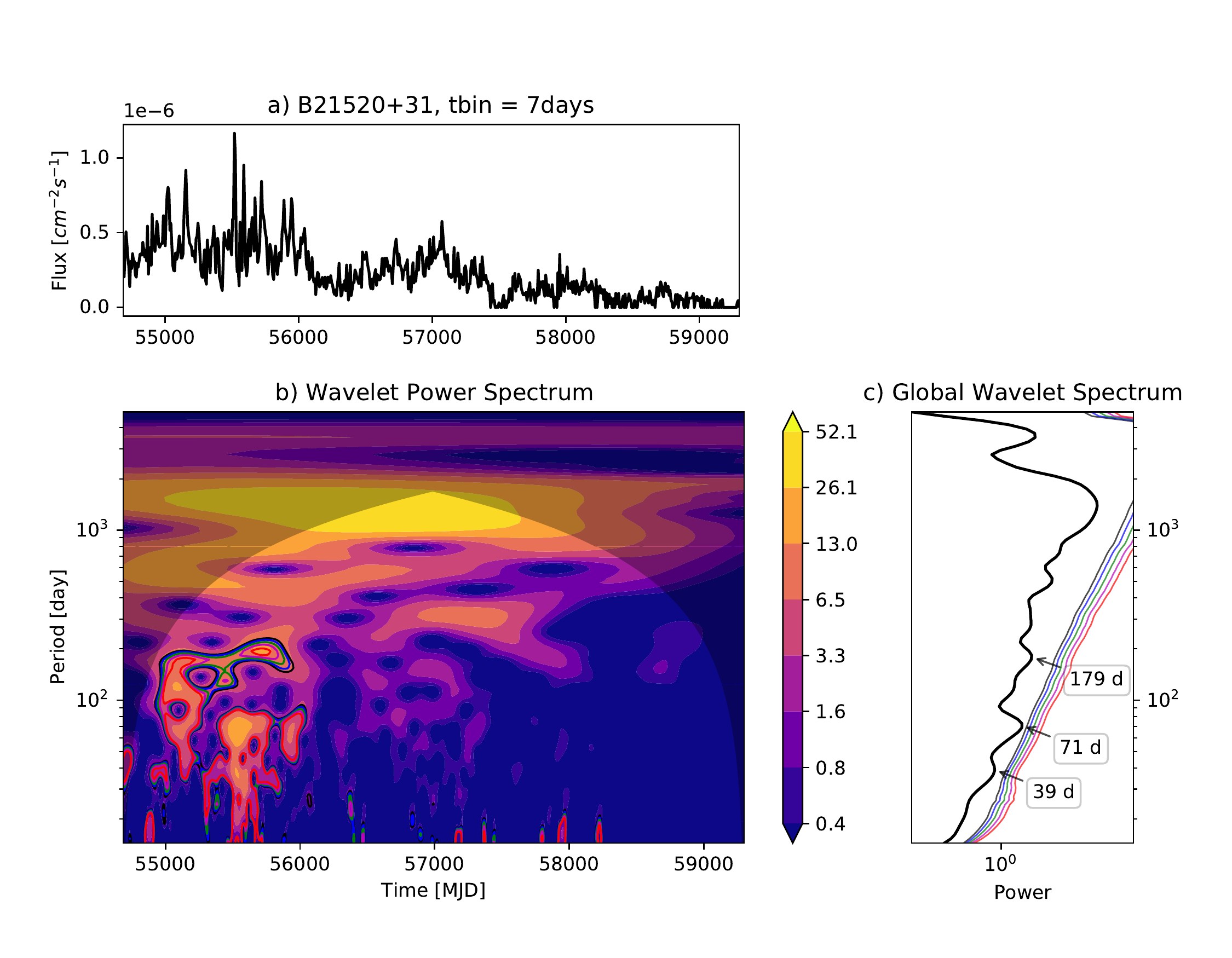}
		\caption{}
	\end{subfigure}
	\hfill
	\begin{subfigure}[b]{0.48\textwidth}  
		\centering 
		\includegraphics[width=\textwidth]{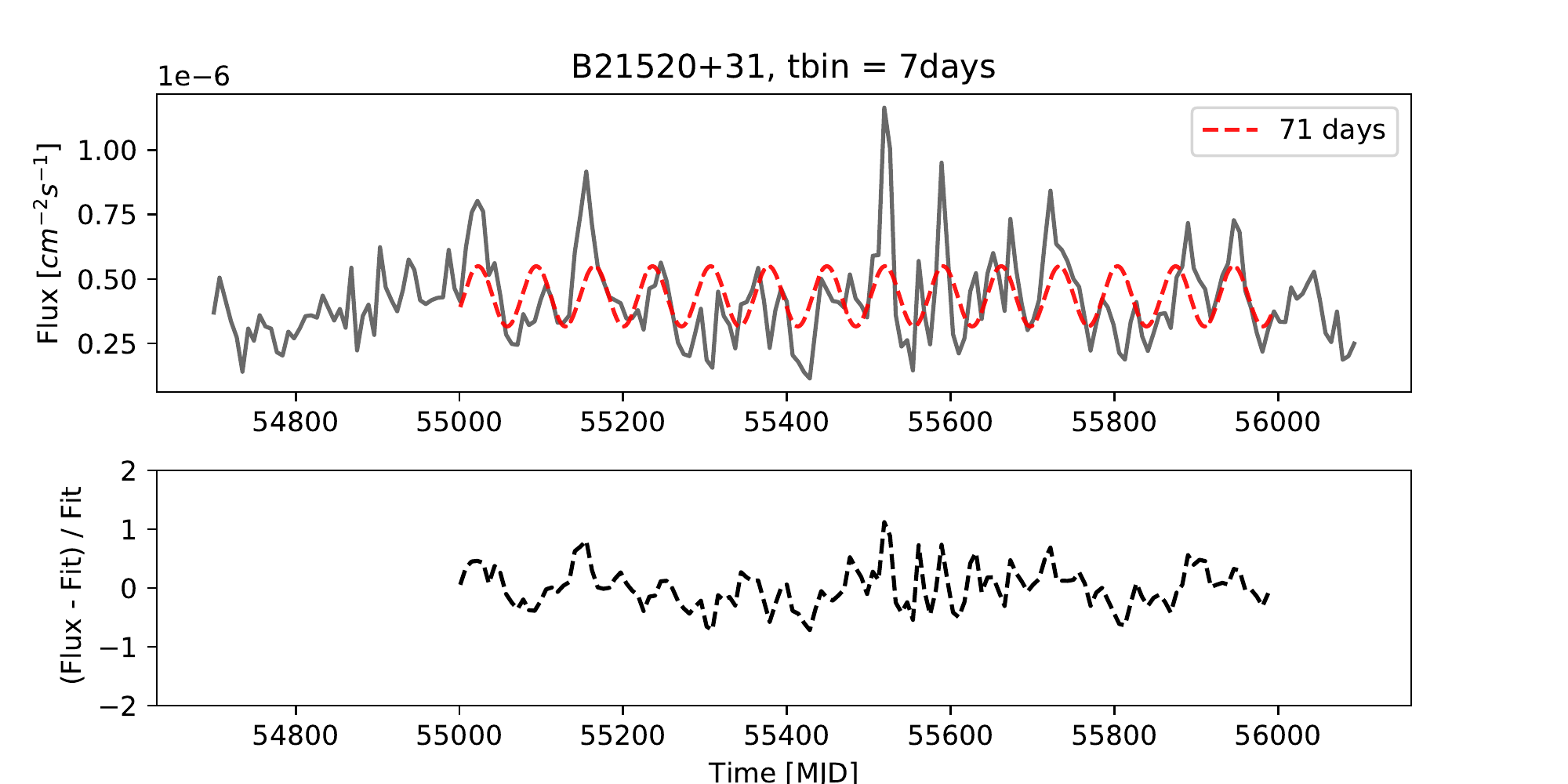}
		\caption{}
	\end{subfigure}
	\vskip\baselineskip

	\begin{subfigure}[b]{0.48\textwidth}  
		\centering 
		\includegraphics[width=\textwidth]{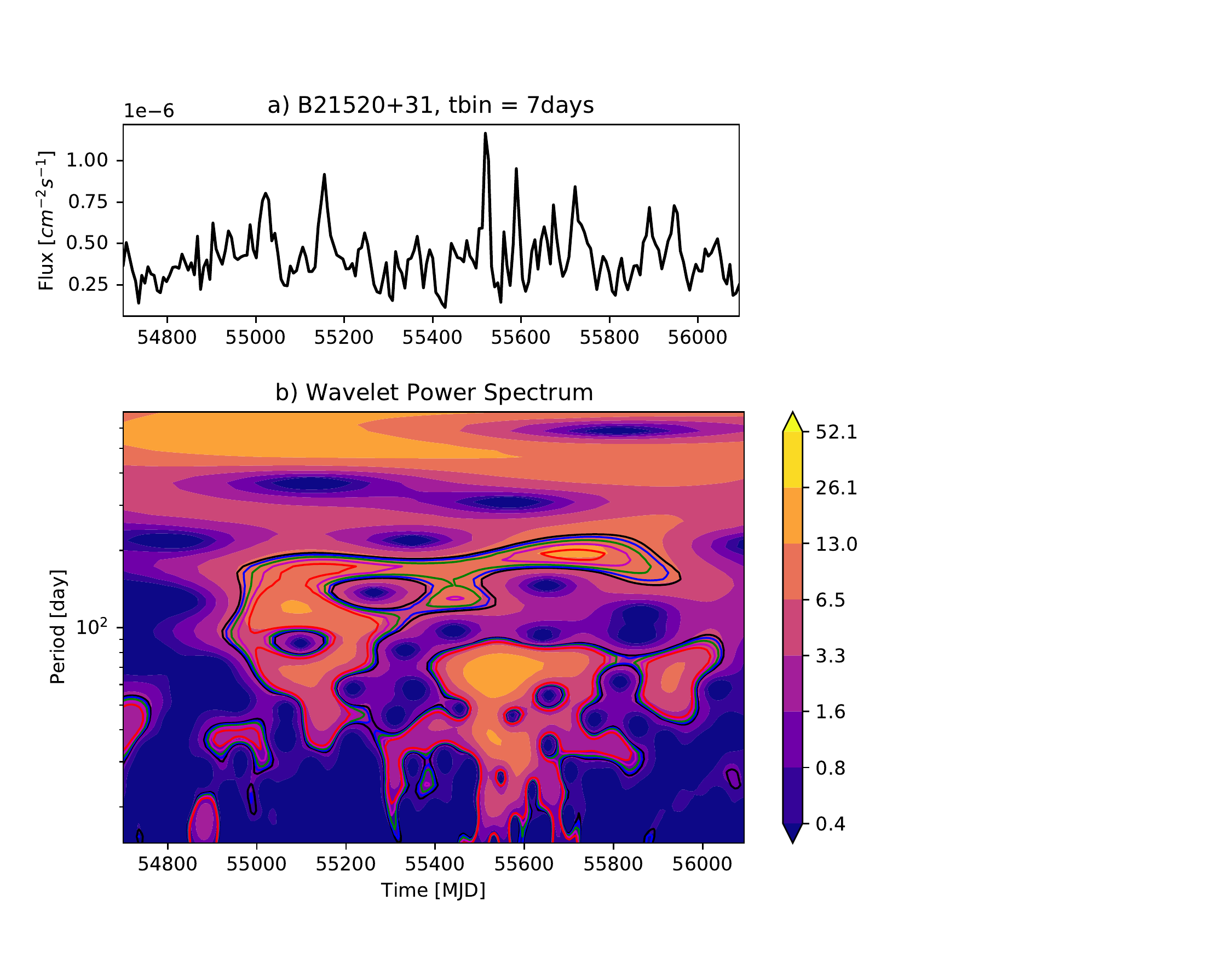}
		\caption{}
	\end{subfigure}
	\hfill
	\begin{subfigure}[b]{0.48\textwidth}  
		\centering 
		\includegraphics[width=\textwidth]{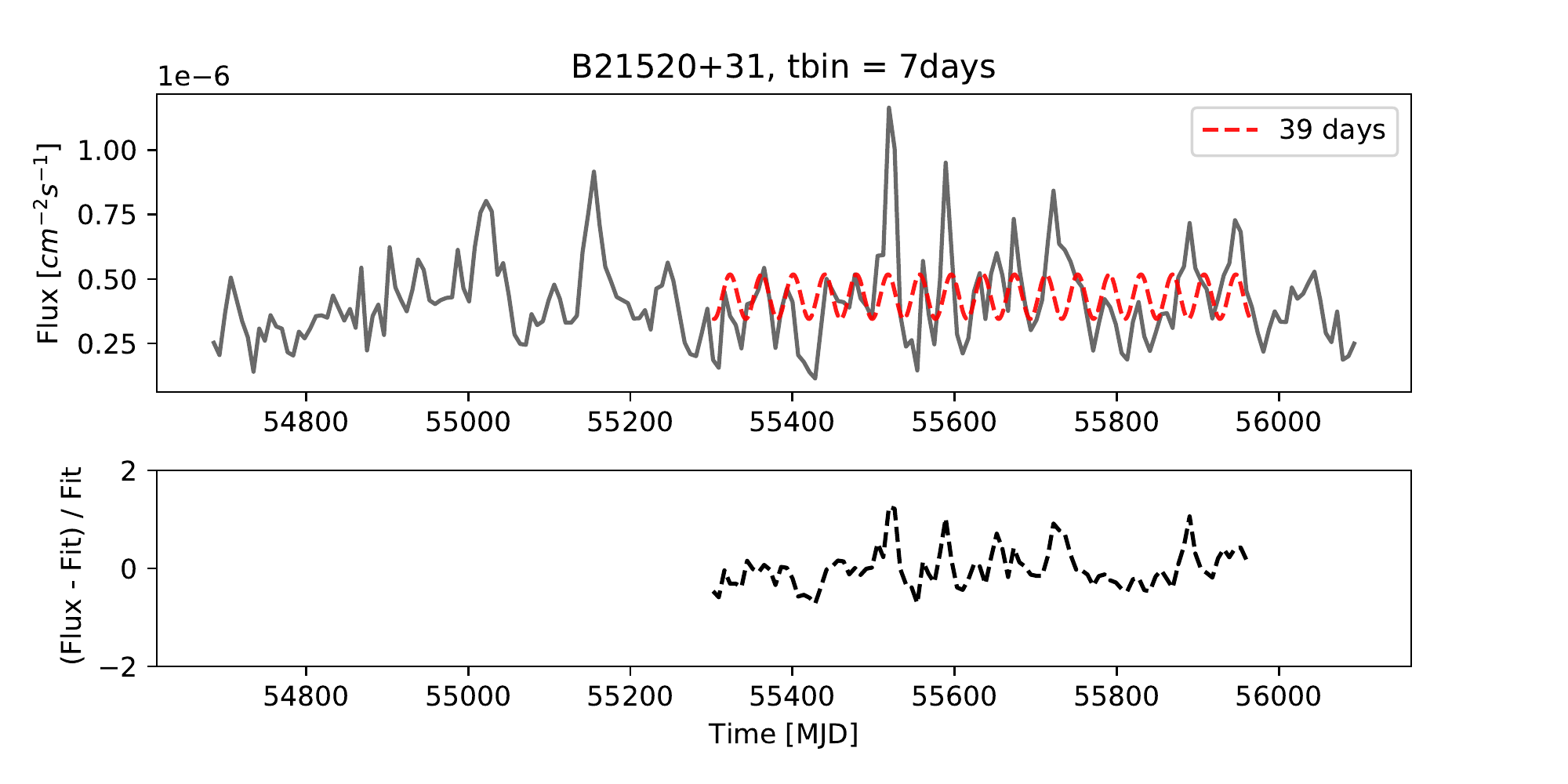}
		\caption{}
	\end{subfigure}
	\caption{Same description as Fig.~\ref{fig:CWT0102} for B2~1520+31 for the upper 4 figures. The third CWT is computed with the weekly binned light curve cut from MJD~54697 to MJD~56100. And on the right, the respective fitted light curve to show the $\sim$39~d period.}
	\label{fig:CWT1520}
\end{figure*}

\begin{figure*}[!htbp]
	\centering
	\begin{subfigure}[b]{0.48\textwidth}
		\centering
		\includegraphics[width=\textwidth]{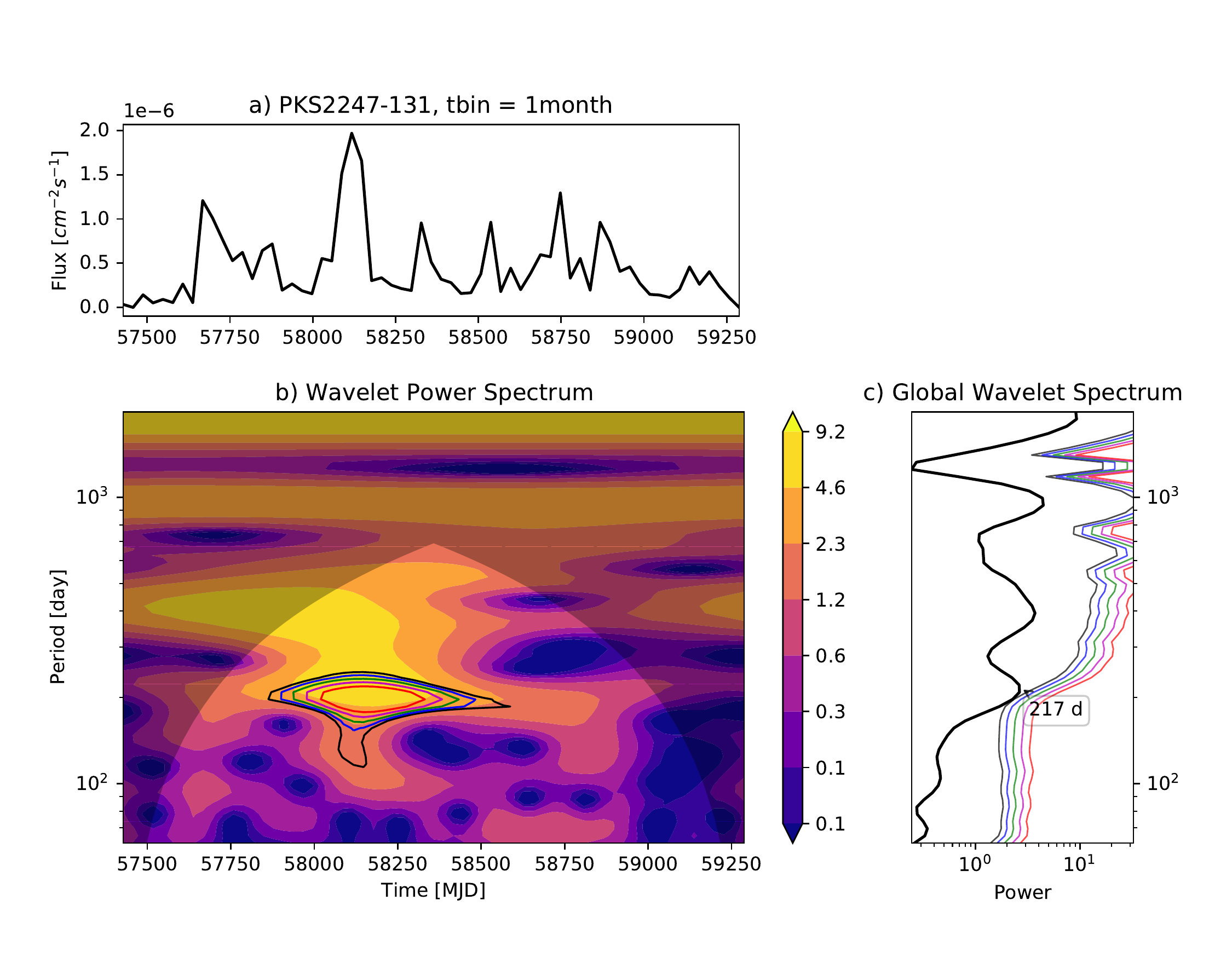}
		\caption{}
	\end{subfigure}
	\hfill
	\begin{subfigure}[b]{0.48\textwidth}
		\centering
		\includegraphics[width=\textwidth]{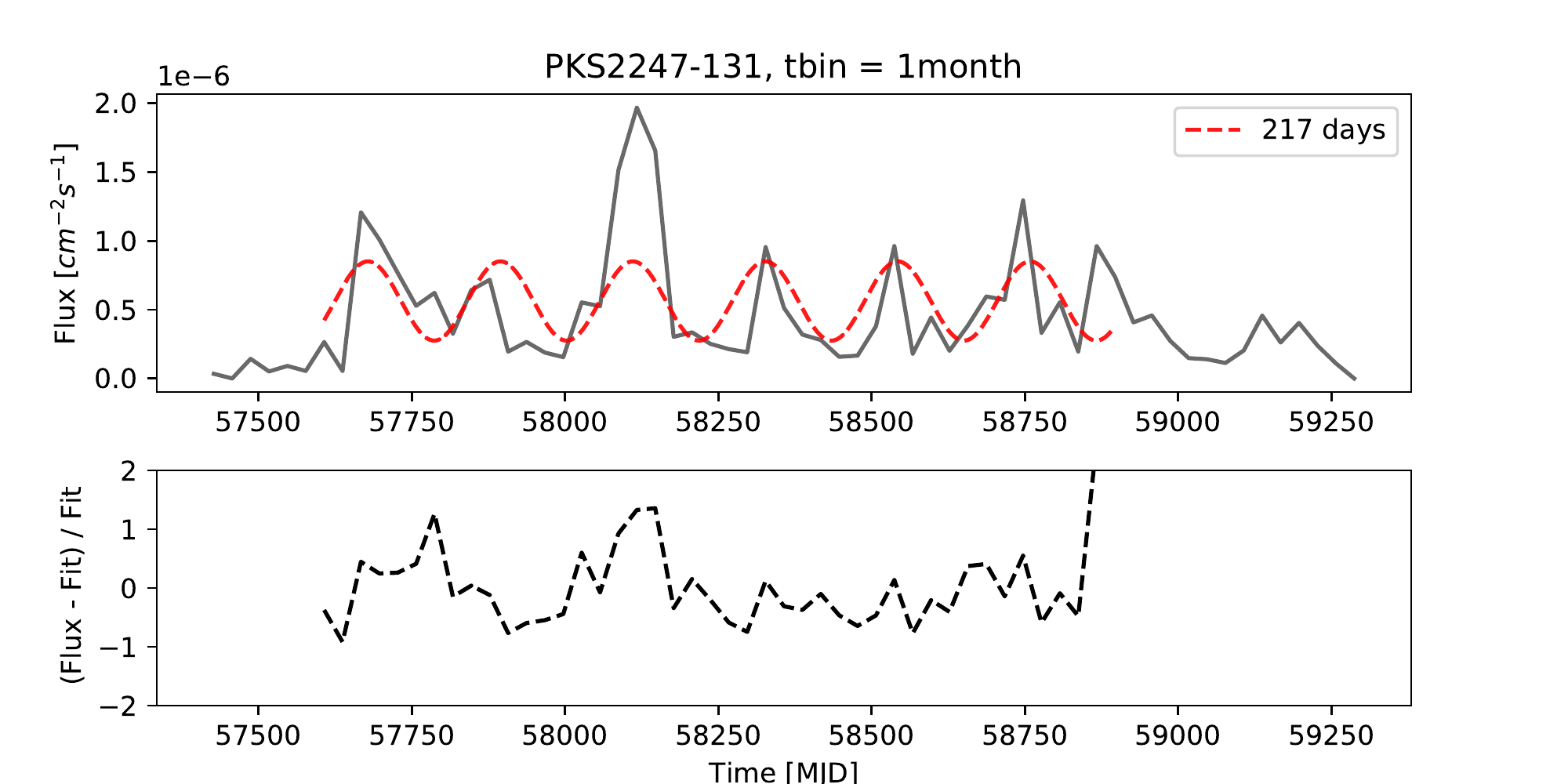}
		\caption{}
	\end{subfigure}
	\vskip\baselineskip
	
	\begin{subfigure}[b]{0.48\textwidth}  
		\centering 
		\includegraphics[width=\textwidth]{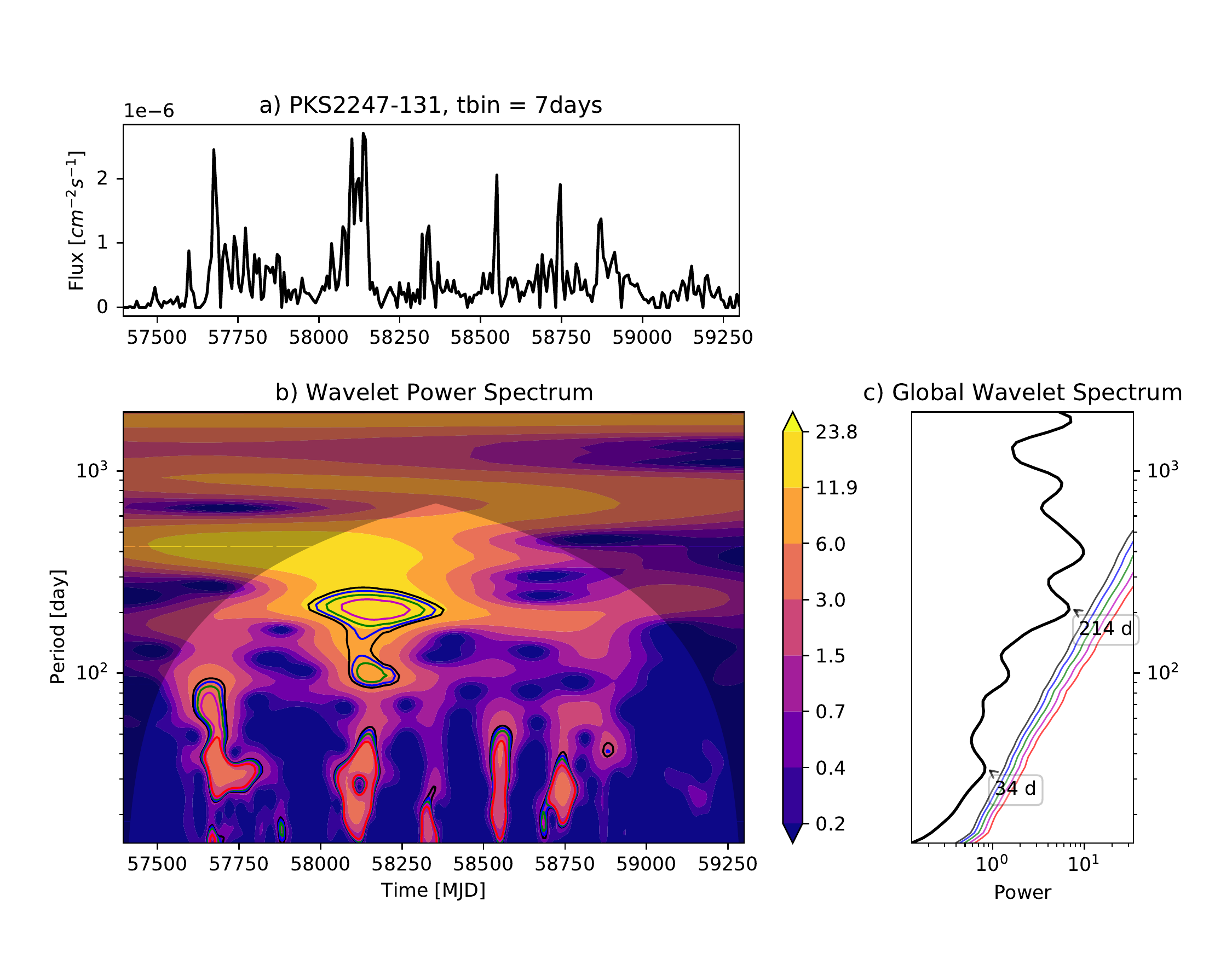}
		\caption{}
	\end{subfigure}
	\hfill
	\begin{subfigure}[b]{0.48\textwidth}  
		\centering 
		\includegraphics[width=\textwidth]{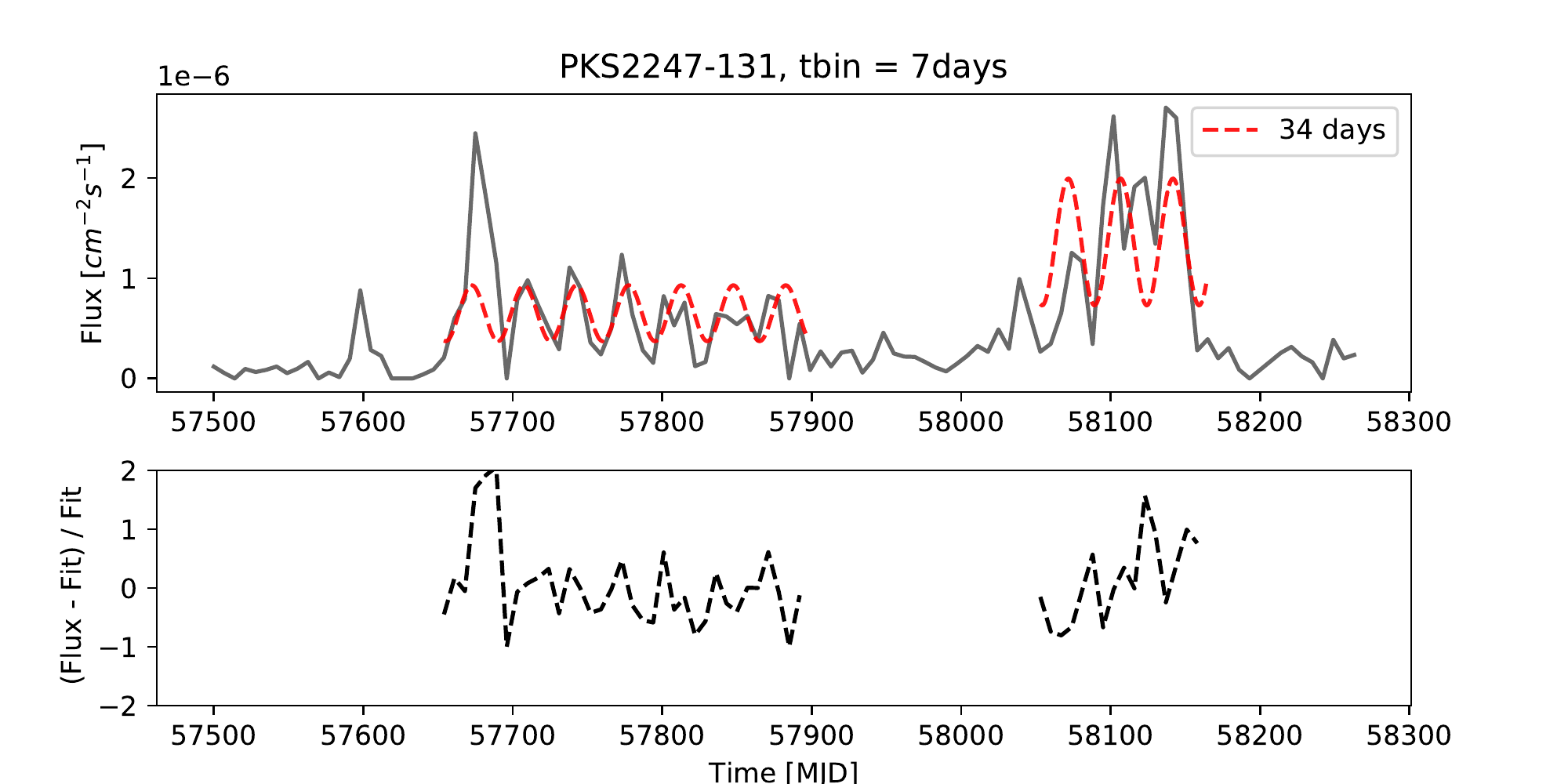}
		\caption{}
	\end{subfigure}
	\vskip\baselineskip
	
	\caption{Same description as Fig.~\ref{fig:CWT0102} for PKS~2247-131. In the bottom fitted light curve we show a possible reappearance of the candidate QPO with $\sim$34~d period at MJD~58050 adding 3 more cycles to the series.}
	\label{fig:CWT2247}
\end{figure*}


We identify a total of 36 QPO candidates in 24 out of the 35 selected sources. We consider candidates that have a post-trial significance larger than 3$\sigma$ in at least one of the two binnings, and that show at least three complete cycles once fitted. Long term QPO candidates with periods longer than and around one year are observed at various significance levels in 5 sources: S5~0716+714, S5~1044+71, Mrk~421, Mrk~501, PKS~2155-304 and CTA~102. On the other hand, month-long QPO candidates with period less than $\sim$300~d can be identified in the following 20 sources: 4C~+01.02, 4C~+28.07, NGC~1275, PKS~0402-362, PKS~0426-380, PKS~0447-439, PKS~0454-234, PKS~0537-441, 1H~1013+498, S5~1044+71, 4C~+21.35, 3C~273, 3C~279, PKS~1424-418, PKS~1510-089, B2~1520+31, CTA~102, PKS~2247-131, 3C~454.3 and PMN~J2345-1555. We report no QPO detection in the $\gamma$-ray emission of sources 3C~66A, PKS~0235+164, 4C~+55.17, B3~1343+451, PKS~1424+240, PKS~1502+106, PG~1553+113, 4C~+38.41, 1ES~1959+650, BL~Lac and PKS~2326-502.\\

The wavelet power spectra of the five more significant sources are shown in Figures~\ref{fig:CWT0102}, \ref{fig:CWT0537}, \ref{fig:CWT1044}, \ref{fig:CWT1520}, and \ref{fig:CWT2247} . Additionally, to help visualizing the QPO candidates identified by the CWT, we fitted periodic functions with periods equal to the ones computed via the CWT, to the \fermilat\ light curves. Here follows a detailed description of these results.\\

\subsection{4C~+01.02}

The wavelet analysis shows two significant QPO candidates in the light curves of 4C~+01.02, a distant FSRQ at z=2.099 (see Fig.~\ref{fig:CWT0102}(A) and (C)). No previous report of QPO analysis is found for this source in the literature. Both QPOs have month-long periods, with the longer one centered at $286\pm55$~d ($268\pm54$~d), and the shorter one at $123\pm26$~d ($122\pm26$~d), in the monthly (weekly) binned light curves. The significance of the longer QPO is over $5\sigma$ in both time binnings, whereas the shorter QPO has a significance $\sim$4.7$\sigma$ for the monthly binned light curve, and larger than $5\sigma$ for the weekly binned light curve. The fitted light curves in Fig.~\ref{fig:CWT0102}(B) and (D) present four complete cycles for the first QPO, starting from around MJD~56900 to MJD~58000 (August 2014 to September 2017), and five complete cycles for the second QPO, between MJD~57300 and MJD~57900 approximately (October 205 and May 2017). Tinier structures also appear at lower periods, although with more vertical and ambiguous shapes, which are inevitably produced by small flares (see for example the signals found between MJD~57500 and MJD~58000 at periods <50~d). We did not include these among our results due to their shortness and unclear shape, but this is a very subjective criteria, and these features could be investigated further with finer time bins.\\  

\subsection{PKS~0537-441}

We identify a significant QPO candidate in this BL Lac object at z=0.892 (see Fig.~\ref{fig:CWT0537}(A) and (C)). The wavelet spectrum shows a clear horizontal feature centered at period $285\pm67$~d ($286\pm73$~d) in the monthly (weekly) binned light curve, at a significance above $5\sigma$ in the CWT map of both time binnings. This result is consistent with \cite{Sandrinelli2016}, who claimed a candidate QPO at $\sim$280~d in the first few years of \fermilat\ data, using the Lomb Scargle periodogram (LSP) technique. We can see four complete cycles of oscillation during a period of high $\gamma$-ray flux starting from the beginning of \fermilat\ observations till around MJD~55700 (May 2011). The cycles are also shown in the fitted light curves in Fig.~\ref{fig:CWT0537}(B) and (D)).\\

\subsection{S5~1044+71}

We identify, for the first time, two highly significant QPO candidates in S5~1044+71, a moderately distant (z=1.15) FSRQ. The two CWTs are shown in Fig.~\ref{fig:CWT1044}(A) and (C). A very evident long term oscillation of period $1134\pm226$~d ($1127\pm224$~d) at a conficence level $\sim$4.9$\sigma$ ($\sim$4.6$\sigma$) for the monthly (weekly) binned light curve appears in the wavelet map, emerging around MJD~56000 (March 2012). Secondly, a short month-long QPO emerged during the last flaring state of the source, between MJD~58650 and MJD~59000 (June 2019 and September 2020), with a period estimated to be $116\pm33$~d ($117\pm38$~d) for the monthly (weekly) binned light curve at a significance more than $5\sigma$. Three complete cycles of the year-long QPO can be observed also in the fitted light curve in Fig.~\ref{fig:CWT1044}(B), which corresponds to the large flaring states of the FSRQ. The last maximum occurred at around MJD~58800 (November 2019), so we expect the next maximum to occur at approximately MJD~59900 (November 2022). The absence of a clear fourth peak at around MJD~55500 might indicate additional modulation.\footnote{During the writing of this manuscript, an analysis of S5~1044+71 has been presented in a pre-print by \cite{Wang22}, claiming a 3 years modulation ($\sim$3.06~$\pm$~0.43~yr) at a significance level of $\sim$3.6$\sigma$. This result is in total agreement with ours.} Regarding the short period QPO, four cycles are indicated in the fitted light curve in Fig.\ref{fig:CWT1044}(D), which are the three narrower flares in the last flaring state and a fourth more suppressed one after MJD~59000. This kind of structured flares are not unique in this last flaring state, but could also appear in the two previous maxima. These signals have however a more complex and vertical shapes, which causes ambiguity. If real, they could also suggest a trend towards longer periods.  One can also notice in the weekly CWT maps hints ($2 \sigma$ at most) of several horizontal features at longer periods, which remain however consistent with noise once corrected for trials. \\

\subsection{B2~1520+31} 
\label{sec:1520}

Three possible month-long QPOs are identified by the CWT of the light curves of this distant (z=1.489) FSRQ, as shown in Fig.~\ref{fig:CWT1520}(A), (C) and (E). The CWT maps of this source are a good example of the inherent difficulty of this analysis technique when dealing with blazar light curves: rapid flares result in vertical structures in the map that overlap with the horizontal bands we are interested into, and require visual inspection of all peaks that appear in the global wavelet spectrum. The first QPO candidate is found in both light curves at around $174\pm48$~d ($179\pm42$~d) for the monthly (weekly) binned light curve with a significance above $5\sigma$ in both cases. This QPO candidate is of particular interest because its period seems to increase with time. The second and third candidate QPOs can be better identified in the CWT map of the weekly binned light curve, with periods of $71\pm15$~d and $39\pm11$~d, both exceeding $5\sigma$ significance. A zoomed-in CWT is shown in Fig.~\ref{fig:CWT1520} (E), which allows us to visualize better the signal of the two very short period QPO candidates. The $71\pm15$~d period is compatible with the one reported by \cite{Gupta19} with a period of $\sim$71~d, by analysing the first 4 years of \textit{Fermi}-LAT data of B2~1520+31 using the LSP and weighted wavelet z-transform (WWZ) techniques. The source has also been studied by \citet{Tarno20}: although some signal can be seen in their analysis at around $\sim$70~d, it remains below the 3$\sigma$ interval they compute.\\

Fitted light curves are presented in the right hand side of Fig.~\ref{fig:CWT1520}. Fig.~\ref{fig:CWT1520} (B) shows a periodic function of $\sim$176~d with six cycles between MJD~54900 and MJD~56000 (March 2009 and March 2012);  in Fig.~\ref{fig:CWT1520} (D) we can see 14 oscillations of $\sim$71~d spanning from MJD~55000 to MJD~56000 (June 2009 to March 2012), and Fig.~\ref{fig:CWT1520} (E) shows a fit with 17 cycles of period $\sim$39~d between MJD~55300 and 55970 (April 2010 and February 2012). A forth QPO could be also considered at a period halfway between $\sim$71~d and $\sim$176~d, which is visible in both the CWT map and the global wavelet spectrum of the weekly binned light curve. Furthermore, the $\sim$39~d period seems to also be present before the above mentioned time span, between around MJD~54800 and MJD~55000. \\

\subsection{PKS~2247-131}
\label{sec:2247}

The wavelet analysis of PKS~2247-131, a BL Lac object at z=0.22, shows two QPO candidates (see Fig.~\ref{fig:CWT2247}(A) and (C)). The first of the QPO candidates is found to be at around $217\pm38$~d ($214\pm43$~d), at above $5\sigma$ confidence level in the CWT of monthly (weekly) binned light curves. This QPO candidate seems to span at least from MJD~57600 to MJD~58500 approximately (July 2016 to January 2019) in the CWT map. Just by looking at the light curve one can notice the tentative periodic oscillations of this source. And this can be confirmed by the fitted function shown in Fig.~\ref{fig:CWT2247}(B), presenting 6 peaks including the one at around MJD~58750.

More interestingly, we identify a QPO candidate in the time interval around MJD~57600 to MJD~57900 approximately (July 2016 to May 2017) with much shorter period. We can appreciate it in Fig.~\ref{fig:CWT2247}(C) which shows the QPO centered around $34\pm14$~d, with a significance larger than $5\sigma$, and only noticeable in the full-range weekly binned light curve. Tentative oscillations can also be seen, with at least 7 full cycles, shown in Fig.~\ref{fig:CWT2247}(D) with the fitted periodic function. In addition, with a more detailed evaluation one can notice that the $\sim$34~d period reappears after MJD~58000 (September 2017). This is also shown in the fitted light curve in (D), with 3 more cycles spanning MJD~58050 to MJD~58170 approximately. \\

In November 2018, two years after the \textit{Fermi}-LAT announcement of detection of this BL Lac type blazar, \cite{Zhou2018} presented the first claim of a month-scale QPO in the $\gamma$-emission of PKS~2247-131. They found a relatively short, month-scale oscillation at period $34.5\pm1.5$~d, which can be indicative of the presence of an SMBH binary in the center of this blazar. This discovery conducted us to explore the CWT of PKS~2247-131 more deeply. Small features like this one also appear in several other sources which are included in \autoref{sec:AAll}, but we do not systematically inspect all of them. The short QPO candidate in our analysis is compatible with the aforementioned one proposed by \cite{Zhou2018}. On the other hand, no previous claims for the longer QPO candidates at periods of around $220$~d have been reported in the literature.\\

\subsection{Other sources}

In the remaining 27 sources showing QPO candidates, we found several candidates compatible with previously reported claims in the literature. The CWT maps and fitted light curves of these AGNs can be found in \autoref{sec:AAll}. 

\begin{itemize}

    \item S5~0716+714: We found a QPO candidate in both the 1-month and 7-days binned light curves of this BL Lac object, centered at $325\pm75$~d ($324\pm77$~d), at $\sim$2.4$\sigma$ ($\sim$3.2$\sigma$) in the wavelet spectrum of the monthly (weekly) binned light curve. This feature only appears in the time interval between MJD~56000 and MJD~57500 approximately (March 2012 to April 2016). Our result is compatible with the analyses by several authors: \cite{Prokhorov17} reported a QPO evidence at $\sim$346~d, \cite{Li18} detected a periodic feature at $\sim$344~d, \cite{Bhatta2020} found a QPO candidate at around 340~d and \cite{Penil20} identified a period at $\sim$0.9~yr. However, \cite{Covino19} and \cite{Zywucka2021} claim the absence of any periodic emission in this object. \cite{Bhatta2020} and \cite{Penil20} also reported a possible QPO at longer period around $1000$~d. In our CWT analysis, a high power spectrum feature does appear and it is centered at around $1000$~d, spanning almost the full time series. However, this result is only significant before applying the trial correction, and is compatible with noise once the ``look-elsewhere effect'' is taken into account. \\

    \item Mrk~421: \cite{Bhatta2020} reported an oscillating $\gamma$-ray emission at $\sim$280~d, compatible with our QPO candidate at $300\pm64$~d ($300\pm65$~d) with significance exceeding $5\sigma$ in both time binnings. This QPO candidate is also not a persistent one, being seen between MJD~55800 and MJD~57000 approximately (August 2011 to December 2014). In our analysis, both maps show a complex structure, with a possible period change at the end. \\
    
    \item Mrk~501: \cite{Bhatta19} found a candidate QPO at period $\sim$230~d, compatible with the one we identify at $315\pm98$~d ($326\pm76$~d) with a significance $\sim$2.9$\sigma$ in the monthly binned light curve, but high significance above $5\sigma$ in the weekly binned one. This QPO candidate is also temporary, emerging from MJD~55800 to MJD~57000 approximately, and reappearing between $\sim$MJD~57800 to $\sim$MJD~58700.\\
    
\end{itemize}

We want to remark the following three sources which also present hints of QPOs compatible with some claims in the literature, although only significant before trial corrections.

\begin{itemize}
    
    \item PKS~1424-418: \cite{Bhatta2020} and \cite{Yang21-1424} claimed a flux oscillation at period $\sim$353~d and $\sim$355~d respectively in the light curve of this FSRQ. The wavelet analysis, however, shows that this QPO candidate is compatible with noise, with a post-trial significance at $\sim$1$\sigma$ for both monthly and weekly binned light curves. On the other hand, we detect a much shorter QPO candidate centered at $\sim$94$\pm25$~d ($90\pm22$~d) in the CWT of the monthly (weekly) binned light curve, at above $5\sigma$ significance. This month-long QPO candidate emerged from MJD~56100 to MJD~56500 approximately (June 2012 to August 2013), showing five complete oscillations in the fitted light curve.\\
    
    \item PG~1553+113: The first detection of a periodic behaviour in the $\gamma$-emission of this high-synchrotron-peak BL Lac object was claimed by \cite{Ackermann15}, with a period of $\sim$798~d. Some more recent studies found a periodic feature at similar values: \cite{Sandrinelli2018} reported two peaks at same frequency $\sim$780$-$810~d, \cite{Covino20} detected a period of $\sim$790~d, \cite{Yang2020} claimed a possible evidence at period $\sim$800~d, and \cite{Penil20} found high significance level at period $\sim$803~d. Our results show that this QPO candidate is significant in pre-trial significance (at a confidence level of $\sim$3$\sigma$ ($\sim$2$\sigma$) for the monthly (weekly) binned light curve), but becomes less than $1\sigma$ after trial correction \citep[consistent with the results by][]{AitBenkhali20}.\\
    
    \item PKS~2155-304: We identify a low-significance QPO candidate for this BL Lac object centered at around $334\pm107$~d ($341\pm106$~d), with a significance $\sim$2.2$\sigma$ ($\sim$3.5$\sigma$) in the monthly (weekly) binned light curve. The QPO candidate clearly shows a time-dependent behaviour, with decreasing frequency in time. This source has been extensively studied in the past and several QPO claims have been made: \cite{Zhang17-2155} who detected a possible QPO at period $\sim$640~d, \cite{Bhatta2020} who found two quasi-periodic features, one at  610$\pm$51~d and the other at $\sim$260~d, \cite{Covino20} who found a QPO at $\sim$610~d although not significant enough, \cite{Penil20} who claim a QPO at $\sim$1.7~yr, \citet{Tarno20} who found a QPO at 610$\pm$42~d, and \cite{Zywucka2021} who detected a QPO at $\sim$612~d. With respect to the results of our analysis, the QPO found by CWT technique is close to the short QPO candidate presented by \cite{Bhatta2020}, whereas the long-term feature at $\sim$600~d, also noticeable in the CWT maps, is not significant after the trial correction, likewise the long $\sim$1000~d QPO candidate in S5~0716+714.\\
    
\end{itemize}

\section{Discussion}
\label{sec:discussion}

We identify a total of 36 QPO candidates in 24 sources, with the longest one found in the light curve of S5~1044+71 at $\sim$1130~d and the shortest in the light curve of PKS~2247-131 at $\sim$34~d. Many of the candidates have a period between one month and one year. This is a new result since not many month-year-long QPOs have been reported in the literature, and might be revealing that middle-term QPOs are actually frequent in nature. On the other hand, only four candidate QPOs with period of around one month are detected in our results. This can be explained by the fact that this kind of short periods, if not lasting continuously in time, would appear in the CWT map as small structures, being difficult to recognize, and to be disentangled from vertical structures produced by flares. A good example is PKS~2247-131, whose analysis was inspired by the work published by \cite{Zhou2018}. More example of short-leaving features with periods of the order of one month can be seen in the CWT maps of B2~1520+31, 1H~1013+498, and 3C~279 (see section~\ref{sec:1520} and \autoref{sec:AAll}). We remind the reader that our threshold to select candidates is for them to have a post-trial significance $>3\sigma$ in at least one of the time-bins, and showing more than three cycles. We cannot exclude then, that some of the candidates are spurious, and coming from the analysis itself.\\

Quasi-periodic modulations observed in the high-energy \gray\ fluxes of AGNs should be related to the relativistic jets launched by these objects or to the process feeding the jet itself. Two main subgroups of QPOs were often discussed in the literature by considering the oscillation period: short intra-day and year-long QPOs. The former are though to be associated to pulsational accretion flow instabilities \citep{Honma1992}, despite that longer periods are observed in magneto-hydrodynamical simulations for slow-spinning SMBH \citep{McKinney2012}. The origin of long-period QPOs, on the other hand, could be related to jet precession, geometry and possibly to the presence of a binary SMBH system \citep[see][and references therein]{Ackermann15}. In simple SMBH systems, jet precession, rotation and helical structure, in the presence of a sufficiently strong magnetic field, yield to observable periodicity from the change of the line of sight. Moreover, these variabilities could also appear in binary SMBH systems due to periodic perturbation of the secondary compact object to the accretion disk and jet. The orbital period of the binary systems is in the range of several years, being compatible with the year-long period QPOs. The most significant, multi-year, QPO candidate in our analysis, seen at more than $4\sigma$ in the blazar S5~1044+71, has a period of about 1100 d (3.5 y), and thus makes this \gray\ source a high-significance SMBH binary candidate in the Universe.\\

No QPOs at an intermediate time scale were reported until the publication of \cite{Zhou2018}, who claimed the detection of $\sim$34.5~d QPO in the light curve of PKS~2247-131. This monthly modulation suggested a helical jet structure due to the short period. However, they also noted that the helical structure could be driven by the orbital period of a secondary SMBH. By considering the time compression due to the high Doppler factor of the emitting region, the observed period will be shortened with respect to the physical period in the host galaxy reference frame, and can still be of the order of orbital periods for close binary SMBHs. In our analysis of PKS~2247-131, an additional QPO ($\sim$200~d) seems to be existing in the form of large $\gamma$-ray flares during the last few years, and could be used to provide further constraints on the model.\\

We highlight here three important findings that emerged from our analysis that might be used as key observables to understand the origin of QPOs in blazar \gray\ light curves:
\begin{itemize}
    \item We do not detect persistent QPOs, that last for the whole observing period. None of the global wavelet power spectra show significant features after trial correction. S5~1044+71 might be considered a persistent QPO candidate if we make the hypothesis that the first maximum is suppressed due to additional modulation. Despite this potential exception, all other QPO candidates in our analysis are transient ones.  \\
    
    \item  We identify some cases where the QPO candidate shows period shifts. This can be seen in the CWT maps of B2~1520+31 and, at lower significance, of PKS~2155-304, where one of the QPO candidates is decreasing its frequency. Hunting for a possible explanation, we might consider changes in the inclination angle of jet precession. But the reasonable timescales involved should not produce a transition as fast as the observed one. A much easier possibility is that we are observing the helical geometry of a jet with an intrinsic opening angle, that naturally leads to a slowdown of the QPO.\\ 
    
    \item Lastly, we see several occurrences of multiple QPO candidates occurring simultaneously, and with harmonic periods. One of the examples is 4C+01.02, where two QPOs with different frequencies overlap in a specified time interval. The longer period is approximately two times the shorter one, suggesting the presence of resonances in the emission. Simultaneous QPO candidates at harmonic ratios (within errors) can be seen in PKS~0402-362, 1H~1013+498, B2~1520+31, and CTA~102. Harmonics are common in QPO analysis of X-ray binaries \citep{Ingramreview}, so it is not surprising to see them also in AGNs. They indicate that the various QPOs share the same origin and are not due to independent physical processes, each one imprinting its particular QPO on the light curve. Theoretical models aimed at explaining QPOs in \gray\ light curves of blazars can thus be further constrained by studying harmonics, and harmonic ratios.  \\
\end{itemize}

Furthermore, no QPO candidates are identified in the $\gamma$-ray emission of 3C~66A, PKS~0235+164, 4C~+55.17, B3~1343+451, PKS~1424+240, PKS~1502+106, PG~1553+113, 4C~+38.41, 1ES~1959+650, BL~Lac and PKS~2326-502. Three of them, PG~1553+113, BL~Lac and 3C~454.3, were previously reported showing periodic oscillation. 

\begin{itemize}
    \item PG~1553+113: We discussed in the previous section that the wavelet analysis do show a increase in power spectrum at a period similar to the one reported by \cite{Ackermann15, Sandrinelli2018, Covino20, Yang2020, Penil20} ($\sim$800~d), but with a very low post-trial significance not reaching $1\sigma$. In order to check the consistency of our result with the literature, we made a test with a reduced light-curve (removing the last cycle). The result shows that although the significance rises a bit, it still remains below $3\sigma$ post-trial. The major effect must come from the trial correction, which indeed, is not done when analyzing a single source. And furthermore, when increasing the duration of the light curve, the number of trials also increases. Before trial correction, we find a significance not much different compared to other works in the literature.\\
     
    \item BL~Lac: \cite{Sandrinelli17, Sandrinelli2018} identified a candidate QPO with period of $\sim$680~d in BL~Lac's light curve. Nevertheless, our results show that this possible QPO is compatible with red noise, as was also found by \citet{Covino19, Penil20}.\\
    
    \item 3C~454.3: \cite{Sakar21} claimed a $>4\sigma$ QPO at period $\sim$47~d analysing the 1-day binned light curve of 3C~454.3, lasting from MJD~56800 to MJD~57250. We can observe in our CWT maps of this source that there is a strong vertical signal before MJD~56000 and no further significant feature appears afterwards. Thus, our CWT maps of 3C~454.3 show no evidence of such a QPO even for the 7-days binned light curve.\\
\end{itemize}

Finally, some other \gray\ AGN have been claimed to show QPOs with periods of the order of years, but were not included in our sample only because they did not pass our original cut on brightness. For completeness, we mention here the cases of PKS~0301-243 \citep{Zhang17-0301}, with a period of $2.1\pm0.3$~yr, PKS~0521-36 \citep{Zhang21-0521}, with a period of about $1.1$~yr, PKS~0601-70 \citep{Zhang20}, with a period of $1.22\pm0.06$~yr, and OJ~287 with a period of about $314$~d \citep{Kushawa20}. \\

We further remind the reader that our work is limited by two main choices: the target list, and the time binning. Extending the target list towards less bright objects is certainly possible, although with a major price: at some point the \fermilat\ light curves of these fainter AGNs will start to show non-detections in individual bins, and the presence of these zeros is problematic for the CWT technique that is sensitive to discontinuities. A simple solution will then be to also investigate longer time bins. Finer time bins could also be investigated for very bright \gray\ flares, giving access to QPO searches on time-scales of days. Both options have a price, that is an increase in the number of trials. As a final caveat, our results and in particular the determination of the significances depend on the Monte-Carlo simulations of artificial light curves. In our study we work under the assumption that the PSDs of the original light curves can be reconstructed by fitting them with a smoothly bended powerlaw function but this choice is far from unique. For a comprehensive study of Fermi-LAT PSD using different methods, see the recent work by \citet{Tarno20}.\\

\section{Conclusions}
\label{sec:conclusions}

The search for QPOs in the light curves of AGNs is a major research topic in astrophysics, providing us additional constraints on the physics of the SMBHs that power these sytems. Long-term QPOs, with periods of the order of months and beyond, are particularly difficult to identify due to the need of highly-sampled and unbiased light curves over long periods of time. The \fermilat\ \gray\ telescope, thanks to its monitoring capabilities is an ideal instrument to perform such a study.\\ 

We analysed 13-years long (from August 2008 to April 2021) \textit{Fermi}-LAT $\gamma$-ray light curve in two different time binnings (7 and 30 days) of 35 bright \gray\ AGNs. By using the CWT technique, we systematically searched for QPO candidates in this data set. In order to compute the confidence levels of the QPO candidates, 10000 simulated artificial light curves are generated for each light curve, and the histograms of global power spectrum at each period scale are fitted to a $\chi^2$ function. We correct for the trial effect in our analysis by estimating the trial number due to the number of sources and the number of time-period bins (in the wavelet power spectrum) or the number of period bins (in the global wavelet spectrum), outside the COI, and following the parameterization given by \cite{Auchere2016}. \\

In this way, 36 QPO candidates in 24 sources are identified (at various significance levels) with periods ranging from one month to several years. Our most significant, multi-year QPO candidate is in the blazar S5~1044+71, with a period of about $1100$~d.\\


We confirm some previously claimed $\gamma$-ray candidate QPOs in the sources PKS~0537-441, B2~1520+31, PKS 2247-131, S5~0716+714, Mrk~421, Mrk~501 and PKS~2155-304, while new possible QPO candidates are identified in 4C~+01.02, 4C~+28.07, NGC~1275, PKS~0402-362, PKS~0426-380, PKS~0447-439, PKS~0454-234, 1H~1013+498, S5~1044+71, 4C~+21.35, 3C~273, 3C~279, PKS~1424-418, PKS~1510-089, B2~1520+31, CTA~102, PKS~2247-131, 3C~454.3 and PMN~J2345-1555. \\

Possible physical origins of these quasi-periodical emissions are the precession of the AGN jet, its helical structure, and changes in the accretion flow. These scenarios can be due to the presence of a second SMBH in the system, opening up the window to study SMBH binaries. Since the orbital periods of the SMBH binaries are of the order of several years, sources showing long-term QPOs are naturally suspected as binary candidates. In particular, we put forward the blazar S5~1044+71 as the most promising SMBH candidate in our sample, due to its high significance QPO candidate with a period of about $1100$~d. Shorter, month-long QPOs, could also be related to close SMBH binaries, once the period is corrected for the Doppler factor of the jet. In our analysis we identify a peculiar behaviour in some QPO candidates, that might hint towards a QPO origin related to the jet geometry, that is a varying (slowing-down) QPO frequency, seen in B2 1520+31. We also put forward the possibility that some simultaneous QPO candidates, such as the ones seen in 4C~+01.02, are indeed harmonics, with a single physical mechanism at the origin of them.\\

With this, we can conclude that the CWT technique is a very powerful tool sensitive to any periodic oscillations in the light curves, considering an appropriate time binning. It has the major advantage over other statistical tools to be sensitive to transient QPO and period-shifting QPOs. However, it also reacts strongly to flares, resulting in vertical features in the map, and is influenced notably by border effects at large periods. This leads to the requirement of a visual inspection of all CWT maps to avoid misleading results. A major point to be highlighted is the trial-correction of significances. Due to the large number of scales probed, the number of sources, and the number of time binnings investigated, the number of trials is pretty large, and a careful correction has to be implemented to avoid false positives. While this correction is clearly mandatory for systematic studies as ours, it is necessary also for studies that analyze a single source without justifying how a particular target has been selected among the \fermilat\ catalog.\\

A natural perspective for a future study would be a multi-wavelength analysis of quasi-periodic emissions from the selected interesting $\gamma$-ray sources. First, the identification of QPOs at multiple wave bands with the same periodicity will automatically boost the significance of the detection, or, if the QPO is only seen in \grays, help identify false positives. Second, and more importantly, multi-wavelength observations will constrain the theoretical models that aim to explain these QPOs. Long-term, unbiased, multi-wavelength monitoring campaigns over several years, as complex and expensive as they might seem, are thus the key to identify AGN QPOs and understand their origin. \\

\begin{acknowledgements}
The authors wish to thank Jonathan Biteau and Pablo Peñil for fruitful inputs and discussions, and Josep Maria Paredes and Bruno Khélifi for comments on the early draft. This work has been possible thanks to the open source code for the simulation of artificial light curves provided by \cite{Emmanoulopoulos13}, its python version provided by \cite{Connolly15} and the open source Python library \texttt{PyCWT} elaborated by \cite{Torrence1998}. M.C. has received financial support through the Postdoctoral Junior Leader Fellowship Programme from la Caixa Banking Foundation, grant No. LCF/BQ/LI18/11630012. N.S. acknowledges the support by the Science Committee of RA, in the frames of the research projects No 20TTCG-1C015 and 21T-1C260.
\end{acknowledgements}

\bibliography{references}{}

\begin{thebibliography}{48}
\expandafter\ifx\csname natexlab\endcsname\relax\def\natexlab#1{#1}\fi

\bibitem[{{Abdollahi} {et~al.}(2020){Abdollahi}, {Acero}, {Ackermann},
  {Ajello}, {Atwood}, {Axelsson}, {Baldini}, {Ballet}, {Barbiellini},
  {Bastieri}, {Becerra Gonzalez}, {Bellazzini}, {Berretta}, {Bissaldi},
  {Blandford}, {Bloom}, {Bonino}, {Bottacini}, {Brandt}, {Bregeon}, {Bruel},
  {Buehler}, {Burnett}, {Buson}, {Cameron}, {Caputo}, {Caraveo}, {Casandjian},
  {Castro}, {Cavazzuti}, {Charles}, {Chaty}, {Chen}, {Cheung}, {Chiaro},
  {Ciprini}, {Cohen-Tanugi}, {Cominsky}, {Coronado-Bl{\'a}zquez}, {Costantin},
  {Cuoco}, {Cutini}, {D'Ammando}, {DeKlotz}, {de la Torre Luque}, {de Palma},
  {Desai}, {Digel}, {Di Lalla}, {Di Mauro}, {Di Venere}, {Dom{\'\i}nguez},
  {Dumora}, {Fana Dirirsa}, {Fegan}, {Ferrara}, {Franckowiak}, {Fukazawa},
  {Funk}, {Fusco}, {Gargano}, {Gasparrini}, {Giglietto}, {Giommi}, {Giordano},
  {Giroletti}, {Glanzman}, {Green}, {Grenier}, {Griffin}, {Grondin}, {Grove},
  {Guiriec}, {Harding}, {Hayashi}, {Hays}, {Hewitt}, {Horan},
  {J{\'o}hannesson}, {Johnson}, {Kamae}, {Kerr}, {Kocevski}, {Kovac'evic'},
  {Kuss}, {Landriu}, {Larsson}, {Latronico}, {Lemoine-Goumard}, {Li},
  {Liodakis}, {Longo}, {Loparco}, {Lott}, {Lovellette}, {Lubrano}, {Madejski},
  {Maldera}, {Malyshev}, {Manfreda}, {Marchesini}, {Marcotulli},
  {Mart{\'\i}-Devesa}, {Martin}, {Massaro}, {Mazziotta}, {McEnery}, {Mereu},
  {Meyer}, {Michelson}, {Mirabal}, {Mizuno}, {Monzani}, {Morselli},
  {Moskalenko}, {Negro}, {Nuss}, {Ojha}, {Omodei}, {Orienti}, {Orlando},
  {Ormes}, {Palatiello}, {Paliya}, {Paneque}, {Pei}, {Pe{\~n}a-Herazo},
  {Perkins}, {Persic}, {Pesce-Rollins}, {Petrosian}, {Petrov}, {Piron}, {Poon},
  {Porter}, {Principe}, {Rain{\`o}}, {Rando}, {Razzano}, {Razzaque}, {Reimer},
  {Reimer}, {Remy}, {Reposeur}, {Romani}, {Saz Parkinson}, {Schinzel},
  {Serini}, {Sgr{\`o}}, {Siskind}, {Smith}, {Spandre}, {Spinelli}, {Strong},
  {Suson}, {Tajima}, {Takahashi}, {Tak}, {Thayer}, {Thompson}, {Tibaldo},
  {Torres}, {Torresi}, {Valverde}, {Van Klaveren}, {van Zyl}, {Wood},
  {Yassine}, \& {Zaharijas}}]{2020ApJS..247...33A}
{Abdollahi}, S., {Acero}, F., {Ackermann}, M., {et~al.} 2020, \apjs, 247, 33

\bibitem[{{Ackermann} {et~al.}(2015){Ackermann}, {Ajello}, {Albert}, {Atwood},
  {Baldini}, {Ballet}, {Barbiellini}, {Bastieri}, {Becerra Gonzalez},
  {Bellazzini}, {Bissaldi}, {Blandford}, {Bloom}, {Bonino}, {Bottacini},
  {Bregeon}, {Bruel}, {Buehler}, {Buson}, {Caliandro}, {Cameron}, {Caputo},
  {Caragiulo}, {Caraveo}, {Cavazzuti}, {Cecchi}, {Chekhtman}, {Chiang},
  {Chiaro}, {Ciprini}, {Cohen-Tanugi}, {Conrad}, {Cutini}, {D'Ammando}, {de
  Angelis}, {de Palma}, {Desiante}, {Di Venere}, {Dom{\'\i}nguez}, {Drell},
  {Favuzzi}, {Fegan}, {Ferrara}, {Focke}, {Fuhrmann}, {Fukazawa}, {Fusco},
  {Gargano}, {Gasparrini}, {Giglietto}, {Giommi}, {Giordano}, {Giroletti},
  {Godfrey}, {Green}, {Grenier}, {Grove}, {Guiriec}, {Harding}, {Hays},
  {Hewitt}, {Hill}, {Horan}, {Jogler}, {J{\'o}hannesson}, {Johnson}, {Kamae},
  {Kuss}, {Larsson}, {Latronico}, {Li}, {Li}, {Longo}, {Loparco}, {Lott},
  {Lovellette}, {Lubrano}, {Magill}, {Maldera}, {Manfreda}, {Max-Moerbeck},
  {Mayer}, {Mazziotta}, {McEnery}, {Michelson}, {Mizuno}, {Monzani},
  {Morselli}, {Moskalenko}, {Murgia}, {Nuss}, {Ohno}, {Ohsugi}, {Ojha},
  {Omodei}, {Orlando}, {Ormes}, {Paneque}, {Pearson}, {Perkins}, {Perri},
  {Pesce-Rollins}, {Petrosian}, {Piron}, {Pivato}, {Porter}, {Rain{\`o}},
  {Rando}, {Razzano}, {Readhead}, {Reimer}, {Reimer}, {Schulz}, {Sgr{\`o}},
  {Siskind}, {Spada}, {Spandre}, {Spinelli}, {Suson}, {Takahashi}, {Thayer},
  {Thompson}, {Tibaldo}, {Torres}, {Tosti}, {Troja}, {Uchiyama}, {Vianello},
  {Wood}, {Wood}, {Zimmer}, {Berdyugin}, {Corbet}, {Hovatta}, {Lindfors},
  {Nilsson}, {Reinthal}, {Sillanp{\"a}{\"a}}, {Stamerra}, {Takalo}, \&
  {Valtonen}}]{Ackermann15}
{Ackermann}, M., {Ajello}, M., {Albert}, A., {et~al.} 2015, \apjl, 813, L41

\bibitem[{{Ait Benkhali} {et~al.}(2020){Ait Benkhali}, {Hofmann}, {Rieger}, \&
  {Chakraborty}}]{AitBenkhali20}
{Ait Benkhali}, F., {Hofmann}, W., {Rieger}, F.~M., \& {Chakraborty}, N. 2020,
  \aap, 634, A120

\bibitem[{{Ajello} {et~al.}(2020){Ajello}, {Angioni}, {Axelsson}, {Ballet},
  {Barbiellini}, {Bastieri}, {Becerra Gonzalez}, {Bellazzini}, {Bissaldi},
  {Bloom}, {Bonino}, {Bottacini}, {Bruel}, {Buson}, {Cafardo}, {Cameron},
  {Cavazzuti}, {Chen}, {Cheung}, {Ciprini}, {Costantin}, {Cutini}, {D'Ammando},
  {de la Torre Luque}, {de Menezes}, {de Palma}, {Desai}, {Di Lalla}, {Di
  Venere}, {Dom{\'\i}nguez}, {Dirirsa}, {Ferrara}, {Finke}, {Franckowiak},
  {Fukazawa}, {Funk}, {Fusco}, {Gargano}, {Garrappa}, {Gasparrini},
  {Giglietto}, {Giordano}, {Giroletti}, {Green}, {Grenier}, {Guiriec},
  {Harita}, {Hays}, {Horan}, {Itoh}, {J{\'o}hannesson}, {Kovac'evic'},
  {Krauss}, {Kreter}, {Kuss}, {Larsson}, {Leto}, {Li}, {Liodakis}, {Longo},
  {Loparco}, {Lott}, {Lovellette}, {Lubrano}, {Madejski}, {Maldera},
  {Manfreda}, {Mart{\'\i}-Devesa}, {Massaro}, {Mazziotta}, {Mereu}, {Meyer},
  {Migliori}, {Mirabal}, {Mizuno}, {Monzani}, {Morselli}, {Moskalenko},
  {Negro}, {Nemmen}, {Nuss}, {Ojha}, {Ojha}, {Omodei}, {Orienti}, {Orlando},
  {Ormes}, {Paliya}, {Pei}, {Pe{\~n}a-Herazo}, {Persic}, {Pesce-Rollins},
  {Petrov}, {Piron}, {Poon}, {Principe}, {Rain{\`o}}, {Rando}, {Rani},
  {Razzano}, {Razzaque}, {Reimer}, {Reimer}, {Schinzel}, {Serini}, {Sgr{\`o}},
  {Siskind}, {Spandre}, {Spinelli}, {Suson}, {Tachibana}, {Thompson}, {Torres},
  {Torresi}, {Troja}, {Valverde}, {van Zyl}, \& {Yassine}}]{4LAC}
{Ajello}, M., {Angioni}, R., {Axelsson}, M., {et~al.} 2020, \apj, 892, 105

\bibitem[{{Alston} {et~al.}(2015){Alston}, {Parker},
  {Markevi{\v{c}}i{\={u}}t{\.{e}}}, {Fabian}, {Middleton}, {Lohfink}, {Kara},
  \& {Pinto}}]{Alston15}
{Alston}, W.~N., {Parker}, M.~L., {Markevi{\v{c}}i{\={u}}t{\.{e}}}, J.,
  {et~al.} 2015, \mnras, 449, 467

\bibitem[{{Atwood} {et~al.}(2009){Atwood}, {Abdo}, {Ackermann}, {Althouse},
  {Anderson}, {Axelsson}, {Baldini}, {Ballet}, {Band}, {Barbiellini},
  {Bartelt}, {Bastieri}, {Baughman}, {Bechtol}, {B{\'e}d{\'e}r{\`e}de},
  {Bellardi}, {Bellazzini}, {Berenji}, {Bignami}, {Bisello}, {Bissaldi},
  {Blandford}, {Bloom}, {Bogart}, {Bonamente}, {Bonnell}, {Borgland},
  {Bouvier}, {Bregeon}, {Brez}, {Brigida}, {Bruel}, {Burnett}, {Busetto},
  {Caliandro}, {Cameron}, {Caraveo}, {Carius}, {Carlson}, {Casandjian},
  {Cavazzuti}, {Ceccanti}, {Cecchi}, {Charles}, {Chekhtman}, {Cheung},
  {Chiang}, {Chipaux}, {Cillis}, {Ciprini}, {Claus}, {Cohen-Tanugi},
  {Condamoor}, {Conrad}, {Corbet}, {Corucci}, {Costamante}, {Cutini}, {Davis},
  {Decotigny}, {DeKlotz}, {Dermer}, {de Angelis}, {Digel}, {do Couto e Silva},
  {Drell}, {Dubois}, {Dumora}, {Edmonds}, {Fabiani}, {Farnier}, {Favuzzi},
  {Flath}, {Fleury}, {Focke}, {Funk}, {Fusco}, {Gargano}, {Gasparrini},
  {Gehrels}, {Gentit}, {Germani}, {Giebels}, {Giglietto}, {Giommi}, {Giordano},
  {Glanzman}, {Godfrey}, {Grenier}, {Grondin}, {Grove}, {Guillemot}, {Guiriec},
  {Haller}, {Harding}, {Hart}, {Hays}, {Healey}, {Hirayama}, {Hjalmarsdotter},
  {Horn}, {Hughes}, {J{\'o}hannesson}, {Johansson}, {Johnson}, {Johnson},
  {Johnson}, {Johnson}, {Kamae}, {Katagiri}, {Kataoka}, {Kavelaars}, {Kawai},
  {Kelly}, {Kerr}, {Klamra}, {Kn{\"o}dlseder}, {Kocian}, {Komin}, {Kuehn},
  {Kuss}, {Landriu}, {Latronico}, {Lee}, {Lee}, {Lemoine-Goumard}, {Lionetto},
  {Longo}, {Loparco}, {Lott}, {Lovellette}, {Lubrano}, {Madejski}, {Makeev},
  {Marangelli}, {Massai}, {Mazziotta}, {McEnery}, {Menon}, {Meurer},
  {Michelson}, {Minuti}, {Mirizzi}, {Mitthumsiri}, {Mizuno}, {Moiseev},
  {Monte}, {Monzani}, {Moretti}, {Morselli}, {Moskalenko}, {Murgia},
  {Nakamori}, {Nishino}, {Nolan}, {Norris}, {Nuss}, {Ohno}, {Ohsugi}, {Omodei},
  {Orlando}, {Ormes}, {Paccagnella}, {Paneque}, {Panetta}, {Parent}, {Pearce},
  {Pepe}, {Perazzo}, {Pesce-Rollins}, {Picozza}, {Pieri}, {Pinchera}, {Piron},
  {Porter}, {Poupard}, {Rain{\`o}}, {Rando}, {Rapposelli}, {Razzano}, {Reimer},
  {Reimer}, {Reposeur}, {Reyes}, {Ritz}, {Rochester}, {Rodriguez}, {Romani},
  {Roth}, {Russell}, {Ryde}, {Sabatini}, {Sadrozinski}, {Sanchez}, {Sander},
  {Sapozhnikov}, {Parkinson}, {Scargle}, {Schalk}, {Scolieri}, {Sgr{\`o}},
  {Share}, {Shaw}, {Shimokawabe}, {Shrader}, {Sierpowska-Bartosik}, {Siskind},
  {Smith}, {Smith}, {Spandre}, {Spinelli}, {Starck}, {Stephens}, {Strickman},
  {Strong}, {Suson}, {Tajima}, {Takahashi}, {Takahashi}, {Tanaka}, {Tenze},
  {Tether}, {Thayer}, {Thayer}, {Thompson}, {Tibaldo}, {Tibolla}, {Torres},
  {Tosti}, {Tramacere}, {Turri}, {Usher}, {Vilchez}, {Vitale}, {Wang},
  {Watters}, {Winer}, {Wood}, {Ylinen}, \& {Ziegler}}]{2009ApJ...697.1071A}
{Atwood}, W.~B., {Abdo}, A.~A., {Ackermann}, M., {et~al.} 2009, \apj, 697, 1071

\bibitem[{{Auch{\`e}re} {et~al.}(2016){Auch{\`e}re}, {Froment}, {Bocchialini},
  {Buchlin}, \& {Solomon}}]{Auchere2016}
{Auch{\`e}re}, F., {Froment}, C., {Bocchialini}, K., {Buchlin}, E., \&
  {Solomon}, J. 2016, \apj, 825, 110

\bibitem[{{Begelman} {et~al.}(1980){Begelman}, {Blandford}, \&
  {Rees}}]{Begelman80}
{Begelman}, M.~C., {Blandford}, R.~D., \& {Rees}, M.~J. 1980, \nat, 287, 307

\bibitem[{{Bhatta}(2019)}]{Bhatta19}
{Bhatta}, G. 2019, \mnras, 487, 3990

\bibitem[{{Bhatta} \& {Dhital}(2020)}]{Bhatta2020}
{Bhatta}, G. \& {Dhital}, N. 2020, \apj, 891, 120

\bibitem[{{Blandford} {et~al.}(2019){Blandford}, {Meier}, \&
  {Readhead}}]{Blandfordreview}
{Blandford}, R., {Meier}, D., \& {Readhead}, A. 2019, \araa, 57, 467

\bibitem[{{Cerruti}(2020)}]{Cerrutireview}
{Cerruti}, M. 2020, Galaxies, 8, 72

\bibitem[{{Connolly}(2015)}]{Connolly15}
{Connolly}, S.~D. 2015, arXiv e-prints, arXiv:1503.06676

\bibitem[{{Covino} {et~al.}(2020){Covino}, {Landoni}, {Sandrinelli}, \&
  {Treves}}]{Covino20}
{Covino}, S., {Landoni}, M., {Sandrinelli}, A., \& {Treves}, A. 2020, \apj,
  895, 122

\bibitem[{{Covino} {et~al.}(2019){Covino}, {Sandrinelli}, \&
  {Treves}}]{Covino19}
{Covino}, S., {Sandrinelli}, A., \& {Treves}, A. 2019, \mnras, 482, 1270

\bibitem[{{Emmanoulopoulos} {et~al.}(2013){Emmanoulopoulos}, {McHardy}, \&
  {Papadakis}}]{Emmanoulopoulos13}
{Emmanoulopoulos}, D., {McHardy}, I.~M., \& {Papadakis}, I.~E. 2013, \mnras,
  433, 907

\bibitem[{{Gierli{\'n}ski} {et~al.}(2008){Gierli{\'n}ski}, {Middleton}, {Ward},
  \& {Done}}]{Gierlinski08}
{Gierli{\'n}ski}, M., {Middleton}, M., {Ward}, M., \& {Done}, C. 2008, \nat,
  455, 369

\bibitem[{{Gupta} {et~al.}(2019){Gupta}, {Tripathi}, {Wiita}, {Kushwaha},
  {Zhang}, \& {Bambi}}]{Gupta19}
{Gupta}, A.~C., {Tripathi}, A., {Wiita}, P.~J., {et~al.} 2019, \mnras, 484,
  5785

\bibitem[{{Honma} {et~al.}(1992){Honma}, {Matsumoto}, \& {Kato}}]{Honma1992}
{Honma}, F., {Matsumoto}, R., \& {Kato}, S. 1992, \pasj, 44, 529

\bibitem[{{Ingram} \& {Motta}(2019)}]{Ingramreview}
{Ingram}, A.~R. \& {Motta}, S.~E. 2019, \nar, 85, 101524

\bibitem[{{Kushwaha} {et~al.}(2020){Kushwaha}, {Sarkar}, {Gupta}, {Tripathi},
  \& {Wiita}}]{Kushawa20}
{Kushwaha}, P., {Sarkar}, A., {Gupta}, A.~C., {Tripathi}, A., \& {Wiita}, P.~J.
  2020, \mnras, 499, 653

\bibitem[{{Li} {et~al.}(2018){Li}, {Jiang}, {Yi}, {Guo}, {Chen}, {Zhang},
  {Gao}, {Lu}, \& {Ren}}]{Li18}
{Li}, H.~Z., {Jiang}, Y.~G., {Yi}, T.~F., {et~al.} 2018, \apss, 363, 45

\bibitem[{{McKinney} {et~al.}(2012){McKinney}, {Tchekhovskoy}, \&
  {Blandford}}]{McKinney2012}
{McKinney}, J.~C., {Tchekhovskoy}, A., \& {Blandford}, R.~D. 2012, \mnras, 423,
  3083

\bibitem[{{Pe{\~n}il} {et~al.}(2020){Pe{\~n}il}, {Dom{\'\i}nguez}, {Buson},
  {Ajello}, {Otero-Santos}, {Barrio}, {Nemmen}, {Cutini}, {Rani},
  {Franckowiak}, \& {Cavazzuti}}]{Penil20}
{Pe{\~n}il}, P., {Dom{\'\i}nguez}, A., {Buson}, S., {et~al.} 2020, \apj, 896,
  134

\bibitem[{Percival(1995)}]{Percival1995}
Percival, D.~P. 1995, Biometrika, 82, 619

\bibitem[{{Prokhorov} \& {Moraghan}(2017)}]{Prokhorov17}
{Prokhorov}, D.~A. \& {Moraghan}, A. 2017, \mnras, 471, 3036

\bibitem[{{Rieger}(2004)}]{Rieger04}
{Rieger}, F.~M. 2004, \apjl, 615, L5

\bibitem[{{Sandrinelli} {et~al.}(2016){Sandrinelli}, {Covino}, \&
  {Treves}}]{Sandrinelli2016}
{Sandrinelli}, A., {Covino}, S., \& {Treves}, A. 2016, \apj, 820, 20

\bibitem[{{Sandrinelli} {et~al.}(2017){Sandrinelli}, {Covino}, {Treves},
  {Lindfors}, {Raiteri}, {Nilsson}, {Takalo}, {Reinthal}, {Berdyugin}, {Fallah
  Ramazani}, {Kadenius}, {Tuominen}, {Kehusmaa}, {Bachev}, \&
  {Strigachev}}]{Sandrinelli17}
{Sandrinelli}, A., {Covino}, S., {Treves}, A., {et~al.} 2017, \aap, 600, A132

\bibitem[{{Sandrinelli} {et~al.}(2018){Sandrinelli}, {Covino}, {Treves},
  {et~al.}}]{Sandrinelli2018}
{Sandrinelli}, A., {Covino}, S., {Treves}, A., {et~al.} 2018, \aap, 615, A118

\bibitem[{{Sarkar} {et~al.}(2021){Sarkar}, {Gupta}, {Chitnis}, \&
  {Wiita}}]{Sakar21}
{Sarkar}, A., {Gupta}, A.~C., {Chitnis}, V.~R., \& {Wiita}, P.~J. 2021, \mnras,
  501, 50

\bibitem[{{Sillanpaa} {et~al.}(1988){Sillanpaa}, {Haarala}, {Valtonen},
  {Sundelius}, \& {Byrd}}]{Silla88}
{Sillanpaa}, A., {Haarala}, S., {Valtonen}, M.~J., {Sundelius}, B., \& {Byrd},
  G.~G. 1988, \apj, 325, 628

\bibitem[{{Sobacchi} {et~al.}(2017){Sobacchi}, {Sormani}, \&
  {Stamerra}}]{Sobacchi17}
{Sobacchi}, E., {Sormani}, M.~C., \& {Stamerra}, A. 2017, \mnras, 465, 161

\bibitem[{{Tarnopolski} {et~al.}(2020){Tarnopolski}, {{\.Z}ywucka},
  {Marchenko}, \& {Pascual-Granado}}]{Tarno20}
{Tarnopolski}, M., {{\.Z}ywucka}, N., {Marchenko}, V., \& {Pascual-Granado}, J.
  2020, \apjs, 250, 1

\bibitem[{{Tavani} {et~al.}(2018){Tavani}, {Cavaliere}, {Munar-Adrover}, \&
  {Argan}}]{Tavani18}
{Tavani}, M., {Cavaliere}, A., {Munar-Adrover}, P., \& {Argan}, A. 2018, \apj,
  854, 11

\bibitem[{{Timmer} \& {Koenig}(1995)}]{Timmer1995}
{Timmer}, J. \& {Koenig}, M. 1995, Astronomy and Astrophysics, 300, 707

\bibitem[{Torrence \& Compo(1998)}]{Torrence1998}
Torrence, C. \& Compo, G.~P. 1998, Bulletin of the American Meteorological
  Society, 79, 61

\bibitem[{{Urry} \& {Padovani}(1995)}]{Urry95}
{Urry}, C.~M. \& {Padovani}, P. 1995, \pasp, 107, 803

\bibitem[{{Valtonen} {et~al.}(2009){Valtonen}, {Nilsson}, {Villforth}, {Lehto},
  {Takalo}, {Lindfors}, {Sillanp{\"a}{\"a}}, {Hentunen}, {Mikkola}, {Zola},
  {Drozdz}, {Koziel}, {Ogloza}, {Kurpinska-Winiarska}, {Siwak}, {Winiarski},
  {Heidt}, {Kidger}, {Pursimo}, {Wu}, {Zhou}, {Sadakane}, {Marchev},
  {Nissinen}, {Niarchos}, {Liakos}, {Gazeas}, {Dogru}, {Poyner}, {Dietrich},
  {Assef}, {Atlee}, {Bird}, {DePoy}, {Eastman}, {Peeples}, {Prieto}, {Watson},
  {Yee}, {Mattingly}, \& {Ohlert}}]{Valtonen09}
{Valtonen}, M.~J., {Nilsson}, K., {Villforth}, C., {et~al.} 2009, \apj, 698,
  781

\bibitem[{{Wang} {et~al.}(2022){Wang}, {Cai}, \& {Fan}}]{Wang22}
{Wang}, G.~G., {Cai}, J.~T., \& {Fan}, J.~H. 2022, arXiv e-prints,
  arXiv:2203.02168

\bibitem[{{Yang} {et~al.}(2021){Yang}, {Cao}, {Zhou}, \& {Qin}}]{Yang21-1424}
{Yang}, J., {Cao}, G., {Zhou}, B., \& {Qin}, L. 2021, \pasp, 133, 024101

\bibitem[{{Yang} {et~al.}(2020){Yang}, {Yan}, {Zhang}, {et~al.}}]{Yang2020}
{Yang}, S., {Yan}, D., {Zhang}, P., {et~al.} 2020, arXiv e-prints,
  arXiv:2011.10186

\bibitem[{{Zhang} {et~al.}(2021){Zhang}, {Yan}, {Zhang}, {Yang}, \&
  {Zhang}}]{Zhang21-0521}
{Zhang}, H., {Yan}, D., {Zhang}, P., {Yang}, S., \& {Zhang}, L. 2021, \apj,
  919, 58

\bibitem[{{Zhang} {et~al.}(2017{\natexlab{a}}){Zhang}, {Yan}, {Liao}, \&
  {Wang}}]{Zhang17-2155}
{Zhang}, P.-F., {Yan}, D.-H., {Liao}, N.-H., \& {Wang}, J.-C.
  2017{\natexlab{a}}, \apj, 835, 260

\bibitem[{{Zhang} {et~al.}(2017{\natexlab{b}}){Zhang}, {Yan}, {Zhou}, {Fan},
  {Wang}, \& {Zhang}}]{Zhang17-0301}
{Zhang}, P.-F., {Yan}, D.-H., {Zhou}, J.-N., {et~al.} 2017{\natexlab{b}}, \apj,
  845, 82

\bibitem[{{Zhang} {et~al.}(2020){Zhang}, {Yan}, {Zhou}, {Wang}, \&
  {Zhang}}]{Zhang20}
{Zhang}, P.-F., {Yan}, D.-H., {Zhou}, J.-N., {Wang}, J.-C., \& {Zhang}, L.
  2020, \apj, 891, 163

\bibitem[{{Zhou} {et~al.}(2018){Zhou}, {Wang}, {Chen}, {et~al.}}]{Zhou2018}
{Zhou}, J., {Wang}, Z., {Chen}, L., {et~al.} 2018, Nature Communications, 9,
  4599

\bibitem[{{{\.Z}ywucka} {et~al.}(2021){{\.Z}ywucka}, {Tarnopolski},
  {Marchenko}, \& {Pascual-Granado}}]{Zywucka2021}
{{\.Z}ywucka}, N., {Tarnopolski}, M., {Marchenko}, V., \& {Pascual-Granado}, J.
  2021, arXiv e-prints, arXiv:2112.01761

\end{thebibliography}
\bibliographystyle{aa}

\appendix

\section{Wavelet Analysis Method}
\label{sec:AWavelet}
\begin{figure}[b!]
	\centering
	\includegraphics[width=0.45\textwidth]{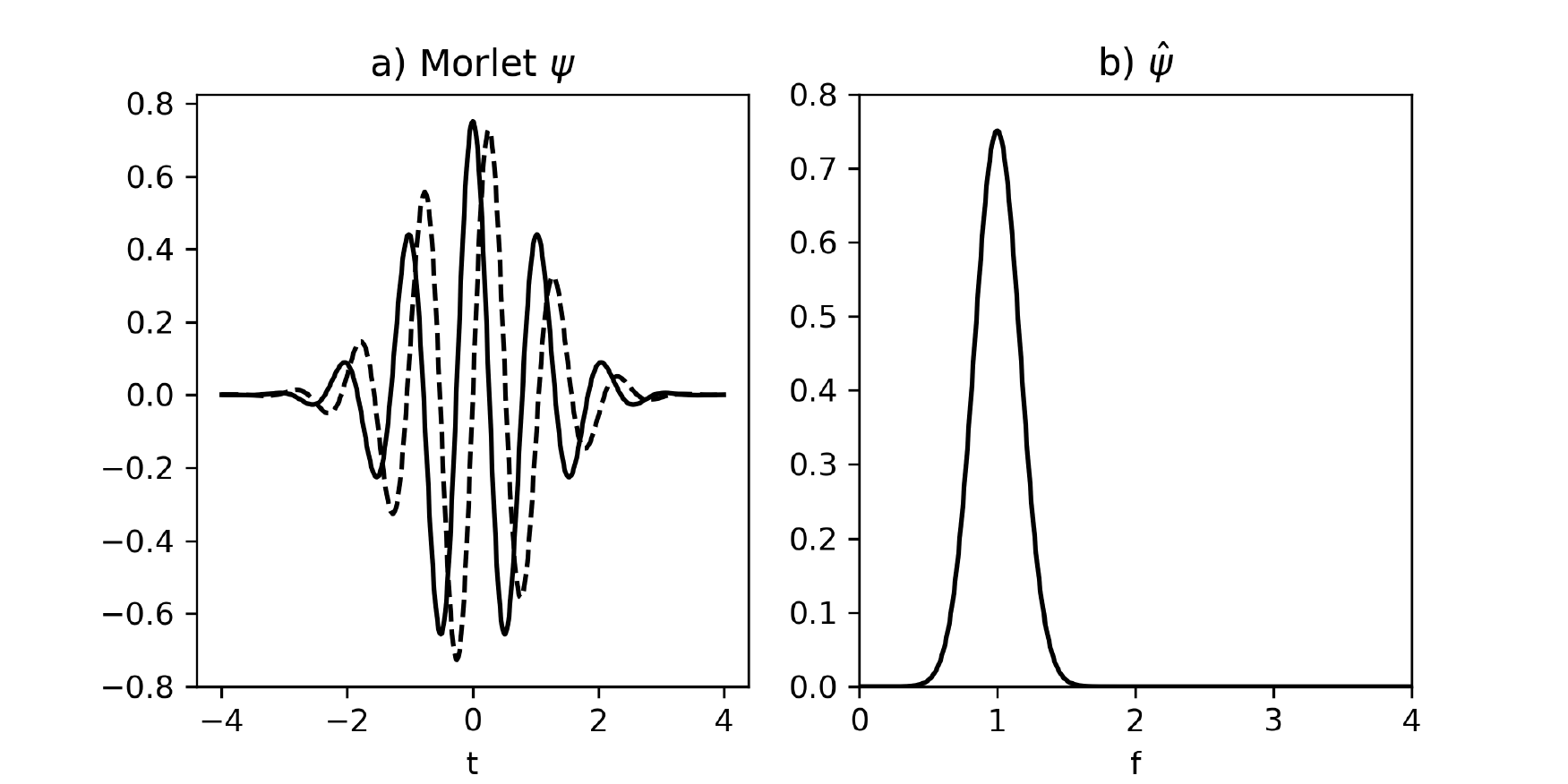}
	\caption{ a) The real part of the Morlet wavelet in solid line and the imaginary part in dashed line. b) Fourier transform of
    the Morlet wavelet.}
	\label{fig:Morlet}
\end{figure}
\vspace*{0.5cm}
\noindent

\begin{figure}[b!]
	\centering
	\includegraphics[width=0.45\textwidth]{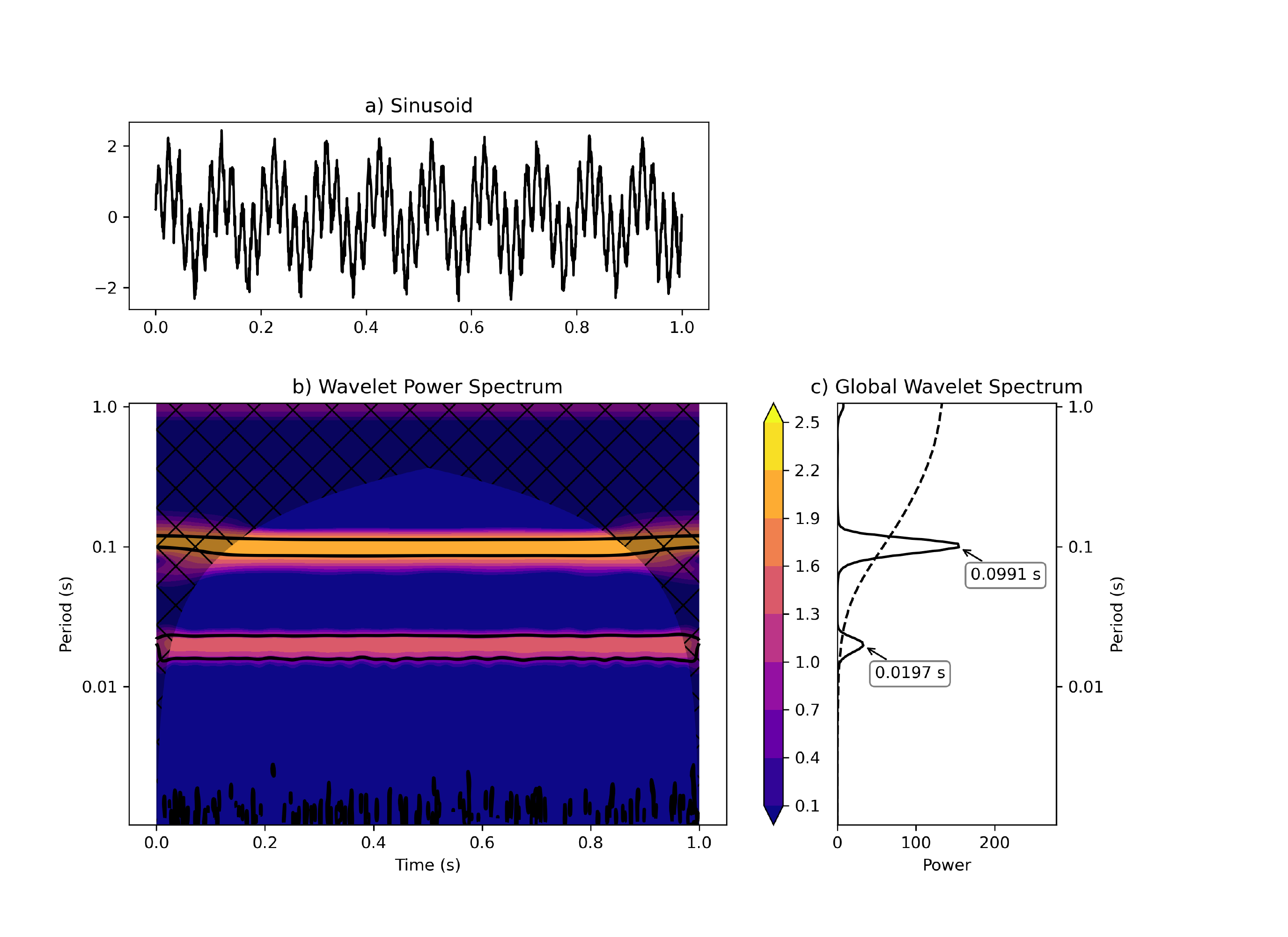}
	\caption{ a) A periodic function with two periods $p_1$ = 0.02 $\mathrm{s}$ and $p_2$ = 0.10 $\mathrm{s}$. b) The corresponding wavelet power spectrum map obtained using \texttt{PyCWT}. The dashed area indicate the COI. And c) the global power spectrum in solid line, and the significance level at 95\% provided by \texttt{PyCWT} in dashed line.}
	\label{fig:Sinusoid}
\end{figure}
\vspace*{0.5cm}
\noindent

\begin{figure}[t!]
	\centering
	\includegraphics[width=0.45\textwidth]{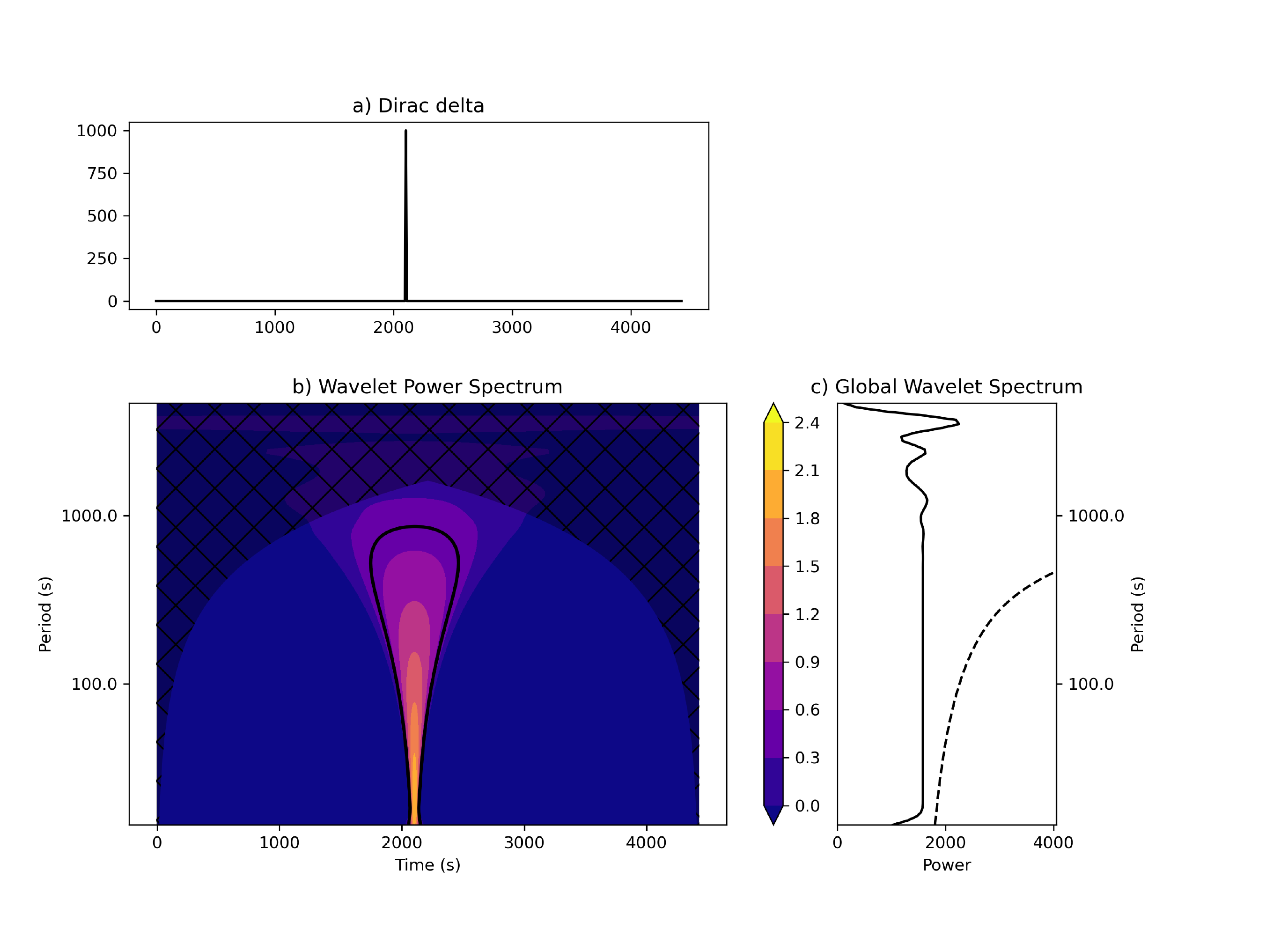}
	\caption{ a) A Dirac delta function. b) The corresponding wavelet power spectrum map obtained using \texttt{PyCWT}. The dashed area indicate the COI. And c) the global power spectrum in solid line, and the significance level at 95\% provided by \texttt{PyCWT} in dashed line.}
	\label{fig:Dirac}
\end{figure}
\vspace*{0.5cm}
\noindent

Fig.~\ref{fig:Morlet} shows the Morlet wavelet used as the mother wavelet in our calculations, and on the right is the Fourier transform of the wavelet. \\

Fig.~\ref{fig:Sinusoid} shows the CWT map and global wavelet spectrum of a periodic function with two periods, 0.1~s and 0.02~s. A random normally distributed error of $\sim$10\% to the amplitude of the function is added, to simulate the error in the light curve data. The maximum values of the global wavelet have a relative error of $\sim$1\% to $\sim$2\%. On the other hand, the broadness of the humps are much larger than the relative error on the maximum, being $\sim$20\% to $\sim$40\% of the real value. Thus, the main contribution to the period uncertainty is the intrinsic broadness of the features. \\

The dashed line in Fig.~\ref{fig:Sinusoid}c) is the 95\% significance level provided by \texttt{PyCWT} which in this work is not used for the light curves of \textit{Fermi}-LAT sources. Instead, we perform a Monte Carlo approach in order to take into account the statistical and variability properties of each light curve, as explained in sections \autoref{sec:AStatistical} and \autoref{sec:ASignificance}. The cross-hatched area in the CWT map shown in Fig.~\ref{fig:Sinusoid}b) corresponds to the cone of influence (COI). The importance of the COI is evident in this figure, where we can visualize the spurious signal at the upper edge, and the first periodicity is clearly distorted when it is within the COI. \\

A second example is shown in the Fig.~\ref{fig:Dirac}, where in this case we can see how the CWT behaves when the input is a Dirac delta function. The vertical shape of the wavelet power spectrum in Fig.~\ref{fig:Dirac}b), with increasing time interval as the scale increases, reveals the response of CWT to flare shaped signals. \\


\section{Statistical and Variability Properties of Artificial Light Curves}
\label{sec:AStatistical}

\begin{figure}[!htbp]
	\centering
	\includegraphics[width=0.45\textwidth]{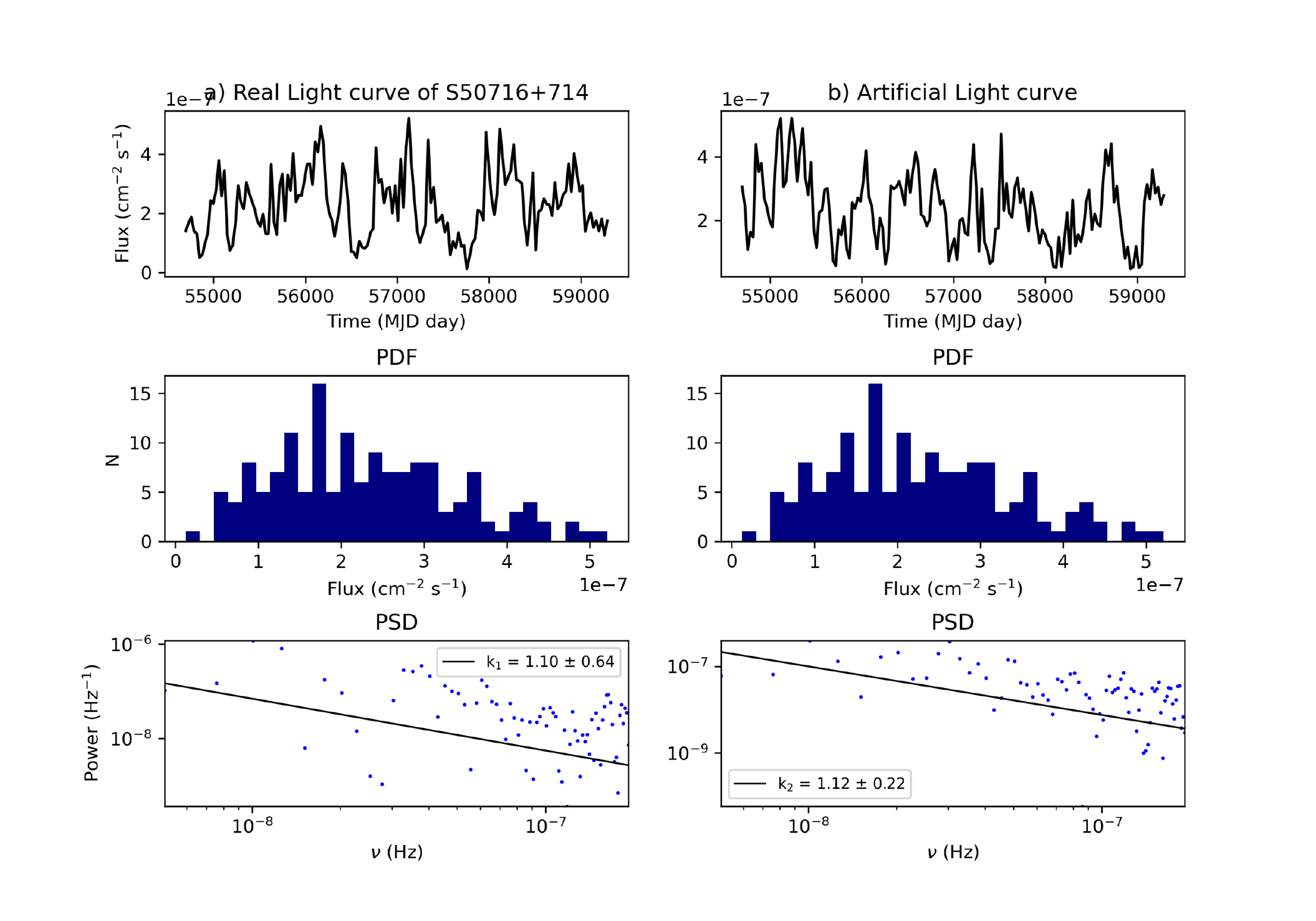}
	\includegraphics[width=0.45\textwidth]{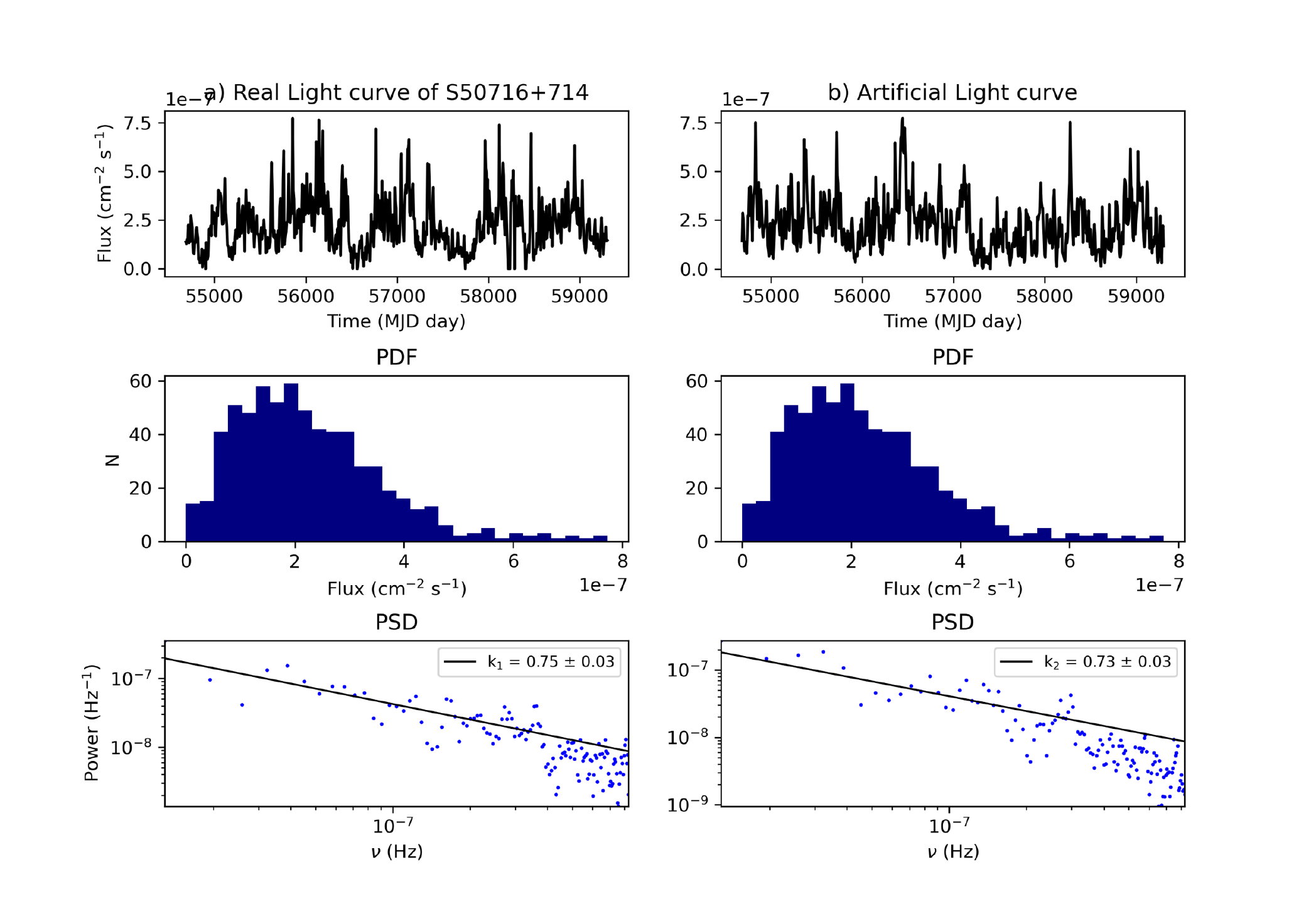}
	\caption{Comparison of the PDF and the PSD of the original light curve of the source S5~0716+714 and the light curve simulated by \cite{Connolly15}, in time bins of 30 (upper) and 7 days (bottom). The solid line in the PSD is a simple power-law fit for a visual check. The actual fitted PSD is a smoothly bending power-law model plus a constant \citep[see equation 2 in][]{Emmanoulopoulos13}.}
	\label{fig:Statistical}
\end{figure}

Artificial light curves computed using the Python code provided by \citet{Connolly15}, following the work of \citet{Emmanoulopoulos13} share both the PDF and the PSD of the original light curve, as can be seen in the Fig.~\ref{fig:Statistical}.

\vspace*{0.5cm}
\noindent

\section{Significance Levels Estimation}
\label{sec:ASignificance}

\begin{figure}[!htbp]
	\centering
	\includegraphics[width=0.45\textwidth]{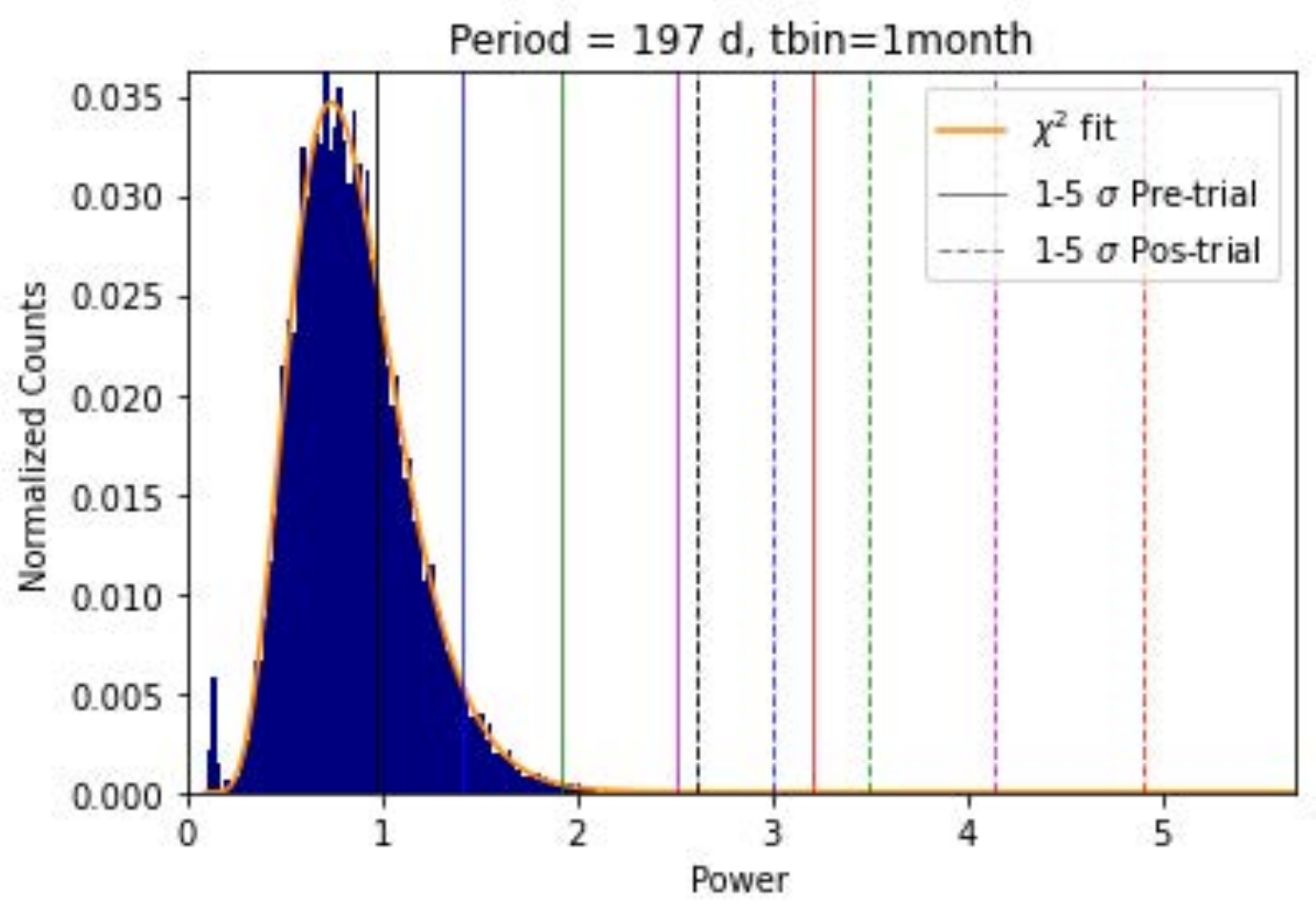}
	\includegraphics[width=0.45\textwidth]{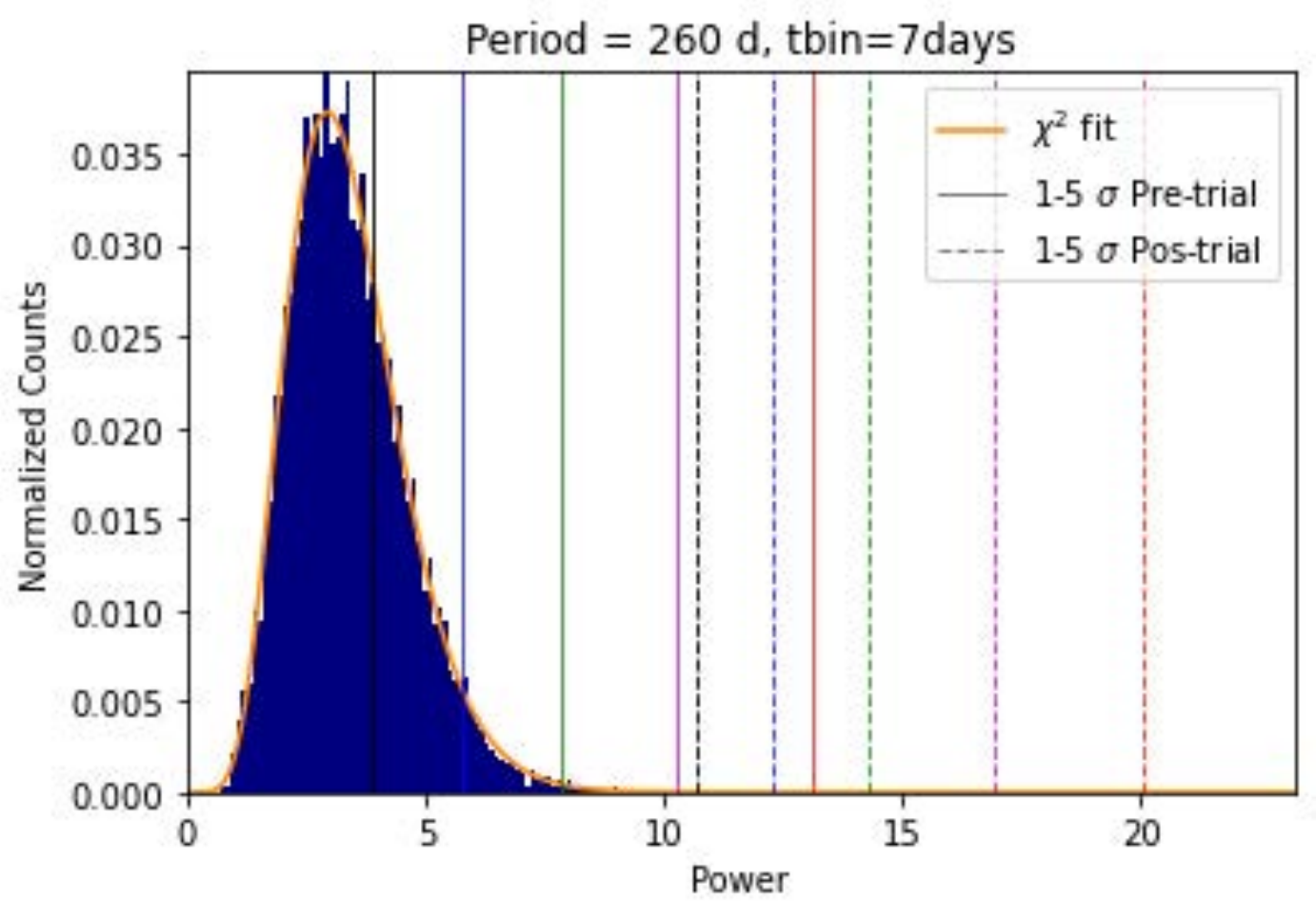}
	\caption{Global power spectrum histogram fitted with a $\chi^2$ function. The upper subfigure is the histogram for the monthly binned light curve of S5~0716+714, at the scale corresponding to period $\sim$197~d, and the bottom subfigure is the histogram for the weekly binned light curve, at the scale corresponding to period $\sim$260~d. Five pre-trial confidence levels are shown in both graphs corresponding to 1 to 5 $\sigma$ significance in solid lines, whereas the post-trial confidence levels are shown in dashed lines.}
	\label{fig:Chi2}
\end{figure}

We fit the global power spectrum histograms, computed using the artificial light curves, for each period with a $\chi^2$ function as shown in example histograms in Fig.~\ref{fig:Chi2}. We also display the five $\sigma$ confidence levels, equivalent to those of a Gaussian distribution, i.e. 68.3\%, 95.4\%, 99.7\%, 99.994\% and 99.99994\%, for both the pre-trial (solid lines) and post-trial (dashed lines) cases.

\vspace*{0.5cm}
\noindent

\section{CWT of All Sources}
\label{sec:AAll}

\begin{figure*}[!htbp]
	\centering
	\begin{subfigure}[b]{0.48\textwidth}   
		\centering 
		\includegraphics[width=\textwidth]{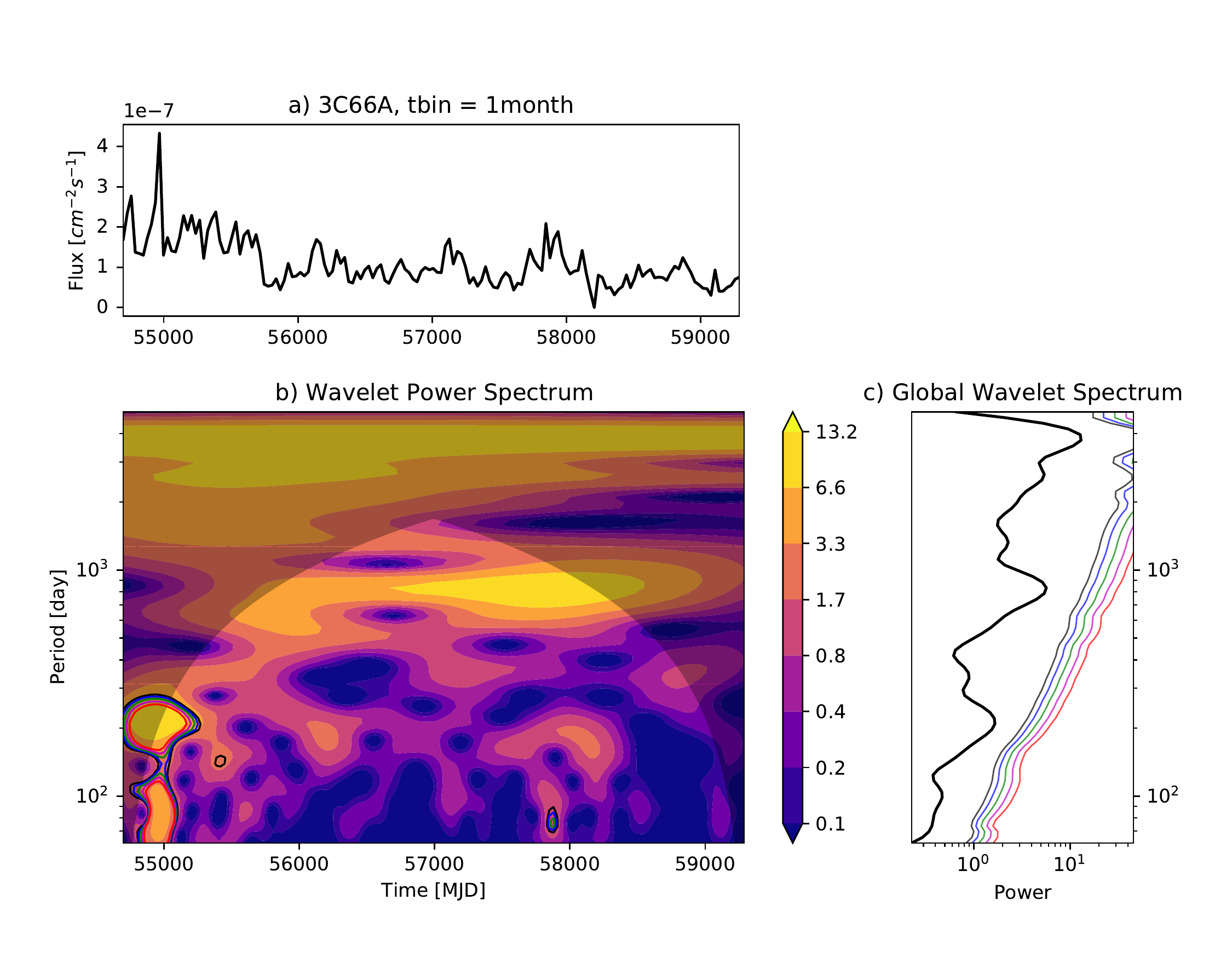}
	\end{subfigure}
	\hfill
	\begin{subfigure}[b]{0.48\textwidth}   
		\centering 
		\includegraphics[width=\textwidth]{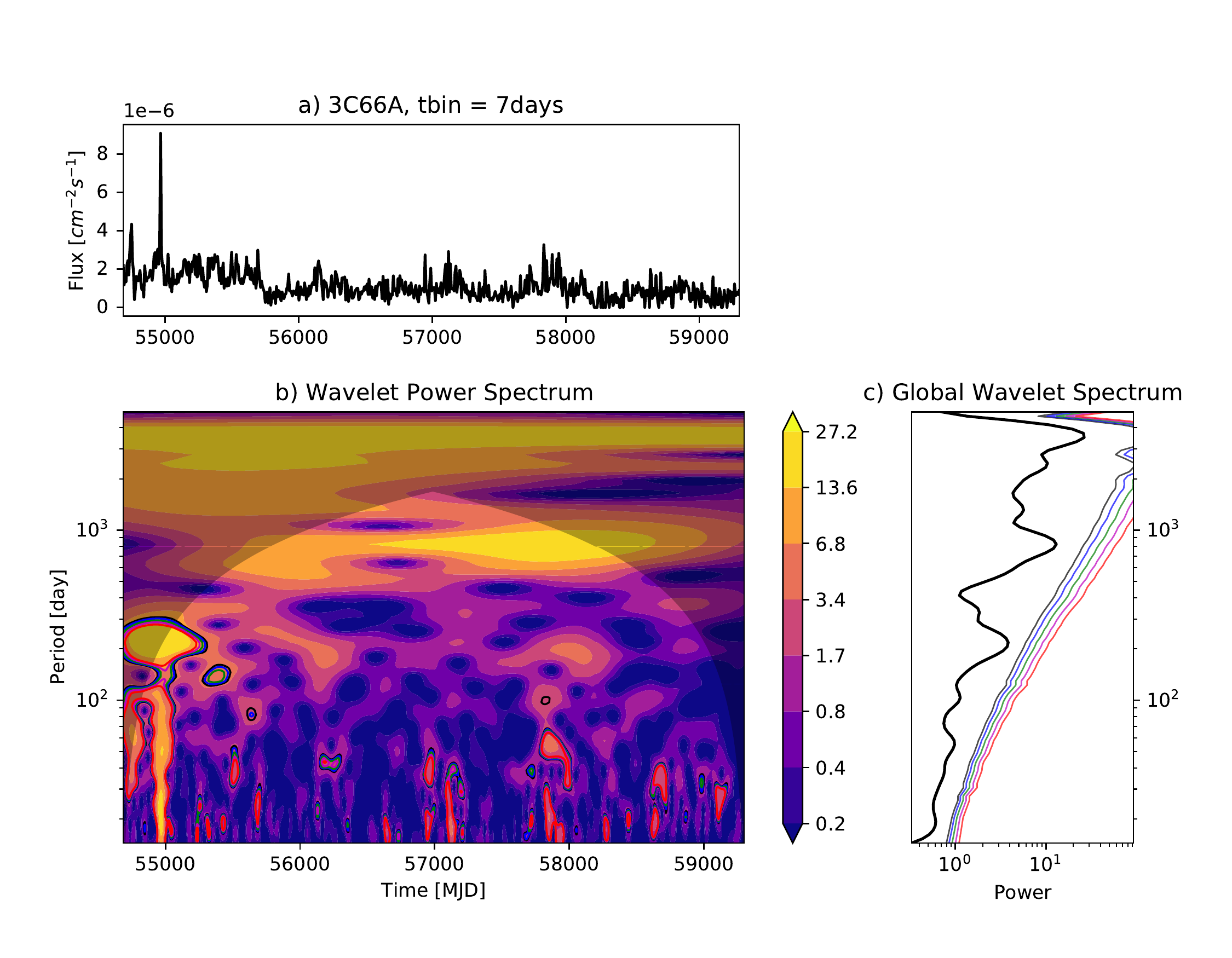}
	\end{subfigure}
	\vskip\baselineskip
	
	\hrule
	
	\begin{subfigure}[b]{0.48\textwidth}
		\centering
		\includegraphics[width=\textwidth]{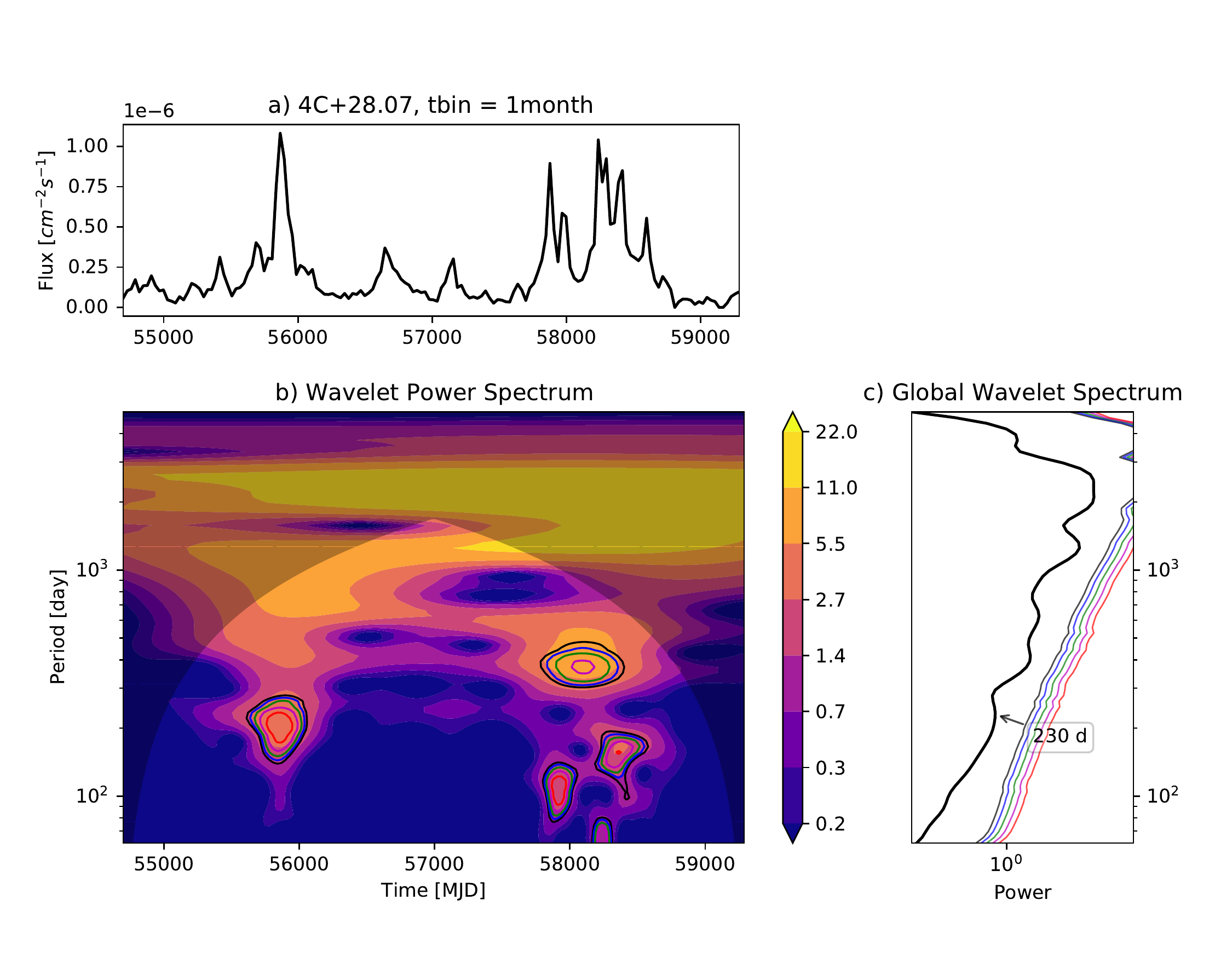}
	\end{subfigure}
	\hfill
	\begin{subfigure}[b]{0.48\textwidth}
		\centering
		\includegraphics[width=\textwidth]{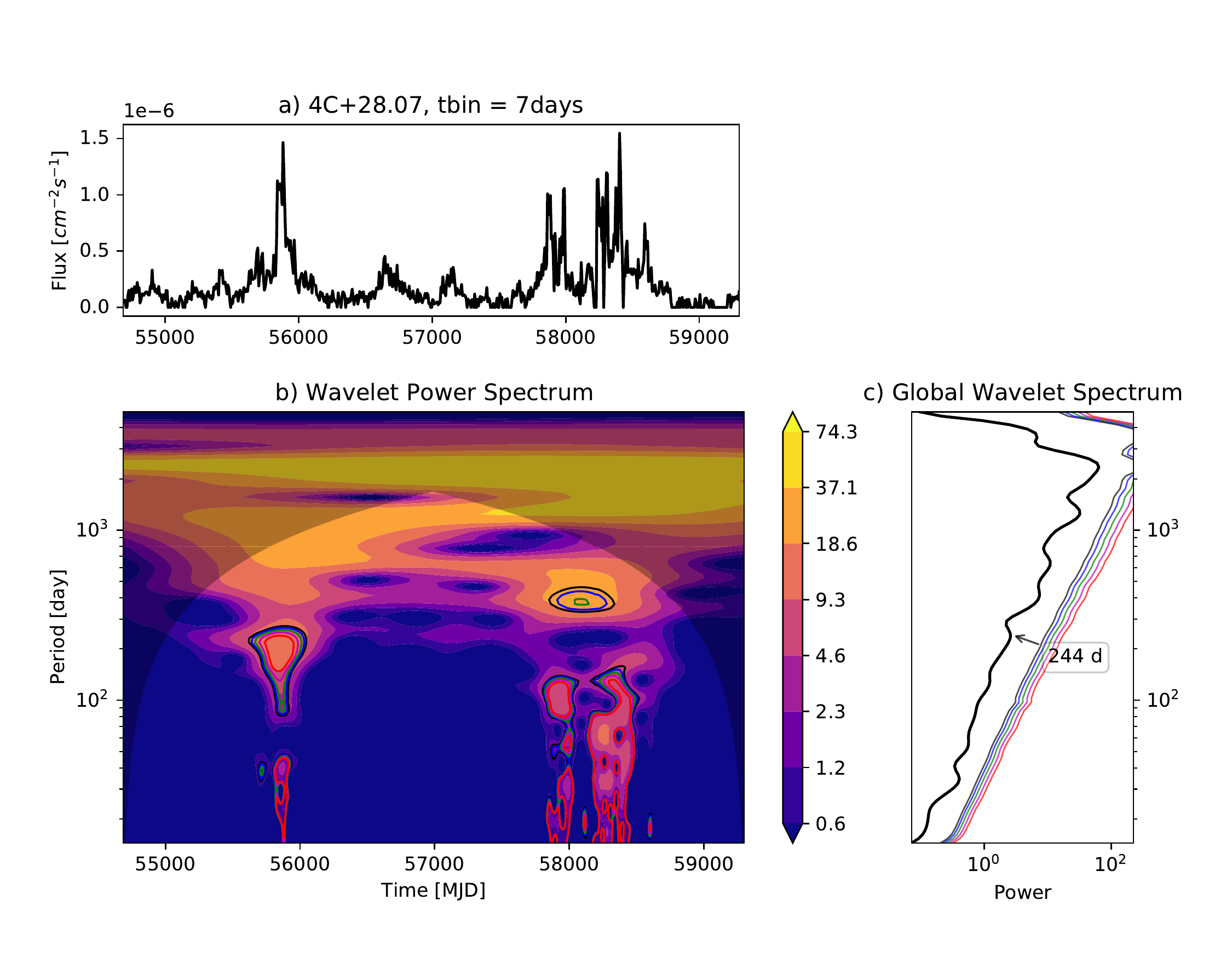}
	\end{subfigure}
	\vskip\baselineskip
	
	\begin{subfigure}[b]{0.48\textwidth}  
		\centering 
		\includegraphics[width=\textwidth]{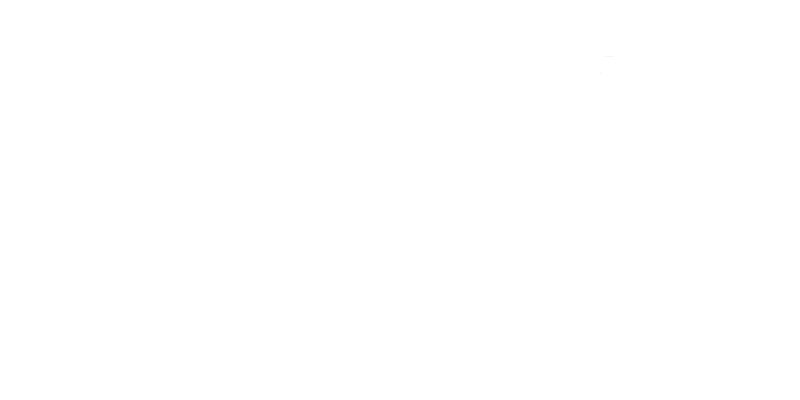}
	\end{subfigure}
	\hfill
	\begin{subfigure}[b]{0.48\textwidth}  
		\centering 
		\includegraphics[width=\textwidth]{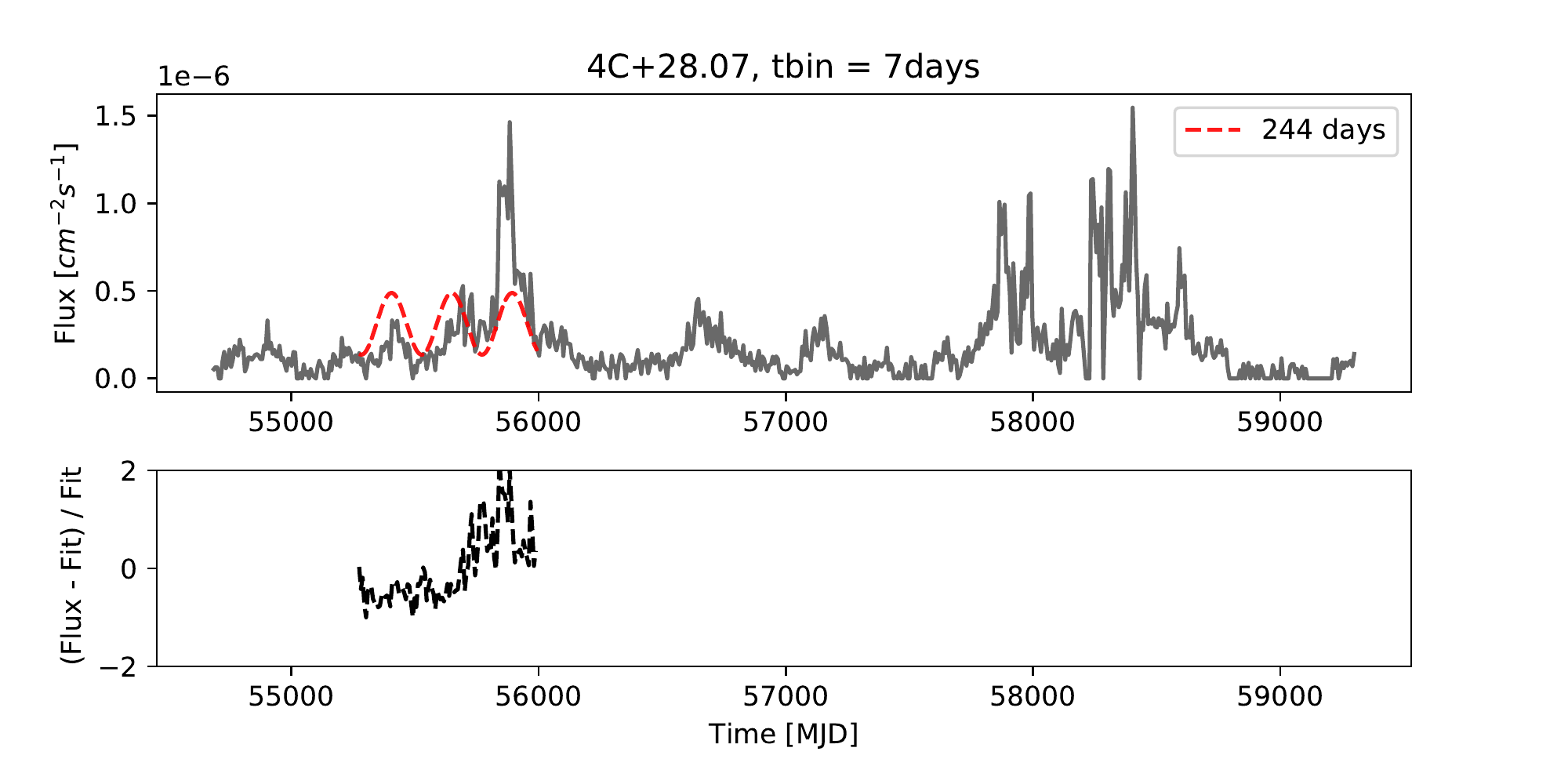}
	\end{subfigure}
	
	\caption{CWT map for monthly binned light curve (left) and weekly binned light curve (right) of  3C~66A and 4C~+28.07, and fitted light curves for 4C~+01.02.}
	\label{fig:CWT1}
\end{figure*}

\begin{figure*}[!htbp]
	\centering
	\begin{subfigure}[b]{0.48\textwidth}   
		\centering 
		\includegraphics[width=\textwidth]{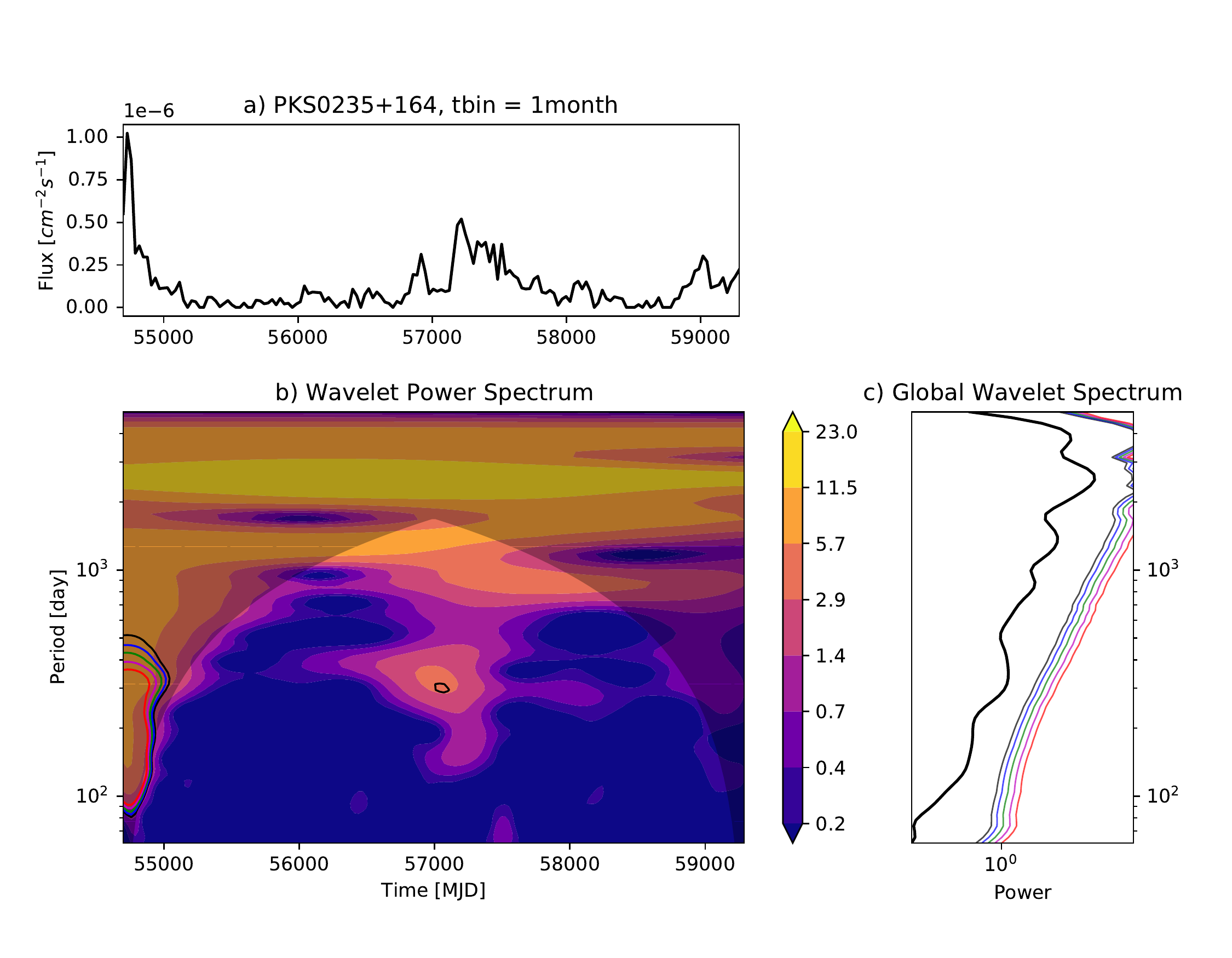}
	\end{subfigure}
	\hfill
	\begin{subfigure}[b]{0.48\textwidth}   
		\centering 
		\includegraphics[width=\textwidth]{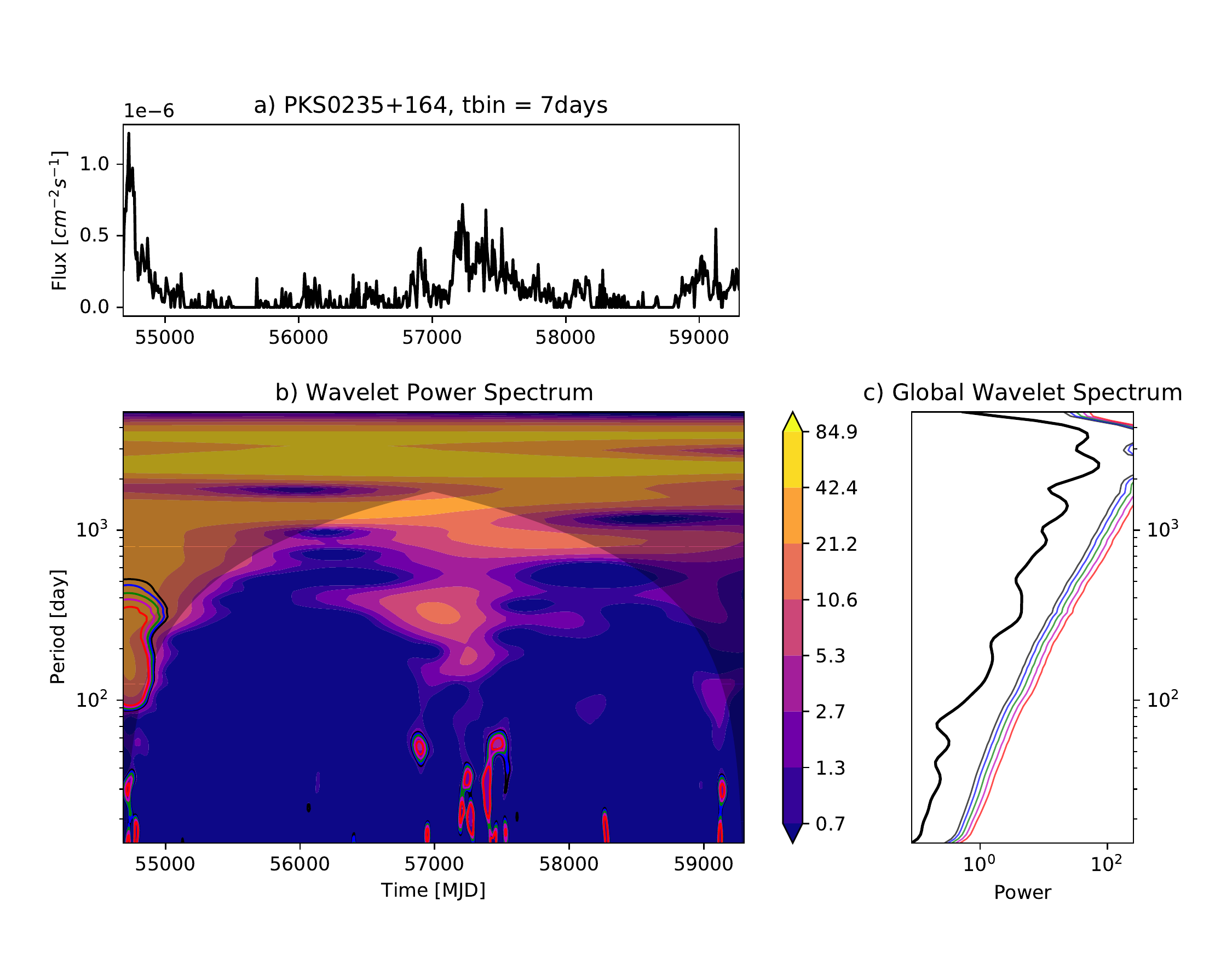}
	\end{subfigure}
	\vskip\baselineskip
	
	\hrule	
	\begin{subfigure}[b]{0.48\textwidth}
		\centering
		\includegraphics[width=\textwidth]{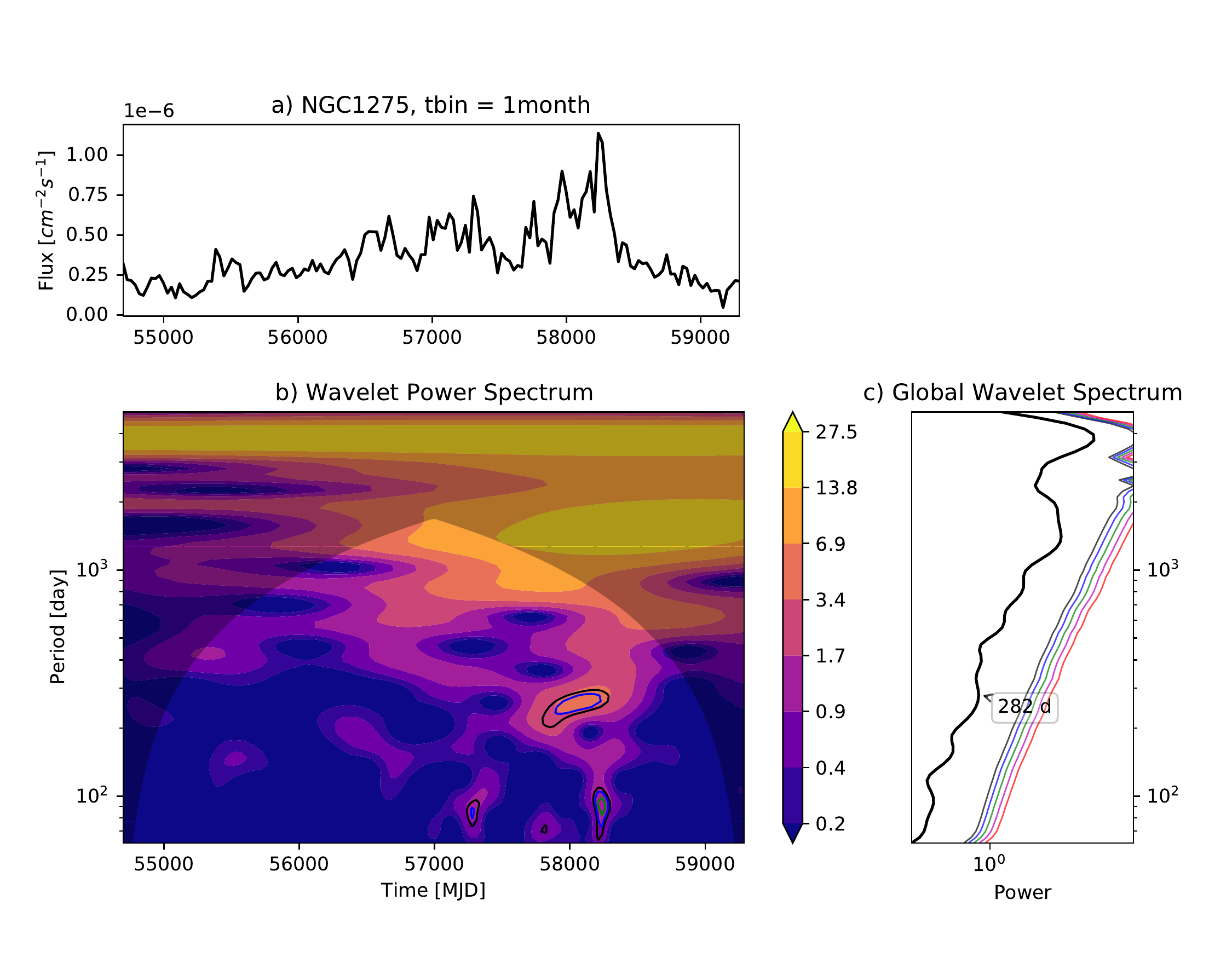}
	\end{subfigure}
	\hfill
	\begin{subfigure}[b]{0.48\textwidth}
		\centering
		\includegraphics[width=\textwidth]{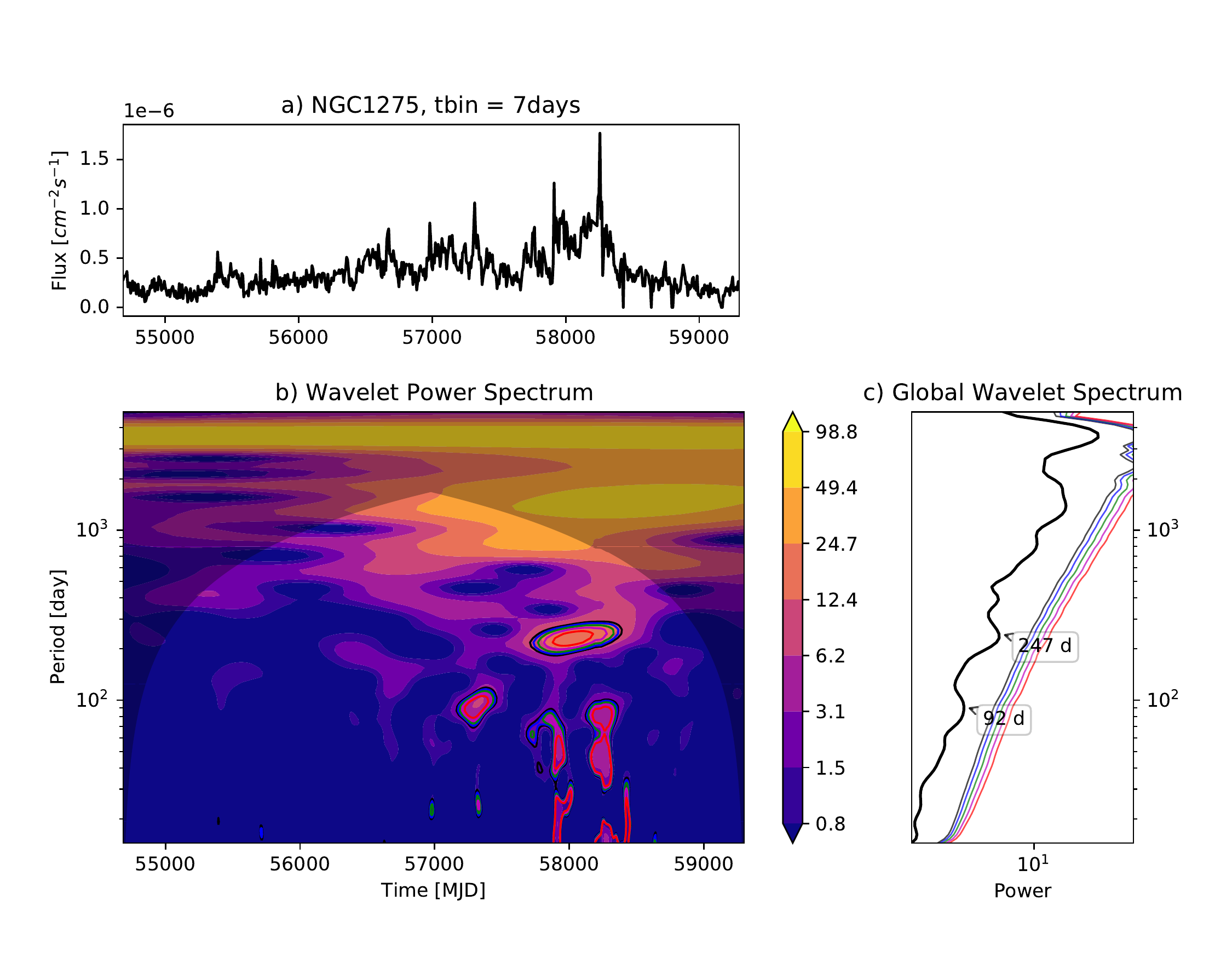}
	\end{subfigure}
	\vskip\baselineskip
	
	\begin{subfigure}[b]{0.48\textwidth}  
		\centering 
		\includegraphics[width=\textwidth]{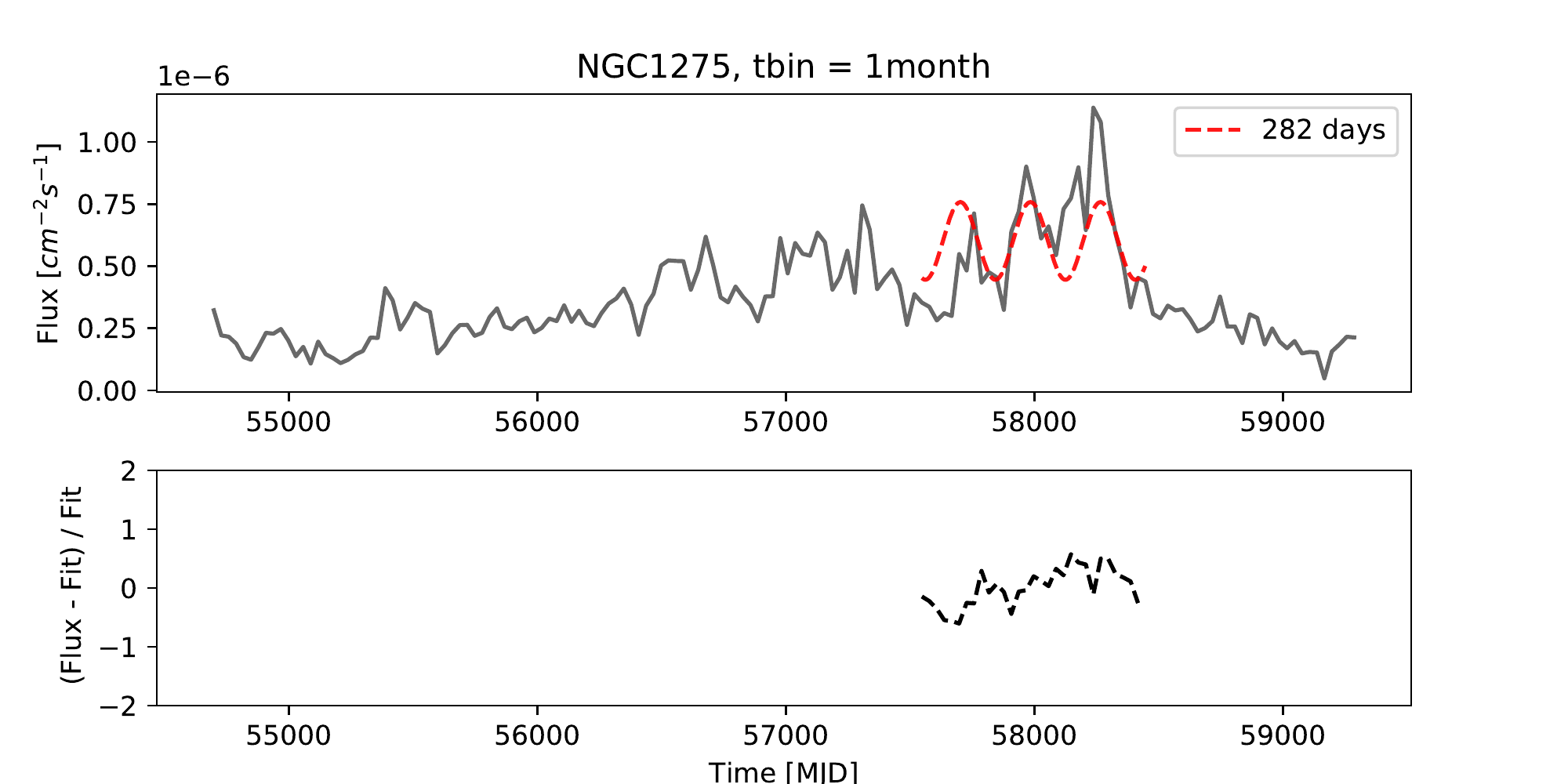}
	\end{subfigure}
	\hfill
	\begin{subfigure}[b]{0.48\textwidth}  
		\centering 
		\includegraphics[width=\textwidth]{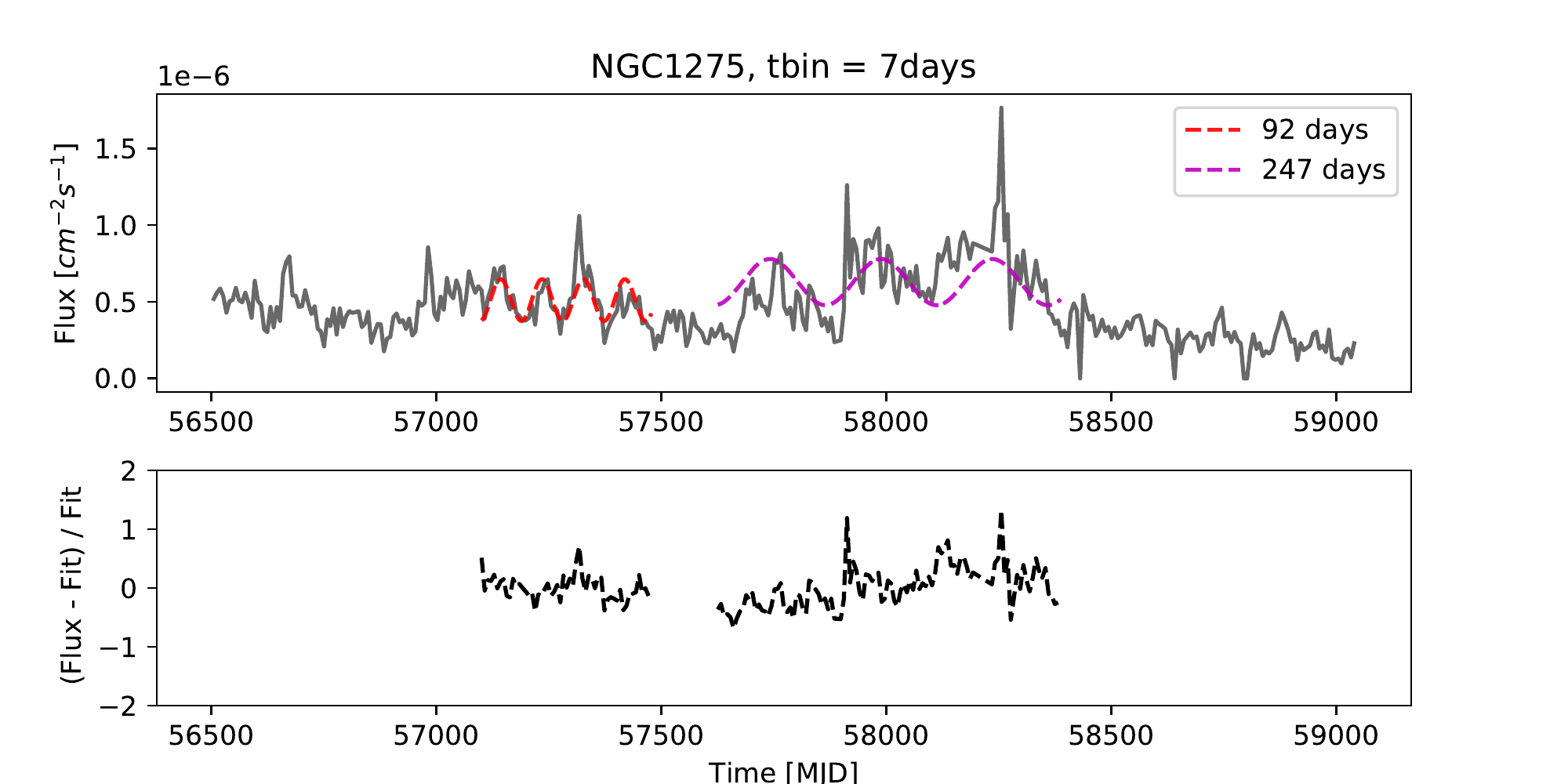}
	\end{subfigure}
	
	\caption{CWT map for monthly binned light curve (left) and weekly binned light curve (right) of PKS~0235+164 and NGC~1275, and fitted light curves for NGC~1275.}
	\label{fig:CWT2}
\end{figure*}

\begin{figure*}[!htbp]
	\centering
	\begin{subfigure}[b]{0.48\textwidth}   
		\centering 
		\includegraphics[width=\textwidth]{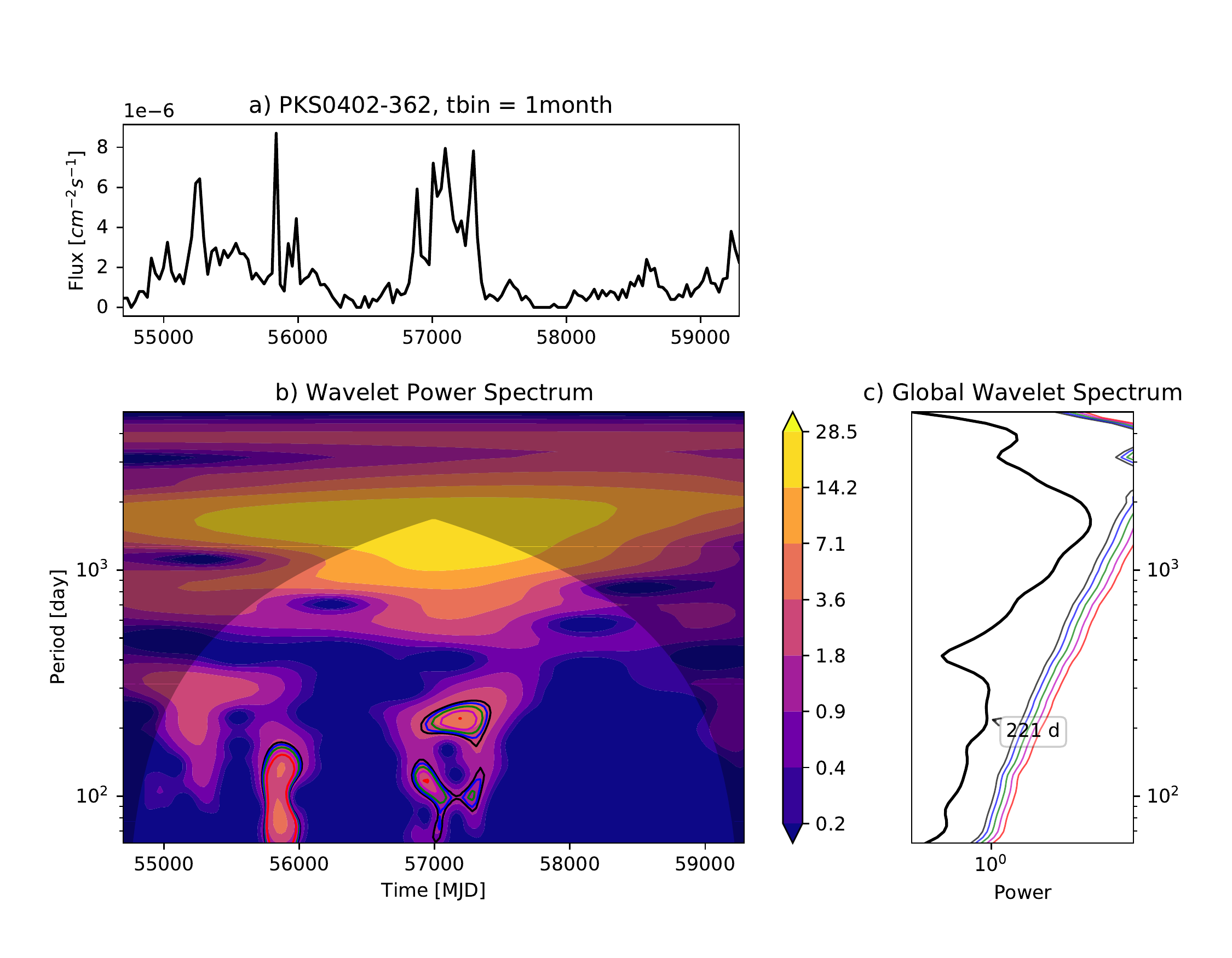}
	\end{subfigure}
	\hfill
	\begin{subfigure}[b]{0.48\textwidth}   
		\centering 
		\includegraphics[width=\textwidth]{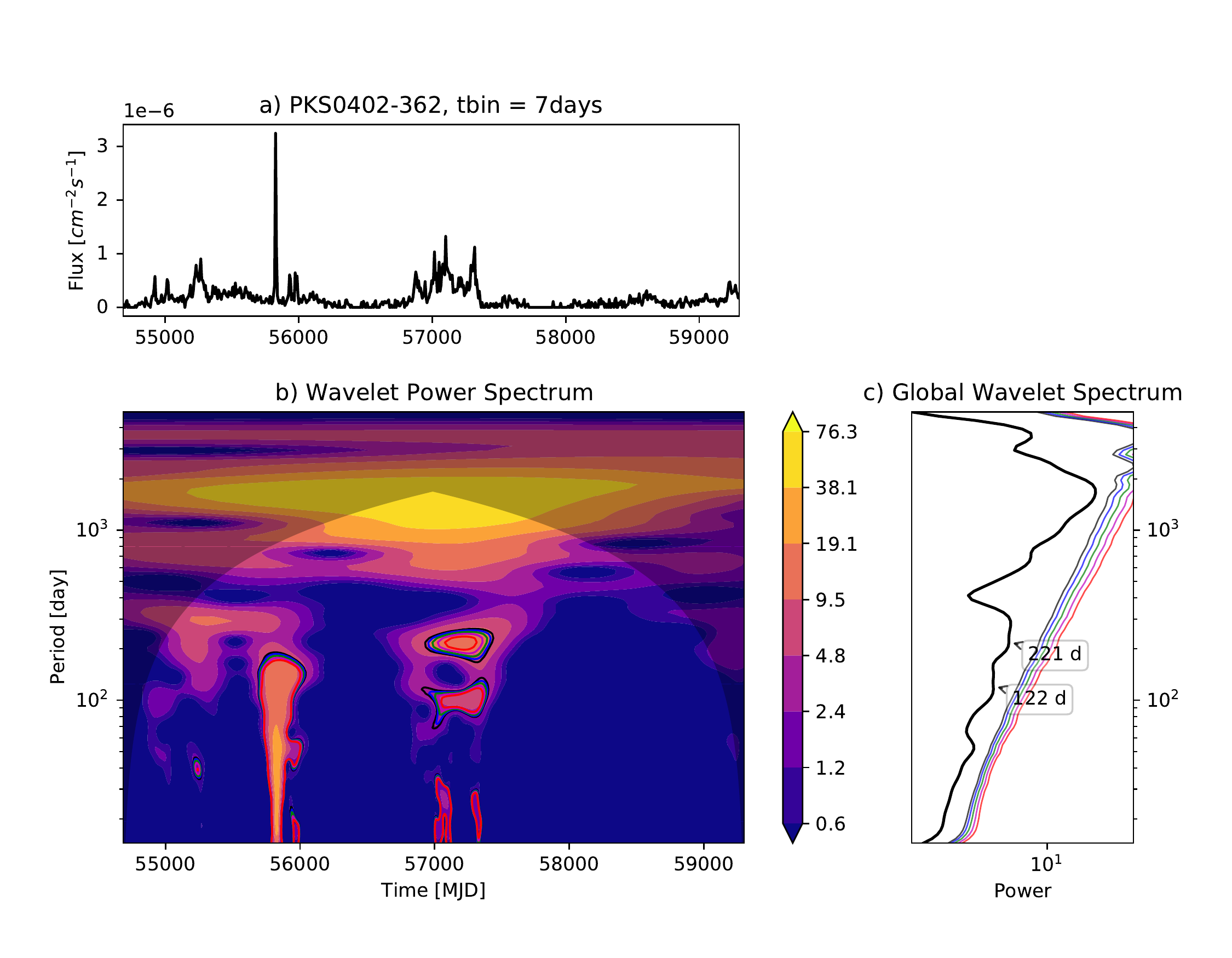}
	\end{subfigure}
	\vskip\baselineskip
	
	\begin{subfigure}[b]{0.48\textwidth}  
		\centering 
		\includegraphics[width=\textwidth]{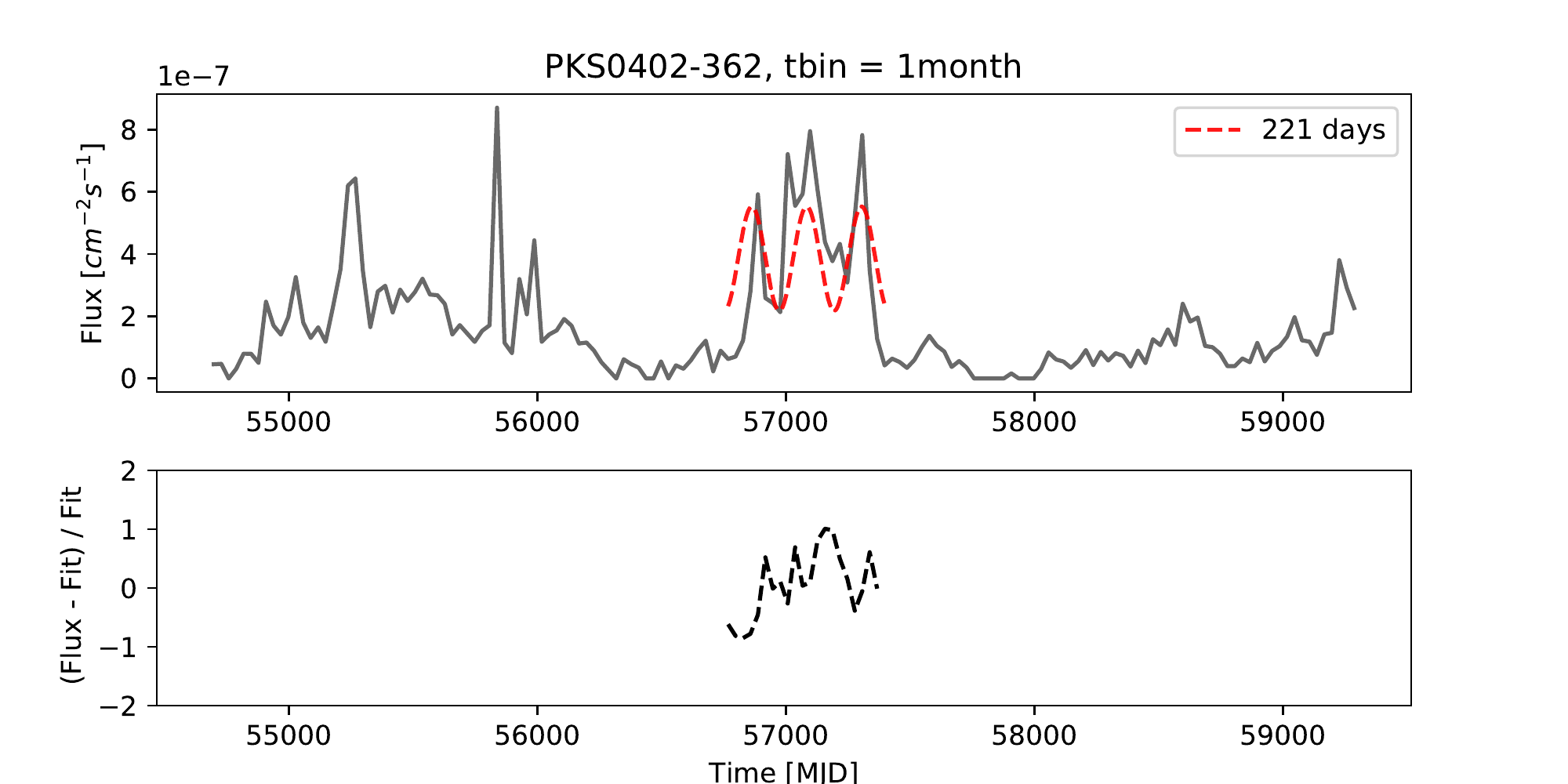}
	\end{subfigure}
	\hfill
	\begin{subfigure}[b]{0.48\textwidth}  
		\centering 
		\includegraphics[width=\textwidth]{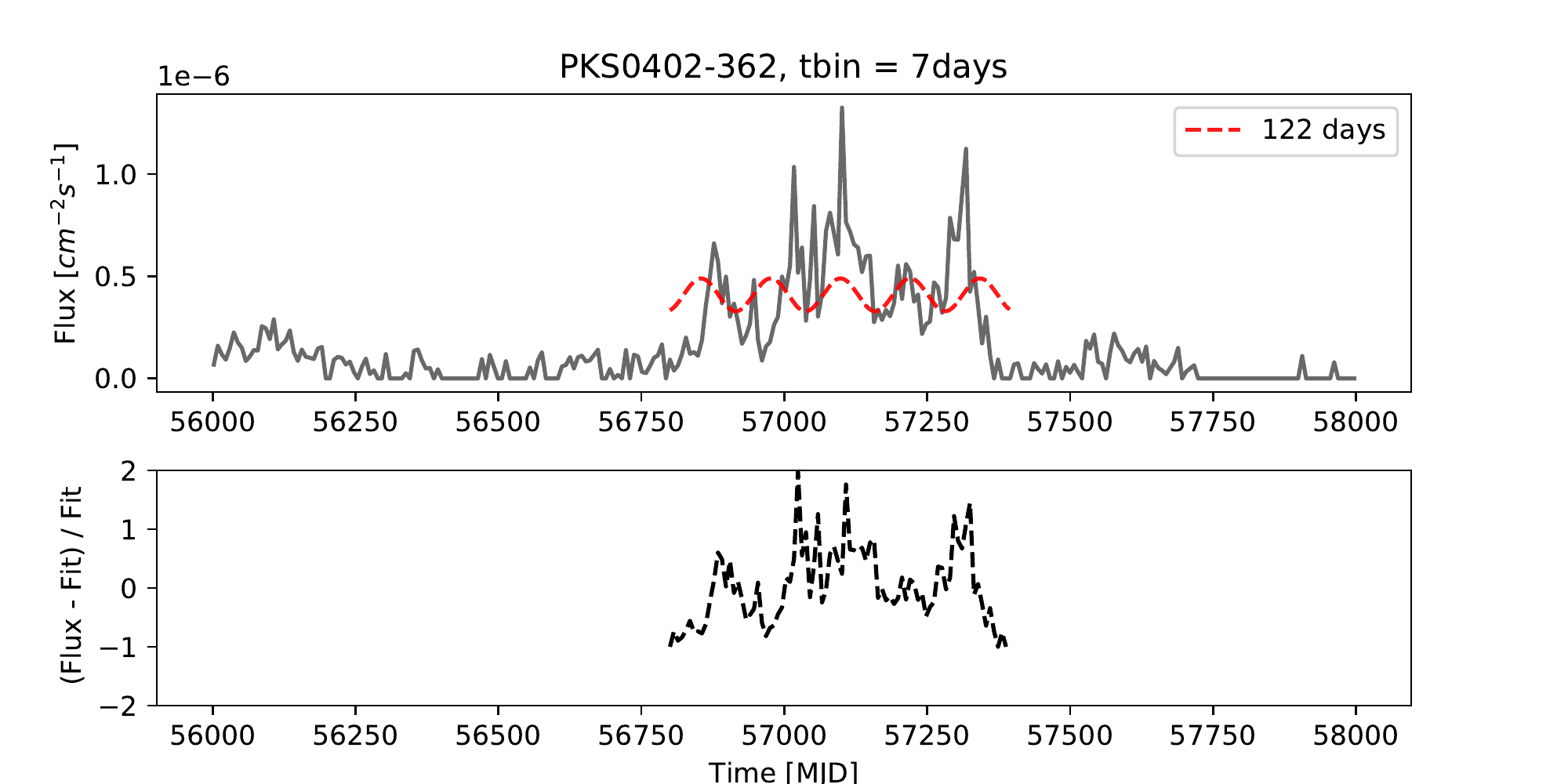}
	\end{subfigure}
	\vskip\baselineskip
	
	\hrule	
	\begin{subfigure}[b]{0.48\textwidth}   
		\centering 
		\includegraphics[width=\textwidth]{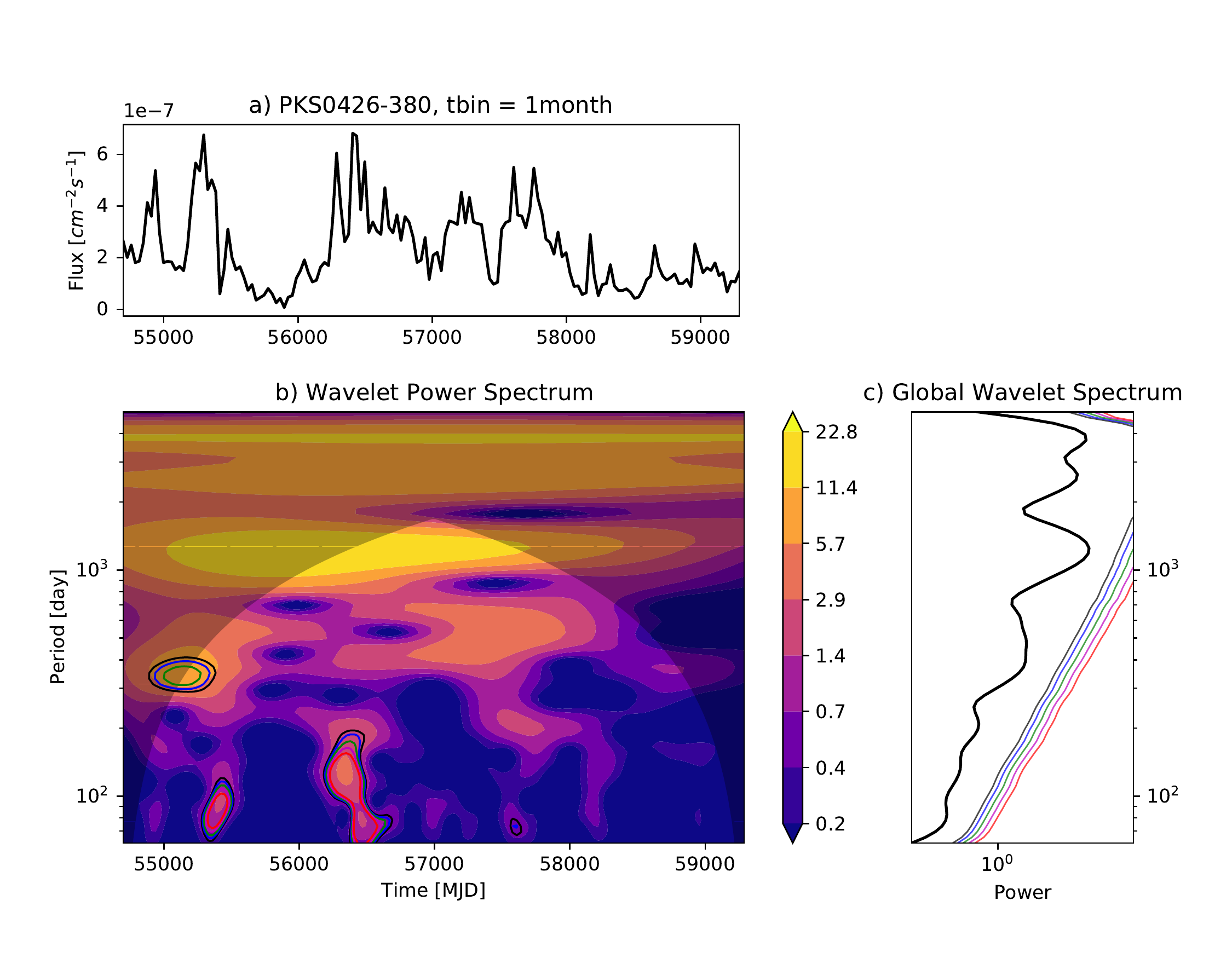}
	\end{subfigure}
	\hfill
	\begin{subfigure}[b]{0.48\textwidth}   
		\centering 
		\includegraphics[width=\textwidth]{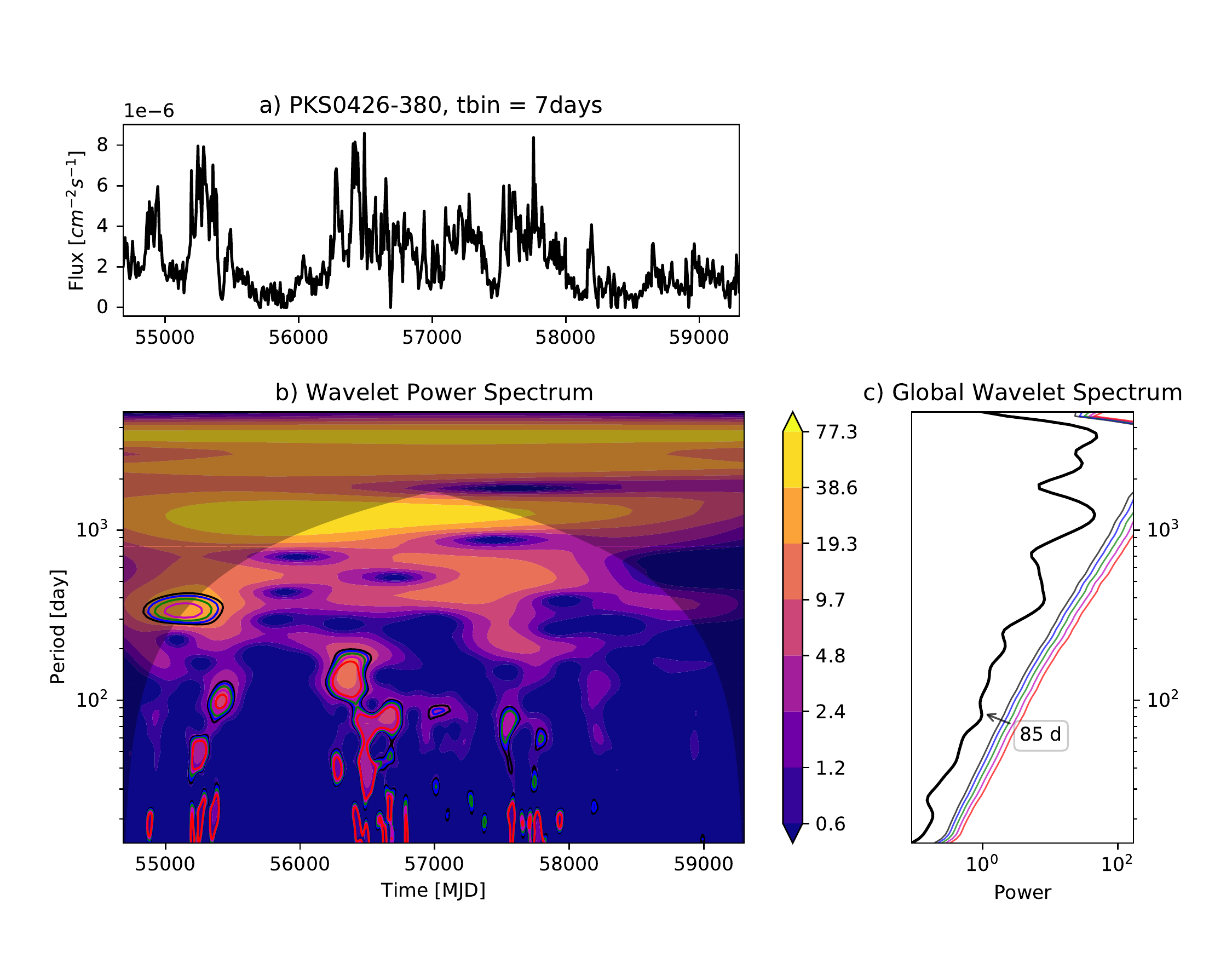}
	\end{subfigure}
	\vskip\baselineskip
	
	\begin{subfigure}[b]{0.48\textwidth}  
		\centering 
		\includegraphics[width=\textwidth]{./Figures/Fit/blanc.png}
	\end{subfigure}
	\hfill
	\begin{subfigure}[b]{0.48\textwidth}  
		\centering 
		\includegraphics[width=\textwidth]{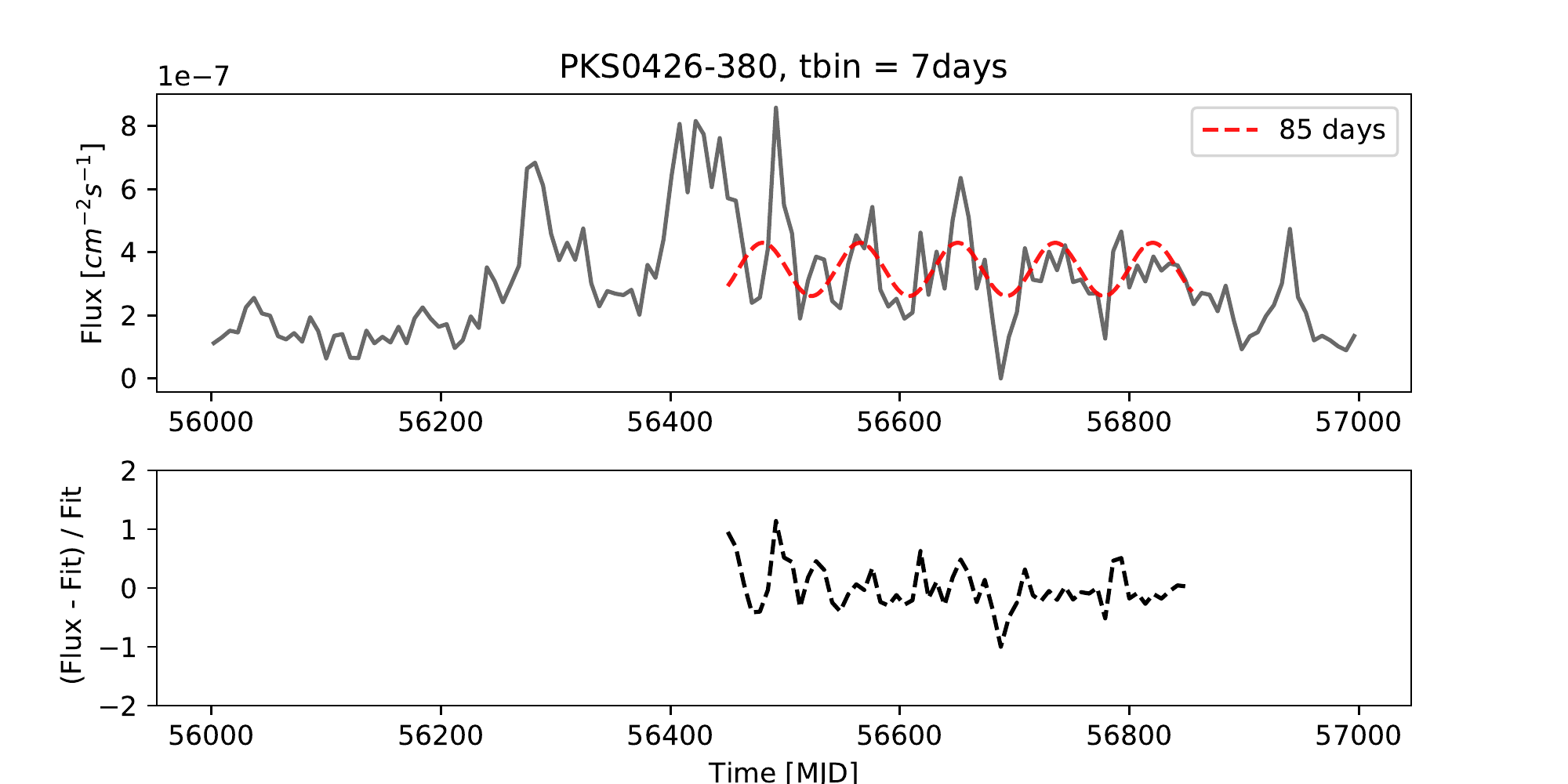}
	\end{subfigure}

	\caption{CWT map for monthly binned light curve (left) and weekly binned light curve (right) of PKS~0402-362 and PKS~0426-380, and the respective fitted light curves.}
	\label{fig:CWT3}
\end{figure*}


\begin{figure*}[!htbp]
	\centering

	\begin{subfigure}[b]{0.48\textwidth}   
		\centering 
		\includegraphics[width=\textwidth]{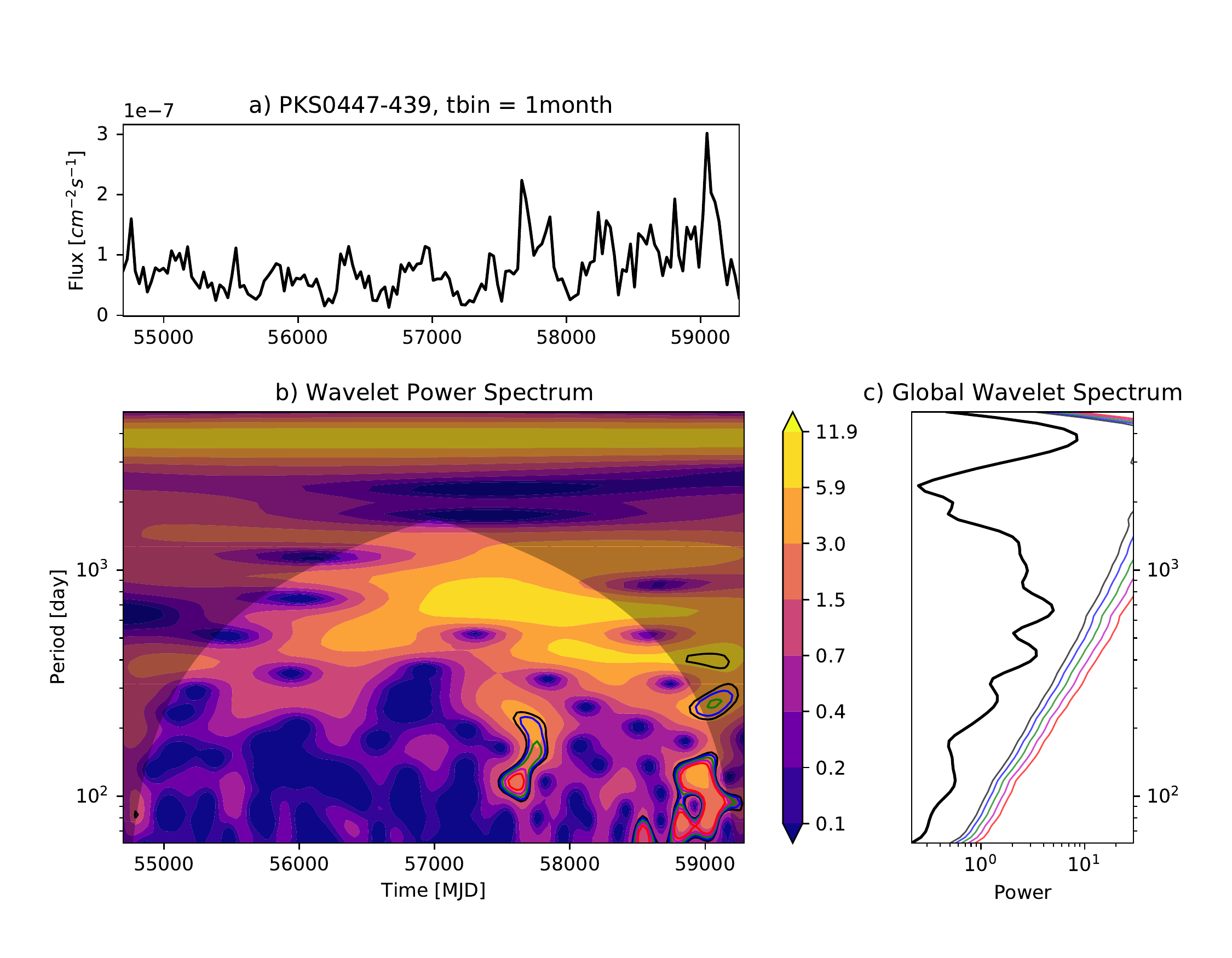}
	\end{subfigure}
	\hfill
	\begin{subfigure}[b]{0.48\textwidth}   
		\centering 
		\includegraphics[width=\textwidth]{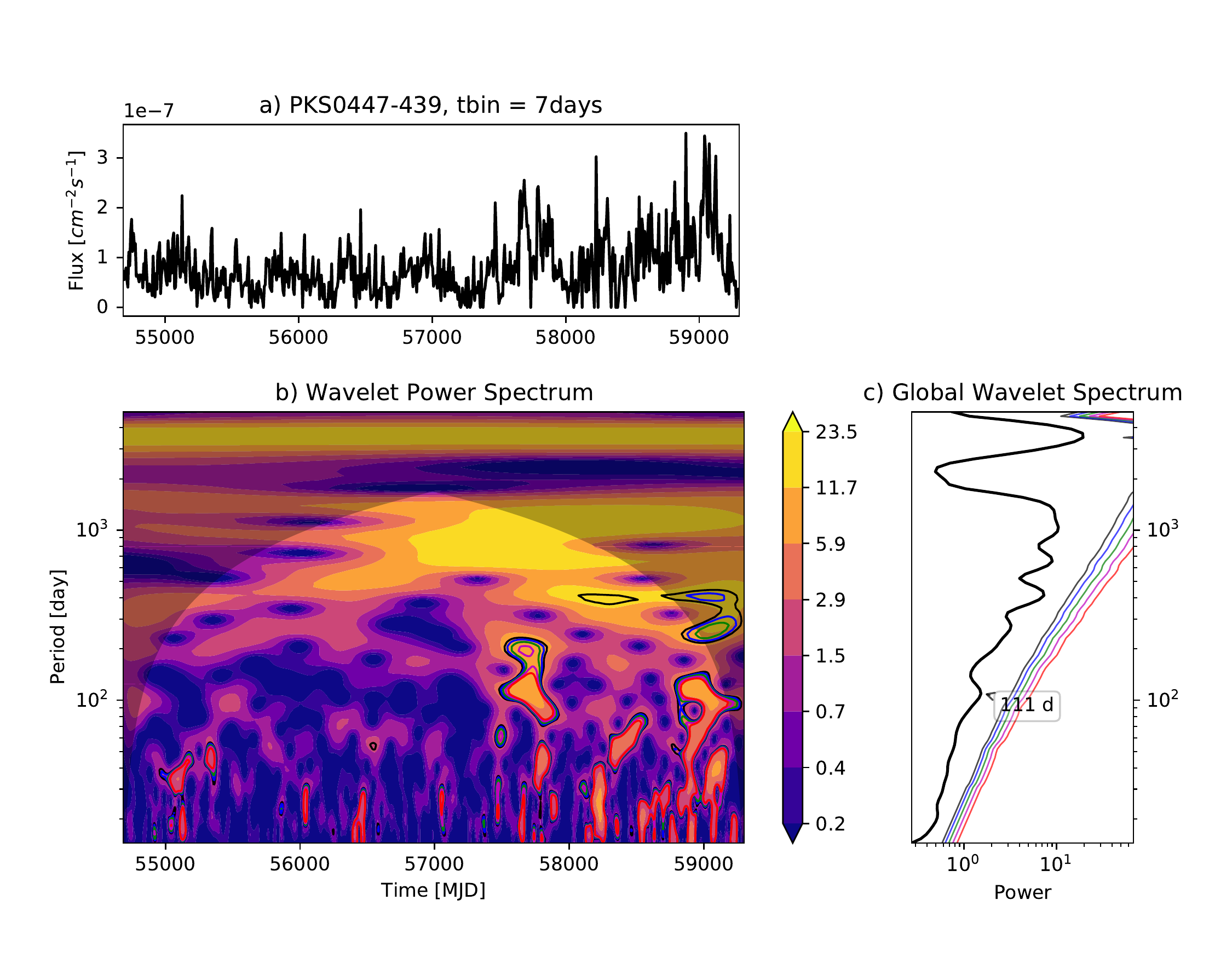}
	\end{subfigure}
	\vskip\baselineskip
	
	\begin{subfigure}[b]{0.48\textwidth}  
		\centering 
		\includegraphics[width=\textwidth]{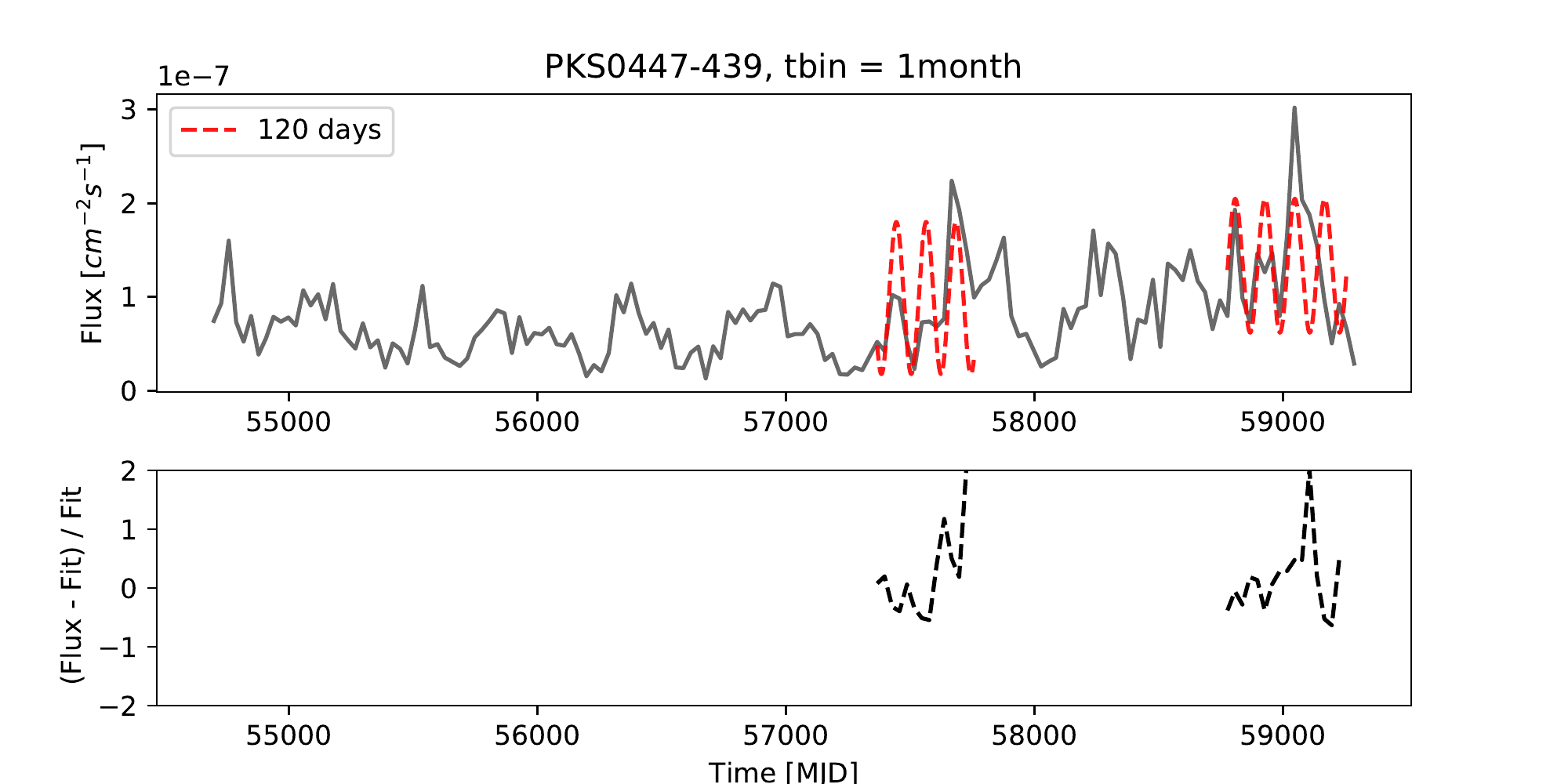}
	\end{subfigure}
	\hfill
	\begin{subfigure}[b]{0.48\textwidth}  
		\centering 
		\includegraphics[width=\textwidth]{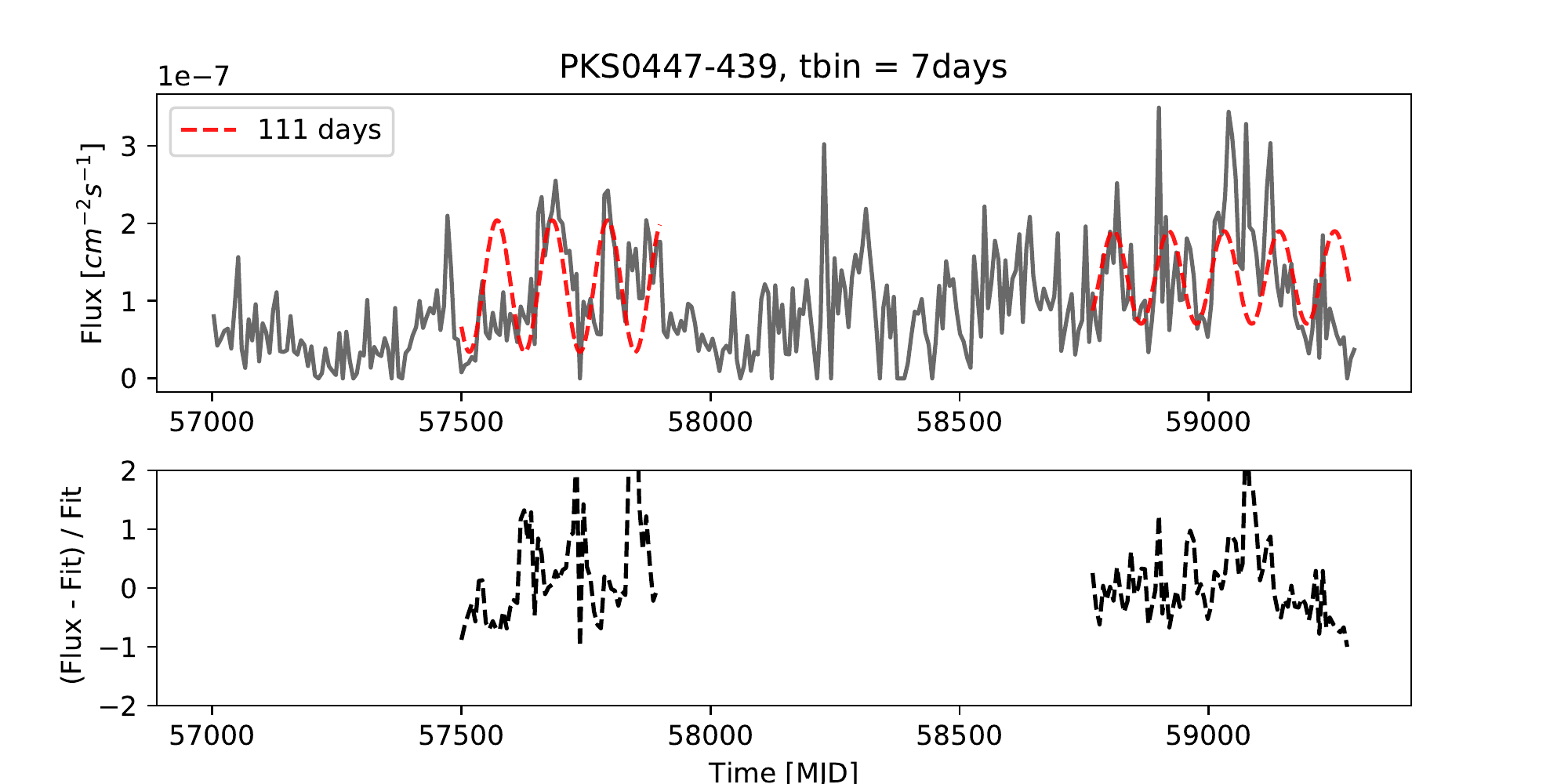}
	\end{subfigure}
	\vskip\baselineskip
	
	\hrule	
	\begin{subfigure}[b]{0.48\textwidth}
		\centering
		\includegraphics[width=\textwidth]{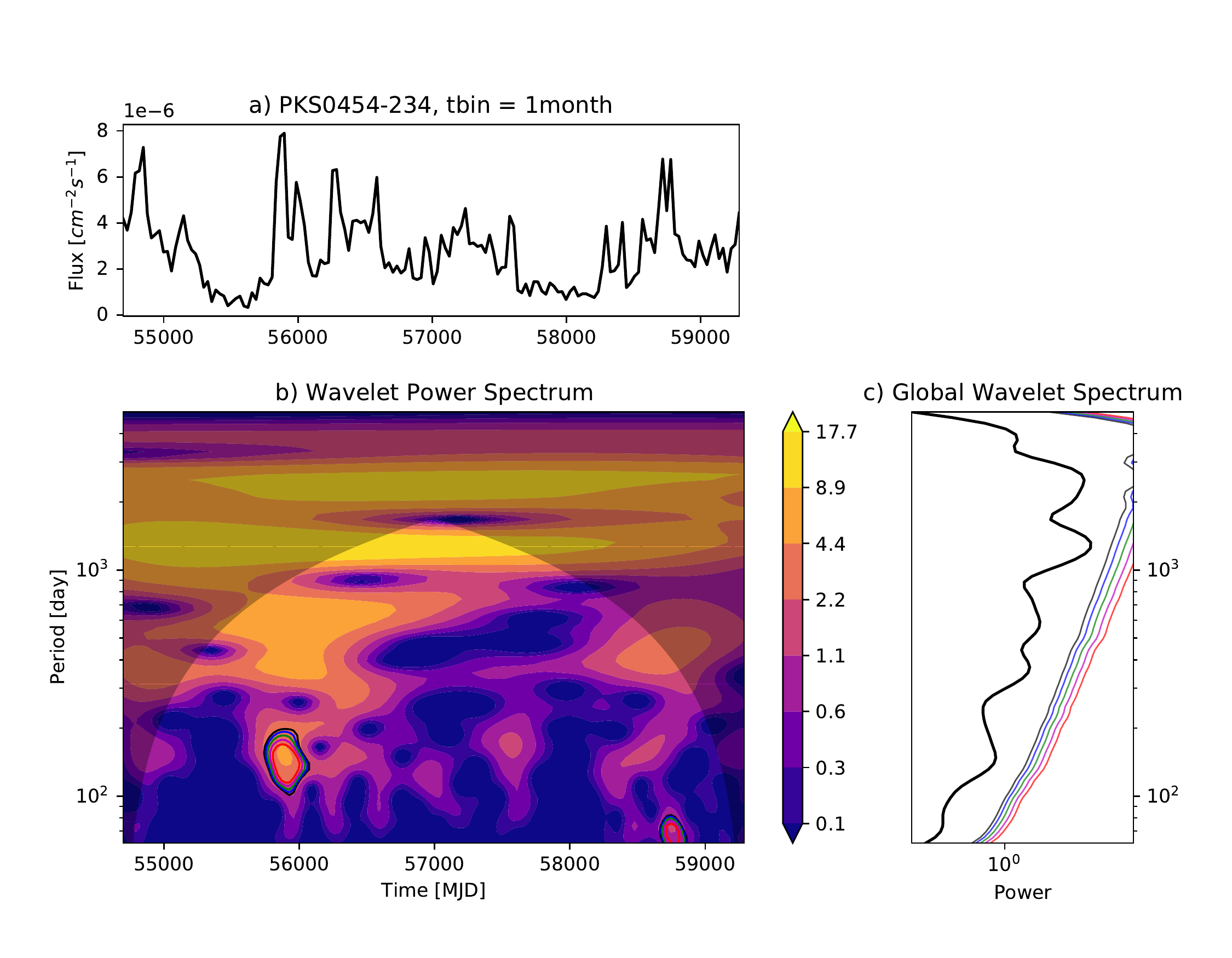}
	\end{subfigure}
	\hfill
	\begin{subfigure}[b]{0.48\textwidth}
		\centering
		\includegraphics[width=\textwidth]{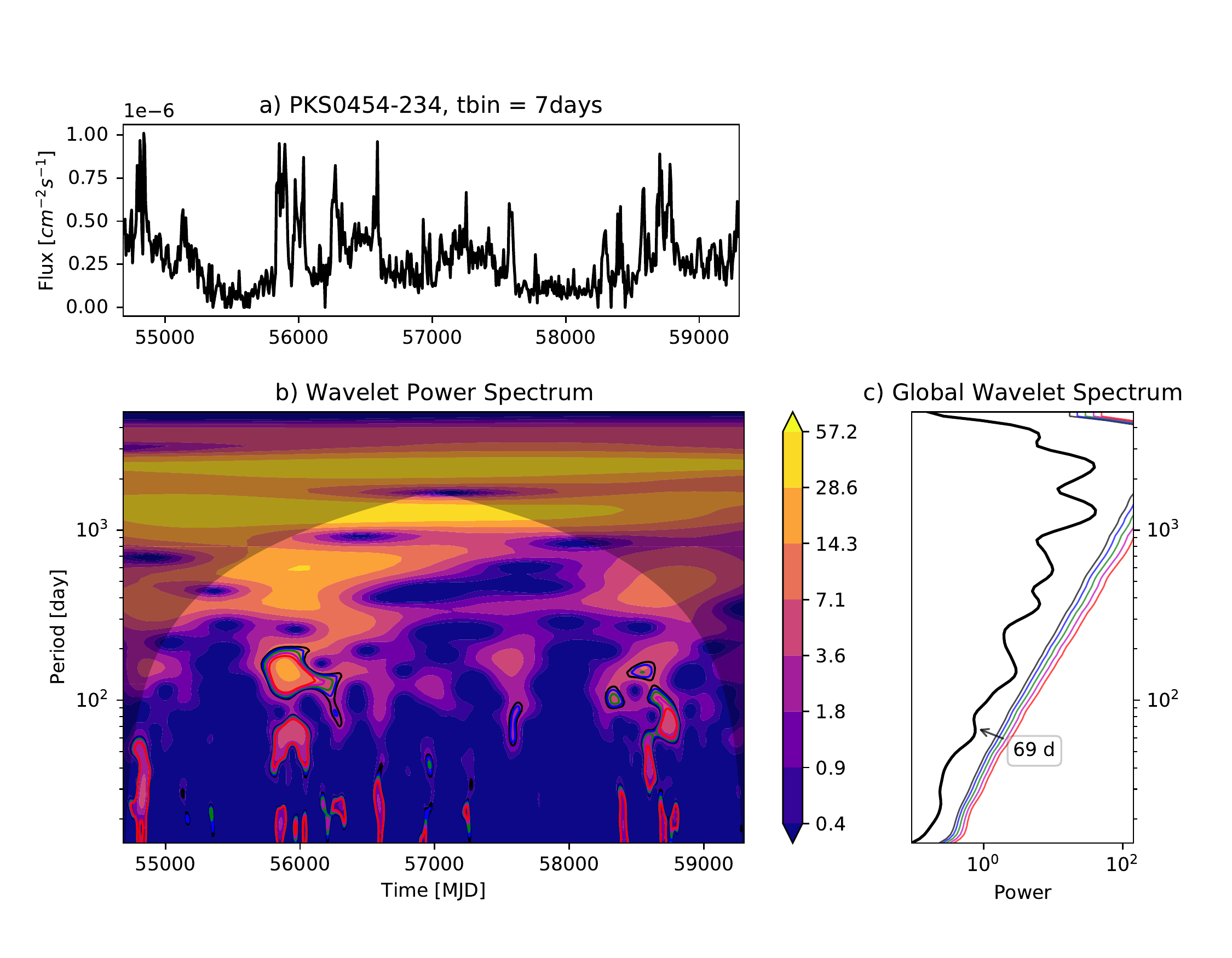}
	\end{subfigure}
	\vskip\baselineskip
	
	\begin{subfigure}[b]{0.48\textwidth}  
		\centering 
		\includegraphics[width=\textwidth]{./Figures/Fit/blanc.png}
	\end{subfigure}
	\hfill
	\begin{subfigure}[b]{0.48\textwidth}  
		\centering 
		\includegraphics[width=\textwidth]{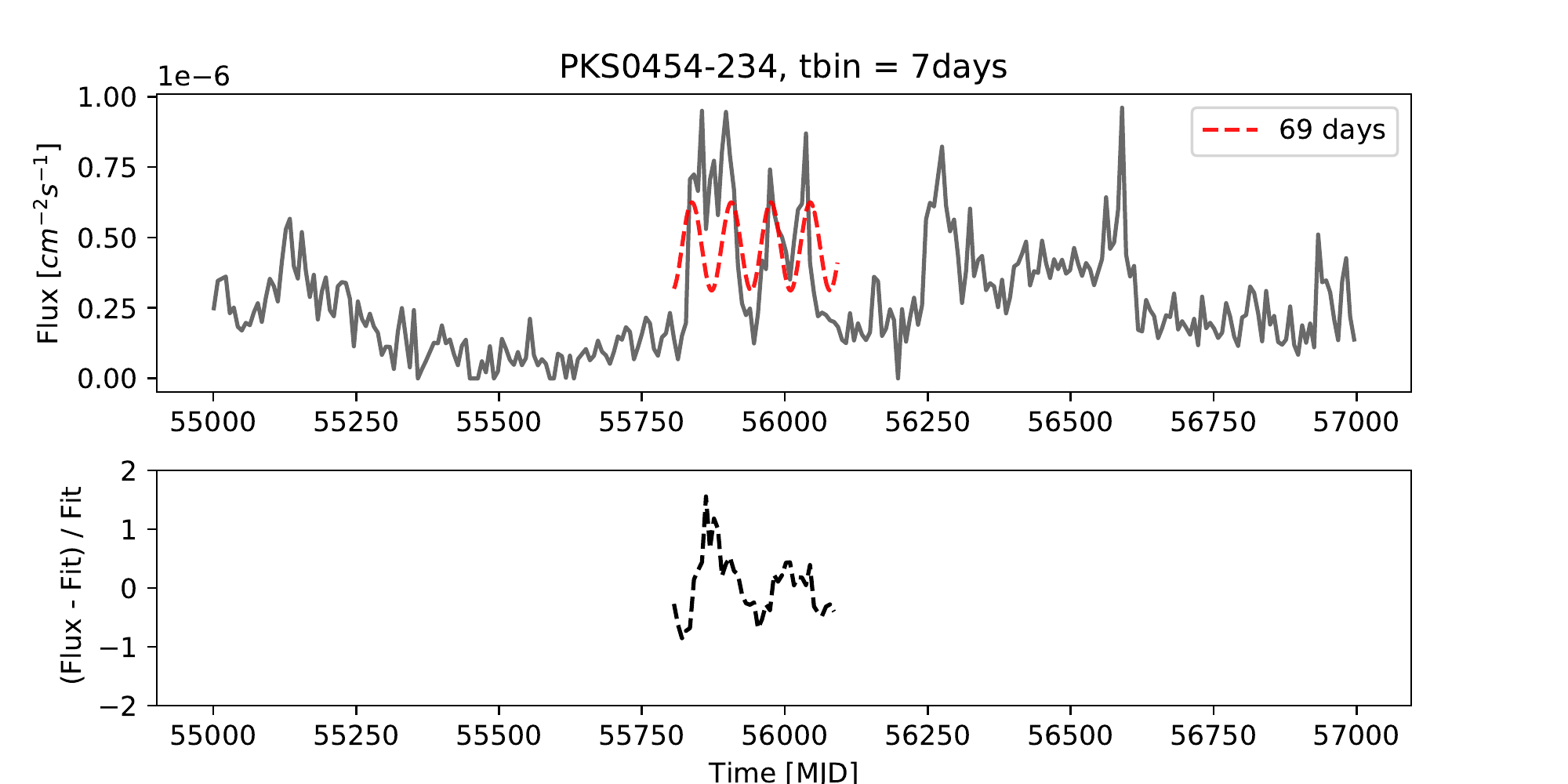}
	\end{subfigure}	
	
	\caption{CWT map for monthly binned light curve (left) and weekly binned light curve (right) of  PKS~0447-439 and PKS~0454-234, and the respective fitted light curves.}
	\label{fig:CWT4}
\end{figure*}

\begin{figure*}[!htbp]
    \centering
	
	\begin{subfigure}[b]{0.48\textwidth}
		\centering
		\includegraphics[width=\textwidth]{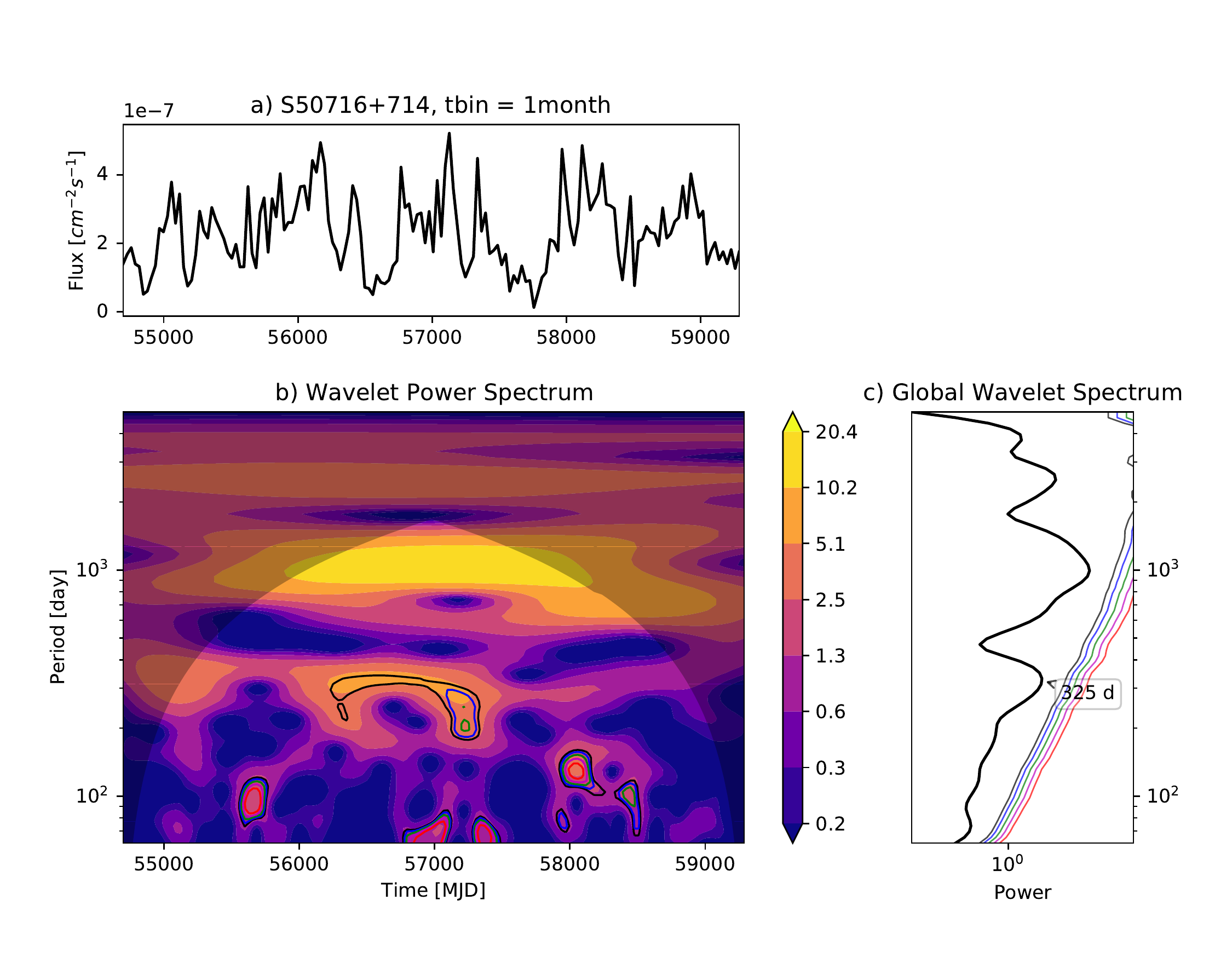}
	\end{subfigure}
	\hfill
	\begin{subfigure}[b]{0.48\textwidth}
		\centering
		\includegraphics[width=\textwidth]{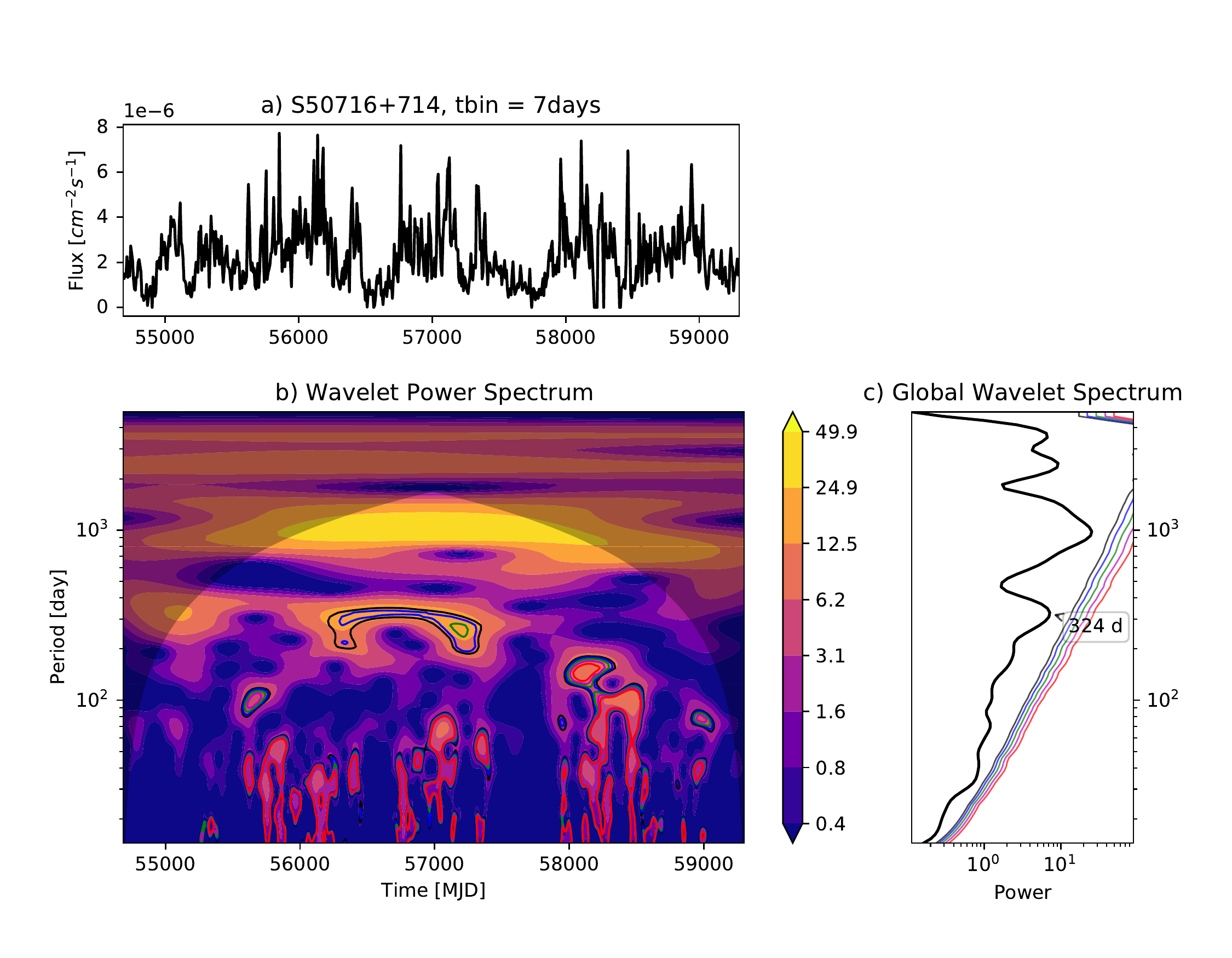}
	\end{subfigure}
	\vskip\baselineskip
	
	\begin{subfigure}[b]{0.48\textwidth}  
		\centering 
		\includegraphics[width=\textwidth]{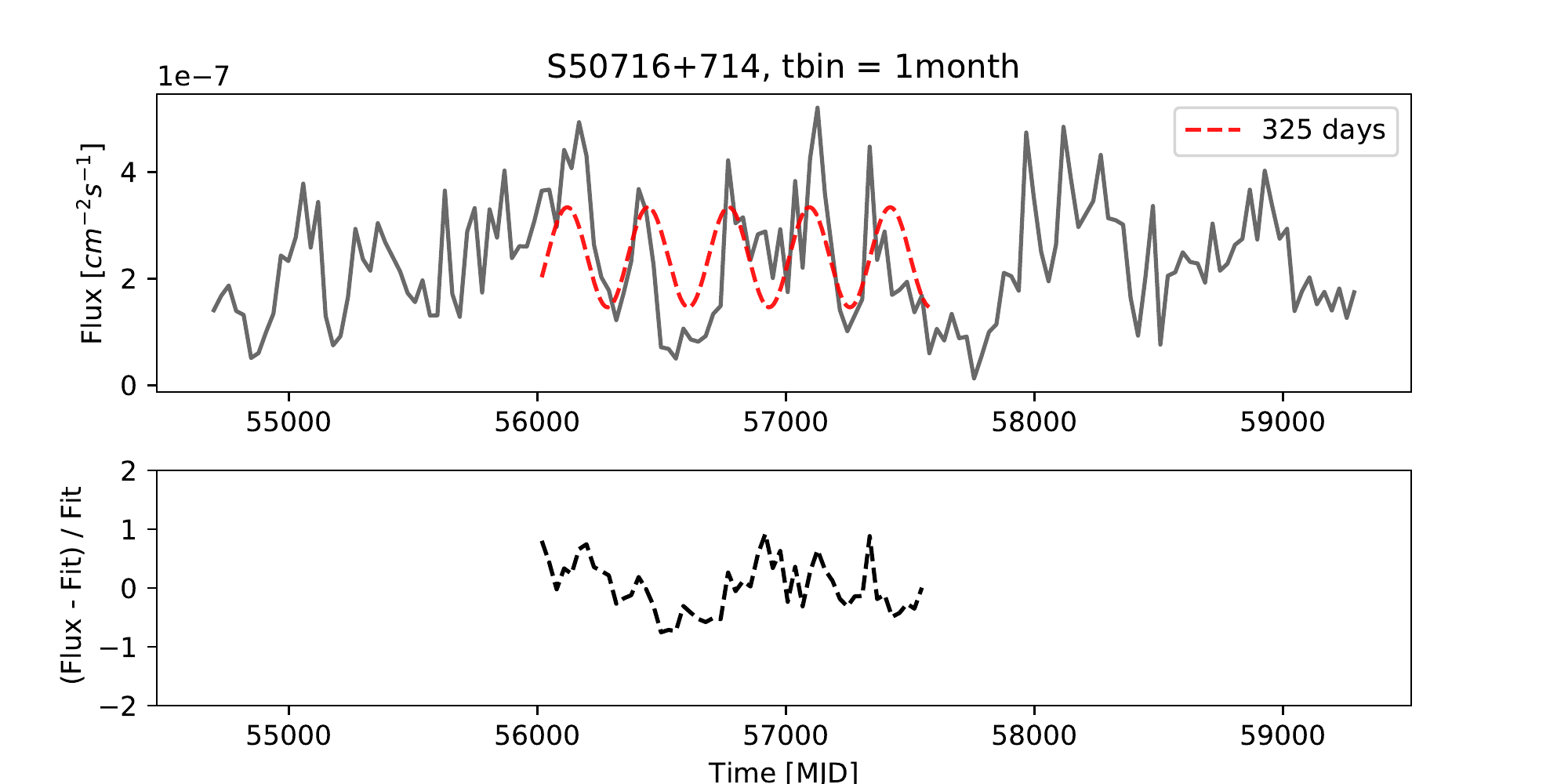}
	\end{subfigure}
	\hfill
	\begin{subfigure}[b]{0.48\textwidth}  
		\centering 
		\includegraphics[width=\textwidth]{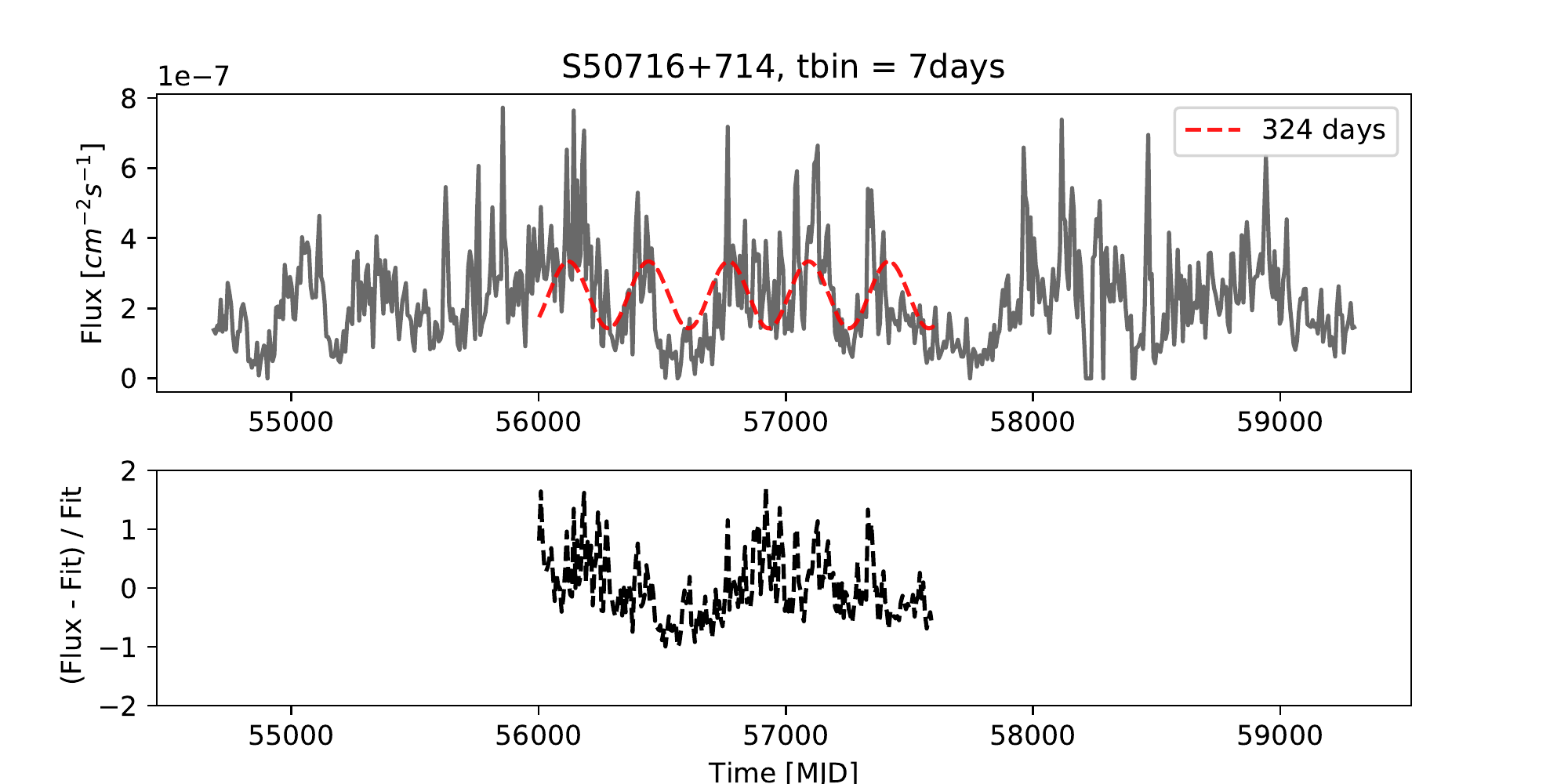}
	\end{subfigure}
	\vskip\baselineskip
	
	\hrule
	
	\centering
	\begin{subfigure}[b]{0.48\textwidth}
		\centering
		\includegraphics[width=\textwidth]{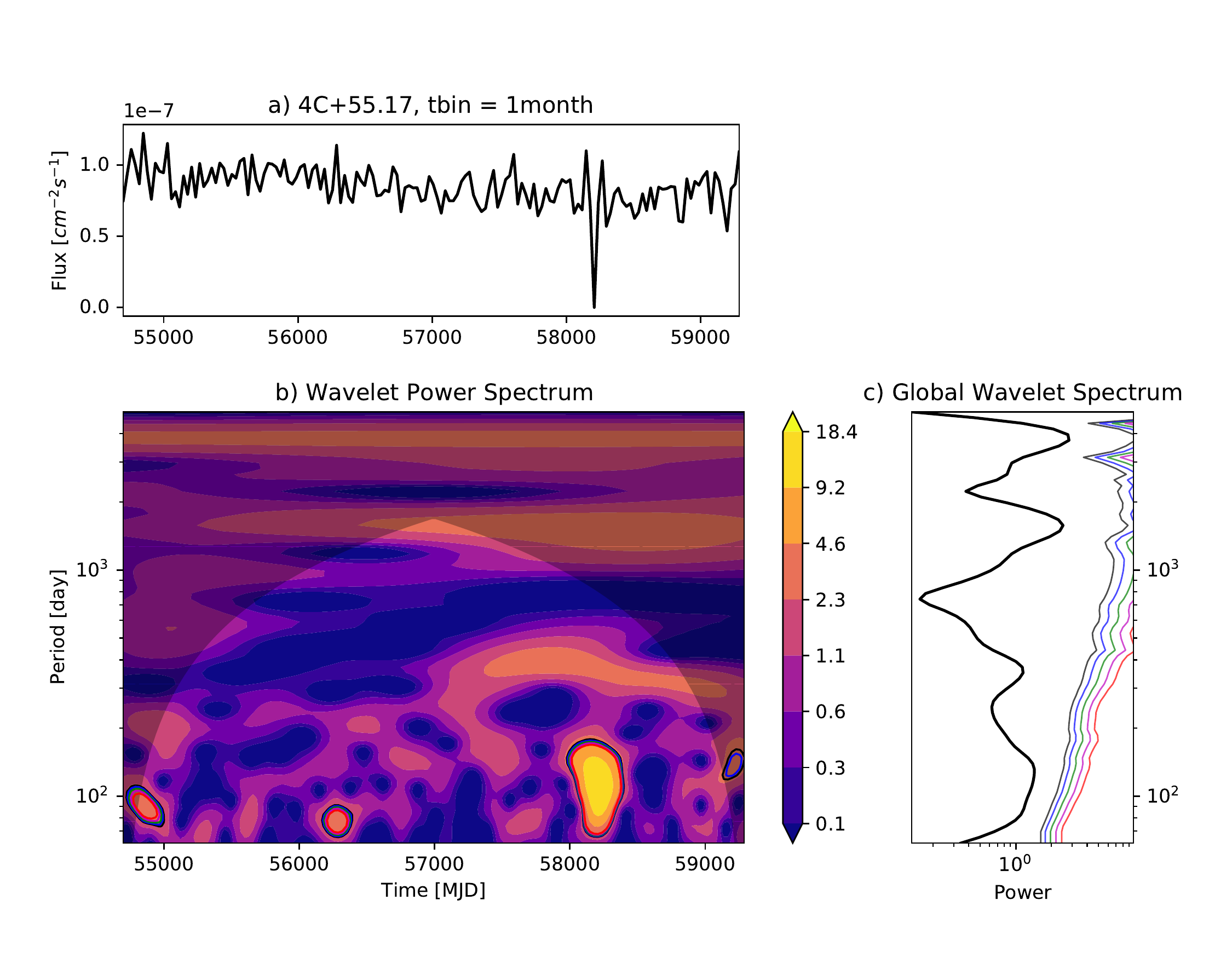}
	\end{subfigure}
	\hfill
	\begin{subfigure}[b]{0.48\textwidth}
		\centering
		\includegraphics[width=\textwidth]{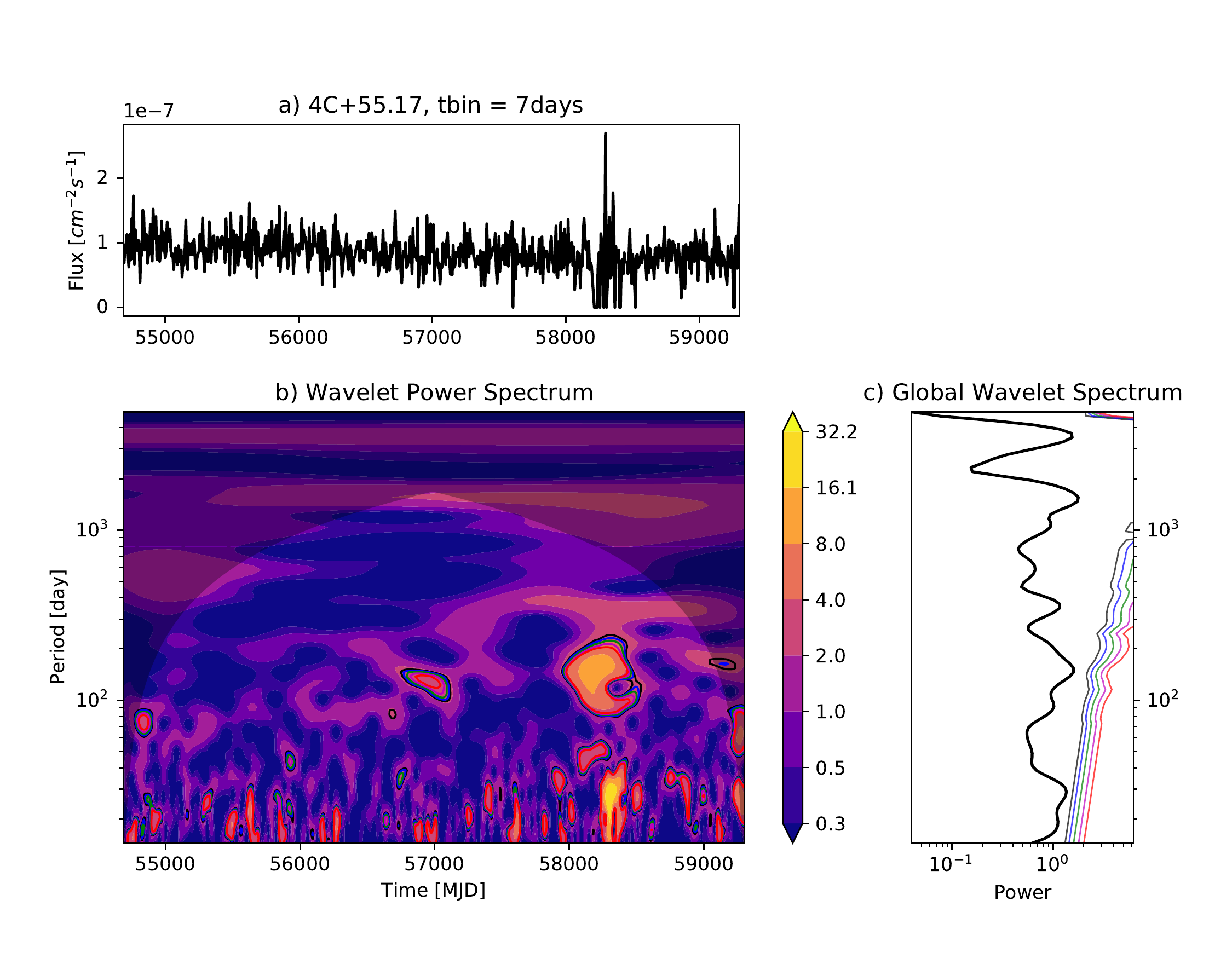}
	\end{subfigure}
	
	\caption{CWT map for monthly binned light curve (left) and weekly binned light curve (right) of S5~0716+714 and 4C~+55.17, and the fitted light curves for S5~0716+714.}
	\label{fig:CWT5}
\end{figure*}


\begin{figure*}[!htbp]
	\begin{subfigure}[b]{0.48\textwidth}
		\centering
		\includegraphics[width=\textwidth]{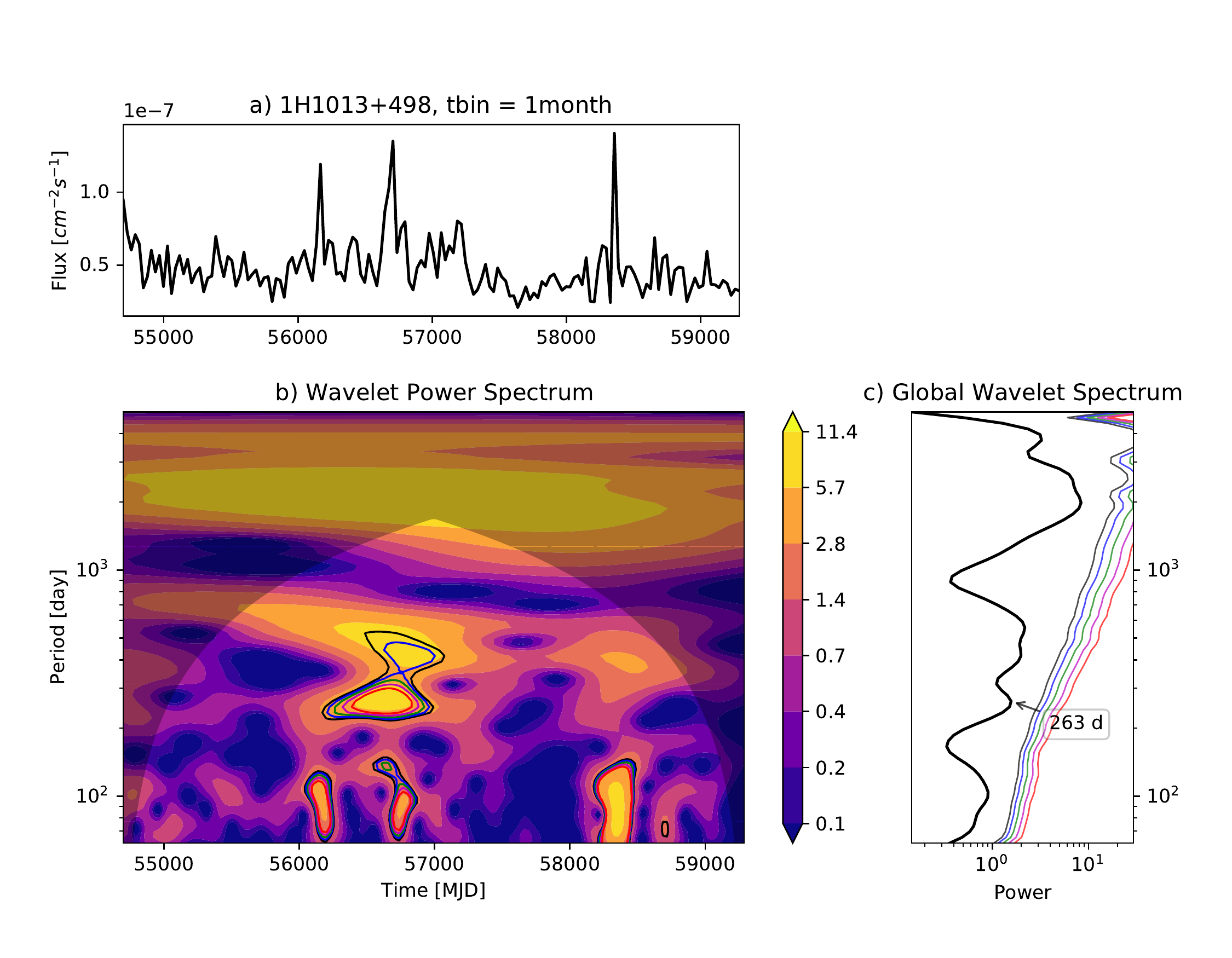}
	\end{subfigure}
	\hfill
	\begin{subfigure}[b]{0.48\textwidth}
		\centering
		\includegraphics[width=\textwidth]{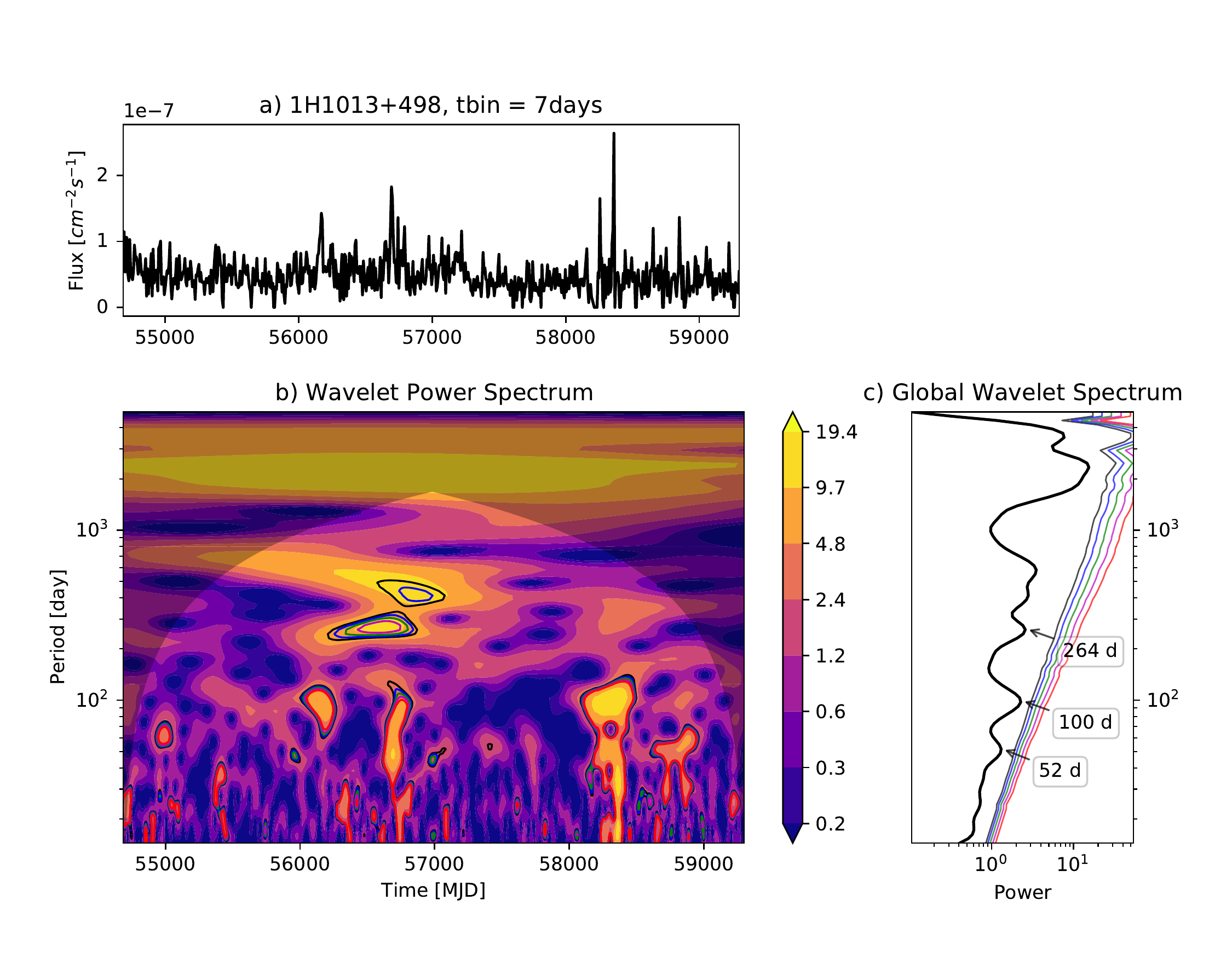}
	\end{subfigure}
	\vskip\baselineskip

	\begin{subfigure}[b]{0.48\textwidth}  
	    \centering 
	    \includegraphics[width=\textwidth]{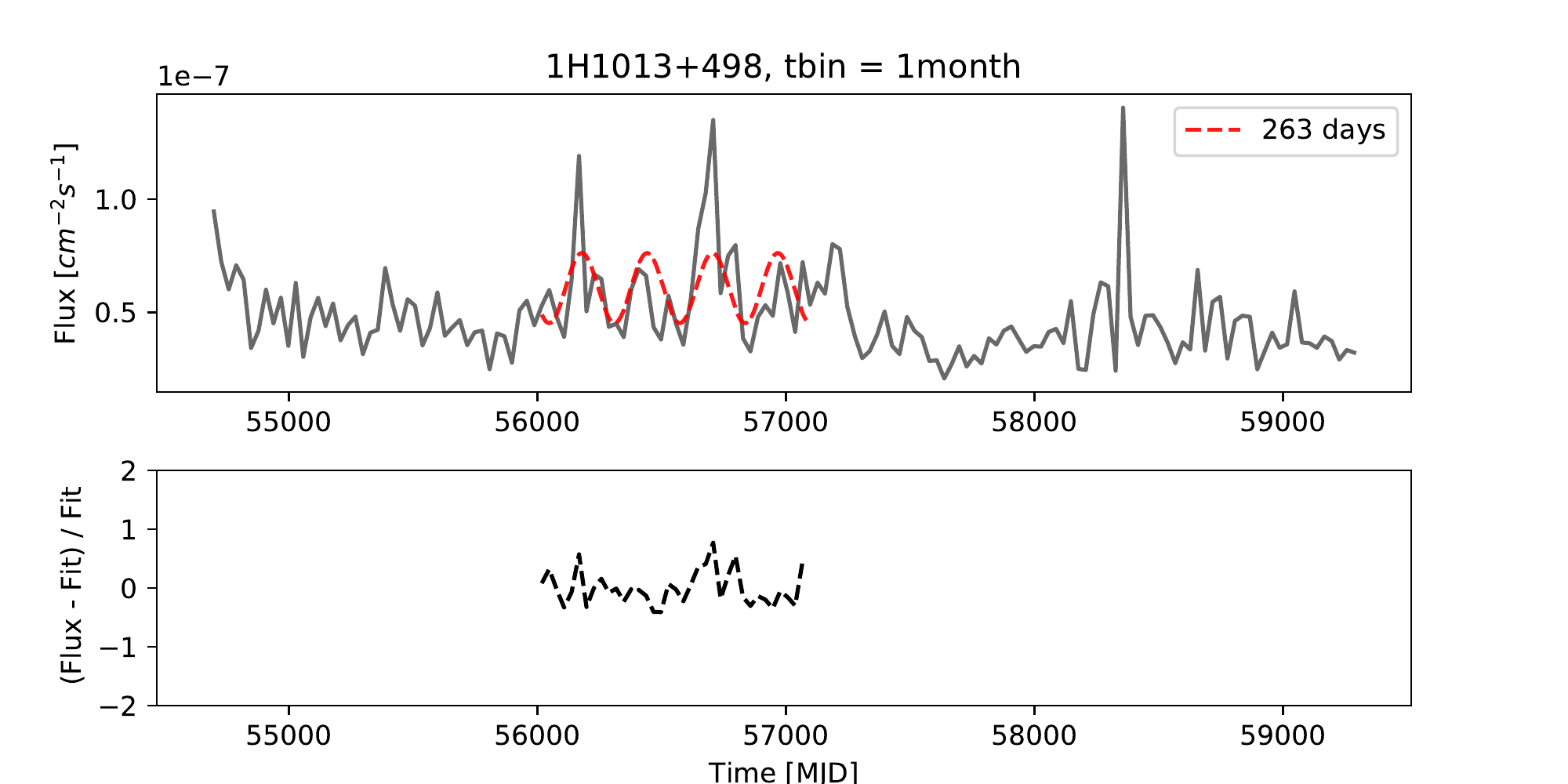}
    \end{subfigure}
	\hfill
	\begin{subfigure}[b]{0.48\textwidth}  
		\centering 
		\includegraphics[width=\textwidth]{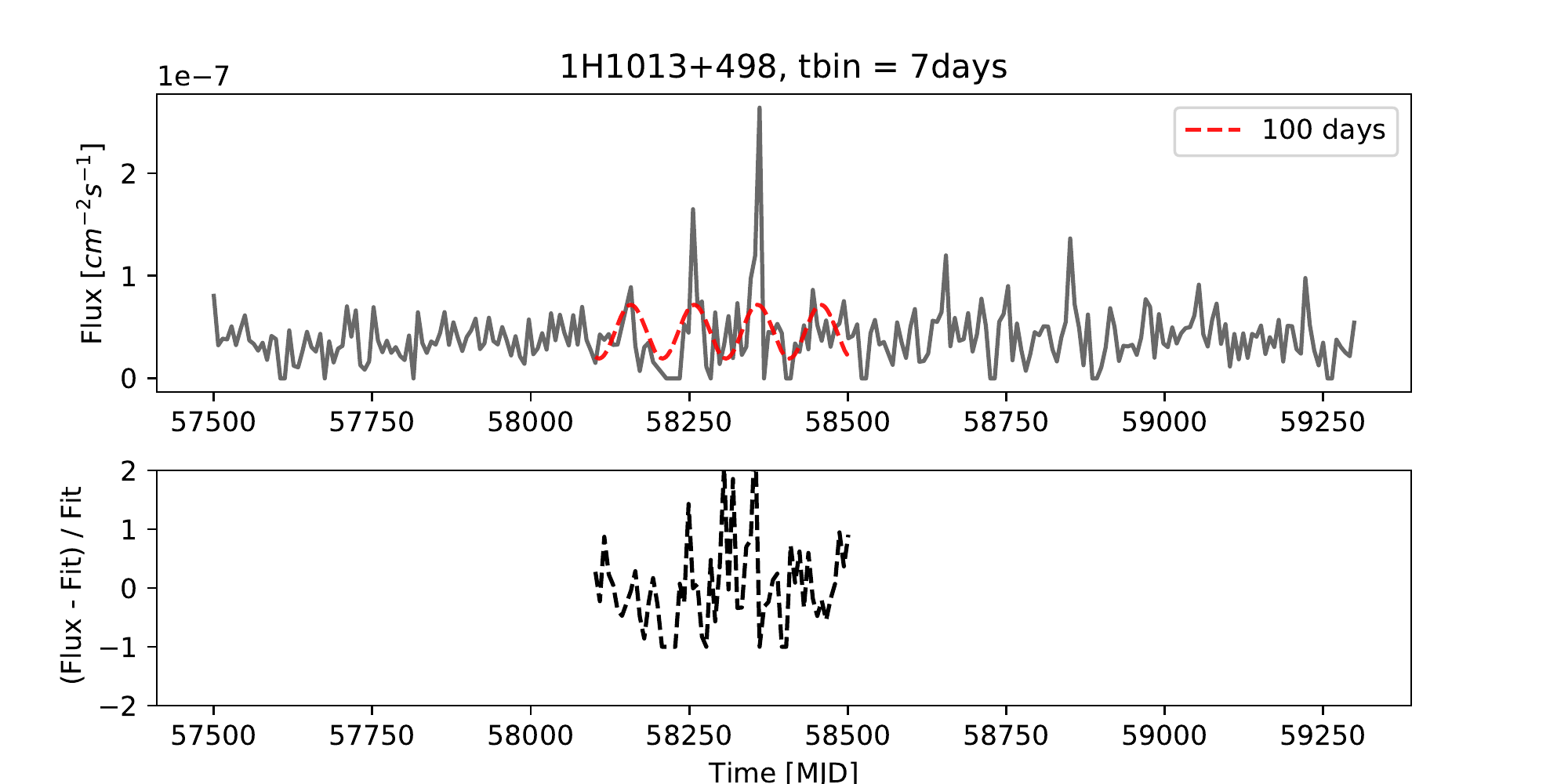}
	\end{subfigure}
	
	\vskip\baselineskip
    
    \begin{subfigure}[b]{0.48\textwidth}  
		\centering 
		\includegraphics[width=\textwidth]{./Figures/Fit/blanc.png}
	\end{subfigure}
	\hfill
	\begin{subfigure}[b]{0.48\textwidth}  
		\centering 
		\includegraphics[width=\textwidth]{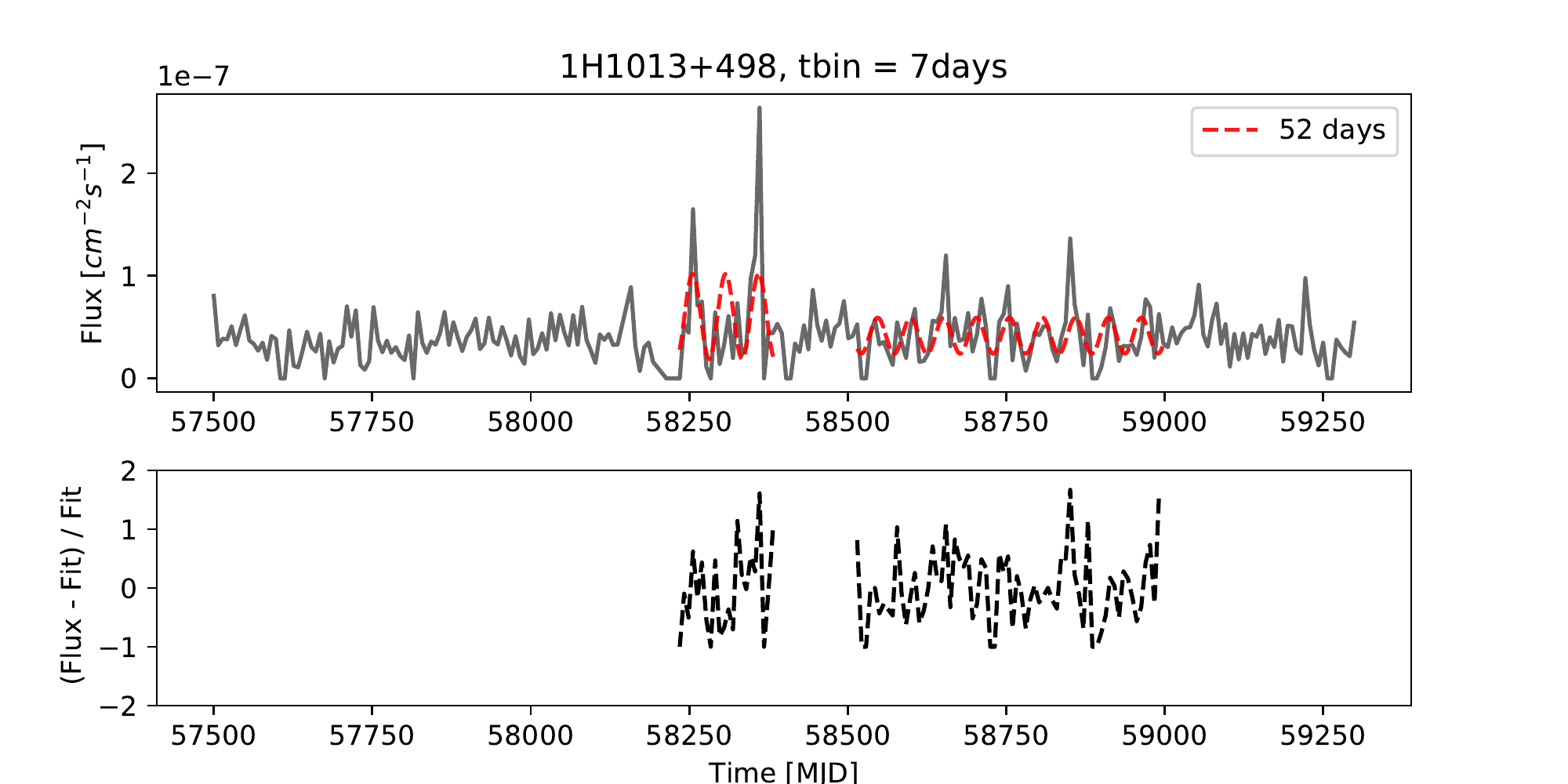}
	\end{subfigure}
	
	\caption{CWT map for monthly binned light curve (left) and weekly binned light curve (right) of 1H~1013+498, and the fitted light curves.}
	\label{fig:CWT6}
\end{figure*}


\begin{figure*}[!htbp]
	\centering
	
	\begin{subfigure}[b]{0.48\textwidth}   
		\centering 
		\includegraphics[width=\textwidth]{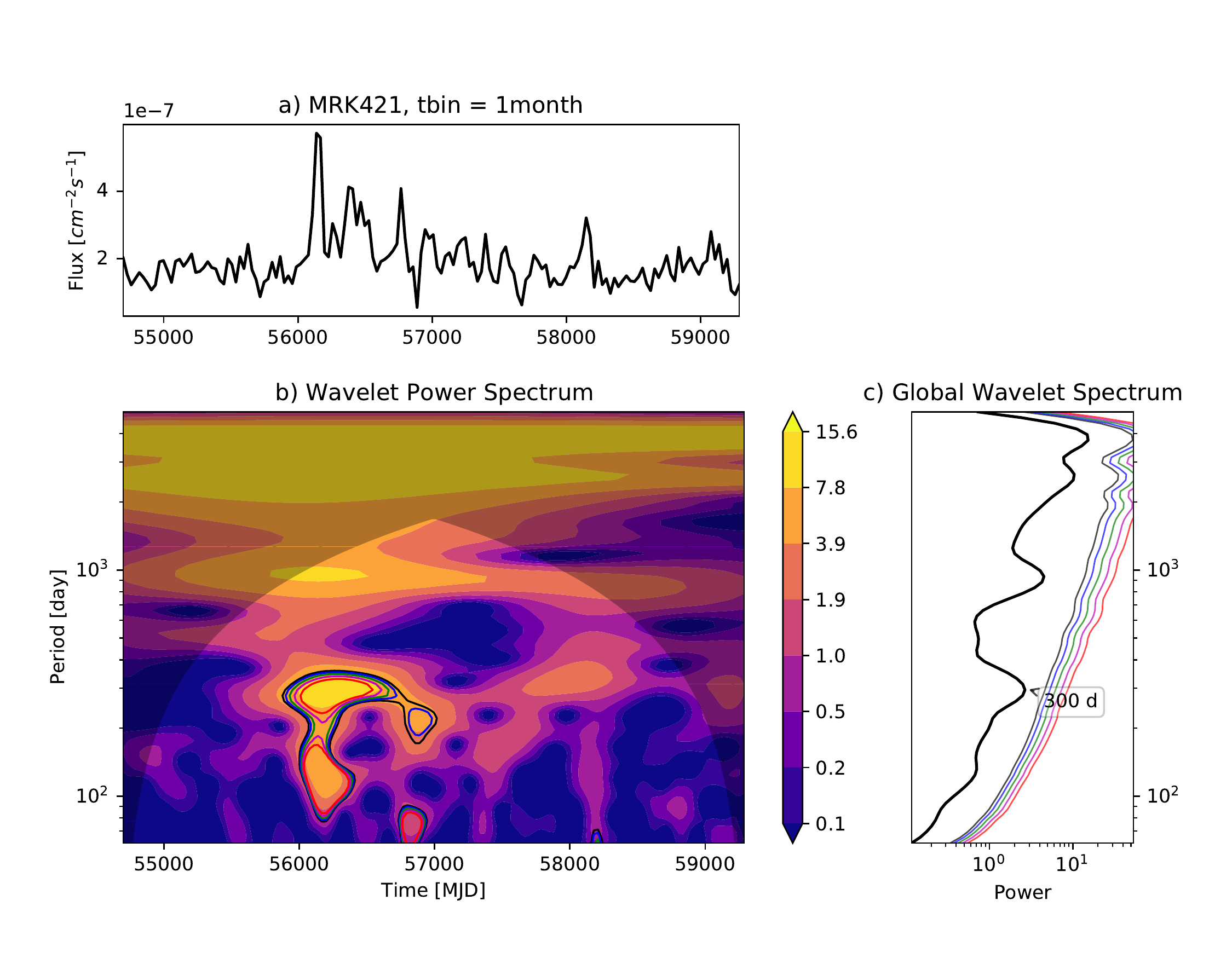}
	\end{subfigure}
	\hfill
	\begin{subfigure}[b]{0.48\textwidth}   
		\centering 
		\includegraphics[width=\textwidth]{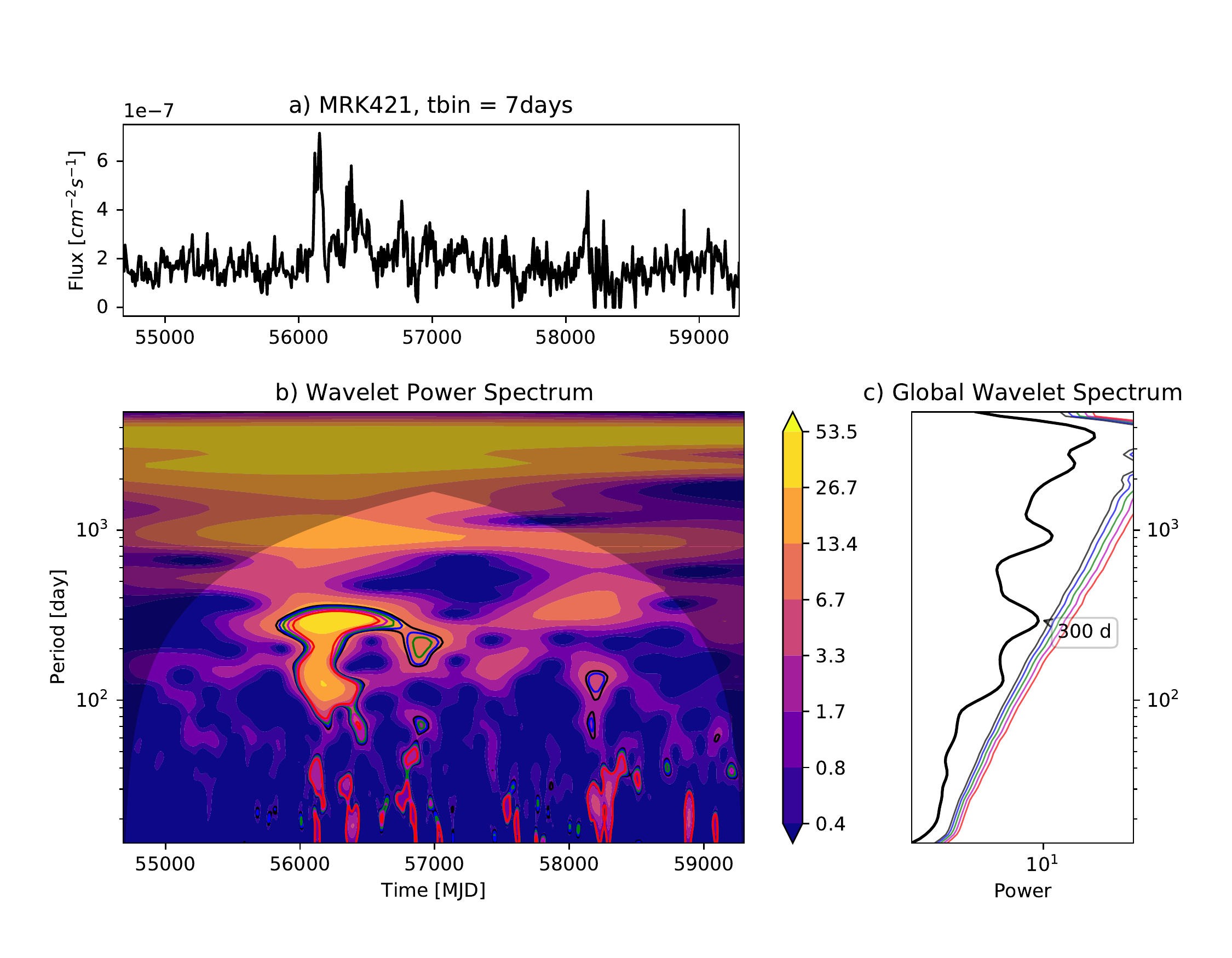}
	\end{subfigure}
	\vskip\baselineskip
	
	\begin{subfigure}[b]{0.48\textwidth}  
		\centering 
		\includegraphics[width=\textwidth]{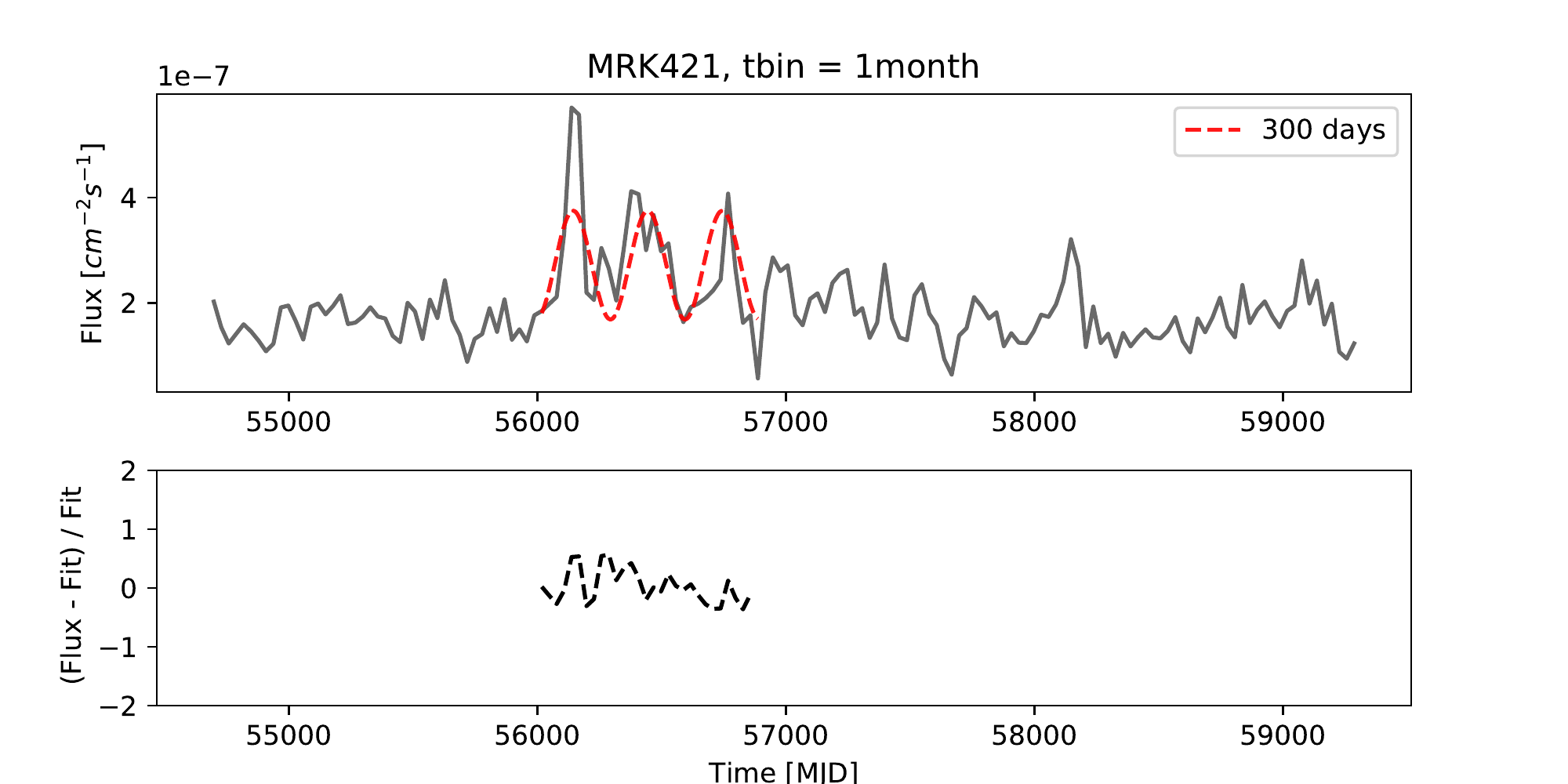}
	\end{subfigure}
	\hfill
	\begin{subfigure}[b]{0.48\textwidth}  
		\centering 
		\includegraphics[width=\textwidth]{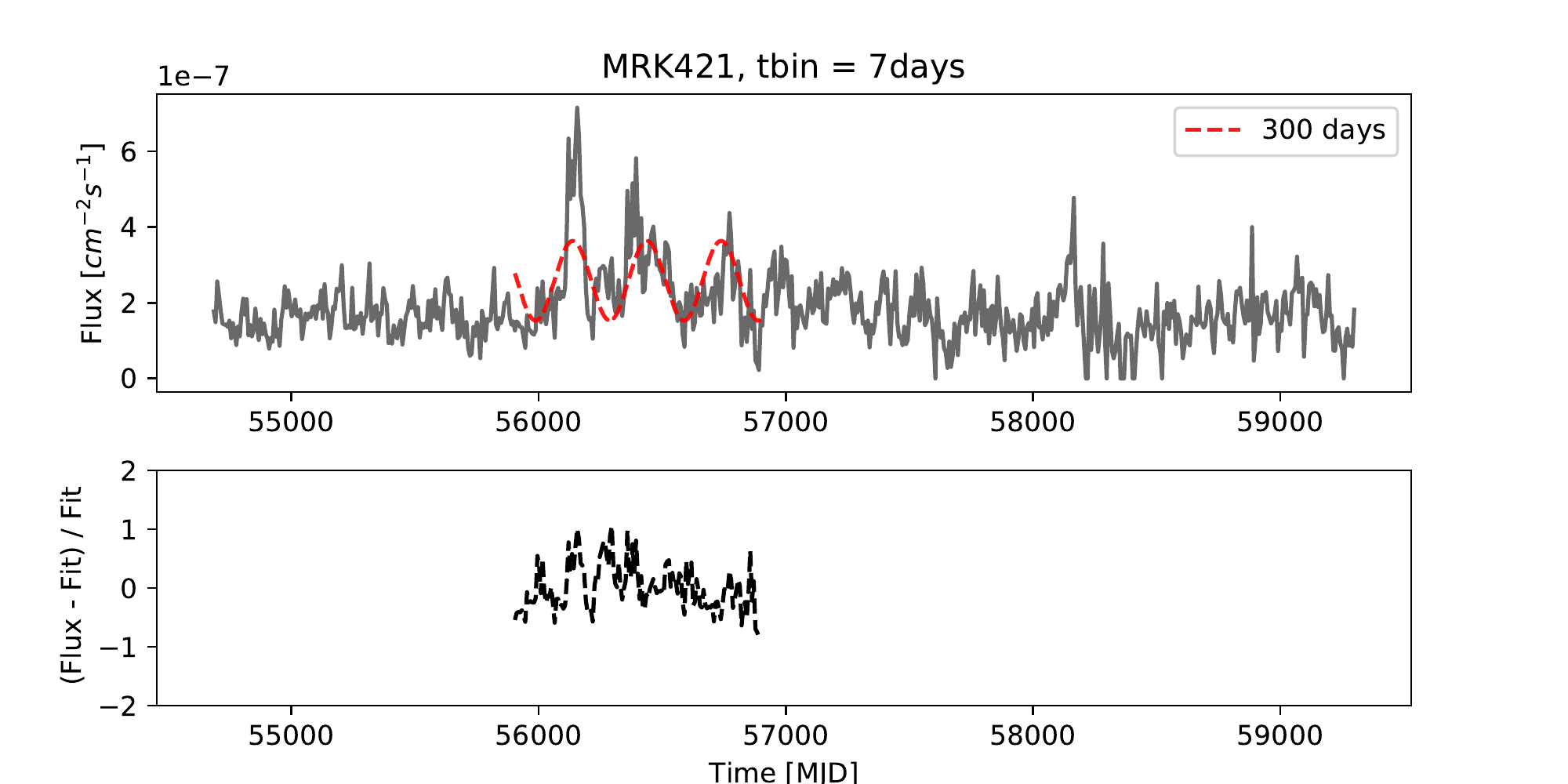}
	\end{subfigure}
    \vskip\baselineskip
	
	\hrule

	\begin{subfigure}[b]{0.48\textwidth}
		\centering
		\includegraphics[width=\textwidth]{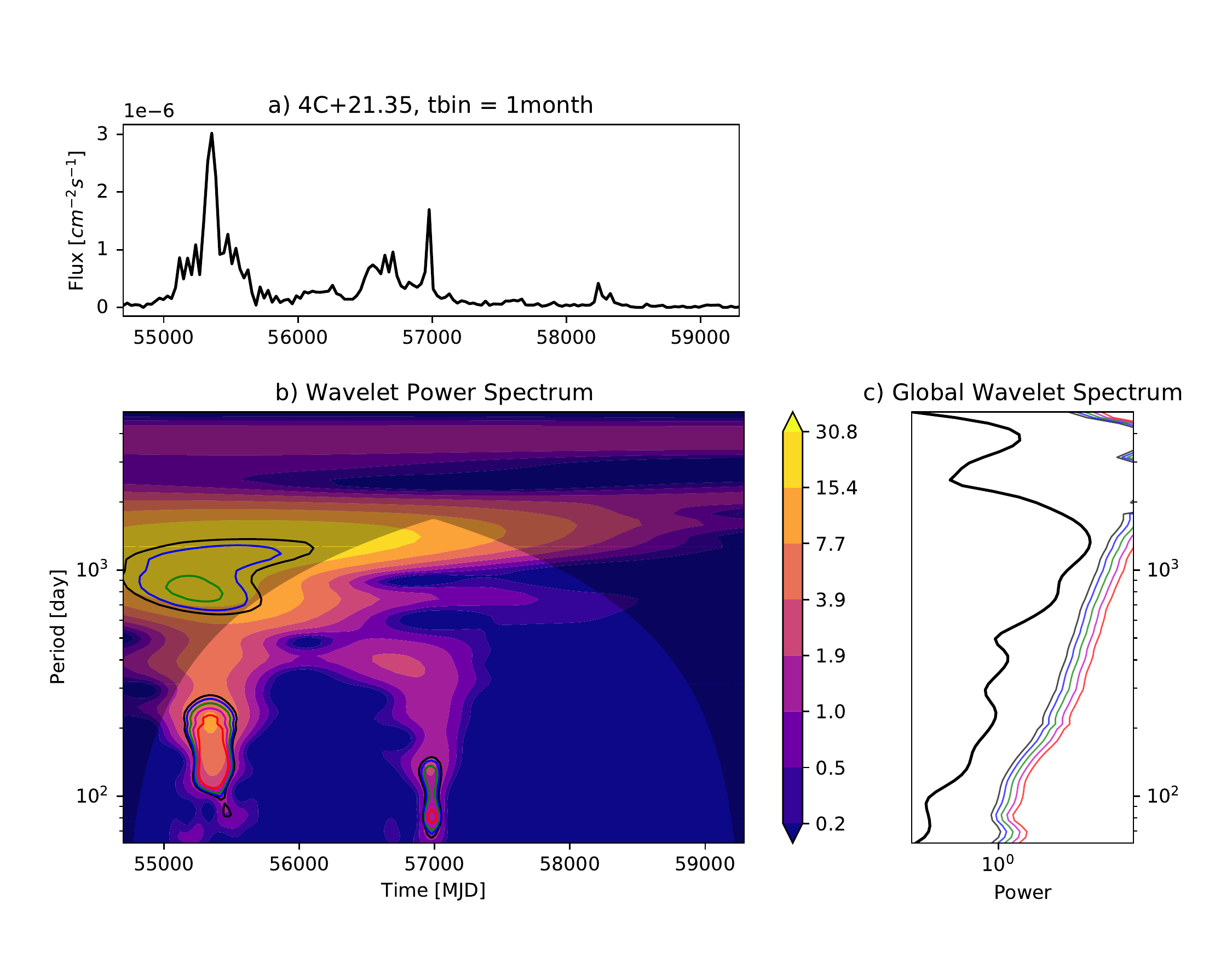}
	\end{subfigure}
	\hfill
	\begin{subfigure}[b]{0.48\textwidth}
		\centering
		\includegraphics[width=\textwidth]{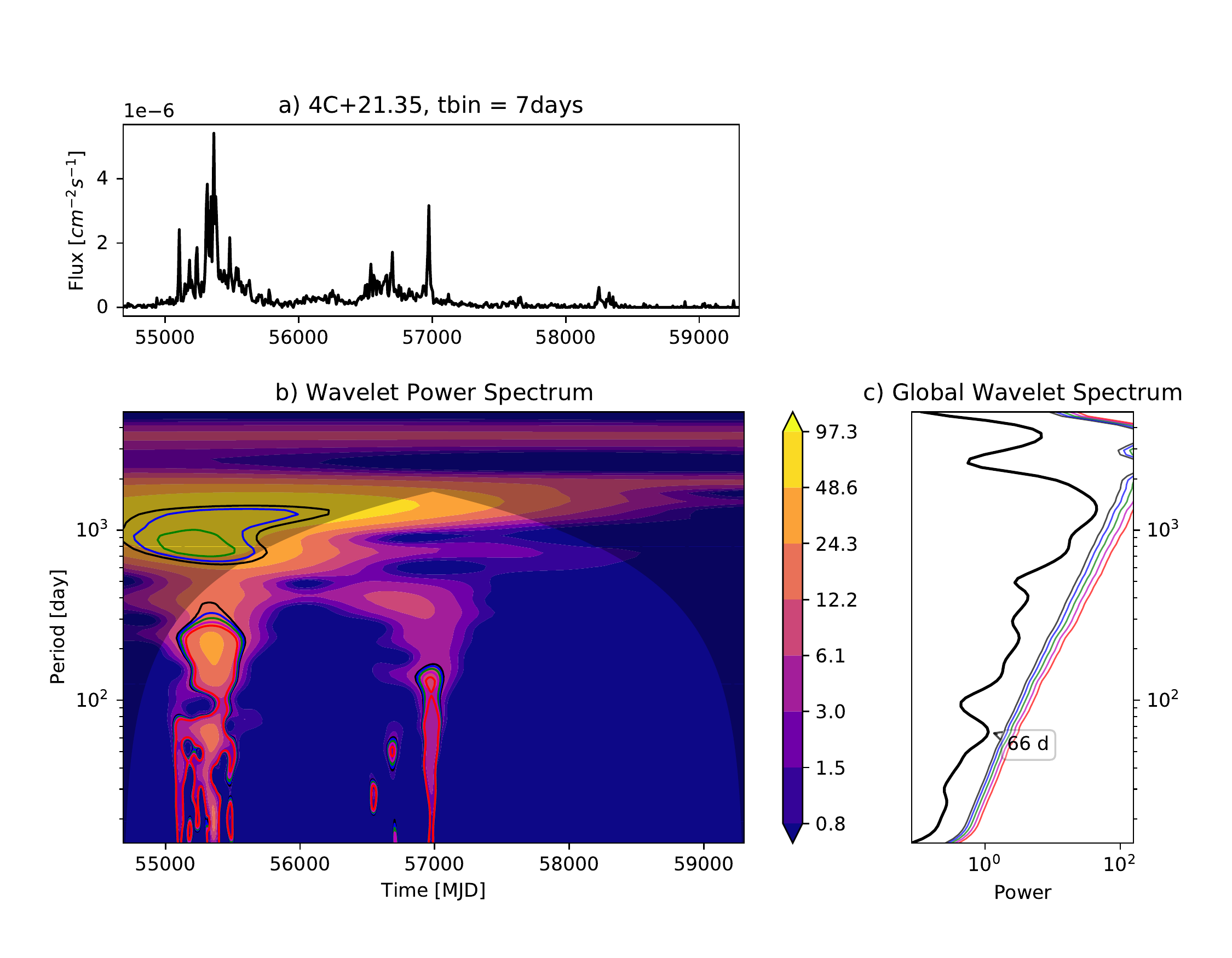}
	\end{subfigure}
	\vskip\baselineskip
	
	\begin{subfigure}[b]{0.48\textwidth}  
		\centering 
		\includegraphics[width=\textwidth]{./Figures/Fit/blanc.png}
	\end{subfigure}
	\hfill
	\begin{subfigure}[b]{0.48\textwidth}  
		\centering 
		\includegraphics[width=\textwidth]{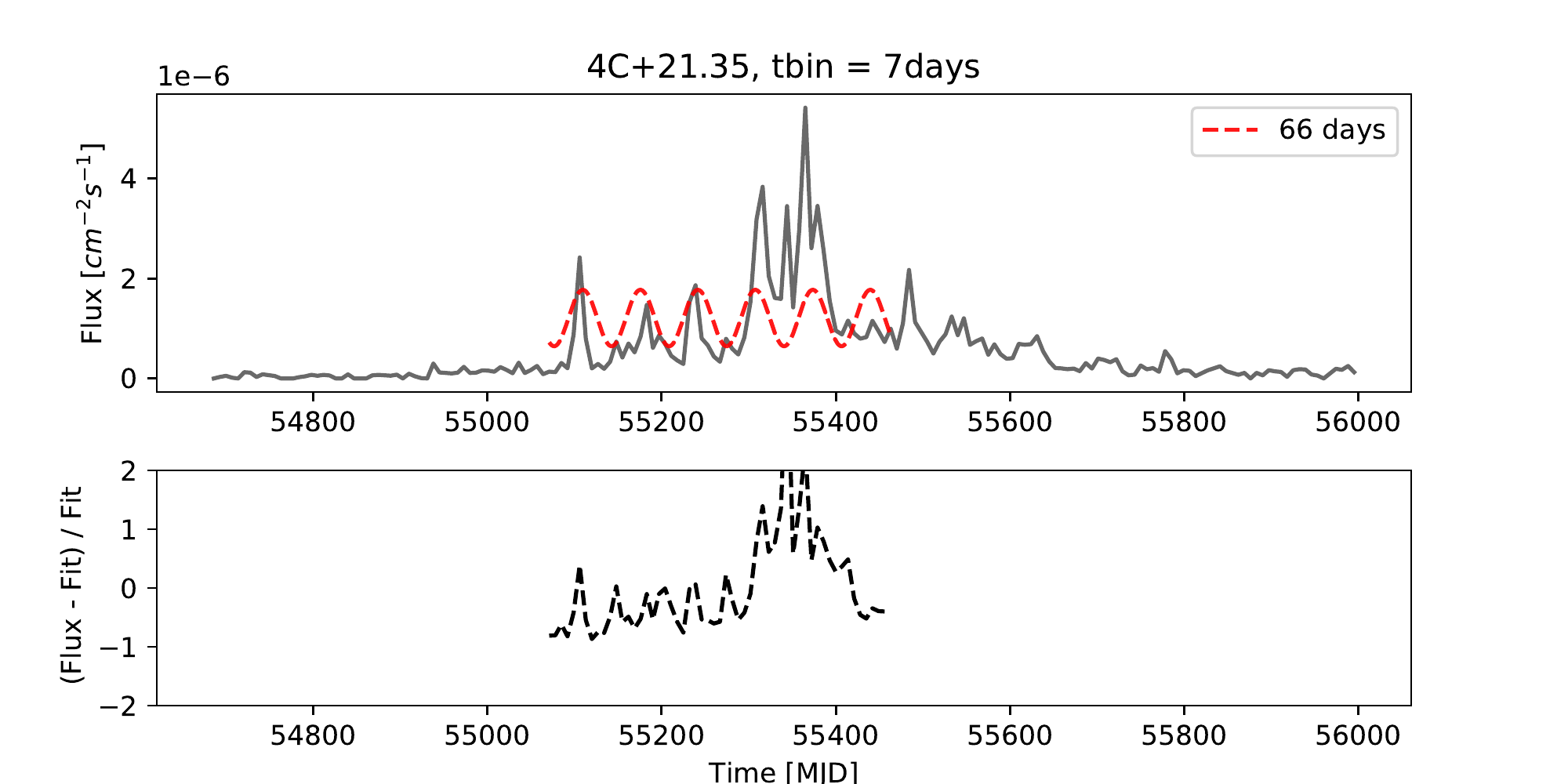}
	\end{subfigure}
	\vskip\baselineskip
	
	\caption{CWT map for monthly binned light curve (left) and weekly binned light curve (right) of MRK~421 and 4C~+21.35, and the respective fitted light curves.}
	\label{fig:CWT7}
\end{figure*}


\begin{figure*}[!htbp]
	\centering
	
	\begin{subfigure}[b]{0.48\textwidth}
		\centering
		\includegraphics[width=\textwidth]{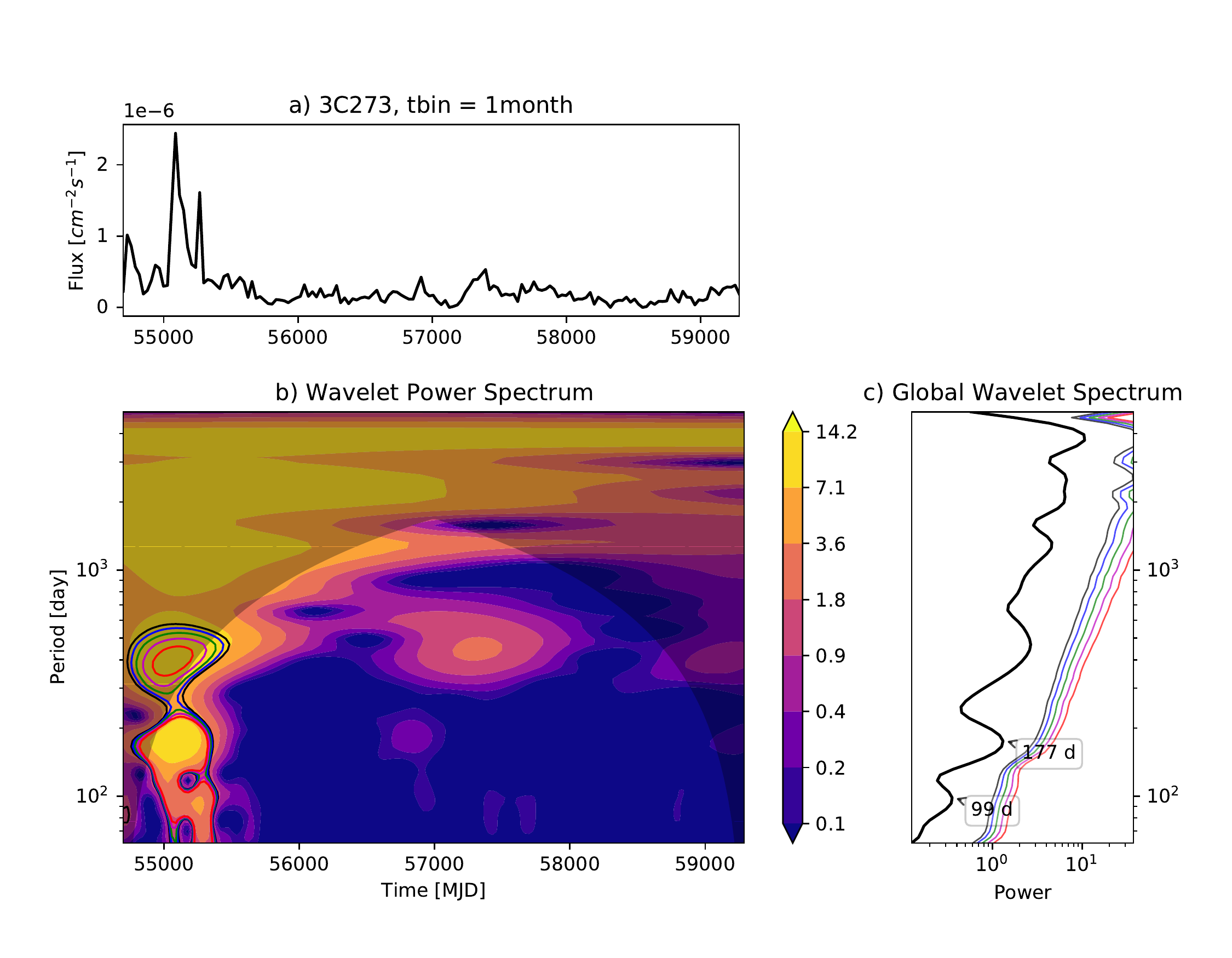}
	\end{subfigure}
	\hfill
	\begin{subfigure}[b]{0.48\textwidth}
		\centering
		\includegraphics[width=\textwidth]{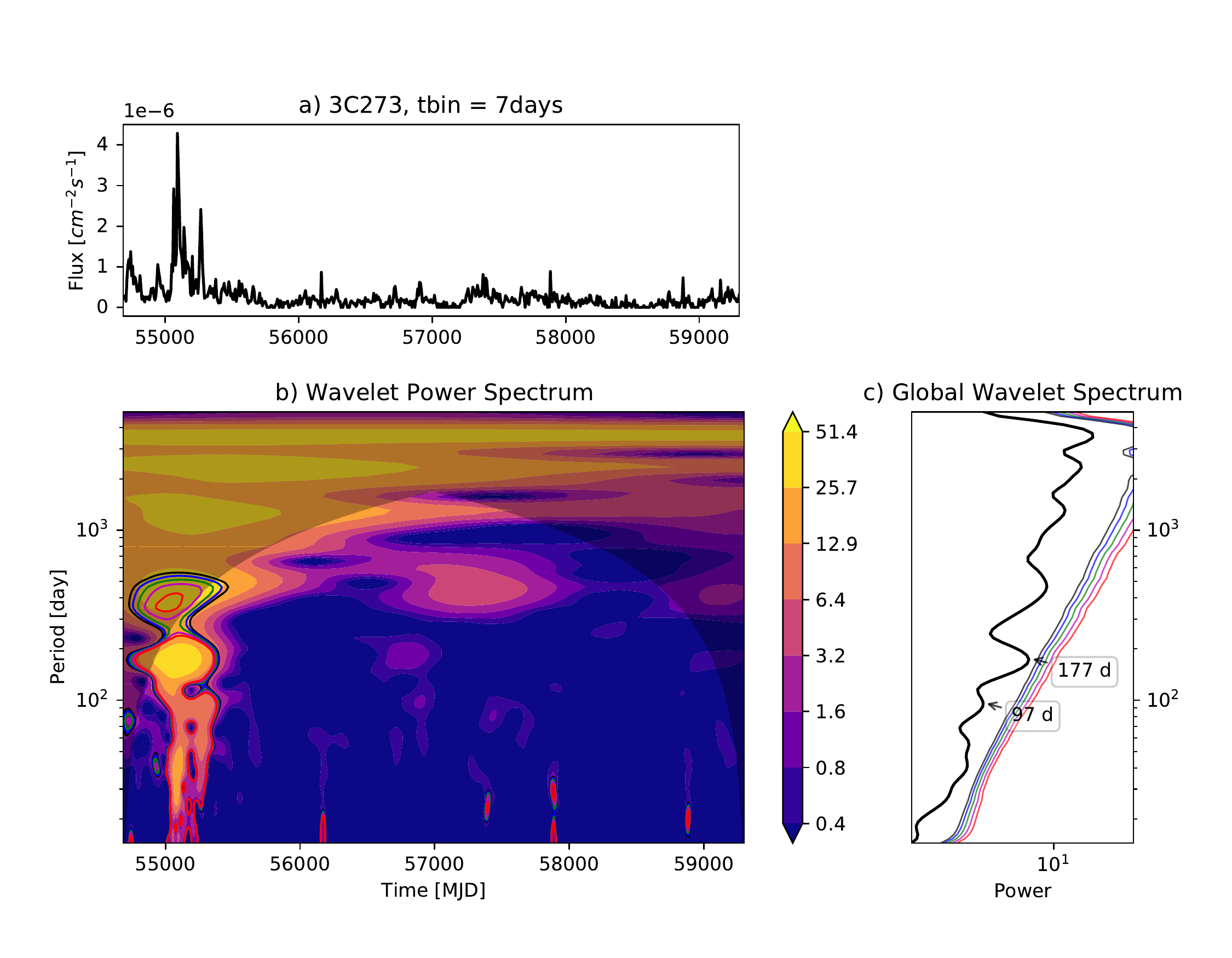}
	\end{subfigure}
	\vskip\baselineskip
	
	\begin{subfigure}[b]{0.48\textwidth}
		\centering
		\includegraphics[width=\textwidth]{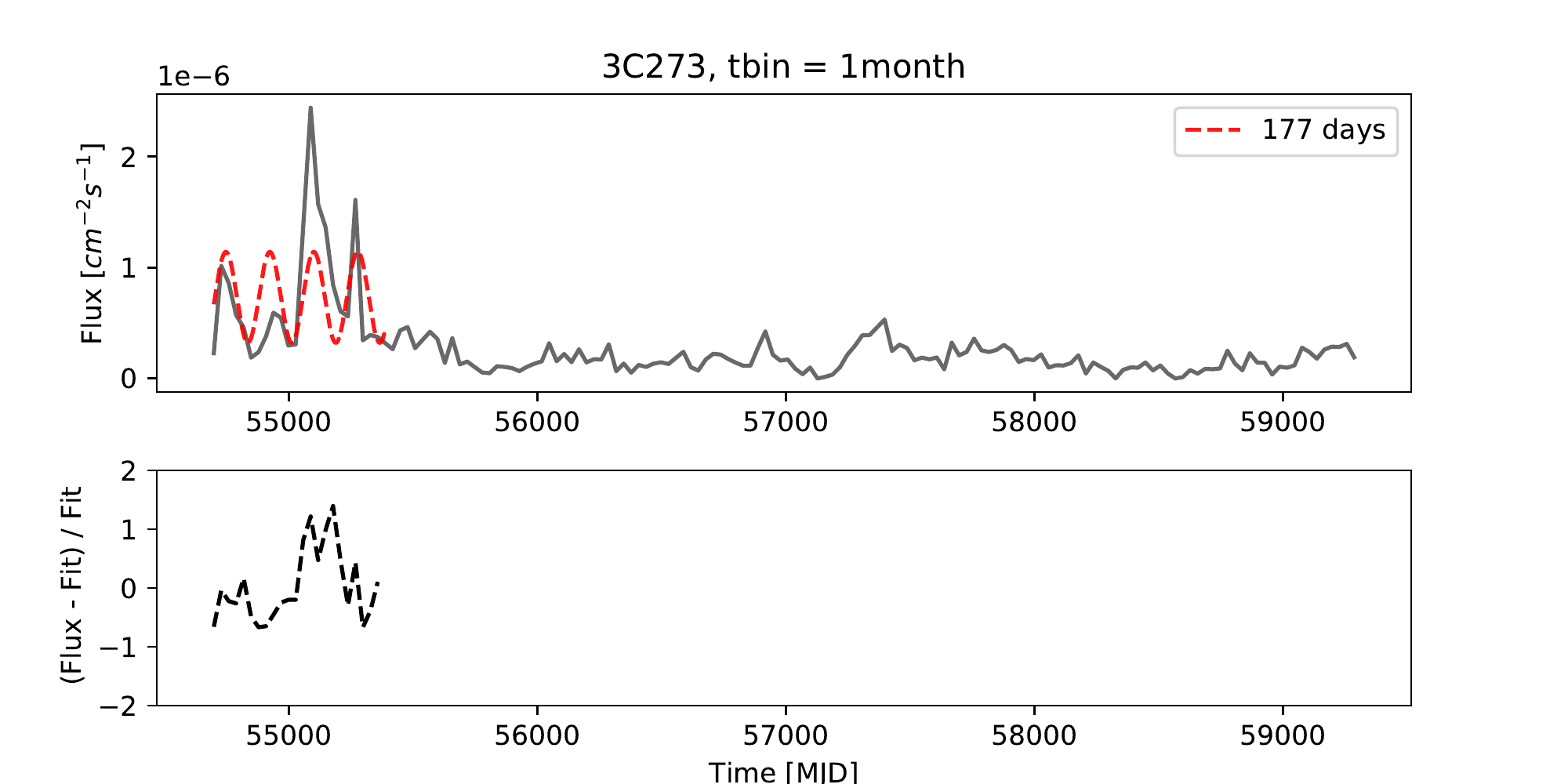}
	\end{subfigure}
	\hfill
	\begin{subfigure}[b]{0.48\textwidth}
		\centering
		\includegraphics[width=\textwidth]{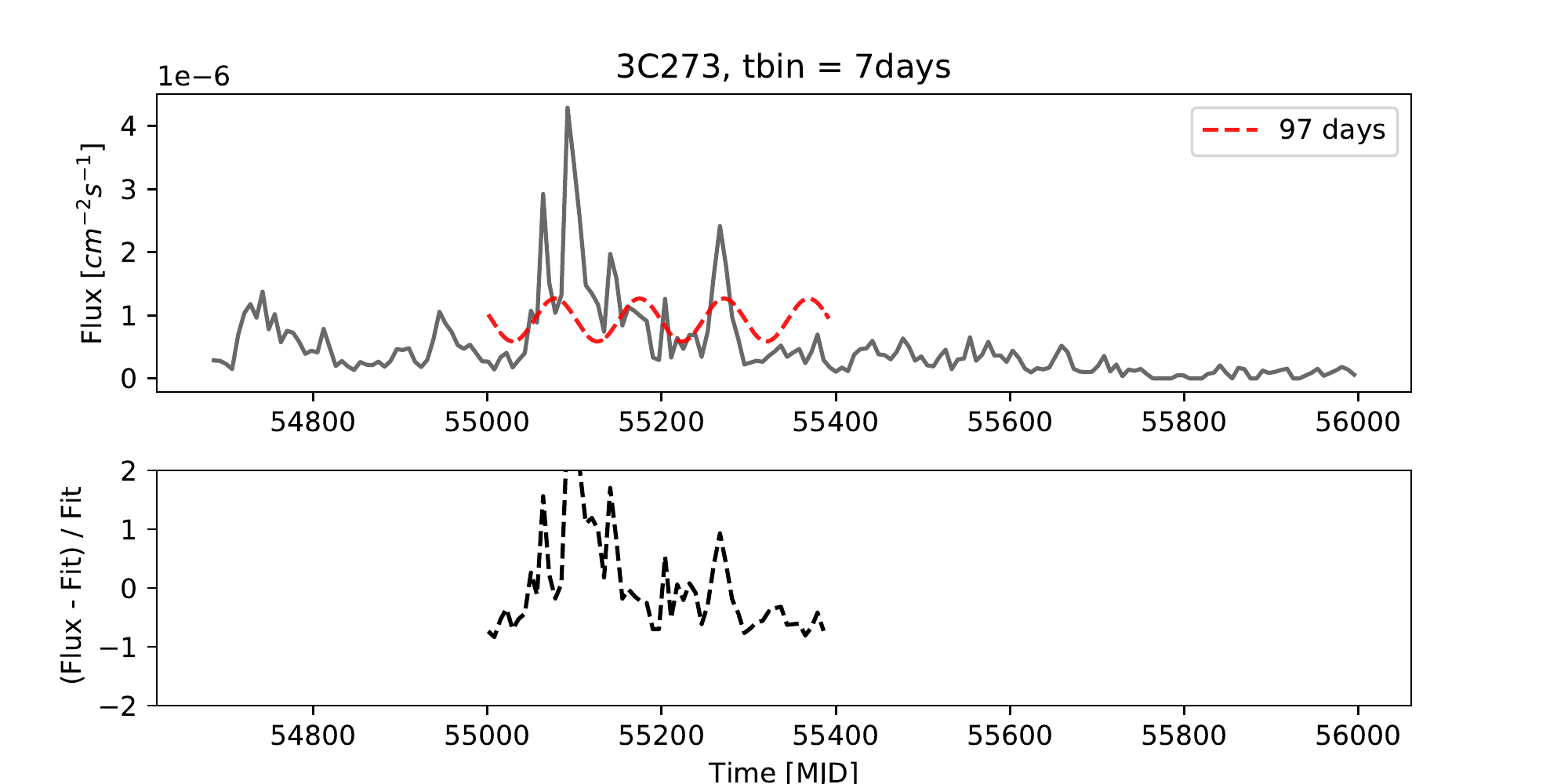}
	\end{subfigure}
	\vskip\baselineskip
	
	\hrule
	
	\begin{subfigure}[b]{0.48\textwidth}   
		\centering 
		\includegraphics[width=\textwidth]{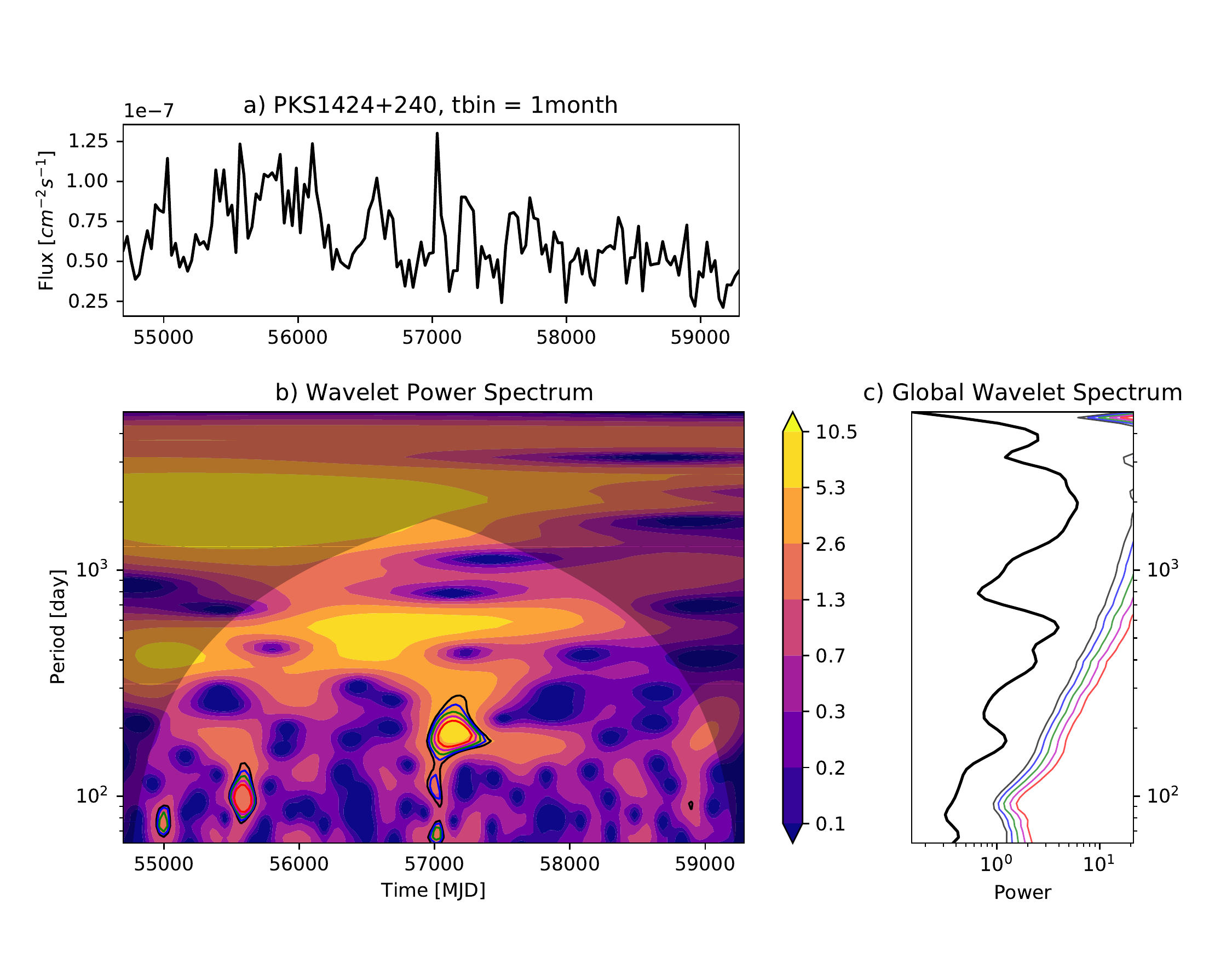}
	\end{subfigure}
	\hfill
	\begin{subfigure}[b]{0.48\textwidth}   
		\centering 
		\includegraphics[width=\textwidth]{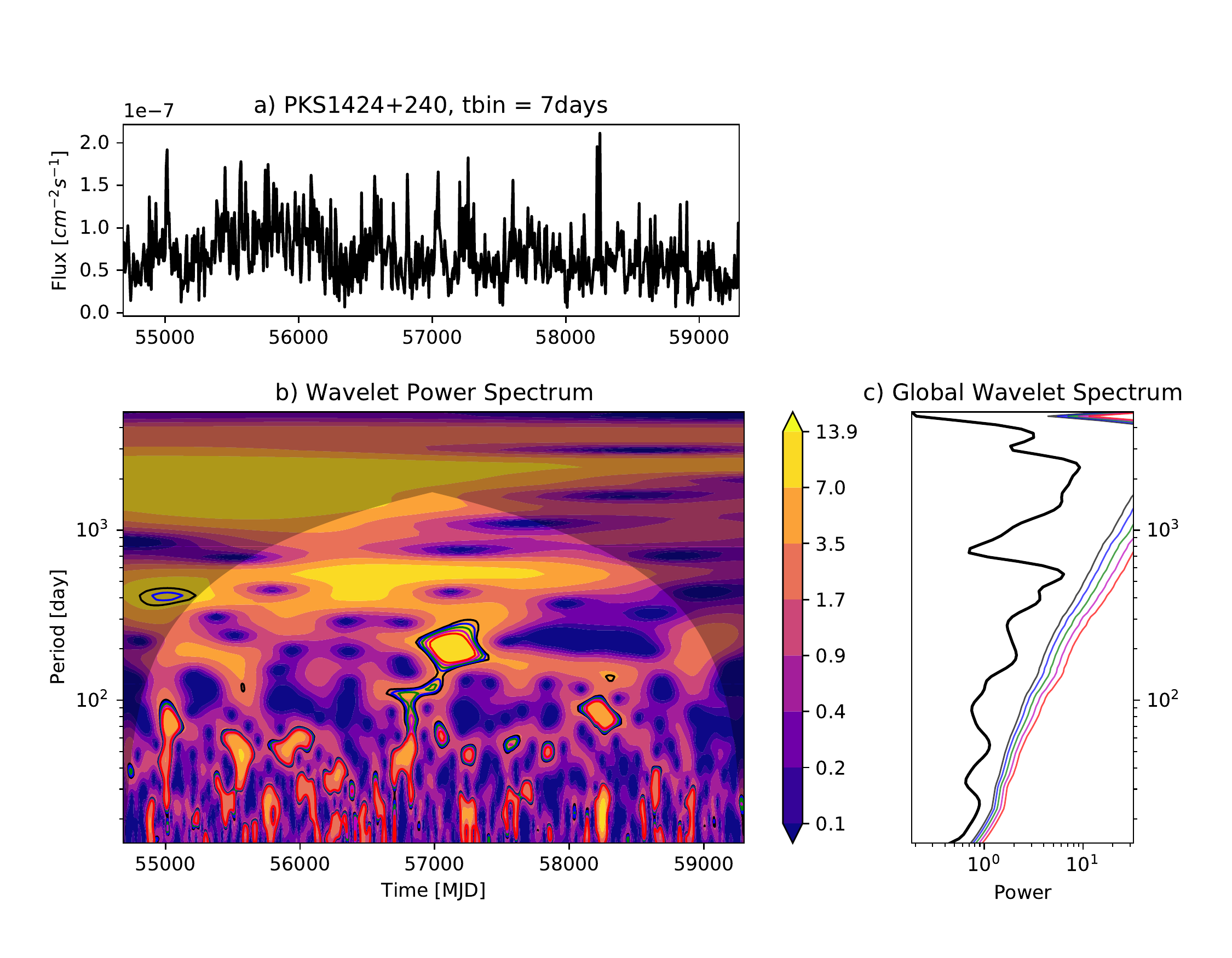}
	\end{subfigure}
	

	\caption{CWT map for monthly binned light curve (left) and weekly binned light curve (right) of 3C~273 and  PKS~1424+240, and the fitted light curves for 3C~273.}
	\label{fig:CWT8}
\end{figure*}


\begin{figure*}[!htbp]
	\centering
	
	\begin{subfigure}[b]{0.48\textwidth}
		\centering
		\includegraphics[width=\textwidth]{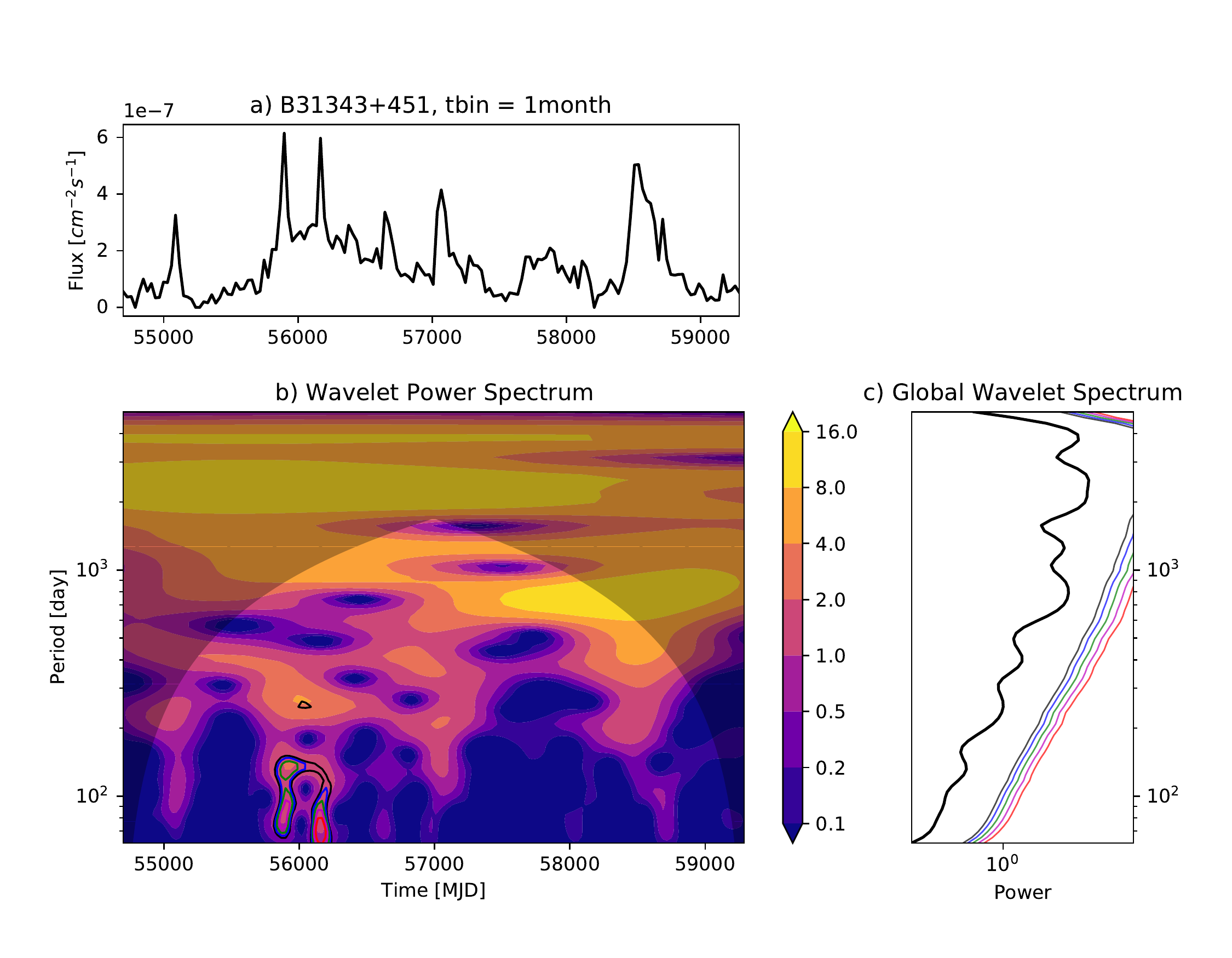}
	\end{subfigure}
	\hfill
	\begin{subfigure}[b]{0.48\textwidth}
		\centering
		\includegraphics[width=\textwidth]{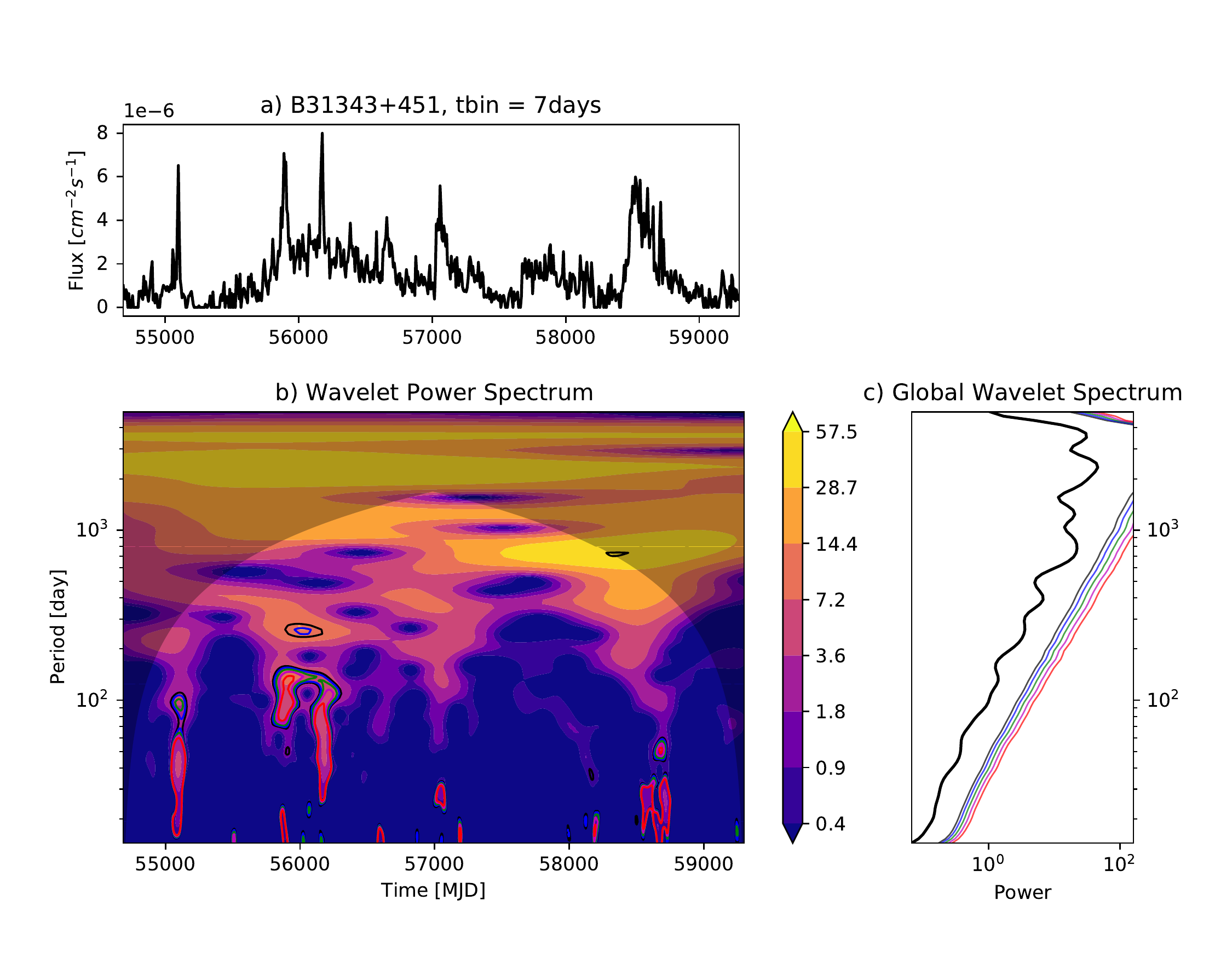}
	\end{subfigure}
	\vskip\baselineskip
	
	\hrule
	
	\begin{subfigure}[b]{0.48\textwidth}   
		\centering 
		\includegraphics[width=\textwidth]{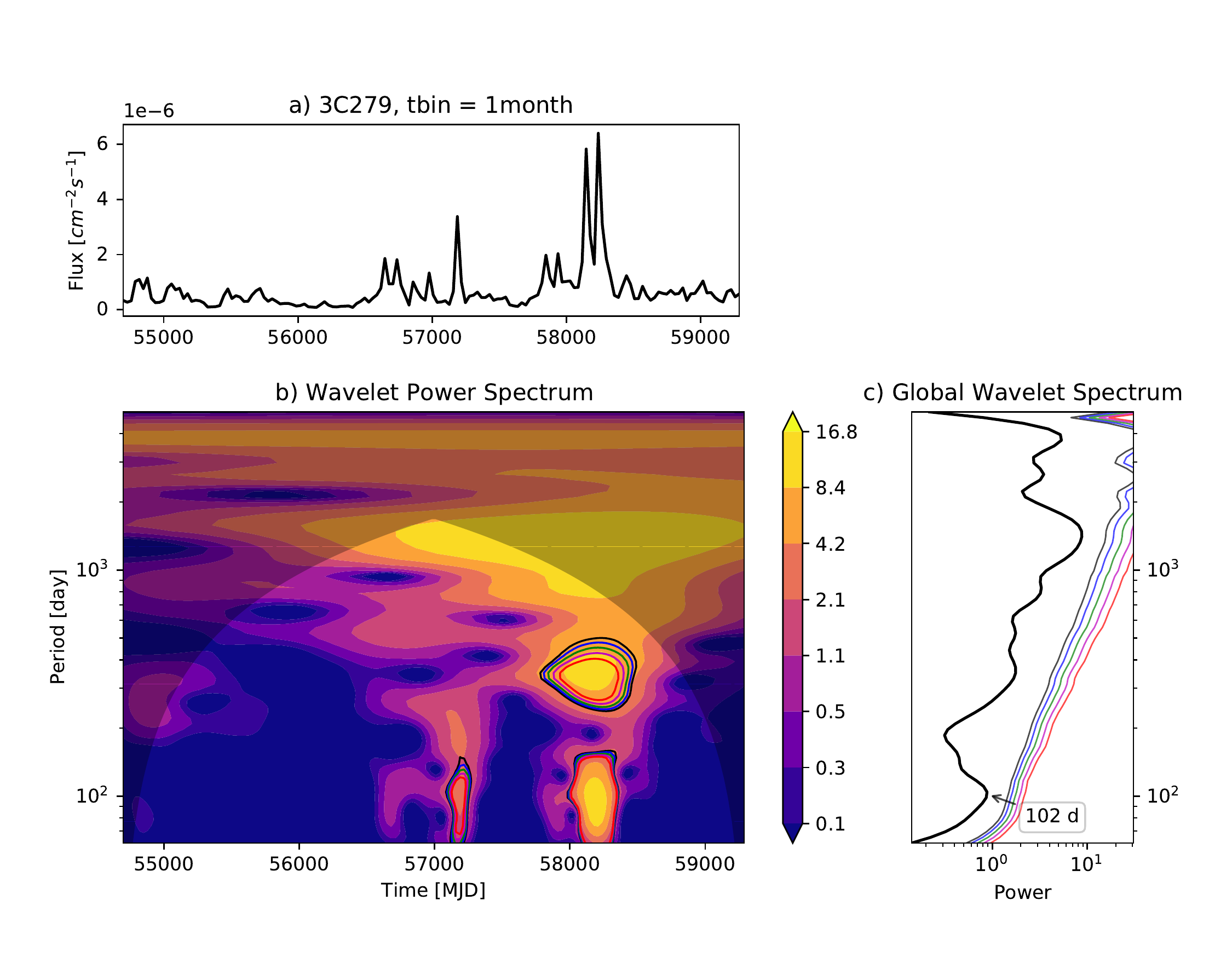}
	\end{subfigure}
	\hfill
	\begin{subfigure}[b]{0.48\textwidth}   
		\centering 
		\includegraphics[width=\textwidth]{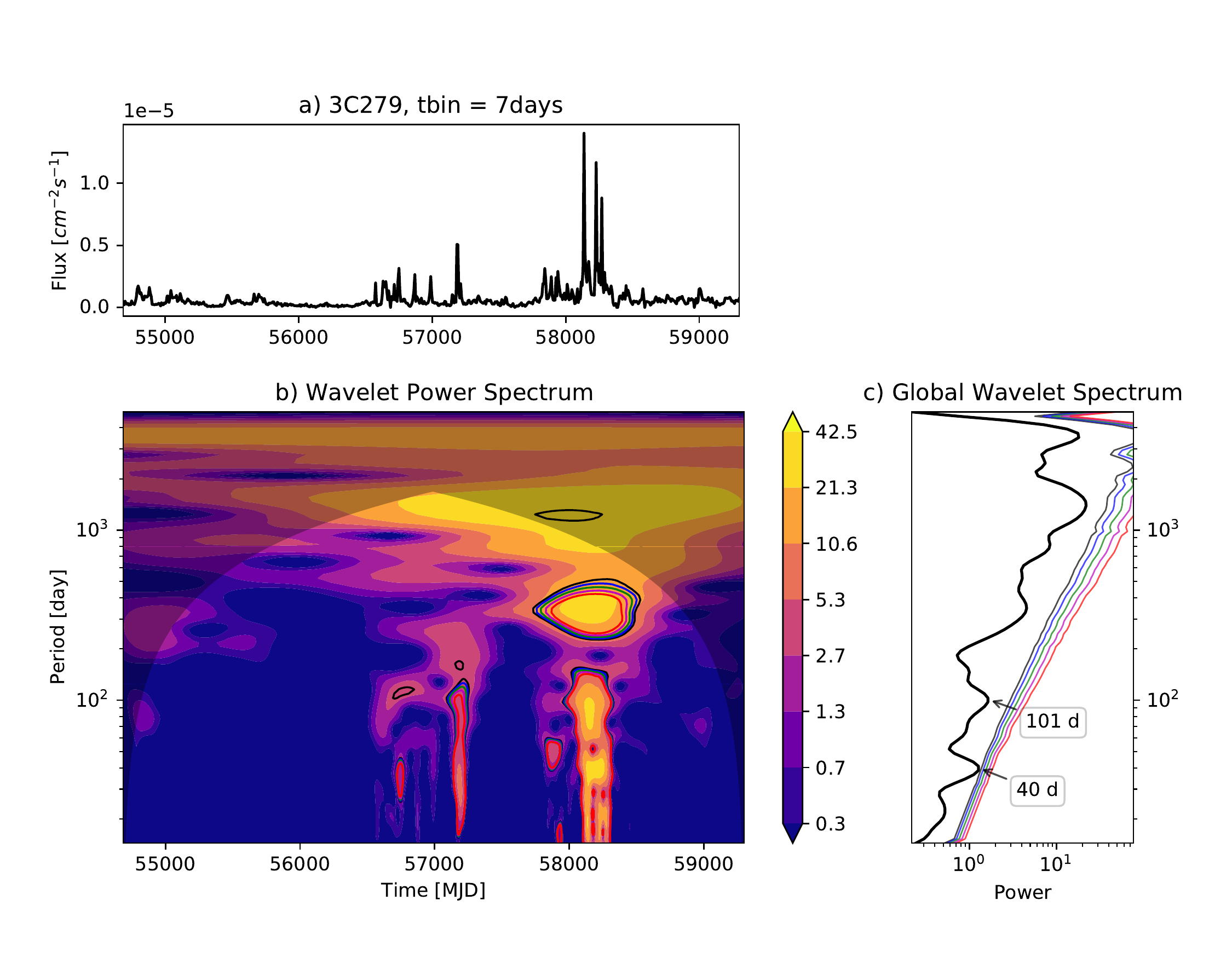}
	\end{subfigure}
	\vskip\baselineskip
	
	\begin{subfigure}[b]{0.48\textwidth}  
		\centering 
		\includegraphics[width=\textwidth]{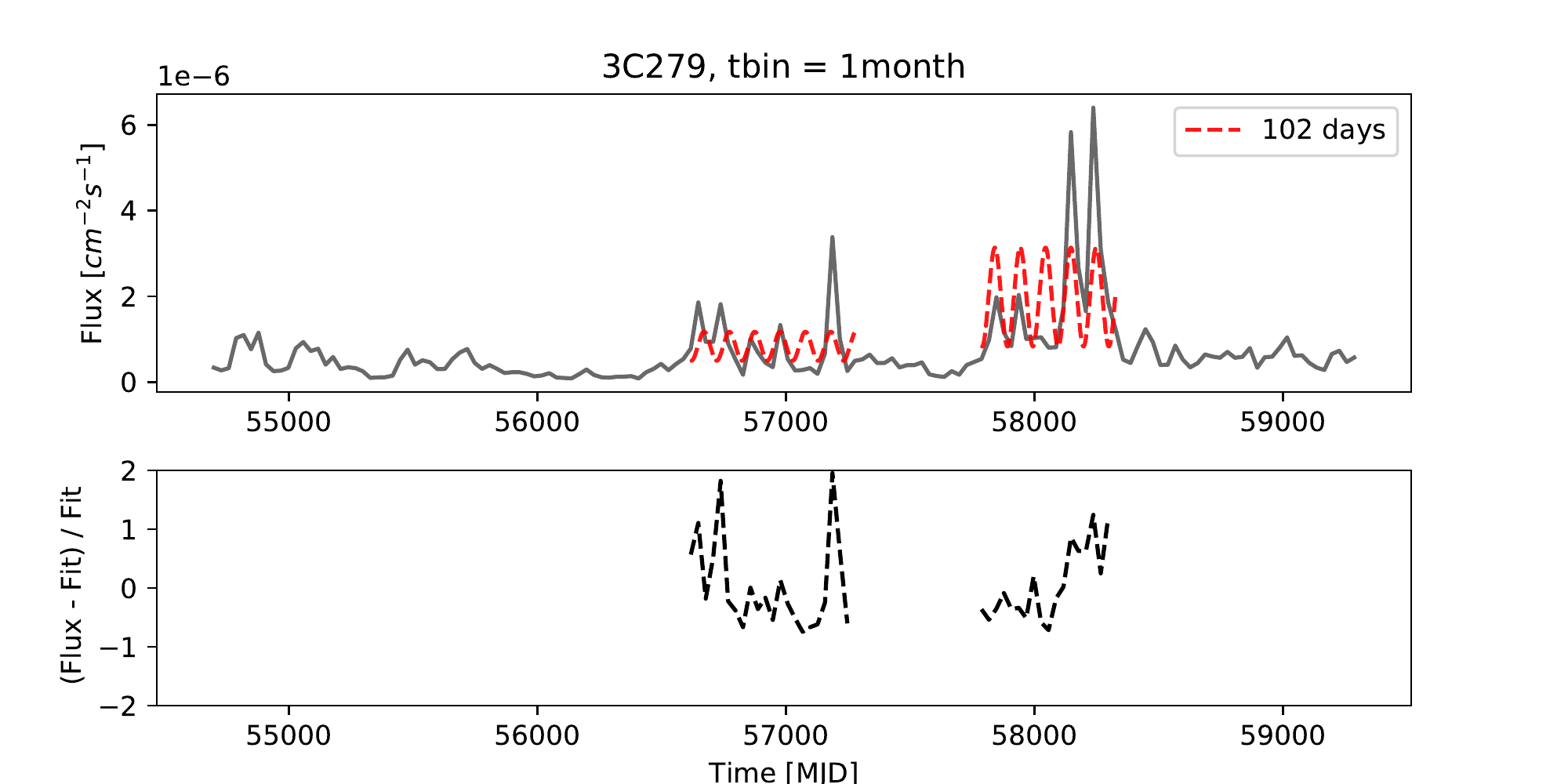}
	\end{subfigure}
	\hfill
	\begin{subfigure}[b]{0.48\textwidth}  
		\centering 
		\includegraphics[width=\textwidth]{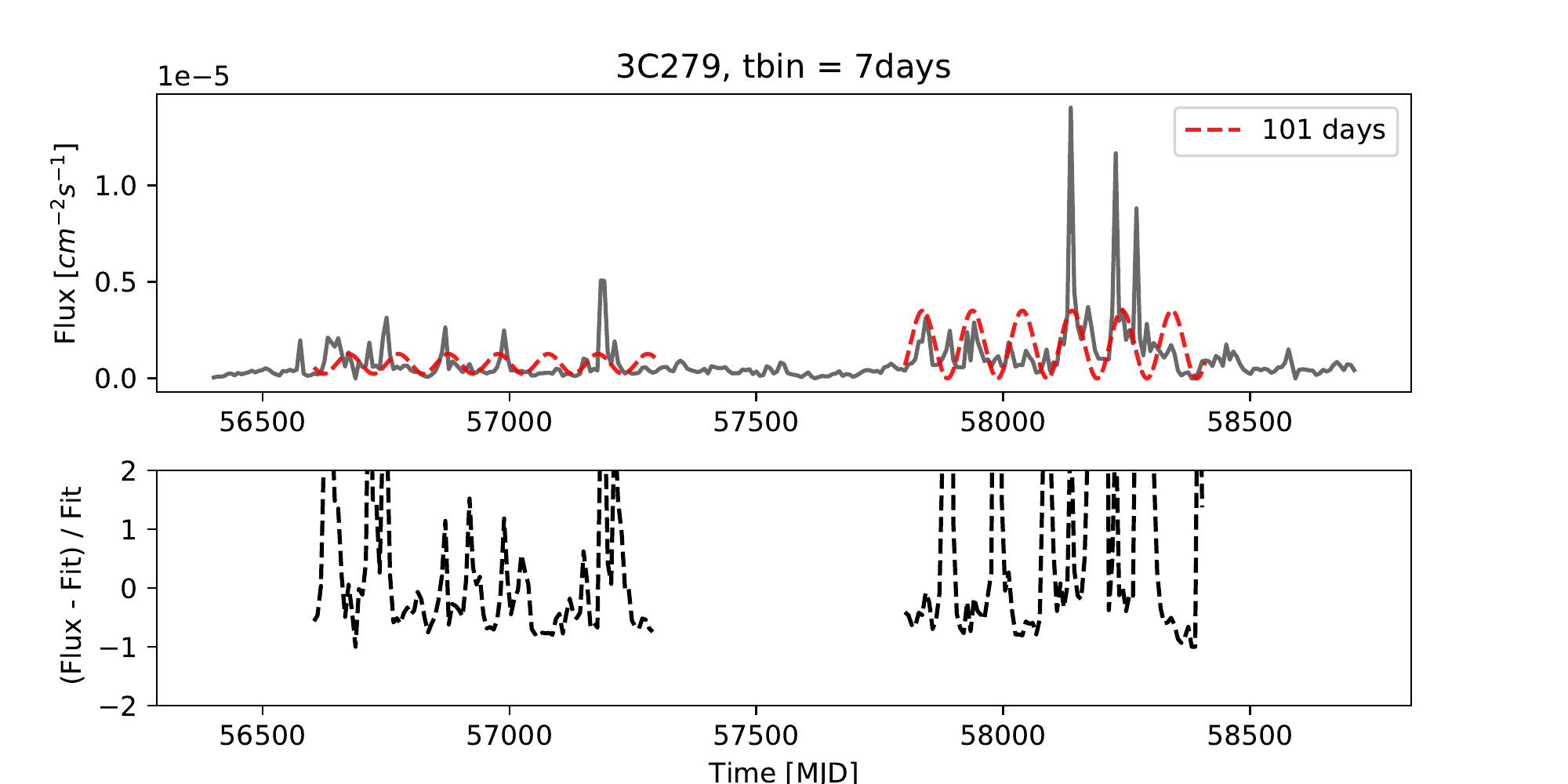}
	\end{subfigure}
	
	\vskip\baselineskip
	
	\begin{subfigure}[b]{0.48\textwidth}  
		\centering 
		\includegraphics[width=\textwidth]{./Figures/Fit/blanc.png}
	\end{subfigure}
	\hfill
	\begin{subfigure}[b]{0.48\textwidth}  
		\centering 
		\includegraphics[width=\textwidth]{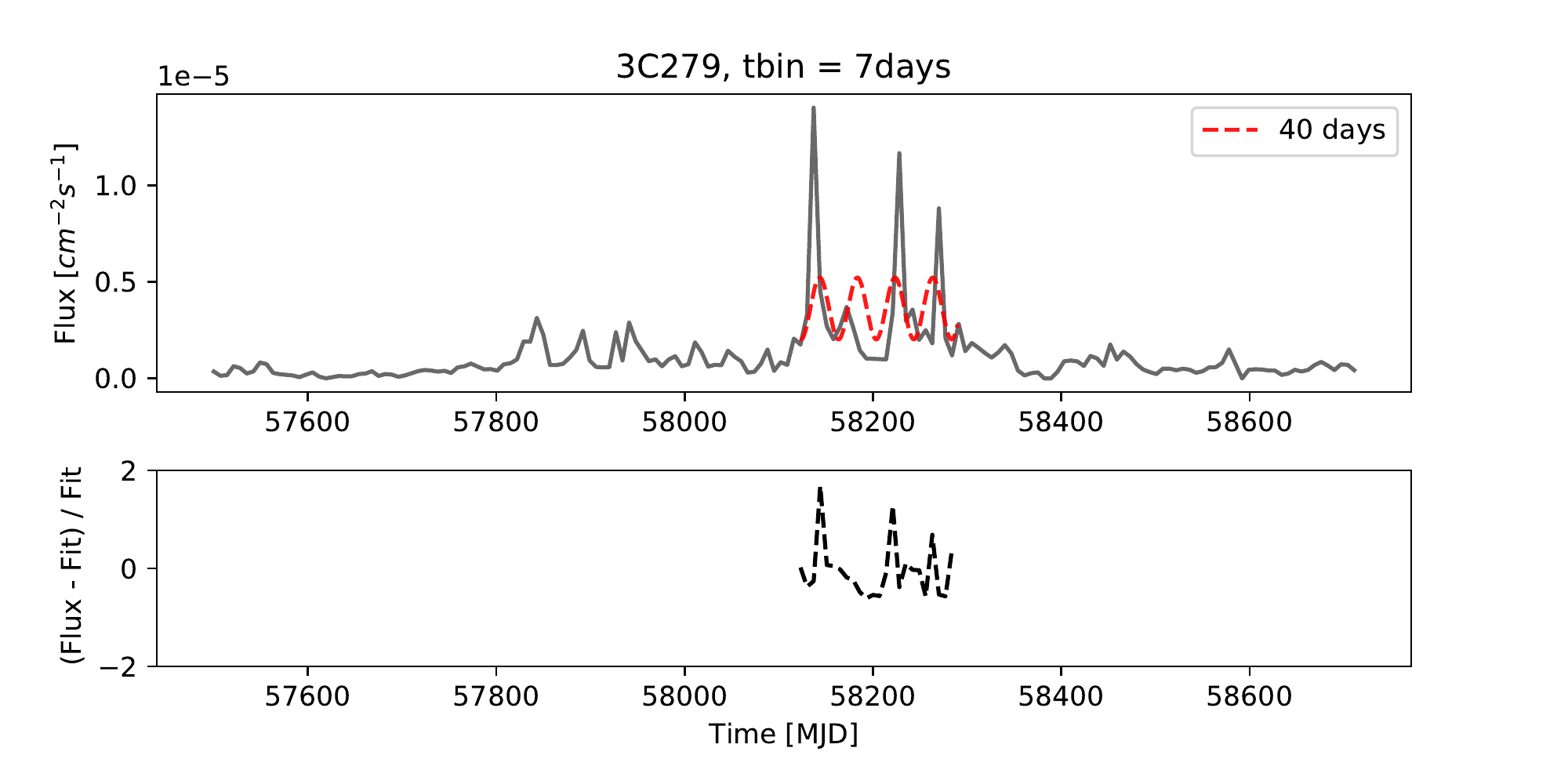}
	\end{subfigure}
	
	\caption{CWT map for monthly binned light curve (left) and weekly binned light curve (right) of B3~1343+451 and 3C~279, and the fitted light curves for 3C~279. The fitted light curve with period $\sim101$~d (third subfigure on the right hand side column) shows six cycles between around MJD~57700 and MJD~58400. However, we can fit six more cycles if we consider the previous rise in power spectrum during MJD~56500 and MJD~57300 approximately. }
	\label{fig:CWT9}
\end{figure*}


\begin{figure*}[!htbp]
	\centering
	\begin{subfigure}[b]{0.48\textwidth}   
		\centering 
		\includegraphics[width=\textwidth]{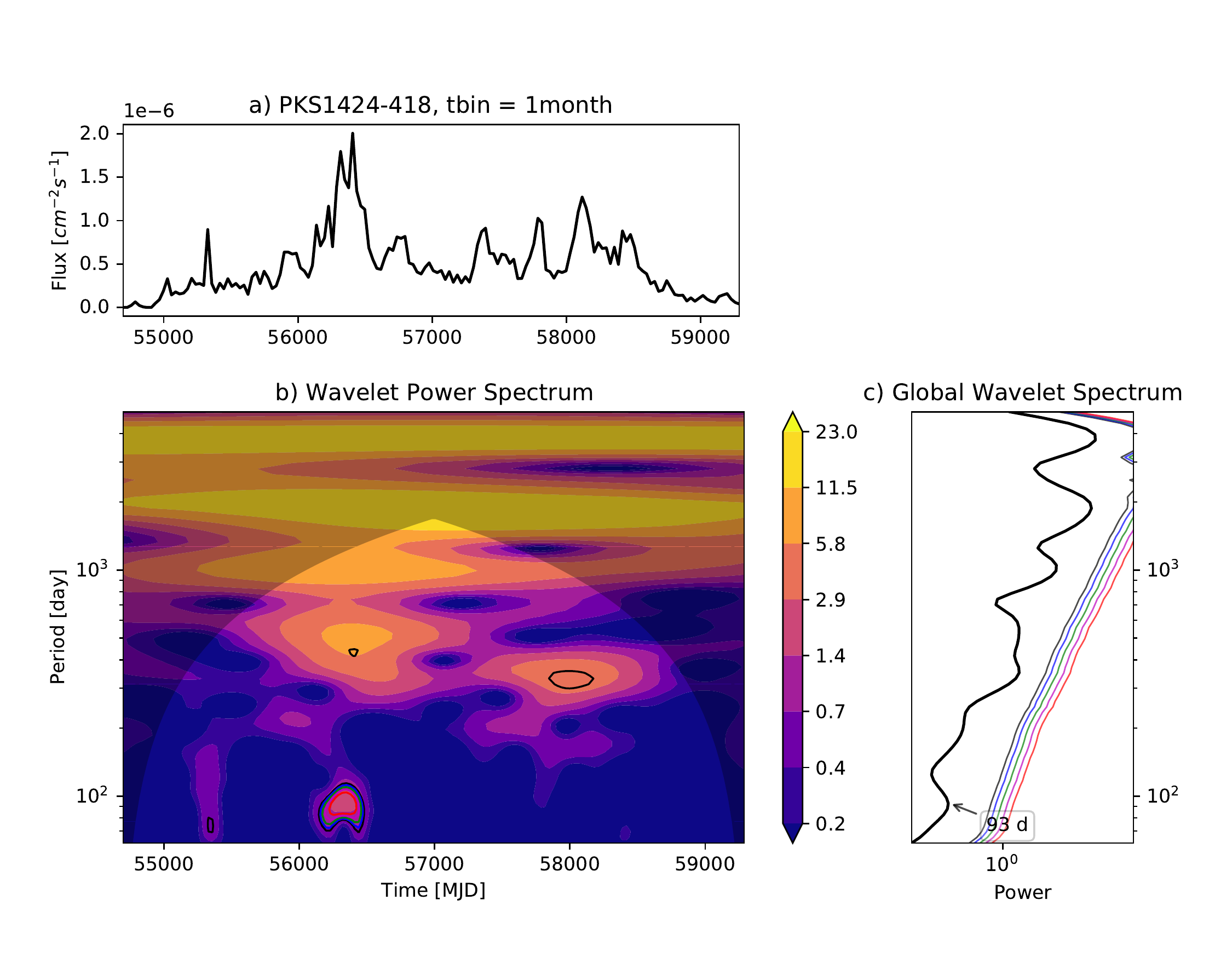}
	\end{subfigure}
	\hfill
	\begin{subfigure}[b]{0.48\textwidth}   
		\centering 
		\includegraphics[width=\textwidth]{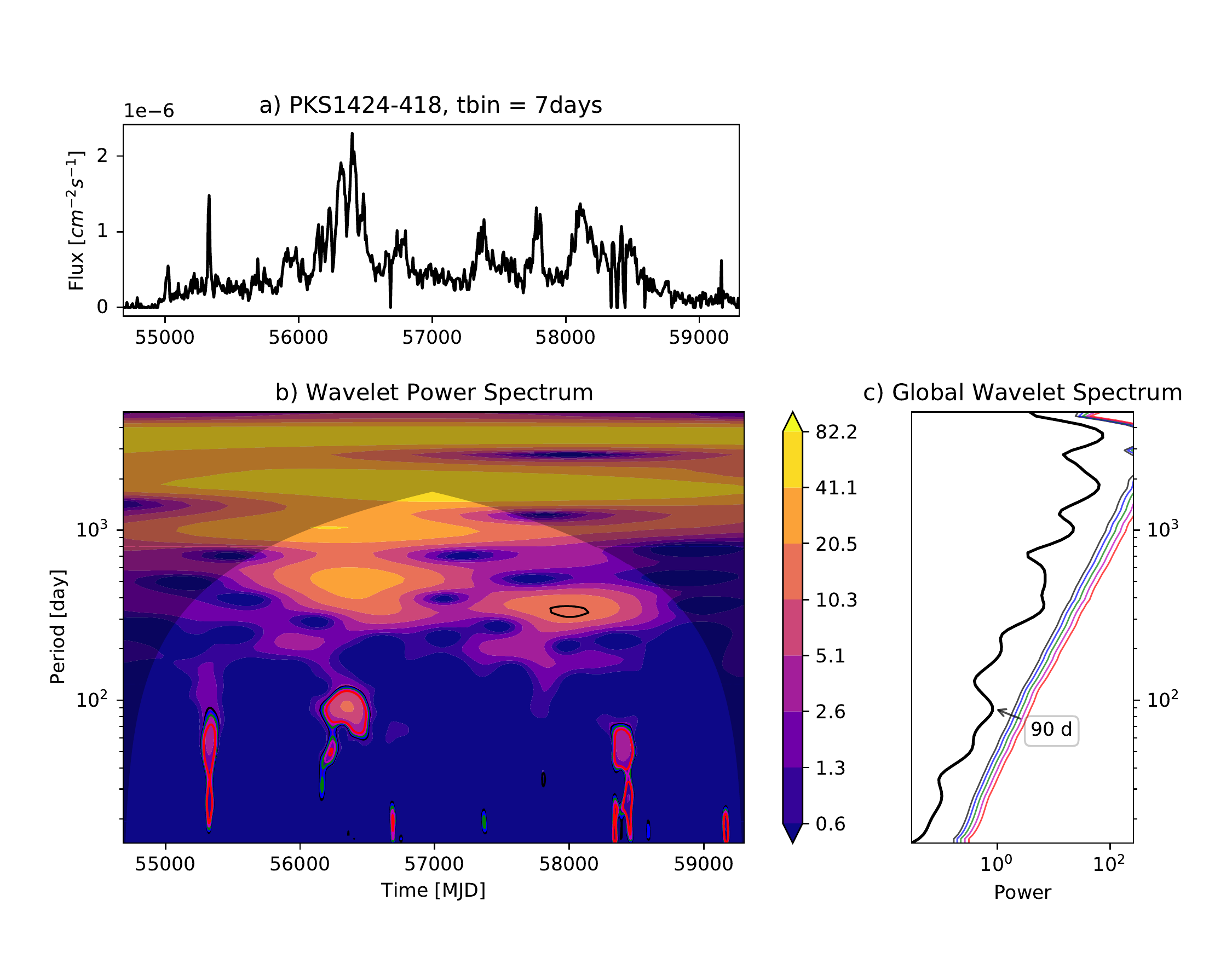}
	\end{subfigure}
	\vskip\baselineskip
	
	\begin{subfigure}[b]{0.48\textwidth}  
		\centering 
		\includegraphics[width=\textwidth]{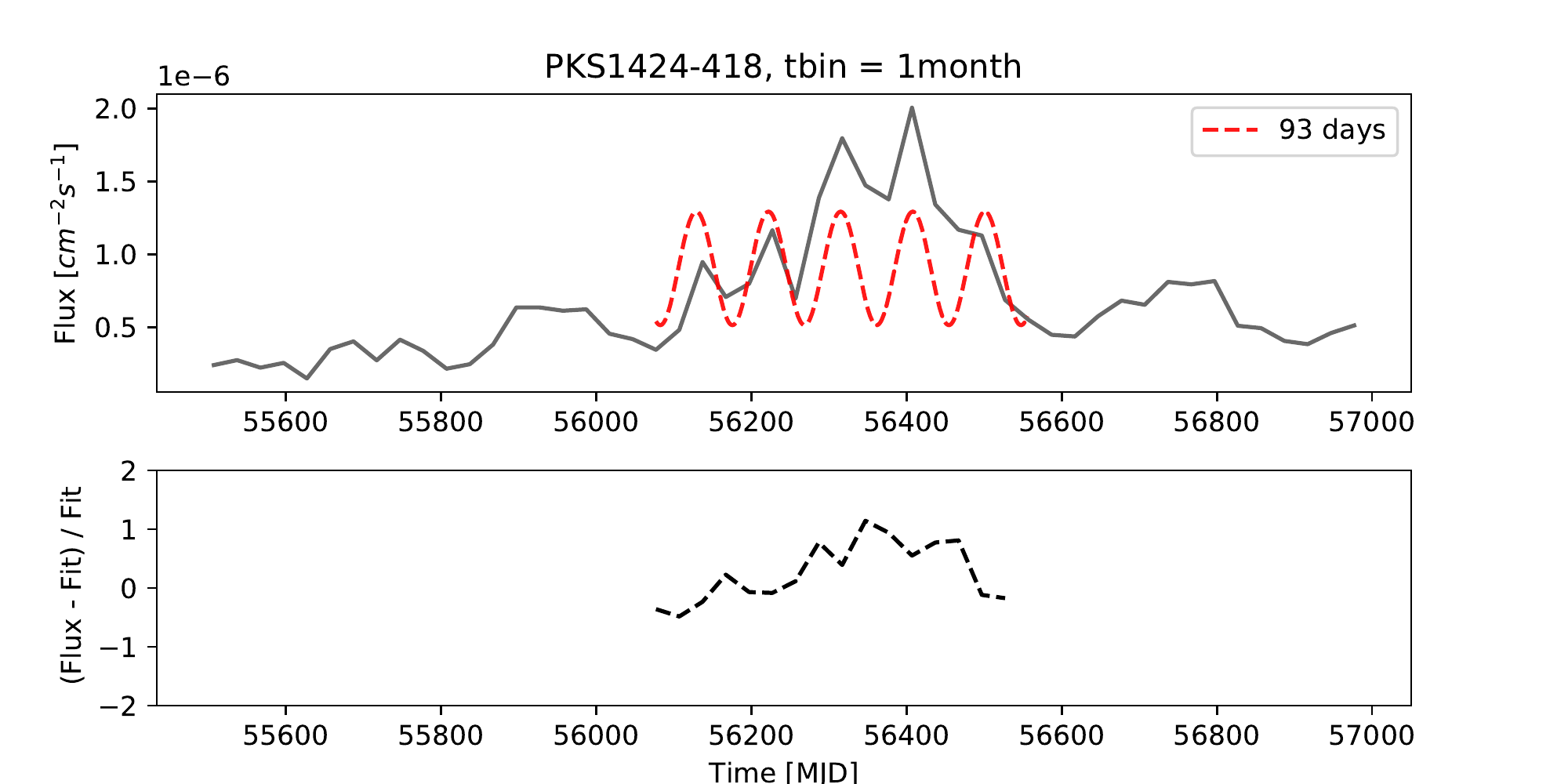}
	\end{subfigure}
	\hfill
	\begin{subfigure}[b]{0.48\textwidth}  
		\centering 
		\includegraphics[width=\textwidth]{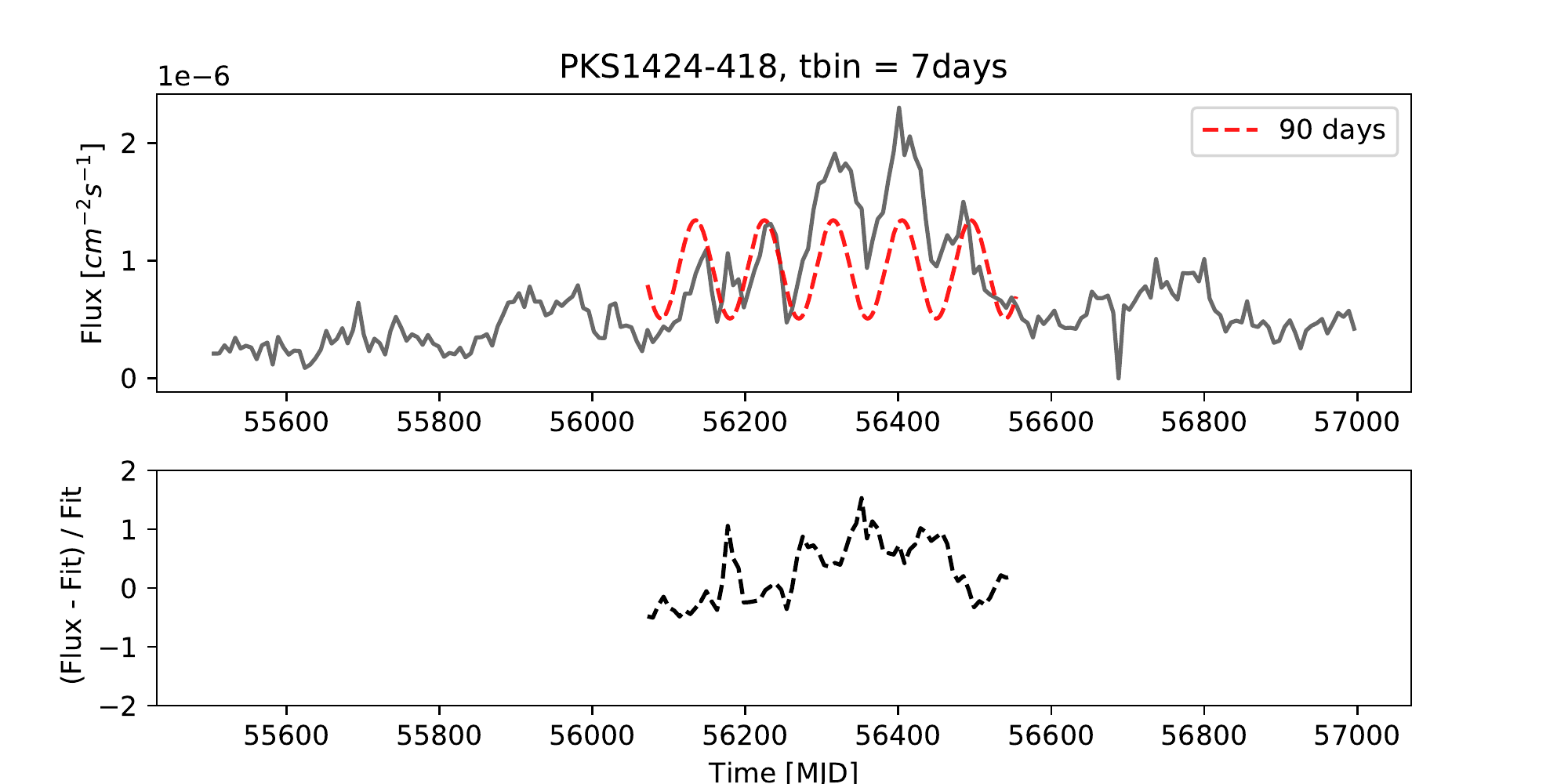}
	\end{subfigure}
	\vskip\baselineskip
	
		\hrule

	\begin{subfigure}[b]{0.48\textwidth}
		\centering
		\includegraphics[width=\textwidth]{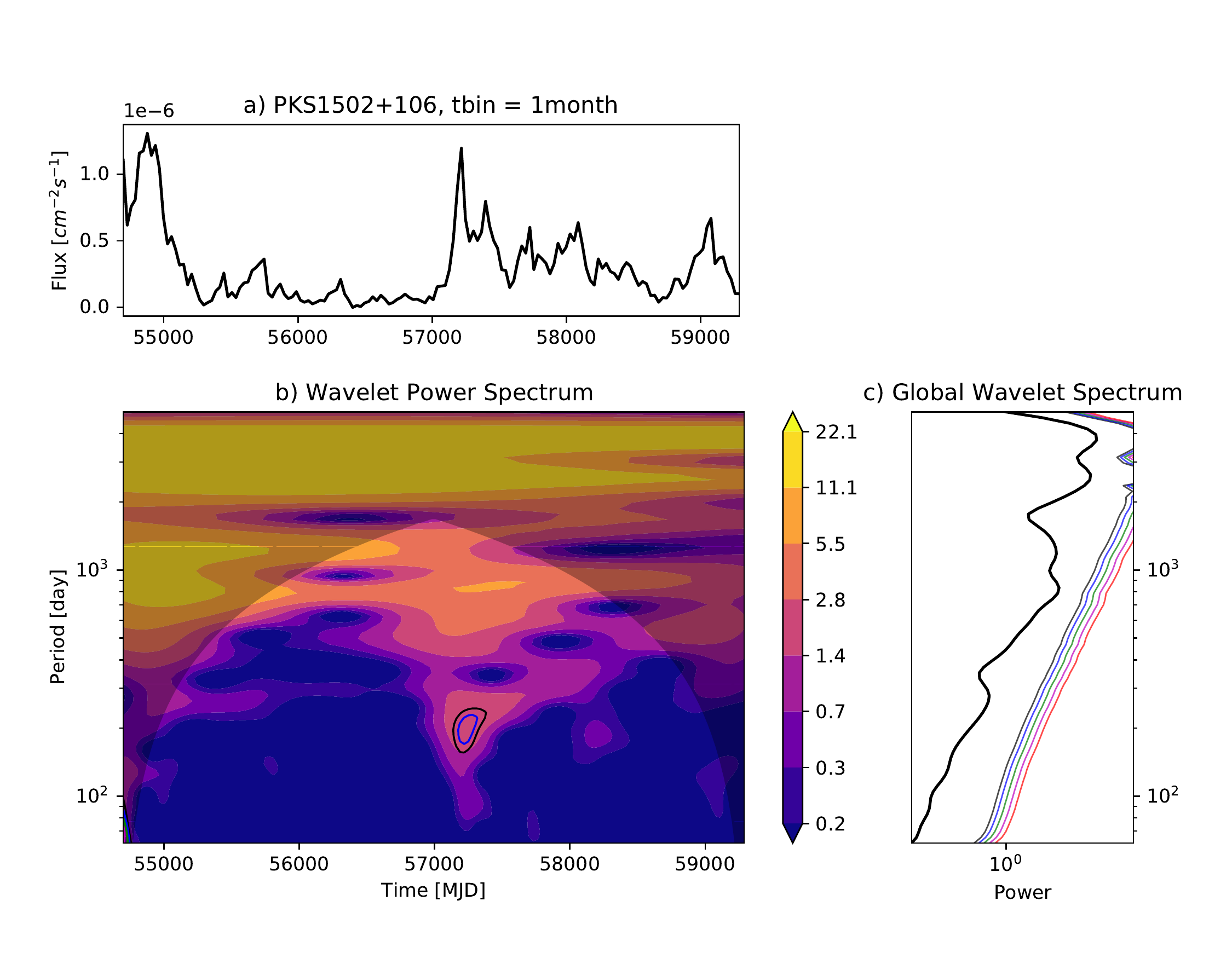}
	\end{subfigure}
	\hfill
	\begin{subfigure}[b]{0.48\textwidth}
		\centering
		\includegraphics[width=\textwidth]{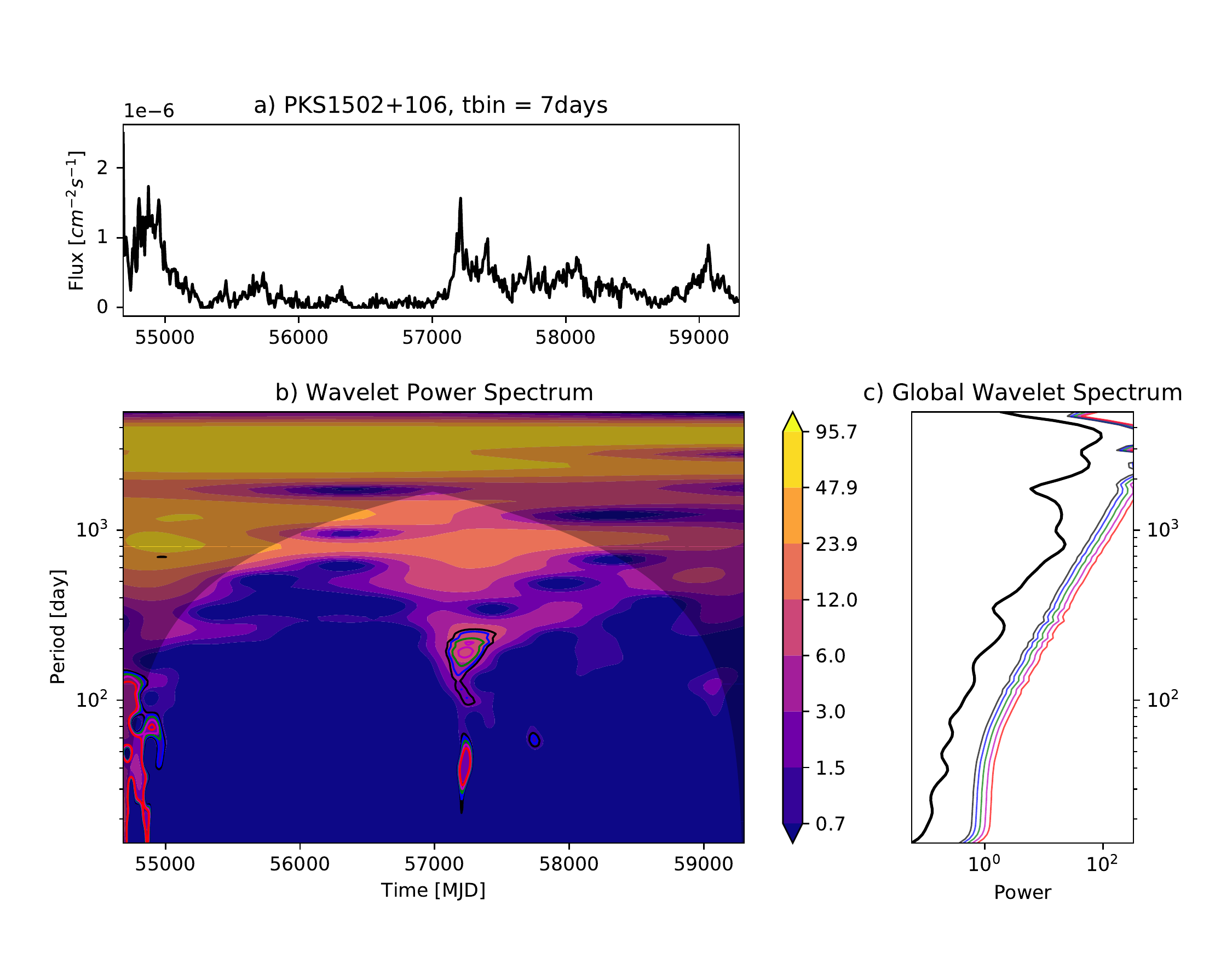}
	\end{subfigure}
	\vskip\baselineskip

	\caption{CWT map for monthly binned light curve (left) and weekly binned light curve (right) of PKS~1424-418 and PKS~1502+106, and the fitted light curves for PKS~1424-418.}
	\label{fig:CWT10}
\end{figure*}


\begin{figure*}[!htbp]
	\centering
	
	\begin{subfigure}[b]{0.48\textwidth}   
		\centering 
		\includegraphics[width=\textwidth]{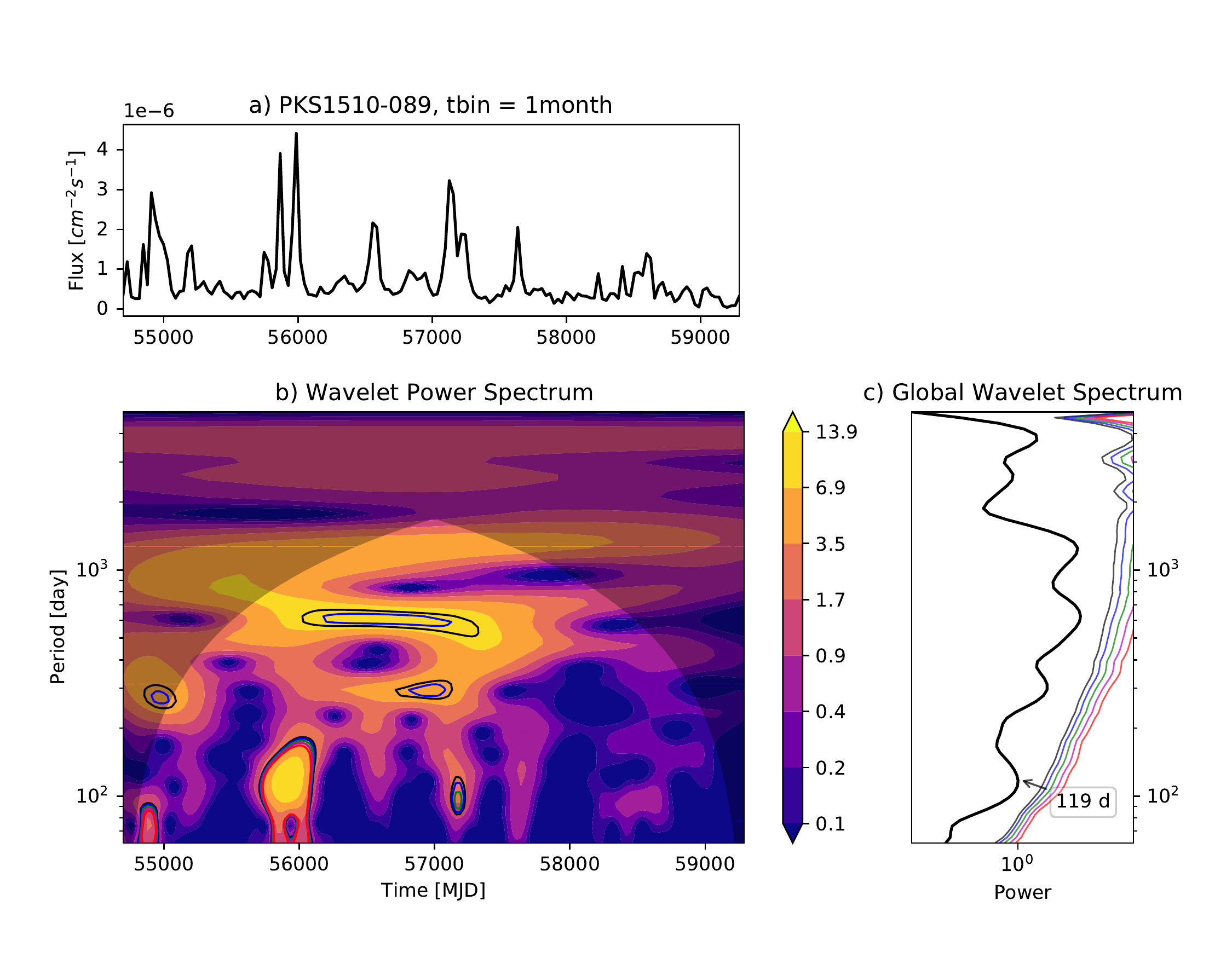}
	\end{subfigure}
	\hfill
	\begin{subfigure}[b]{0.48\textwidth}   
		\centering 
		\includegraphics[width=\textwidth]{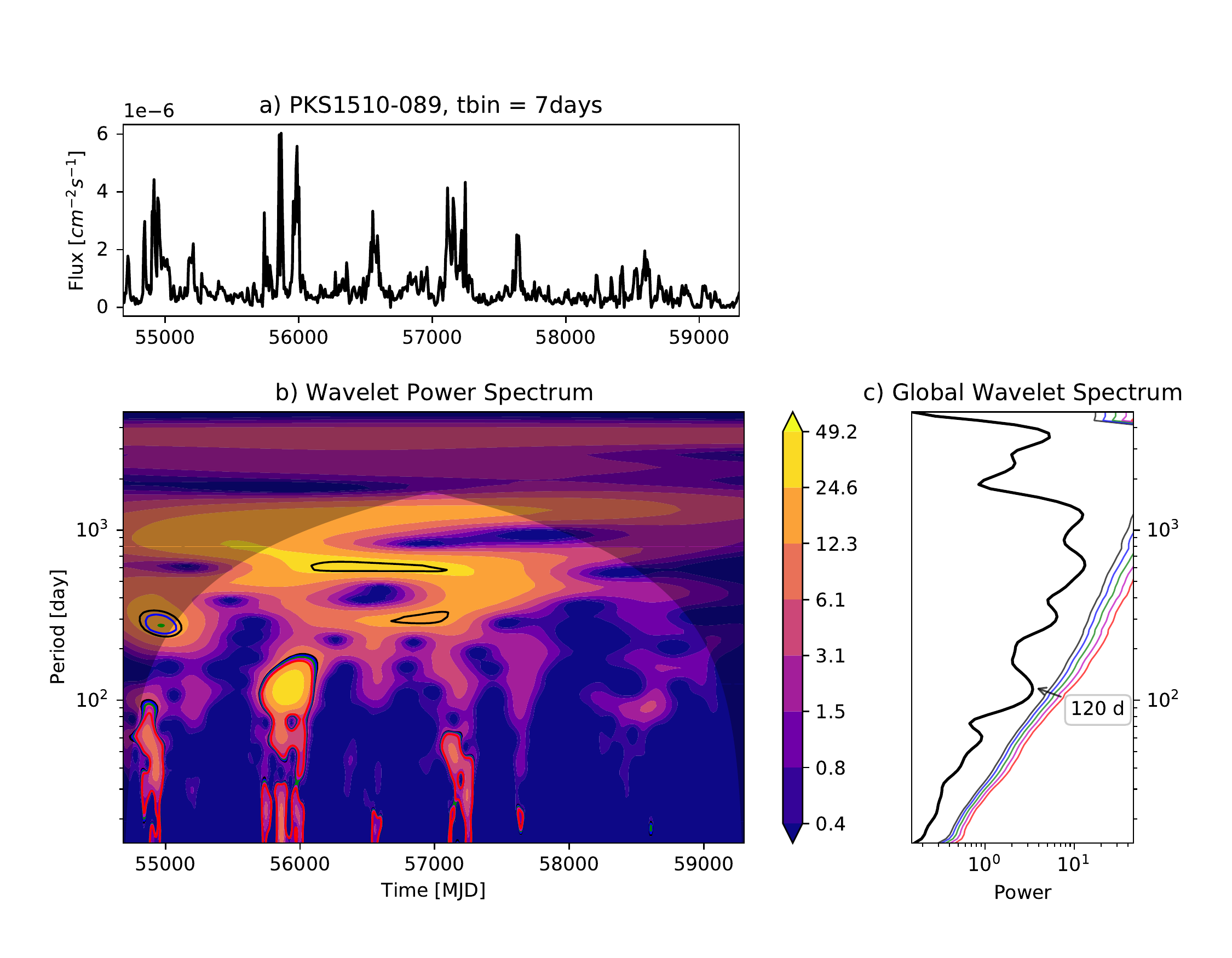}
	\end{subfigure}
	\vskip\baselineskip
	
	\begin{subfigure}[b]{0.48\textwidth}  
		\centering 
		\includegraphics[width=\textwidth]{./Figures/Fit/blanc.png}
	\end{subfigure}
	\hfill
	\begin{subfigure}[b]{0.48\textwidth}  
		\centering 
		\includegraphics[width=\textwidth]{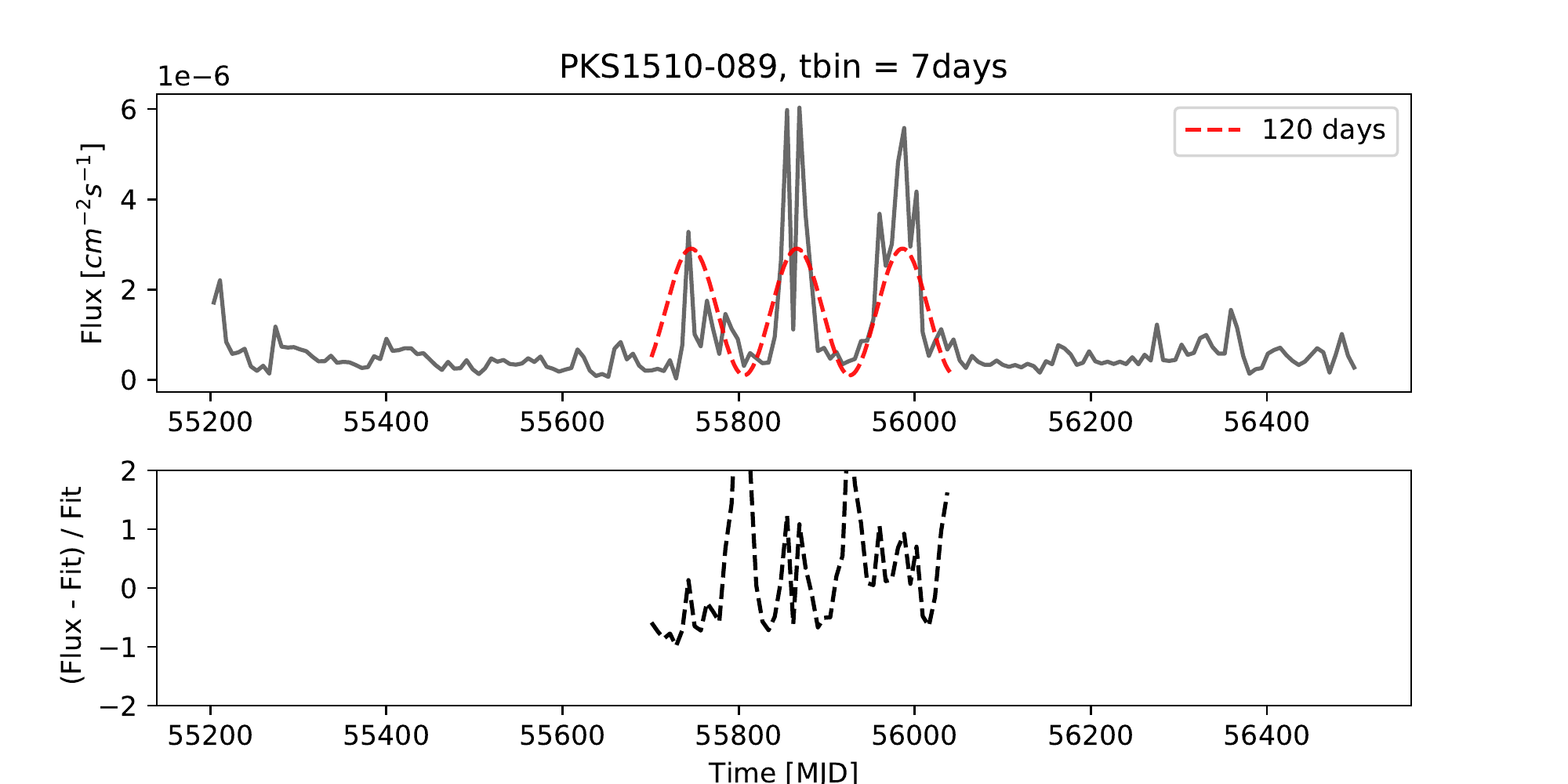}
	\end{subfigure}
	
	\vskip\baselineskip

    \hrule
    \begin{subfigure}[b]{0.48\textwidth}
		\centering
		\includegraphics[width=\textwidth]{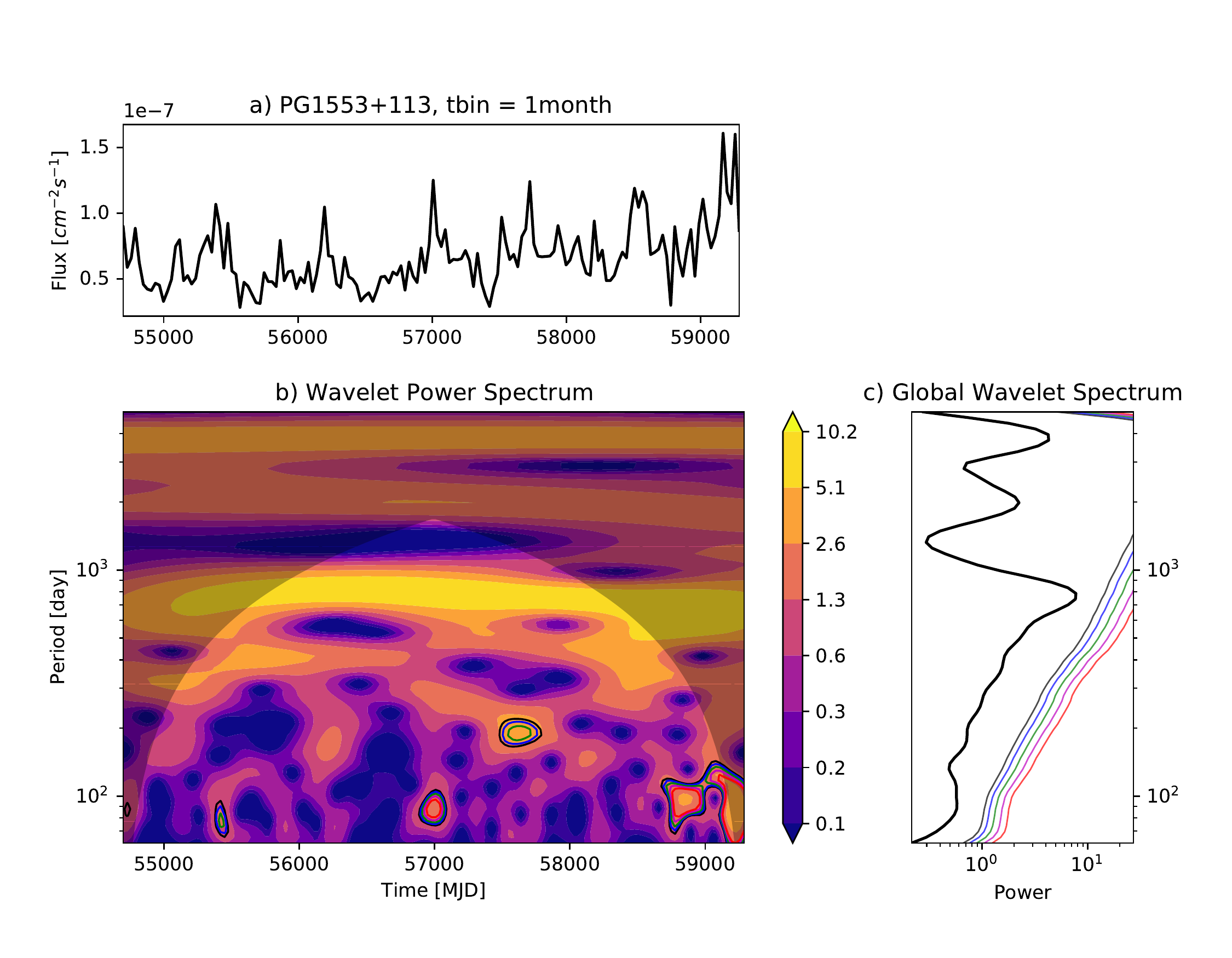}
	\end{subfigure}
	\hfill
	\begin{subfigure}[b]{0.48\textwidth}
		\centering
		\includegraphics[width=\textwidth]{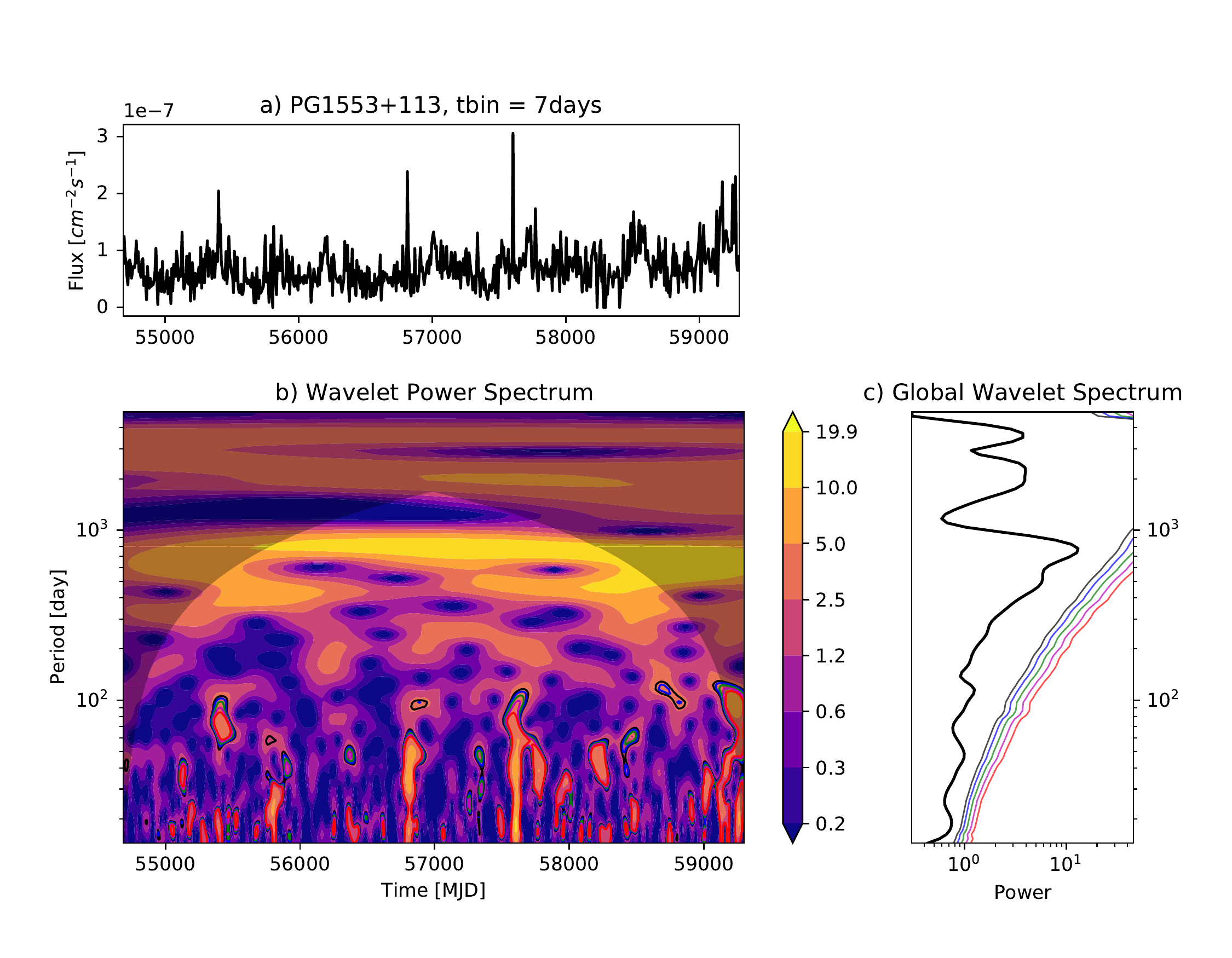}
	\end{subfigure}

	\caption{CWT map for monthly binned light curve (left) and weekly binned light curve (right) of PKS~1510-089 and PG~1553+113, and the fitted light curve for PKS~1510-089.}
	\label{fig:CWT11}
\end{figure*}


\begin{figure*}[!htbp]
	\centering
	\begin{subfigure}[b]{0.48\textwidth}   
		\centering 
		\includegraphics[width=\textwidth]{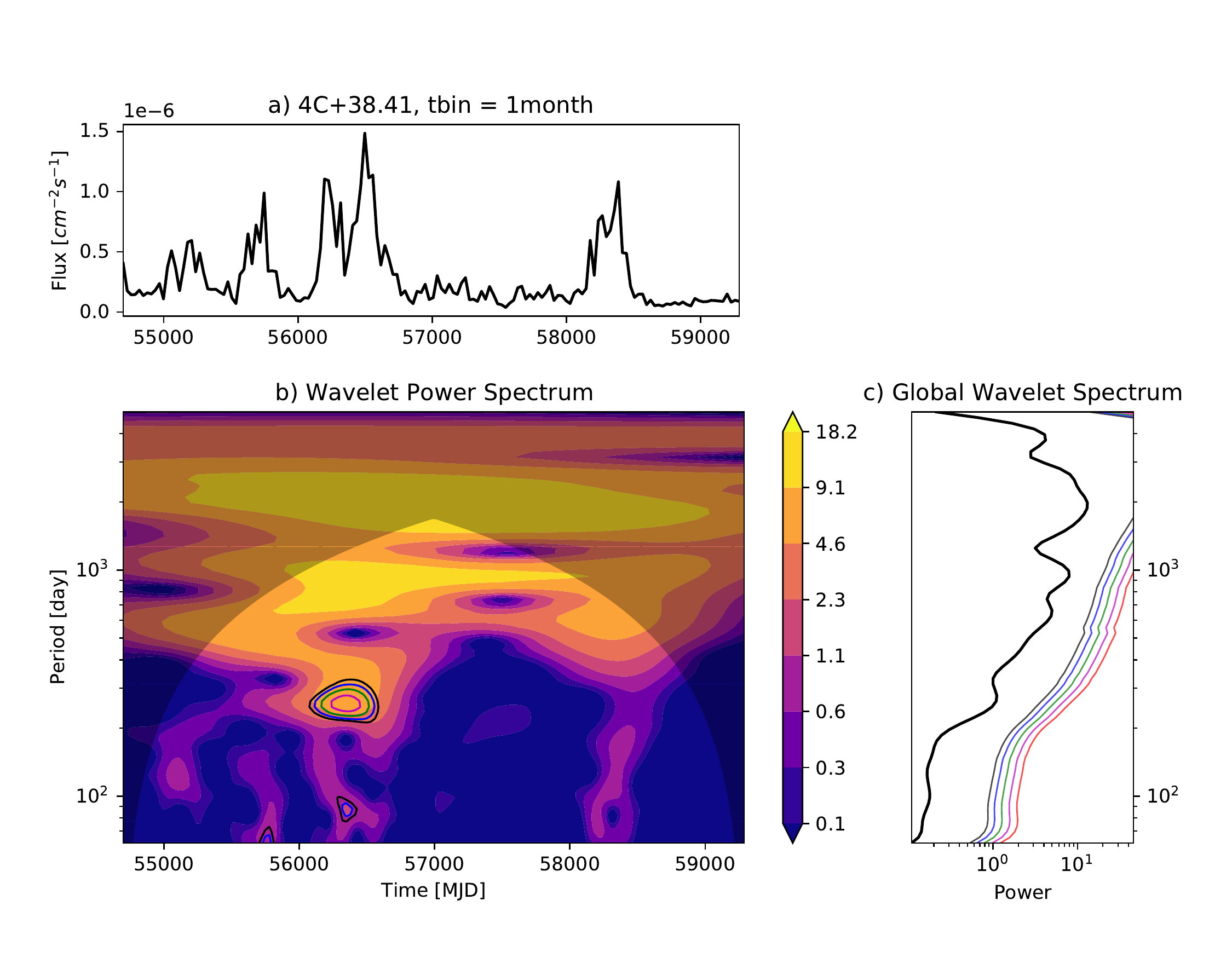}
	\end{subfigure}
	\hfill
	\begin{subfigure}[b]{0.48\textwidth}   
		\centering 
		\includegraphics[width=\textwidth]{./Figures/Figures_AuchereTrials/4FGLJ1635.2+3808_1month_10000_cwt.pdf}
	\end{subfigure}
	
	
	\vskip\baselineskip

    \hrule
	
	\begin{subfigure}[b]{0.48\textwidth}   
		\centering 
		\includegraphics[width=\textwidth]{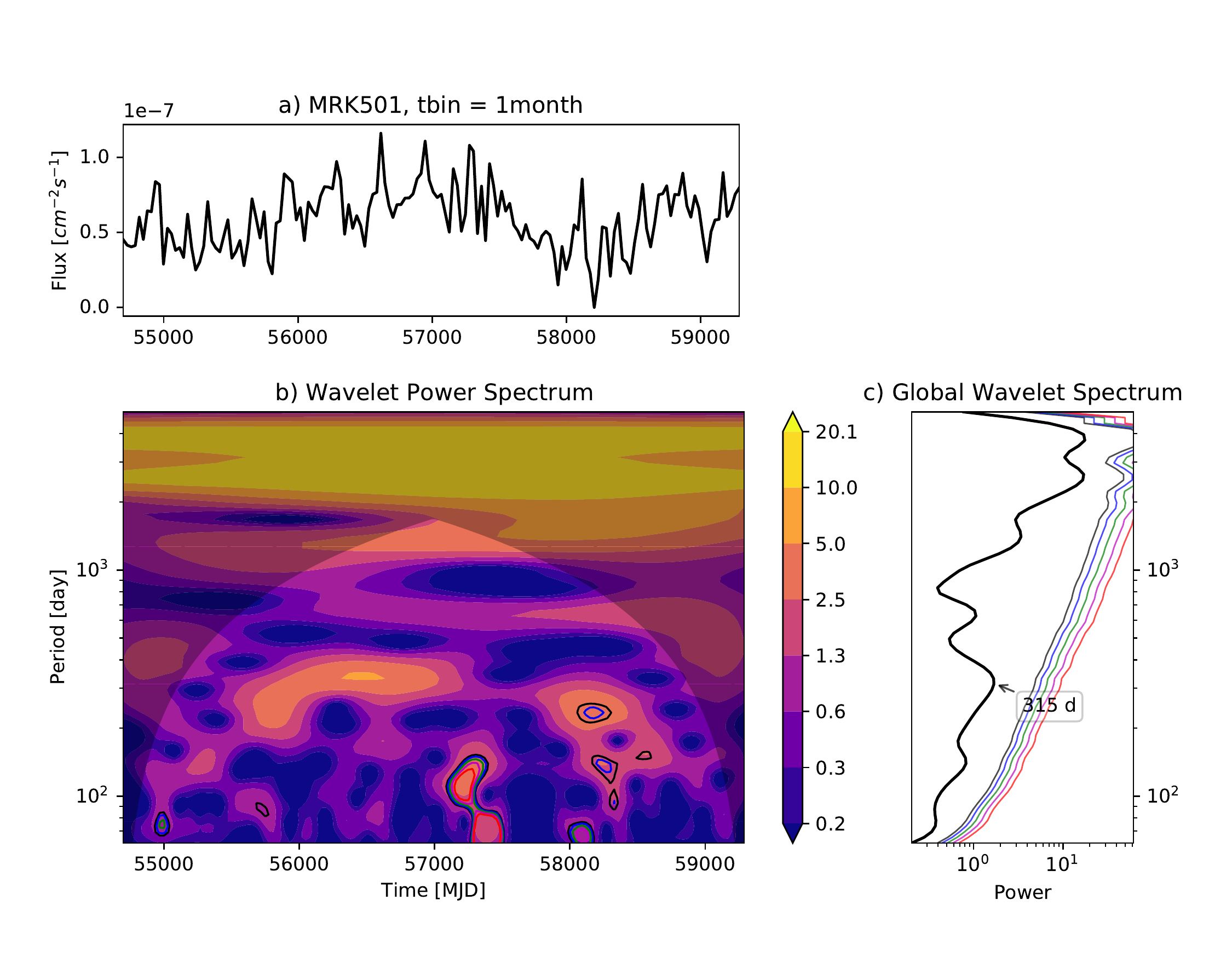}
	\end{subfigure}
	\hfill
	\begin{subfigure}[b]{0.48\textwidth}   
		\centering 
		\includegraphics[width=\textwidth]{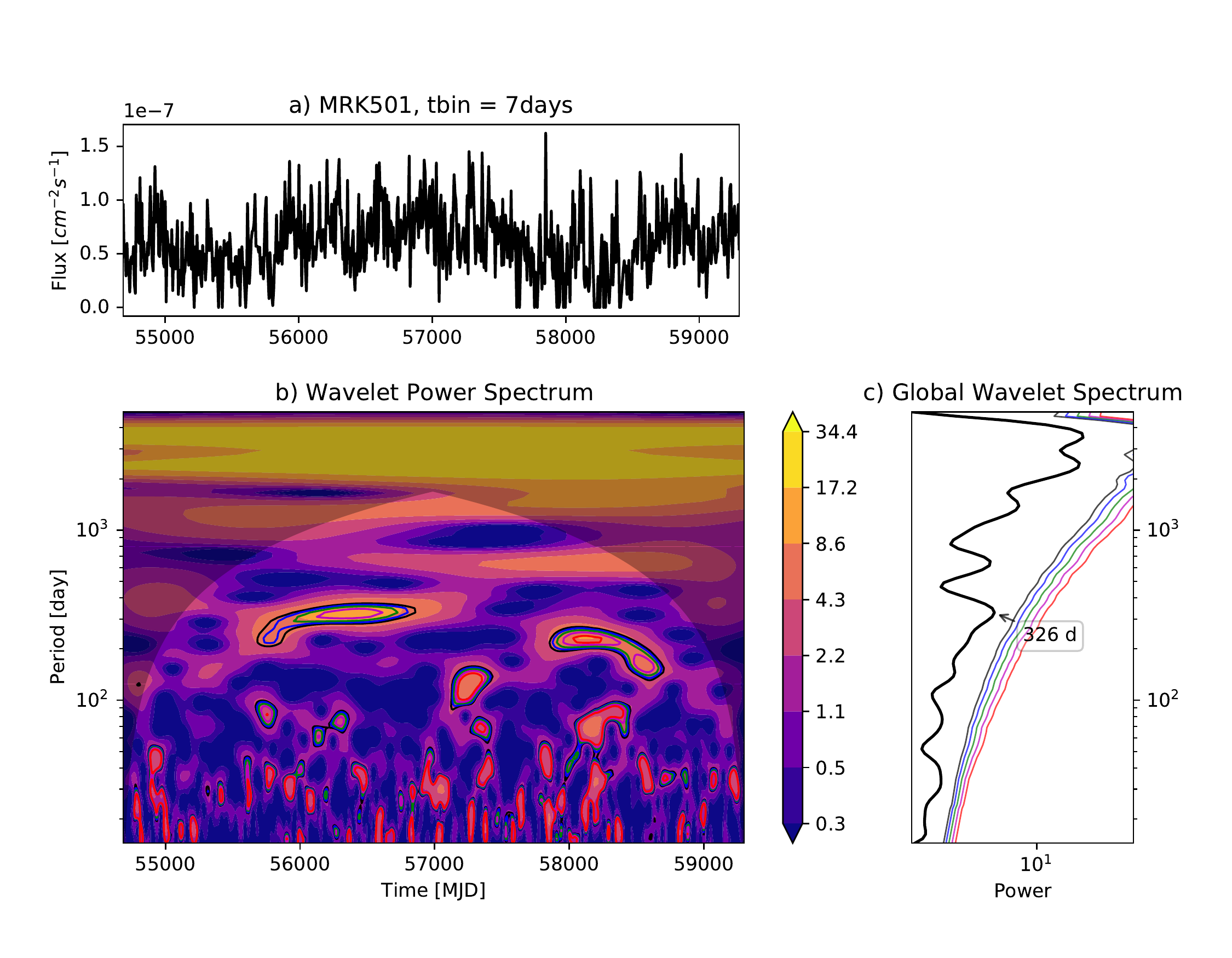}
	\end{subfigure}
	\vskip\baselineskip
	
	\begin{subfigure}[b]{0.48\textwidth}  
		\centering 
		\includegraphics[width=\textwidth]{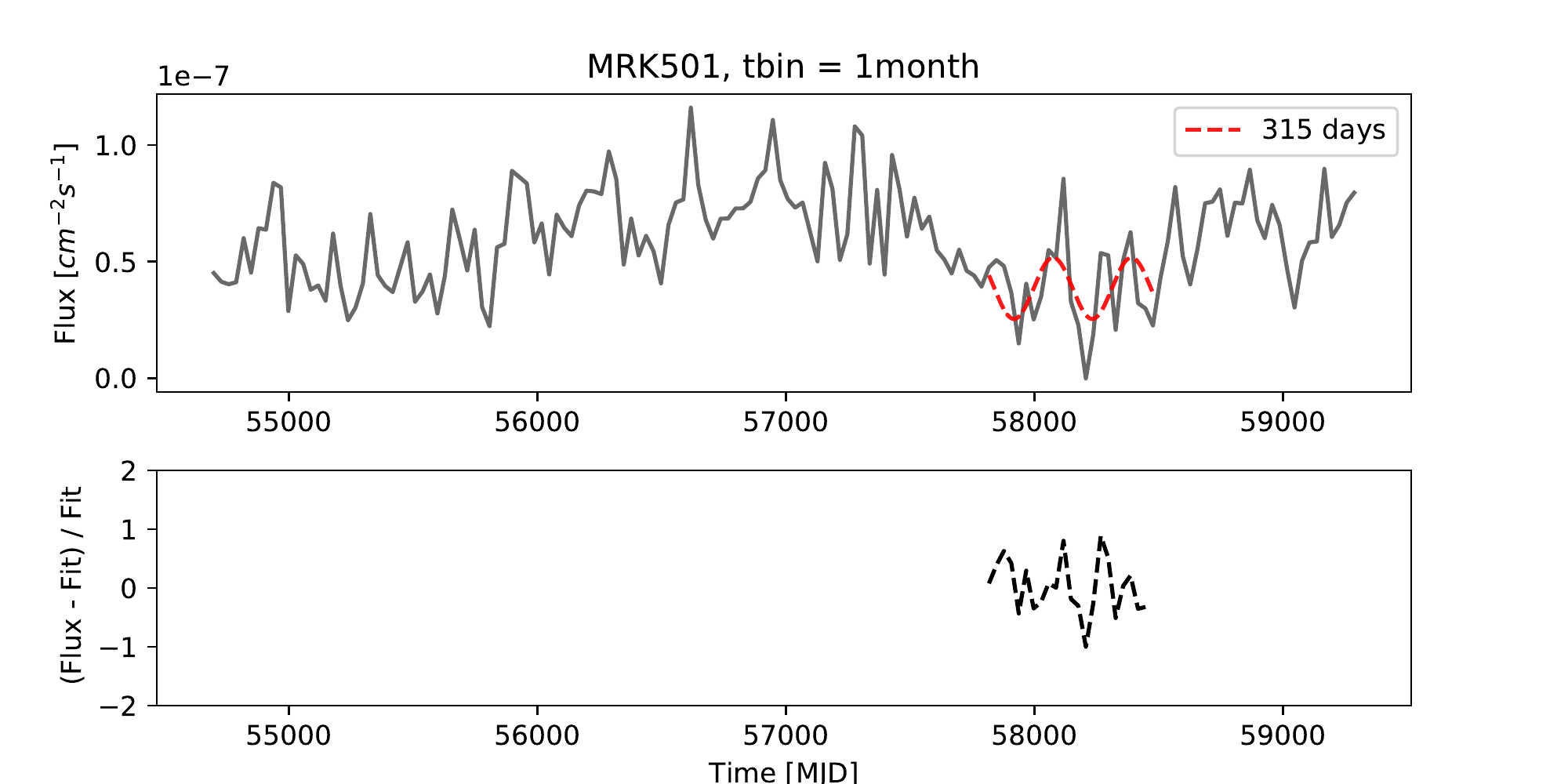}
	\end{subfigure}
	\hfill
	\begin{subfigure}[b]{0.48\textwidth}  
		\centering 
		\includegraphics[width=\textwidth]{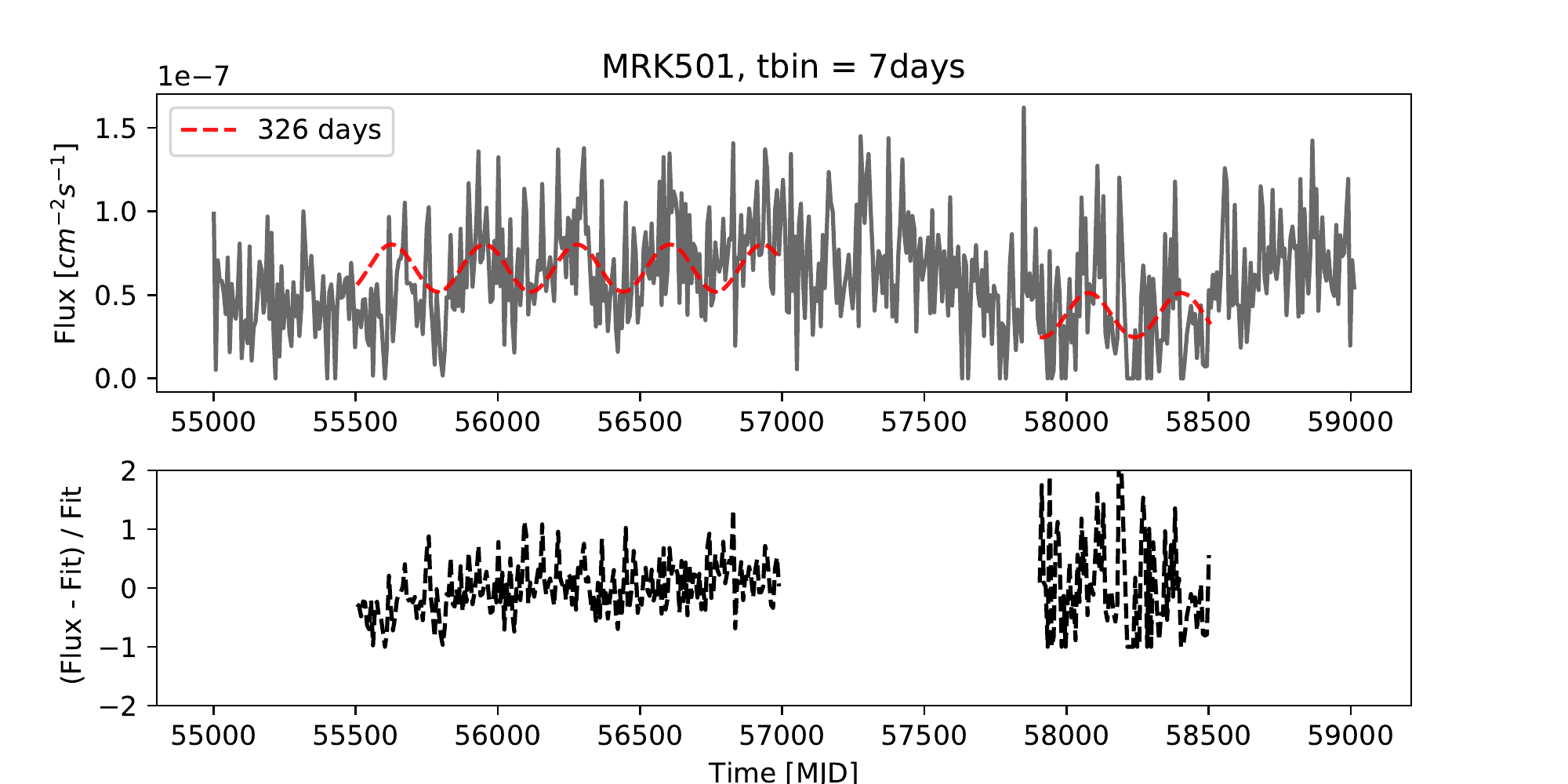}
	\end{subfigure}
	
	\caption{CWT map for monthly binned light curve (left) and weekly binned light curve (right) of 4C~~+38.41 and Mrk~501, and the fitted light curves for Mrk~501.}
	\label{fig:CWT12}
\end{figure*}


\begin{figure*}[!htbp]
	\centering

	\begin{subfigure}[b]{0.48\textwidth}
		\centering
		\includegraphics[width=\textwidth]{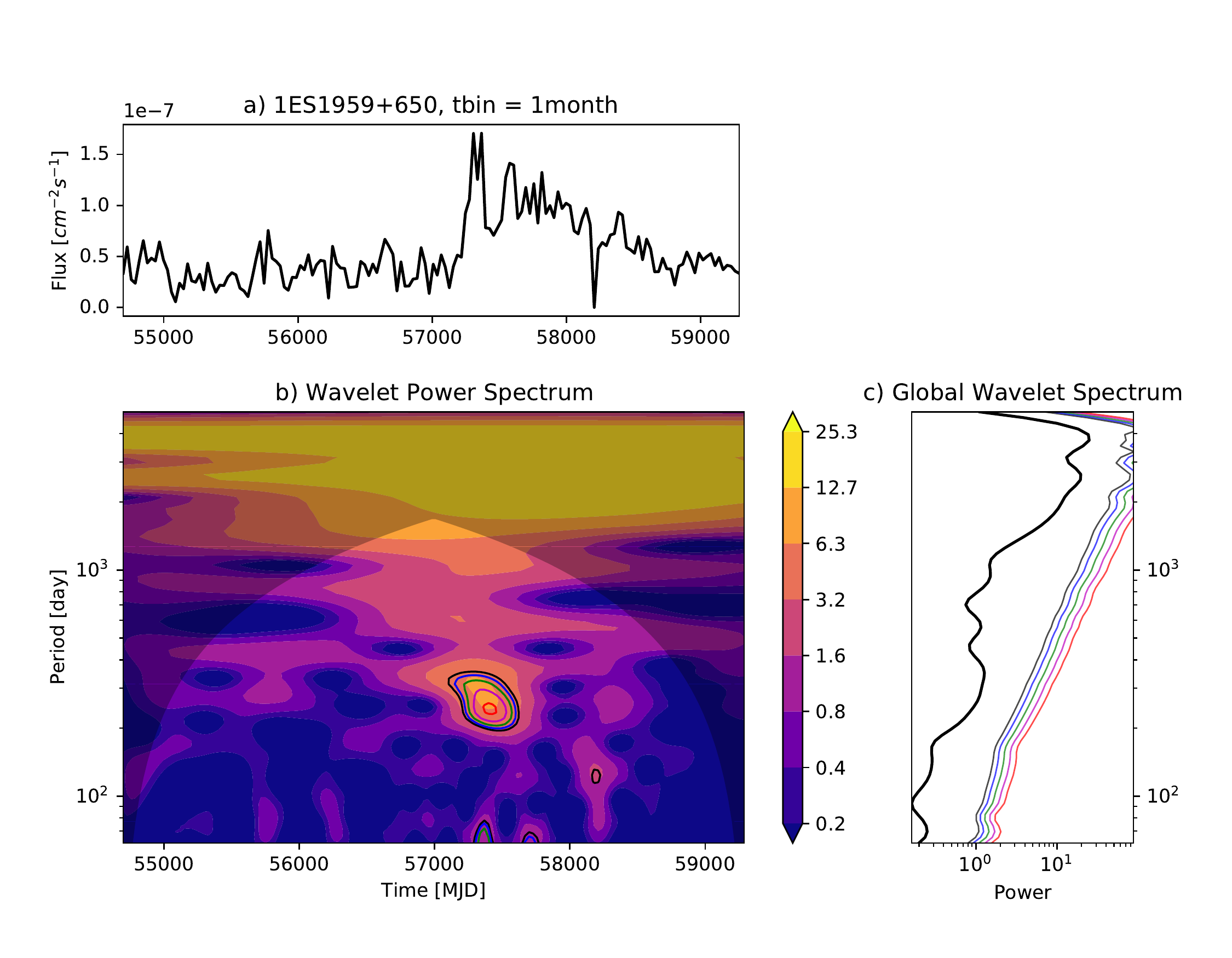}
	\end{subfigure}
	\hfill
	\begin{subfigure}[b]{0.48\textwidth}
		\centering
		\includegraphics[width=\textwidth]{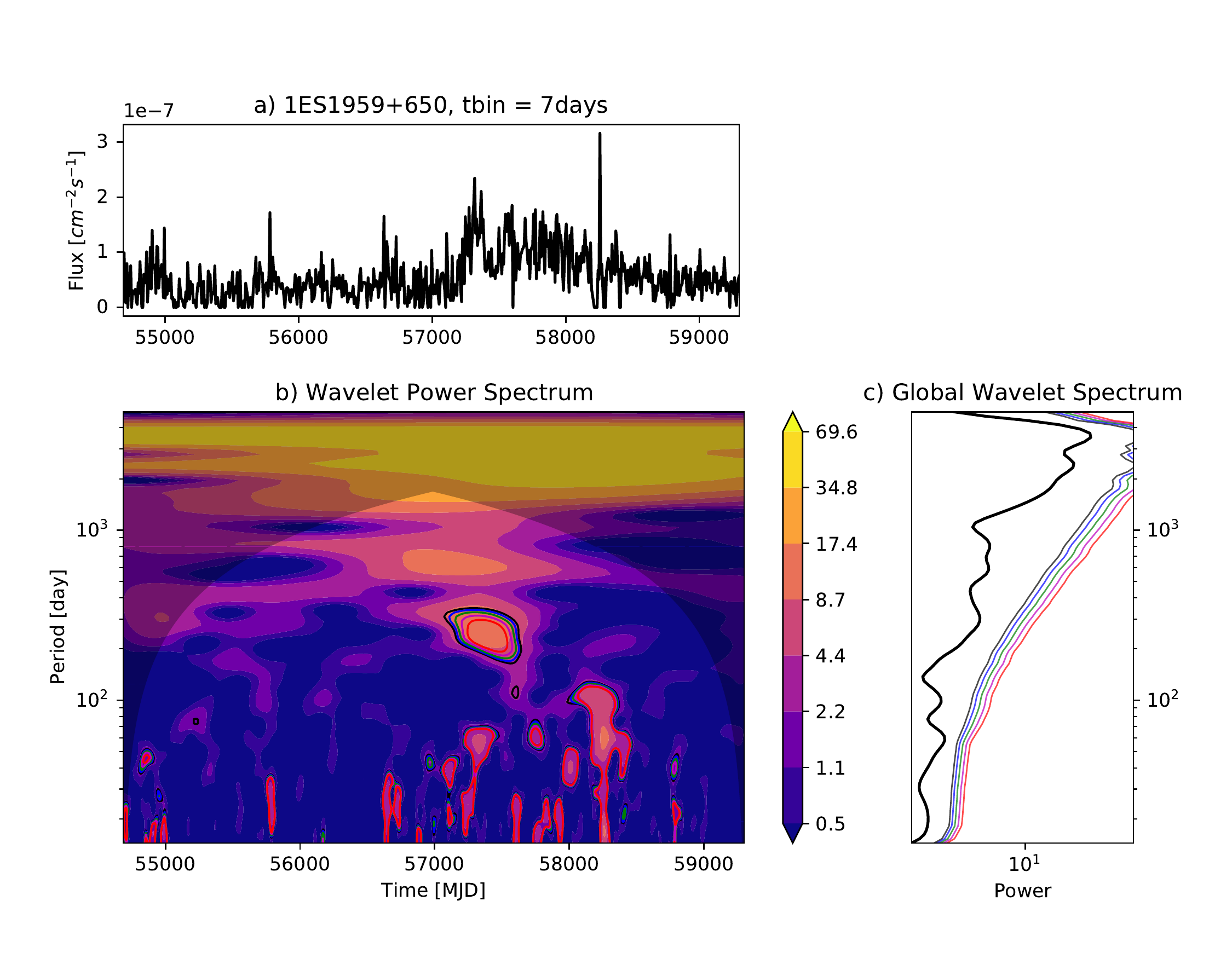}
	\end{subfigure}
	
	\vskip\baselineskip
	
	\hrule
	
	\begin{subfigure}[b]{0.48\textwidth}
		\centering
		\includegraphics[width=\textwidth]{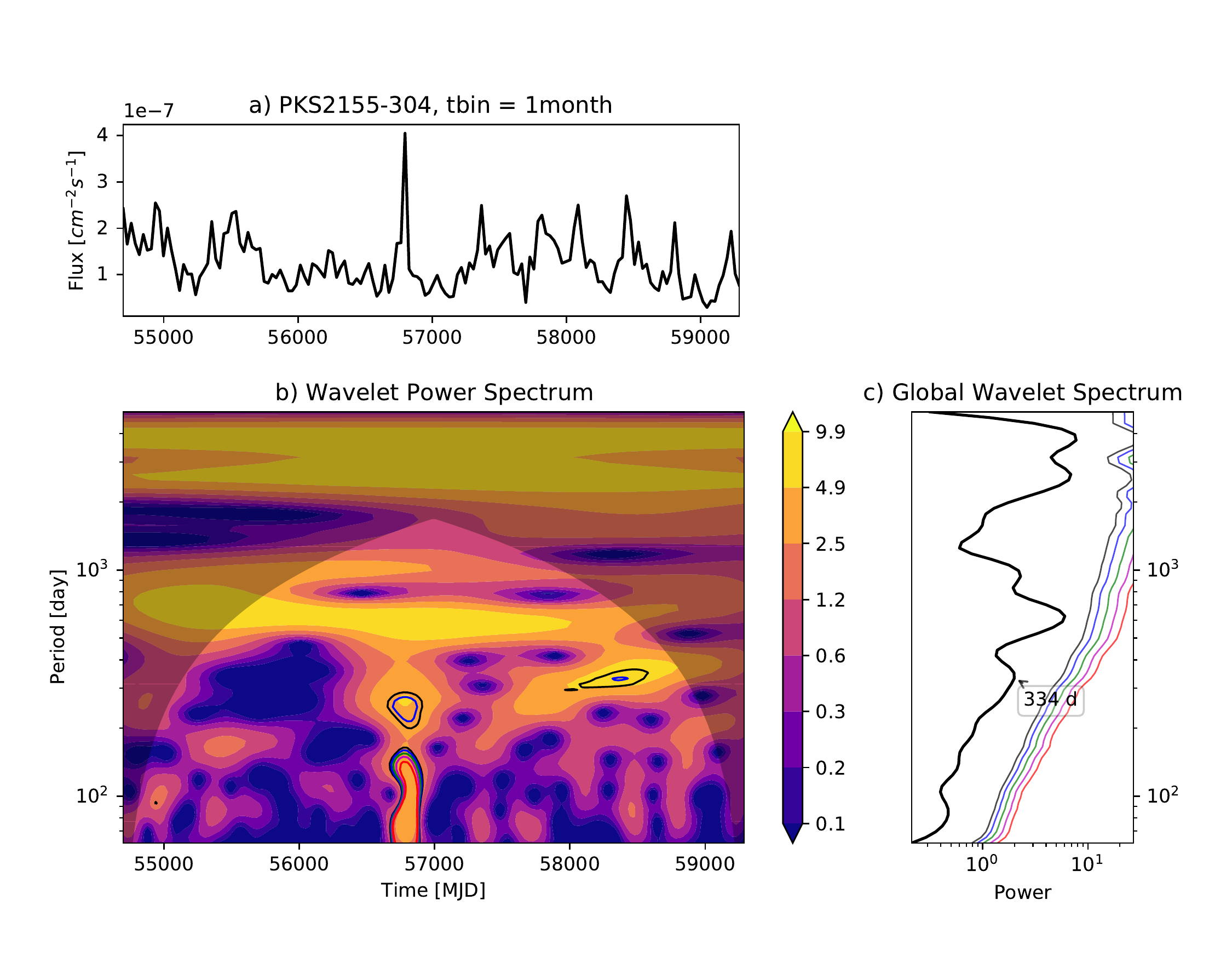}
	\end{subfigure}
	\hfill
	\begin{subfigure}[b]{0.48\textwidth}
		\centering
		\includegraphics[width=\textwidth]{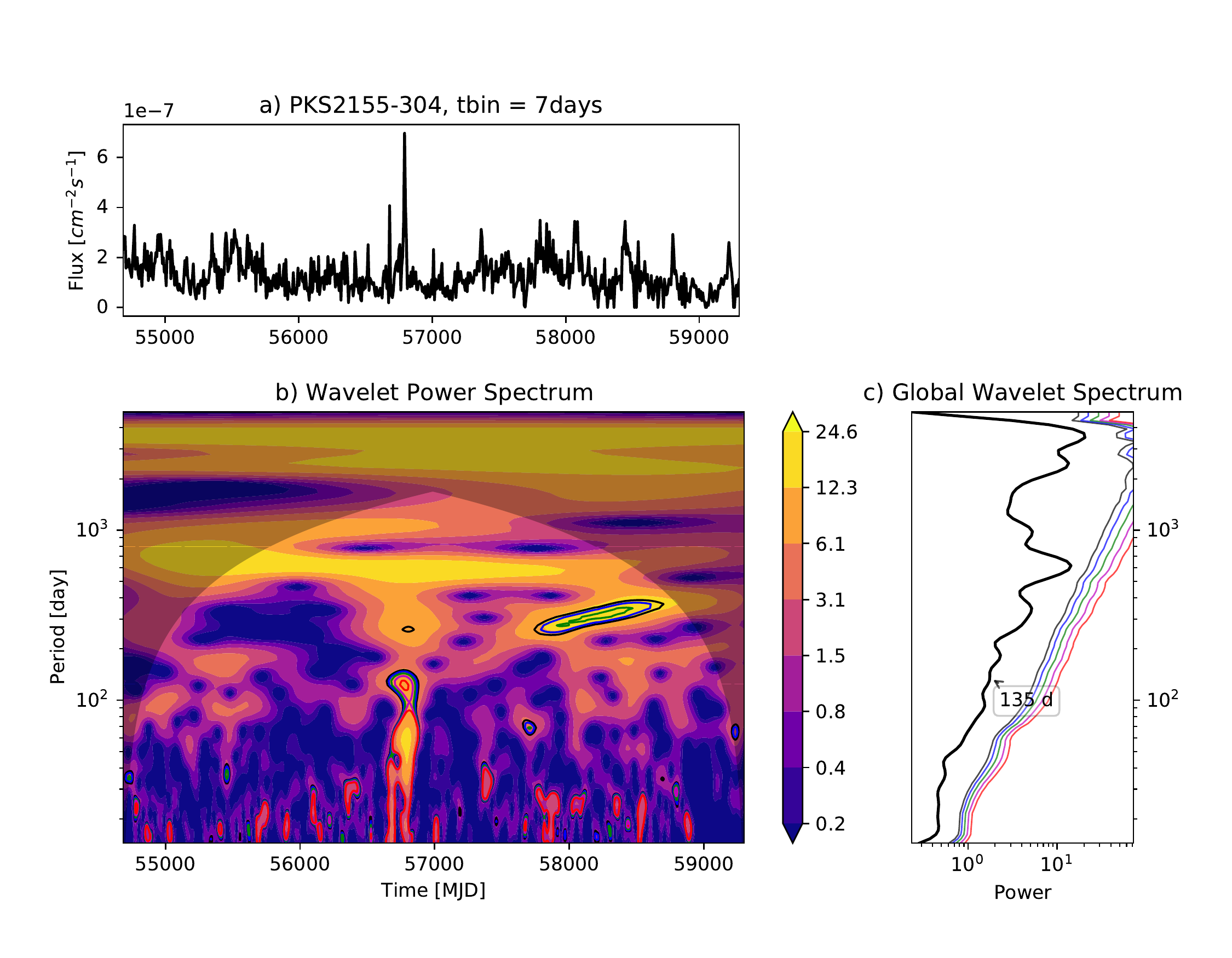}
	\end{subfigure}
	\vskip\baselineskip
	
	\begin{subfigure}[b]{0.48\textwidth}   
		\centering 
		\includegraphics[width=\textwidth]{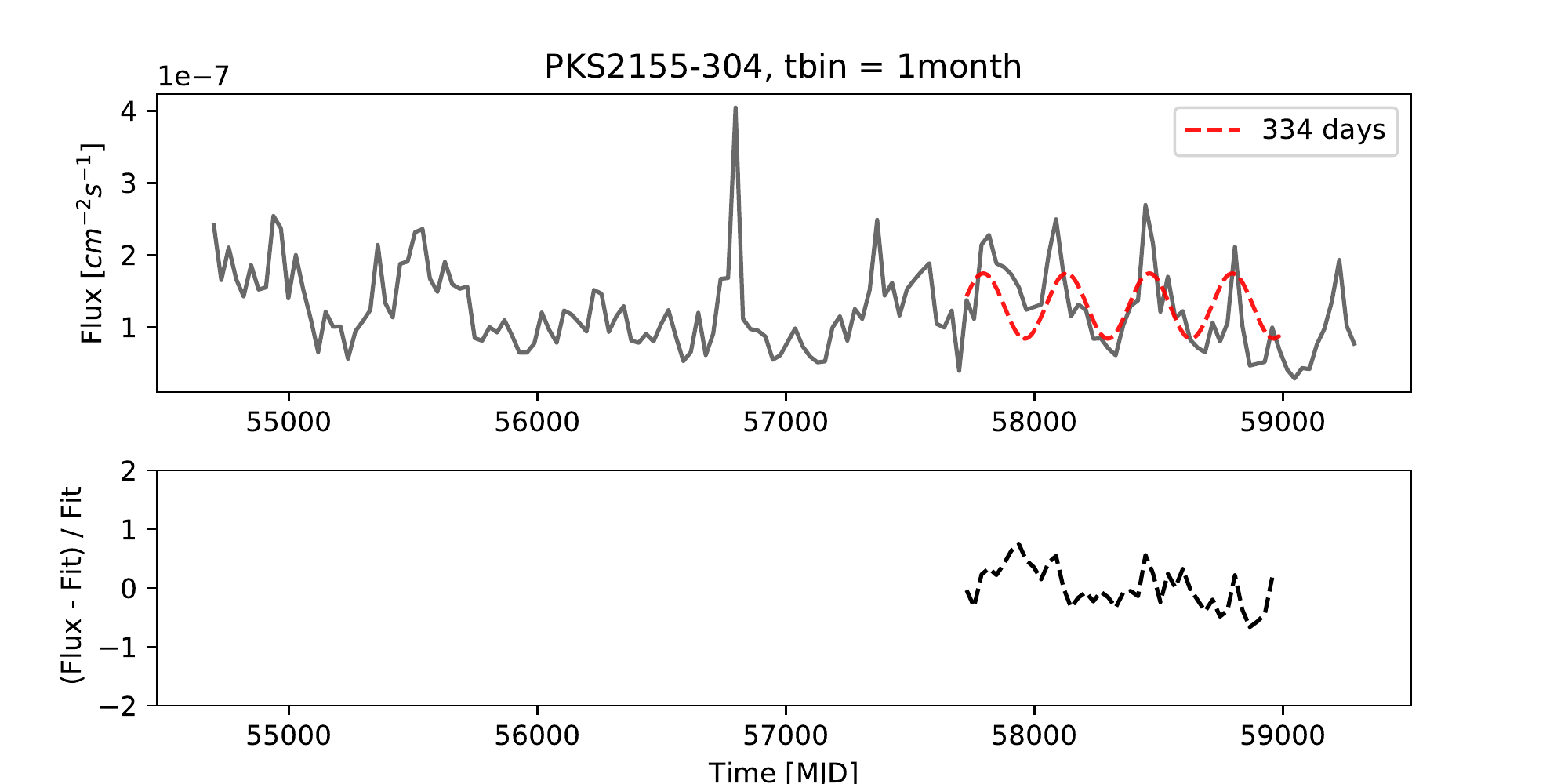}
	\end{subfigure}
	\hfill
	\begin{subfigure}[b]{0.48\textwidth}   
		\centering 
		\includegraphics[width=\textwidth]{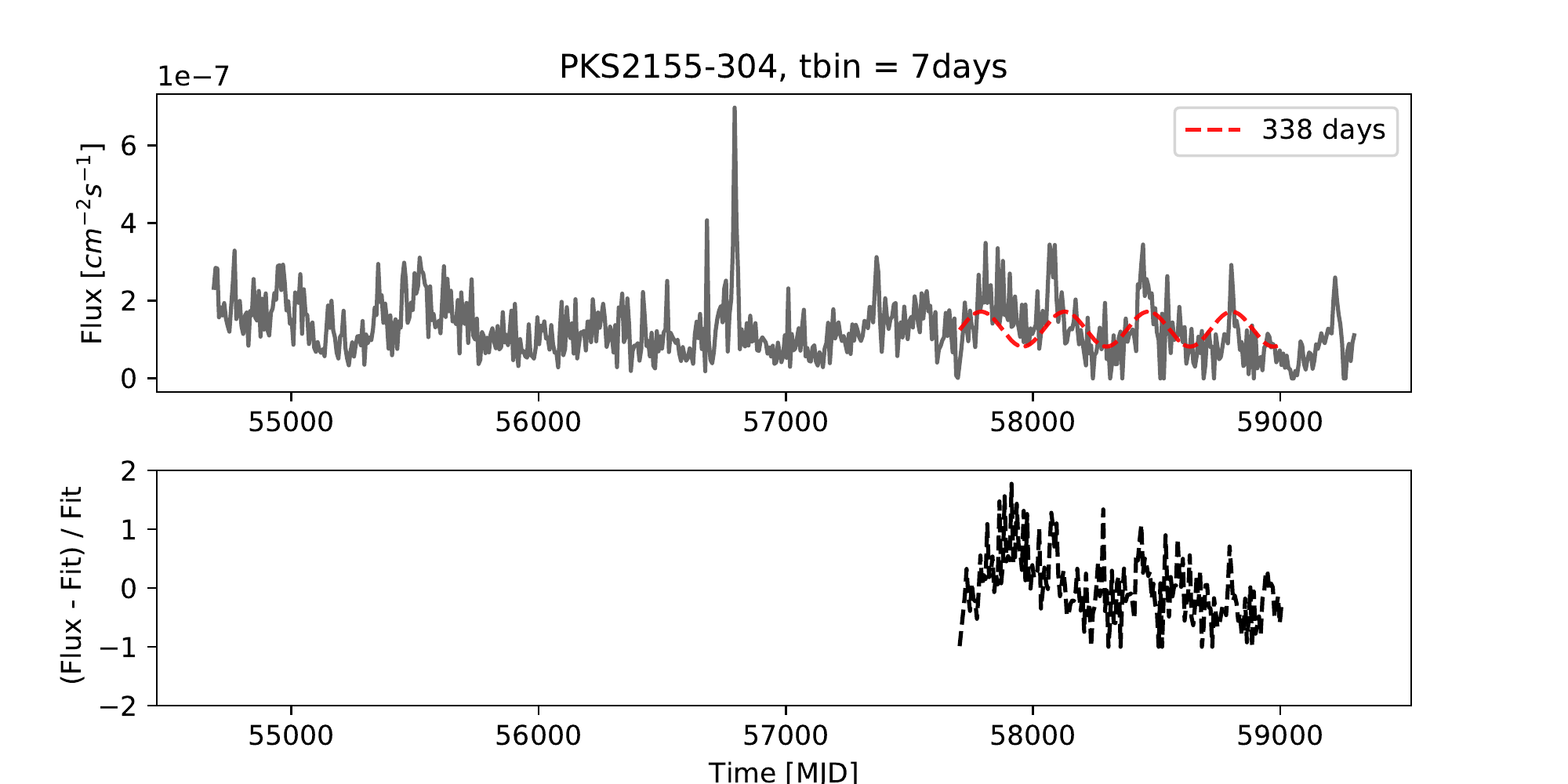}
	\end{subfigure}
	
	\caption{CWT map for monthly binned light curve (left) and weekly binned light curve (right) of 1E~1959+650 and PKS~2155-304, and the fitted light curves for PKS~2155-304.}
	\label{fig:CWT13}
\end{figure*}


\begin{figure*}[!htbp]
	\centering

	\begin{subfigure}[b]{0.48\textwidth}  
    	\centering 
    	\includegraphics[width=\textwidth]{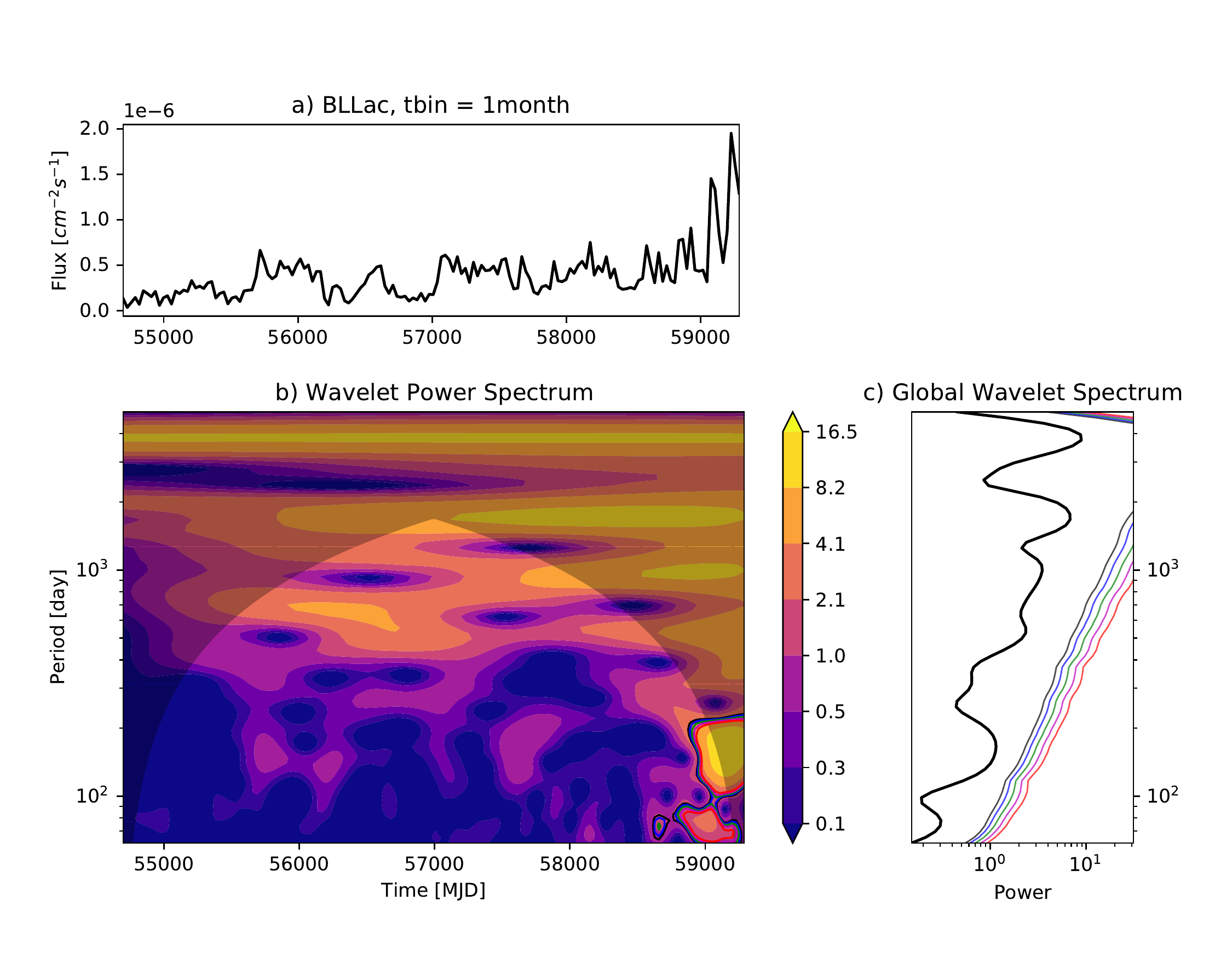}
	\end{subfigure}
	\hfill
	\begin{subfigure}[b]{0.48\textwidth}  
		\centering 
		\includegraphics[width=\textwidth]{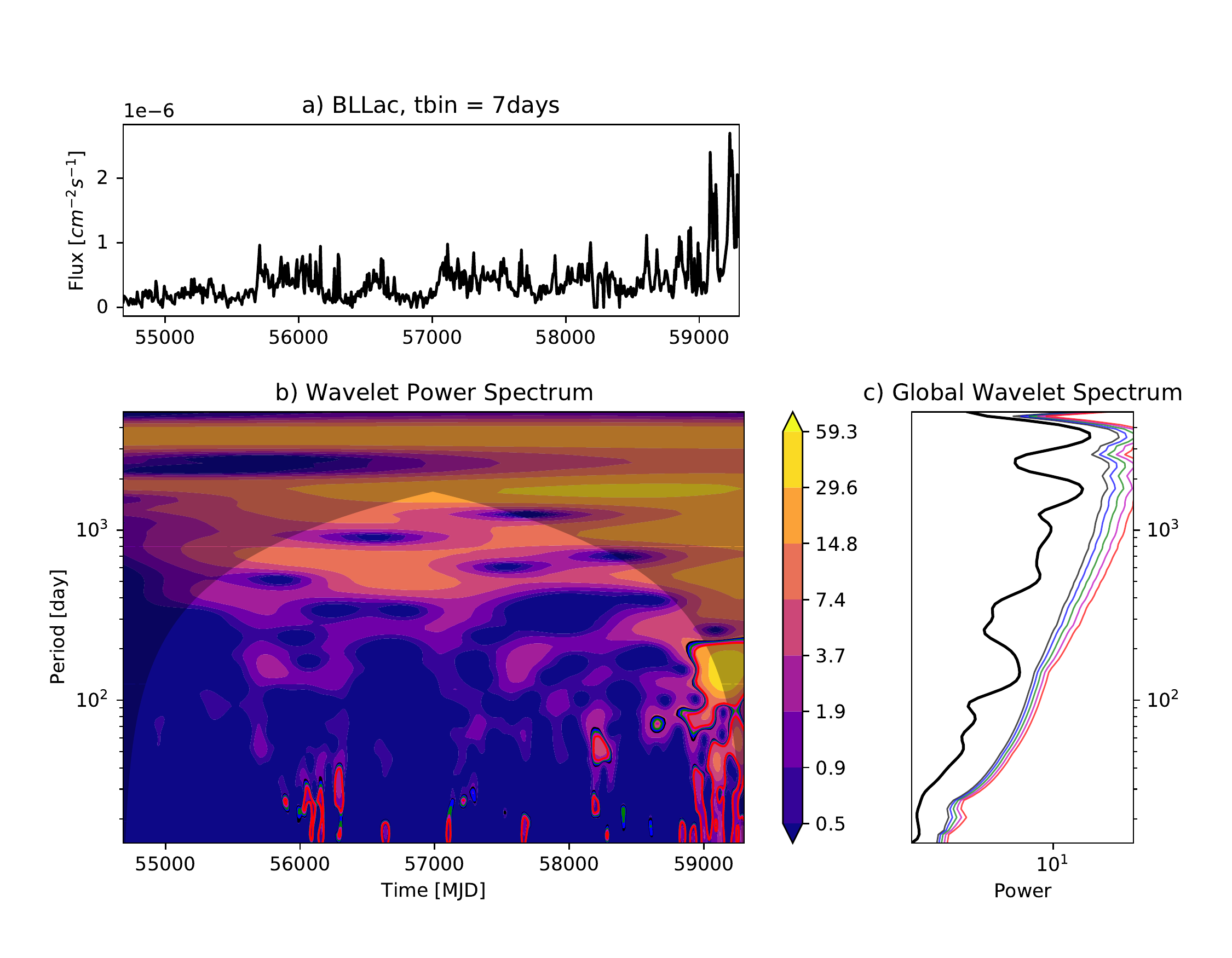}
	\end{subfigure}
	\vskip\baselineskip
	
	\hrule
	
	\begin{subfigure}[b]{0.48\textwidth}
		\centering
		\includegraphics[width=\textwidth]{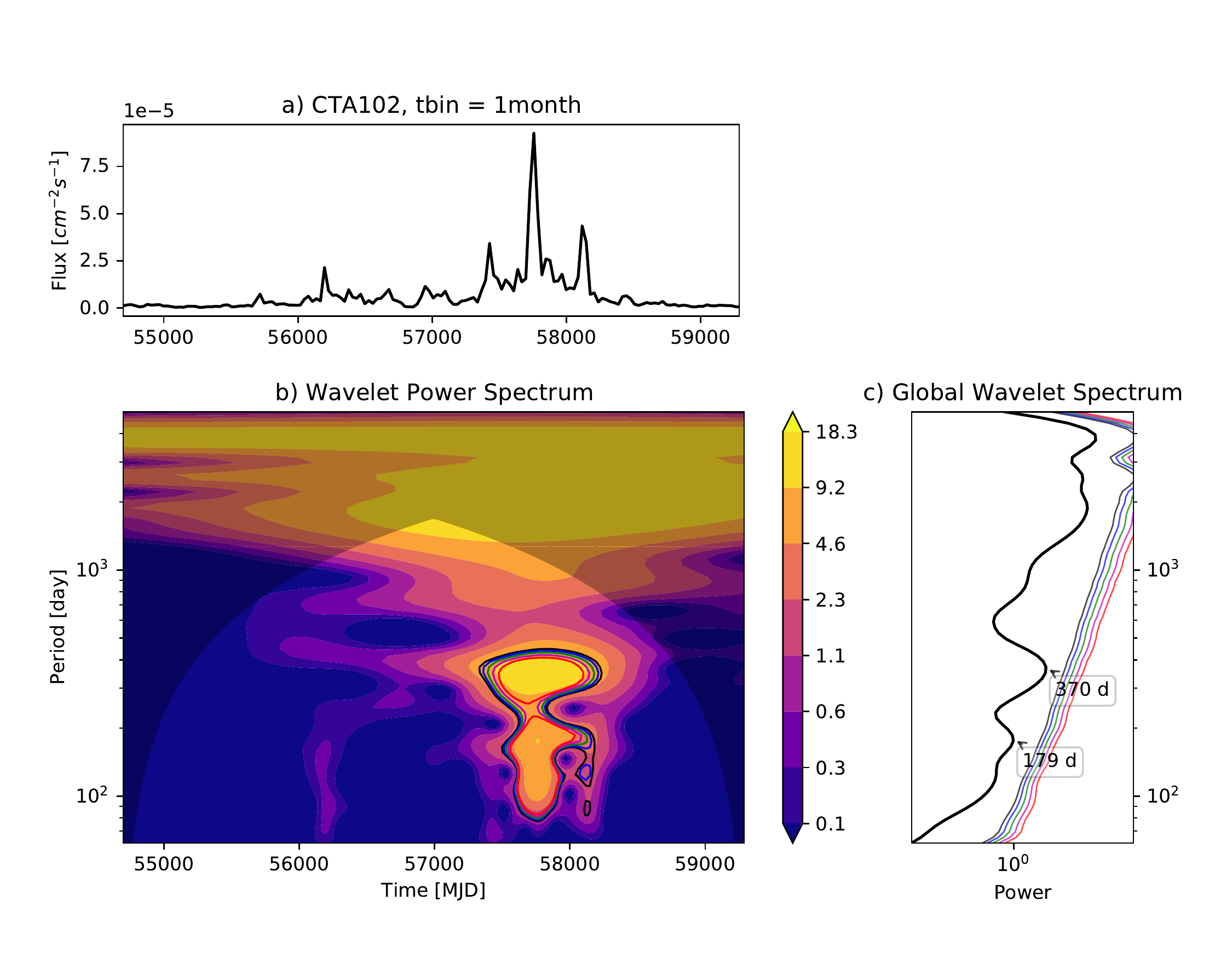}
	\end{subfigure}
	\hfill
	\begin{subfigure}[b]{0.48\textwidth}
		\centering
		\includegraphics[width=\textwidth]{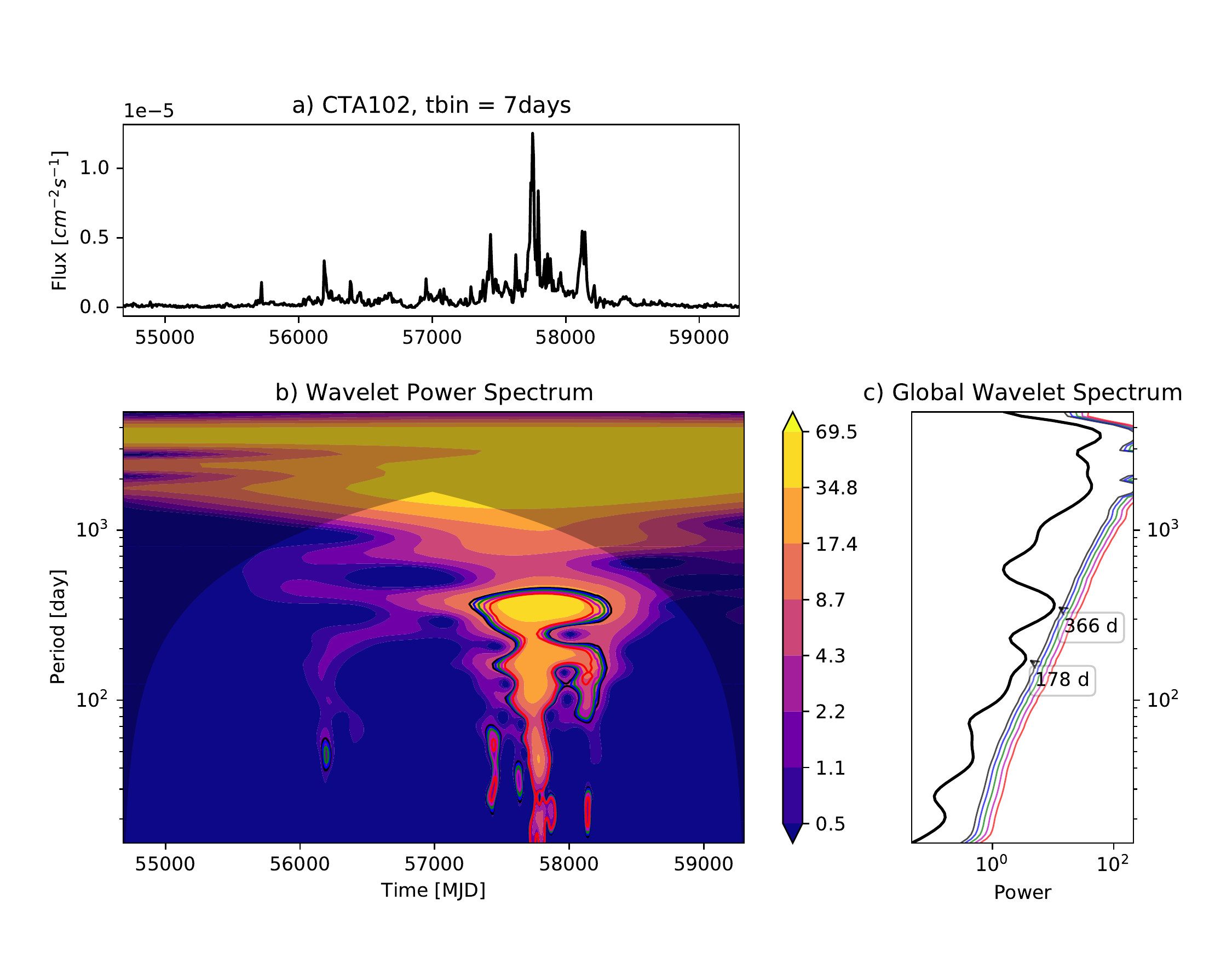}
	\end{subfigure}
	\vskip\baselineskip
	
	\begin{subfigure}[b]{0.48\textwidth}   
		\centering 
		\includegraphics[width=\textwidth]{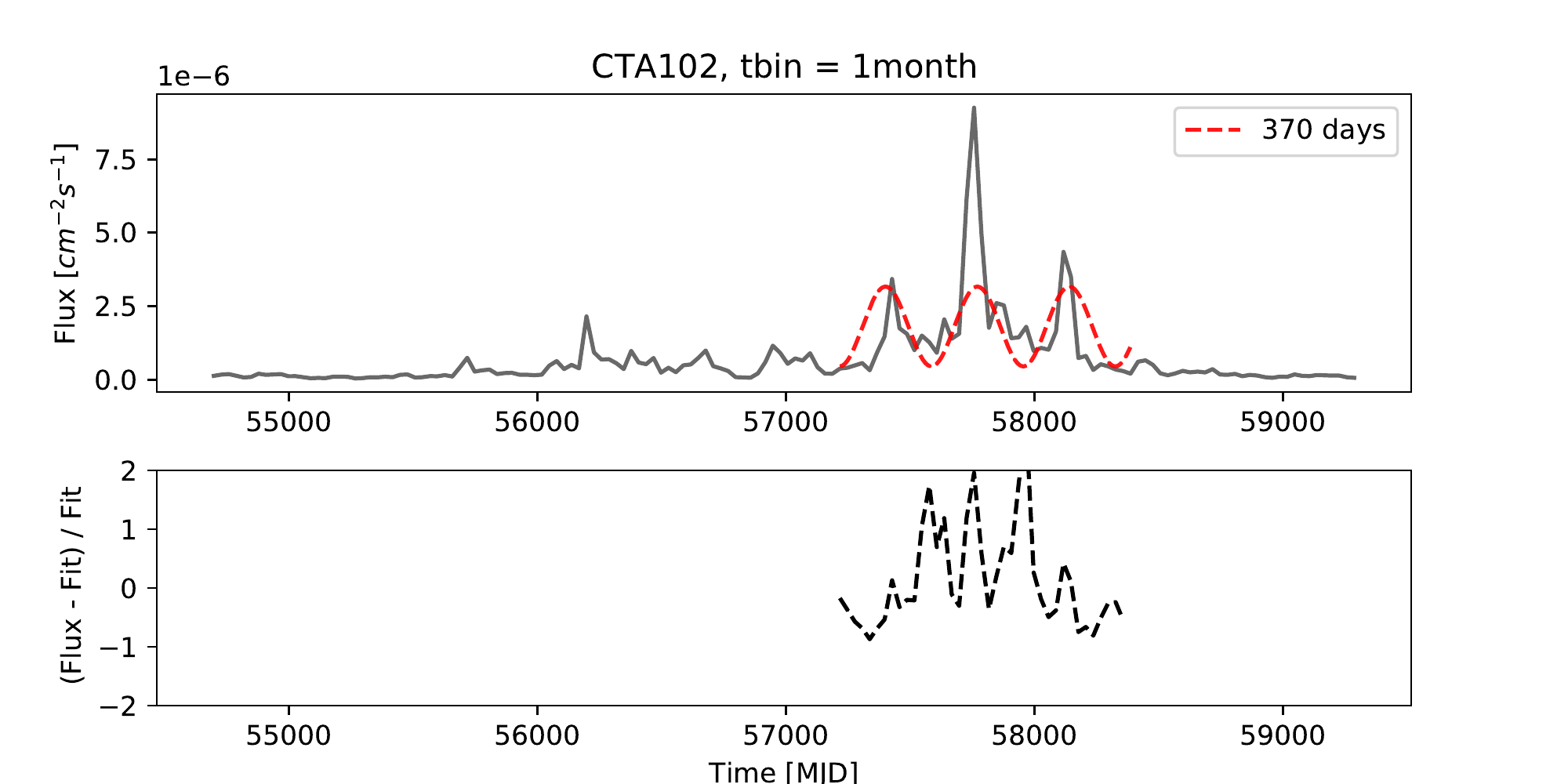}
	\end{subfigure}
	\hfill
	\begin{subfigure}[b]{0.48\textwidth}   
		\centering 
		\includegraphics[width=\textwidth]{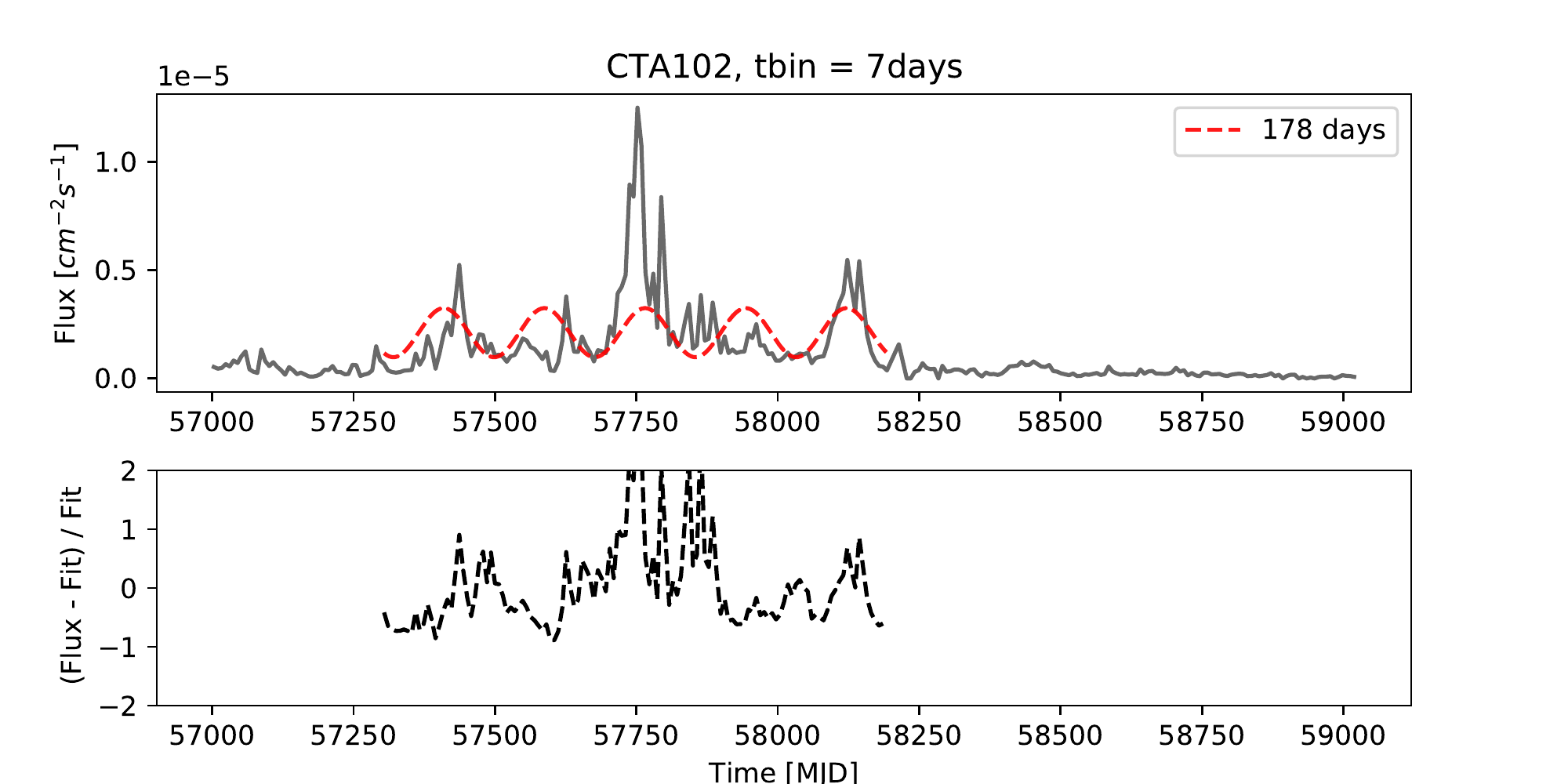}
	\end{subfigure}
	
	\caption{CWT map for monthly binned light curve (left) and weekly binned light curve (right) of BL~Lac and CTA~102, and the fitted light curves for CTA~102.}
	\label{fig:CWT14}
\end{figure*}


\begin{figure*}[!htbp]
	\centering
    
    \begin{subfigure}[b]{0.48\textwidth}  
		\centering 
		\includegraphics[width=\textwidth]{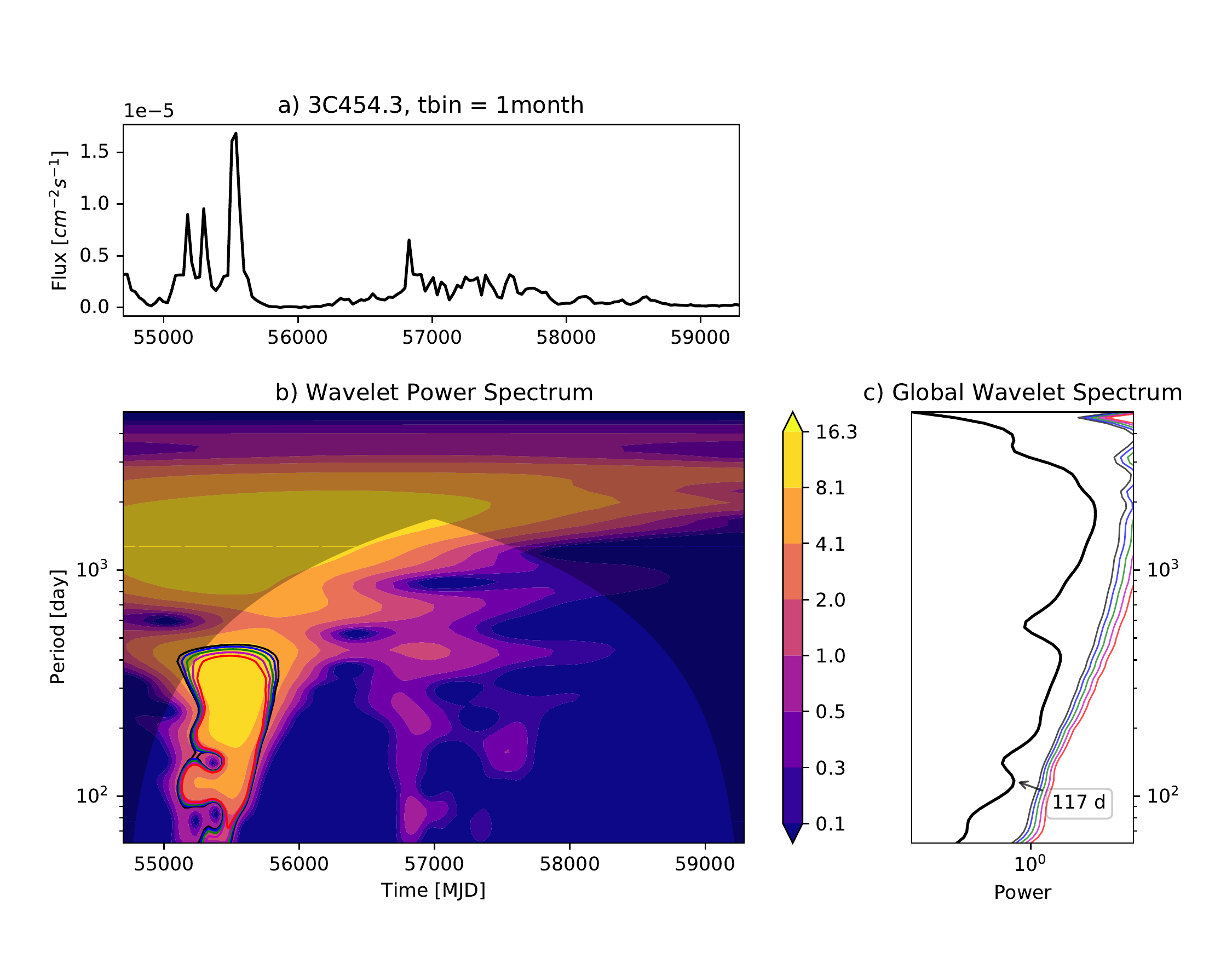}
	\end{subfigure}
	\hfill
	\begin{subfigure}[b]{0.48\textwidth}  
		\centering 
		\includegraphics[width=\textwidth]{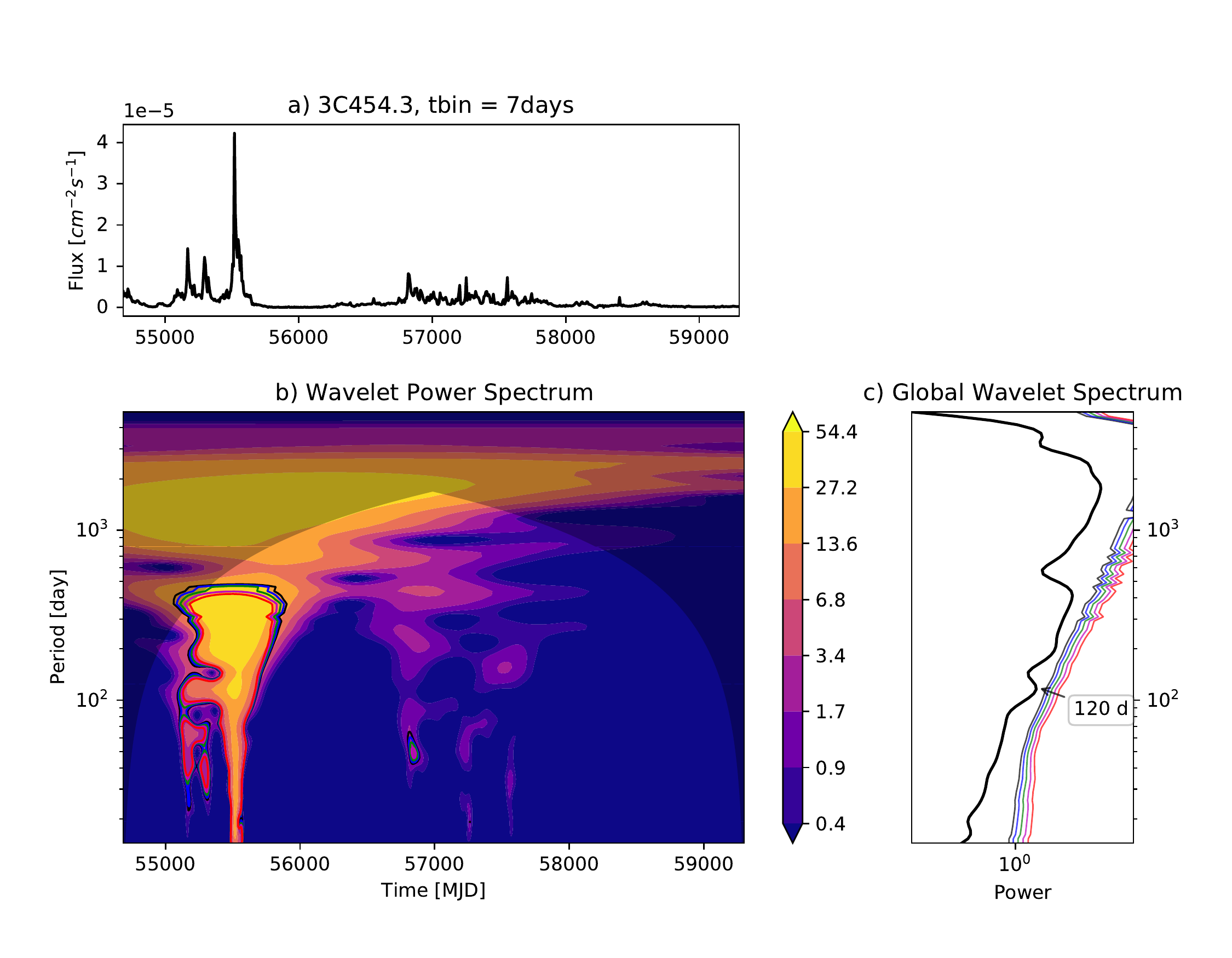}
	\end{subfigure}
    \vskip\baselineskip
	
	\begin{subfigure}[b]{0.48\textwidth}   
		\centering 
		\includegraphics[width=\textwidth]{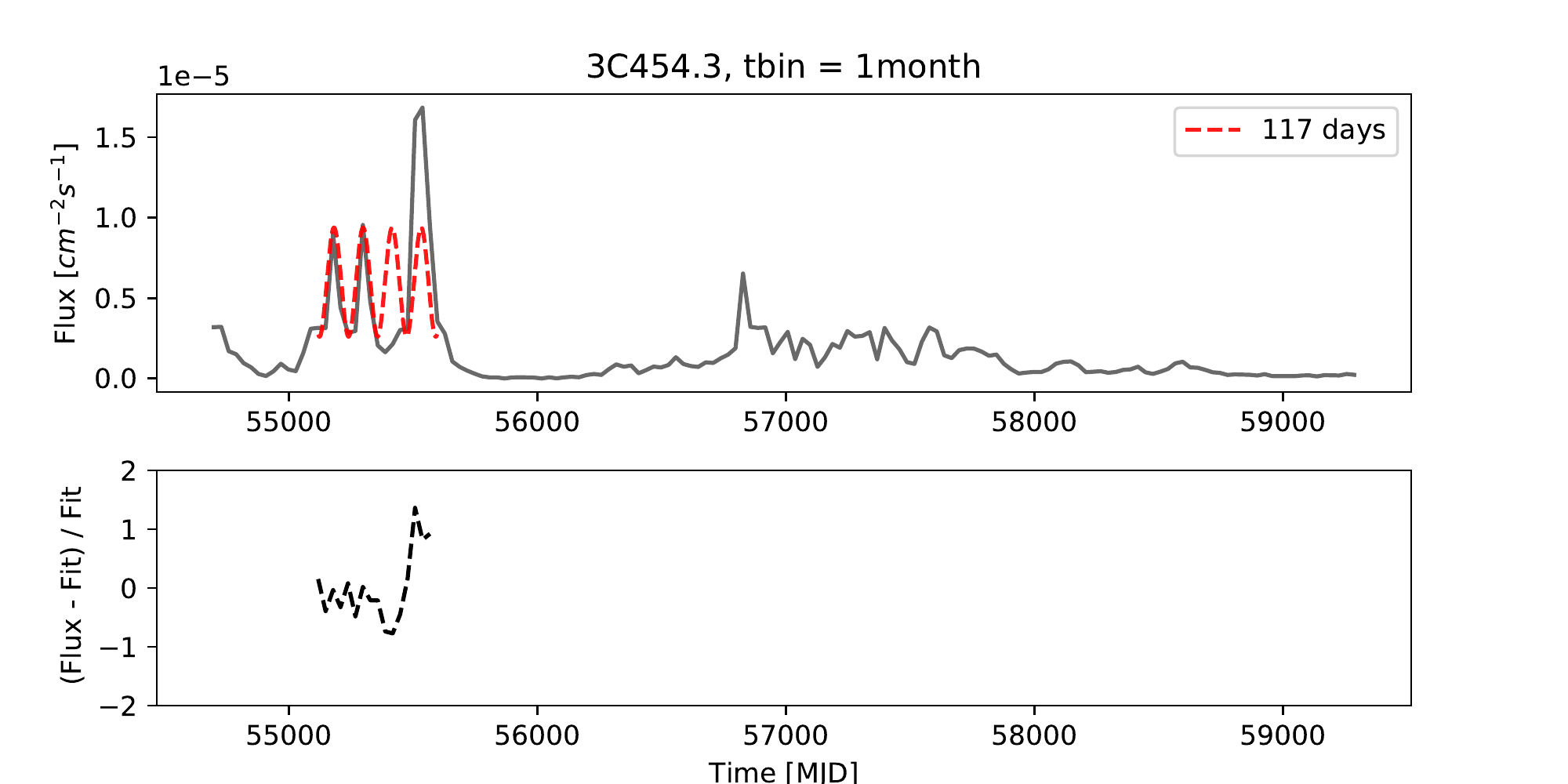}
	\end{subfigure}
	\hfill
	\begin{subfigure}[b]{0.48\textwidth}   
		\centering 
		\includegraphics[width=\textwidth]{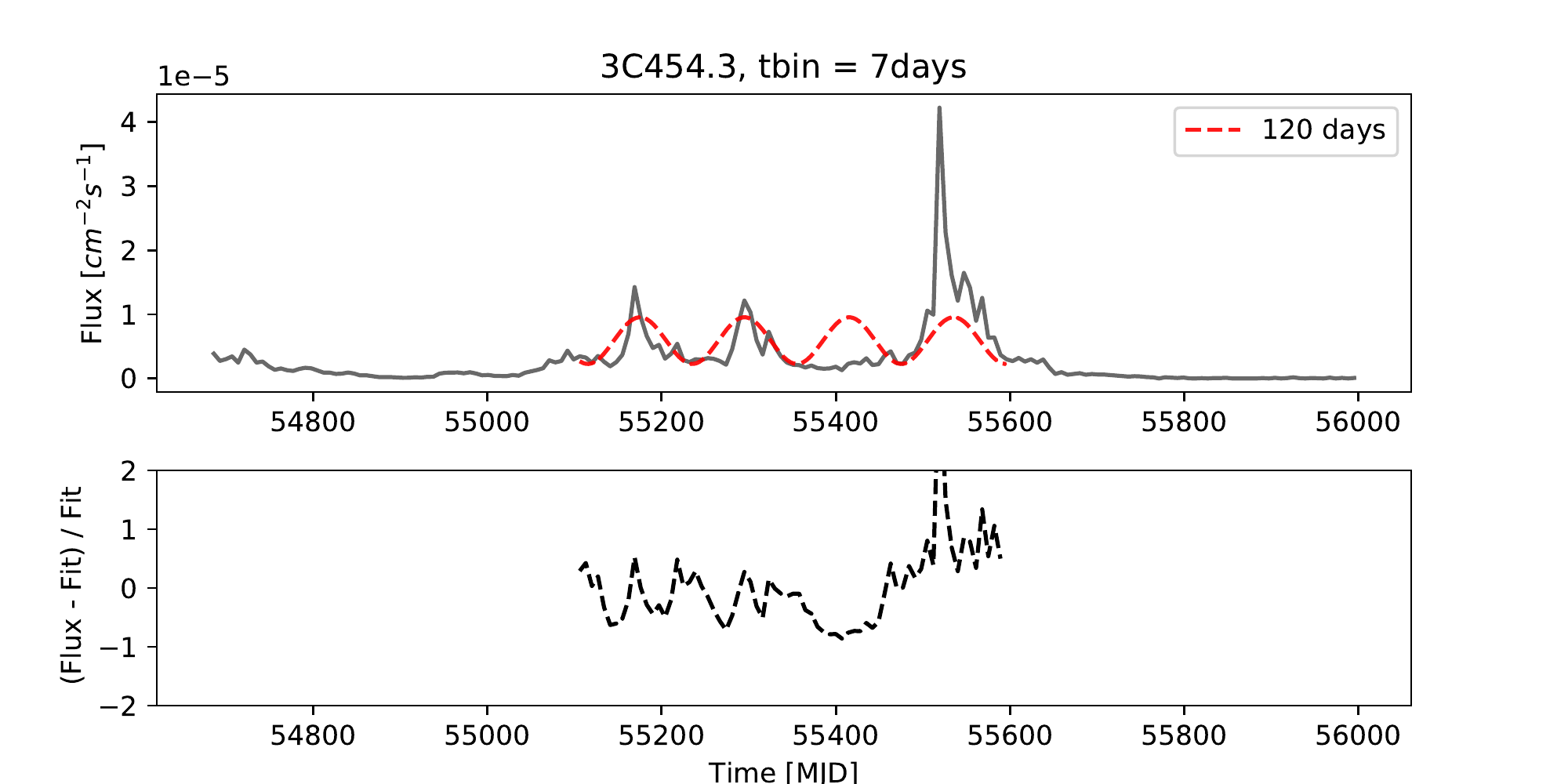}
	\end{subfigure}
	\vskip\baselineskip

    \hrule	
	
	\begin{subfigure}[b]{0.48\textwidth}   
		\centering 
		\includegraphics[width=\textwidth]{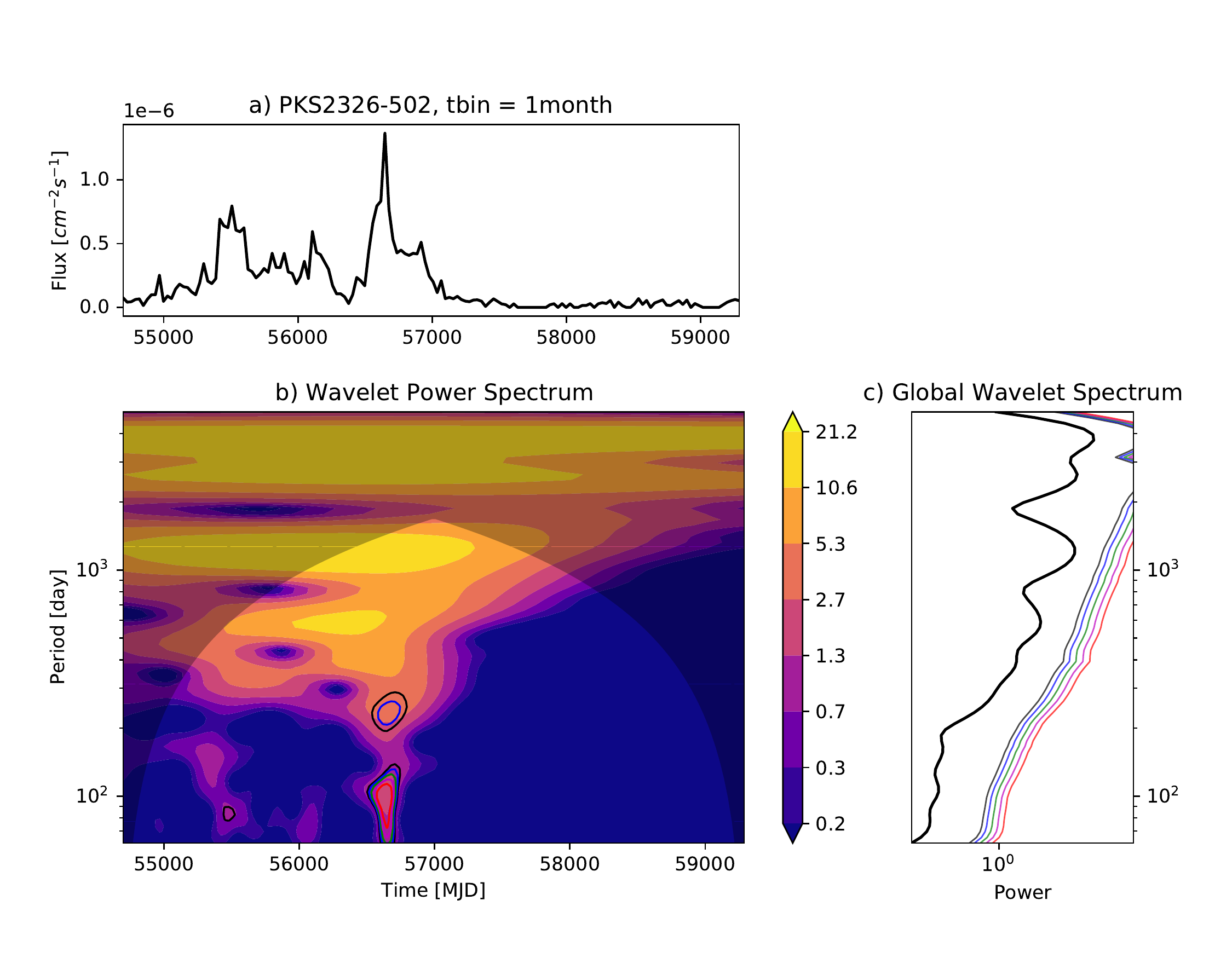}
	\end{subfigure}
	\hfill
	\begin{subfigure}[b]{0.48\textwidth}   
		\centering 
		\includegraphics[width=\textwidth]{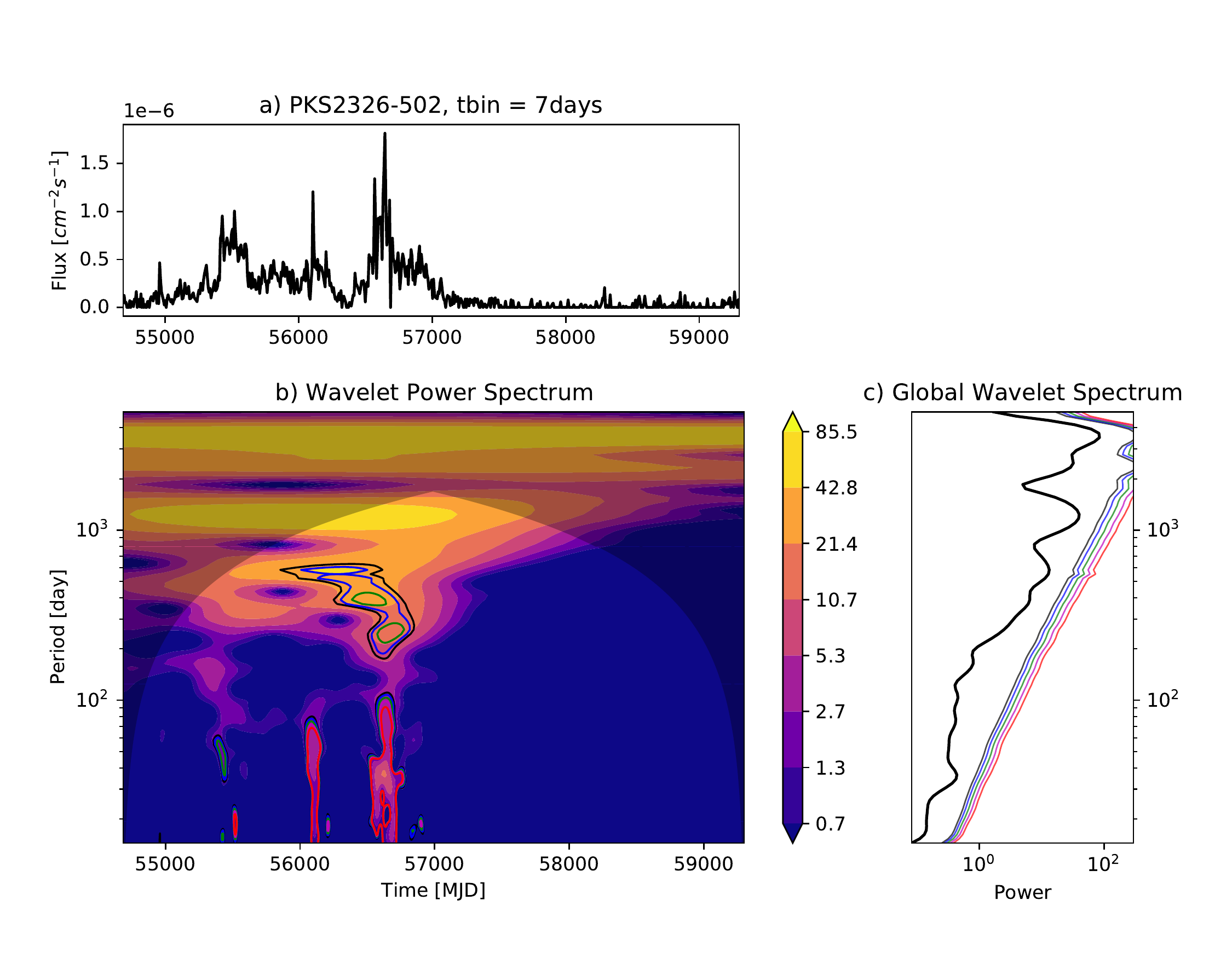}
	\end{subfigure}

	\caption{CWT map for monthly binned light curve (left) and weekly binned light curve (right) of 3C~454.3 and PKS~2326-502, and the fitted light curves for 3C~454.3.}
	\label{fig:CWT15}
\end{figure*}


\begin{figure*}[!htbp]
	\centering
	
	\begin{subfigure}[b]{0.48\textwidth}   
		\centering 
		\includegraphics[width=\textwidth]{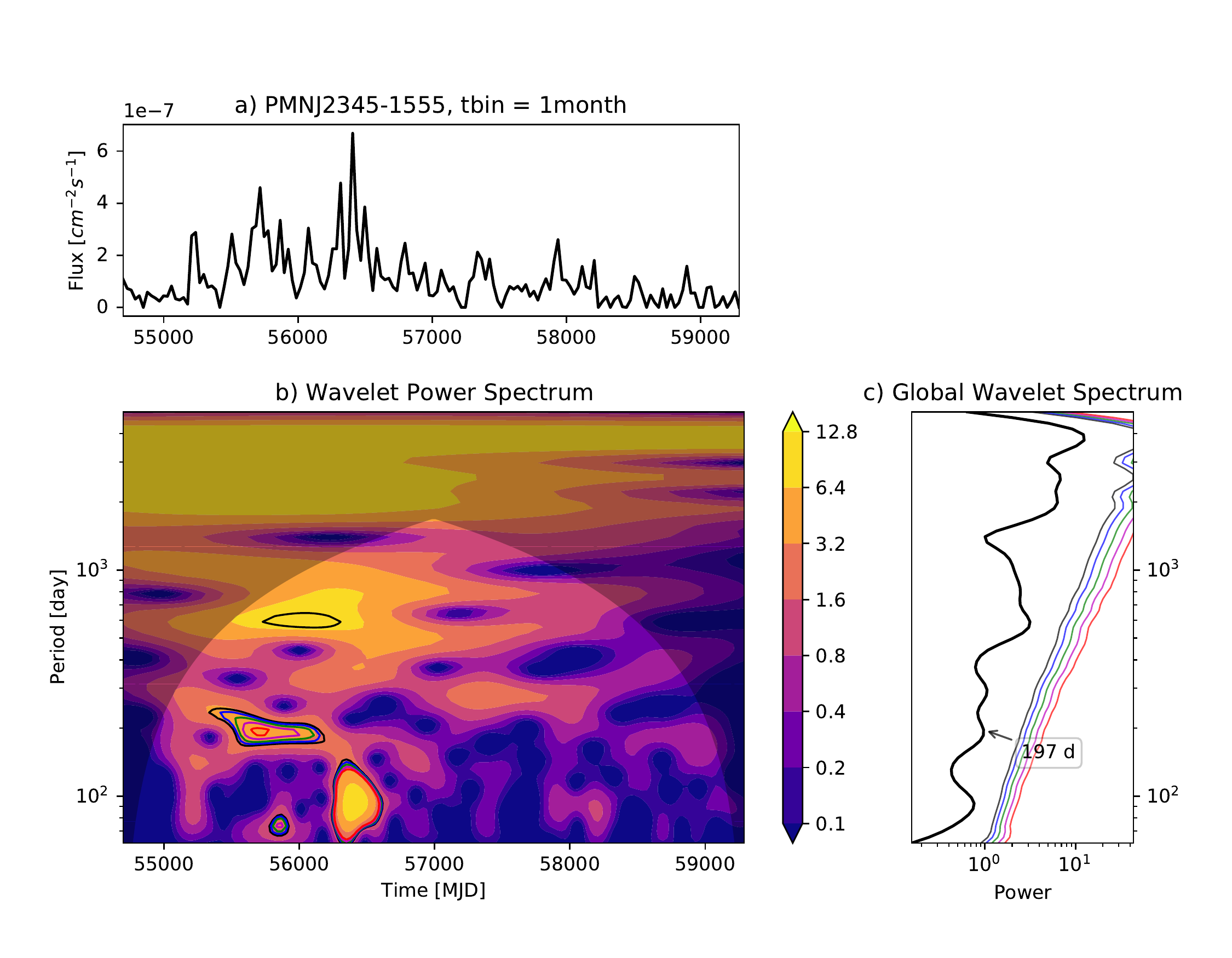}
	\end{subfigure}
	\hfill
	\begin{subfigure}[b]{0.48\textwidth}   
		\centering 
		\includegraphics[width=\textwidth]{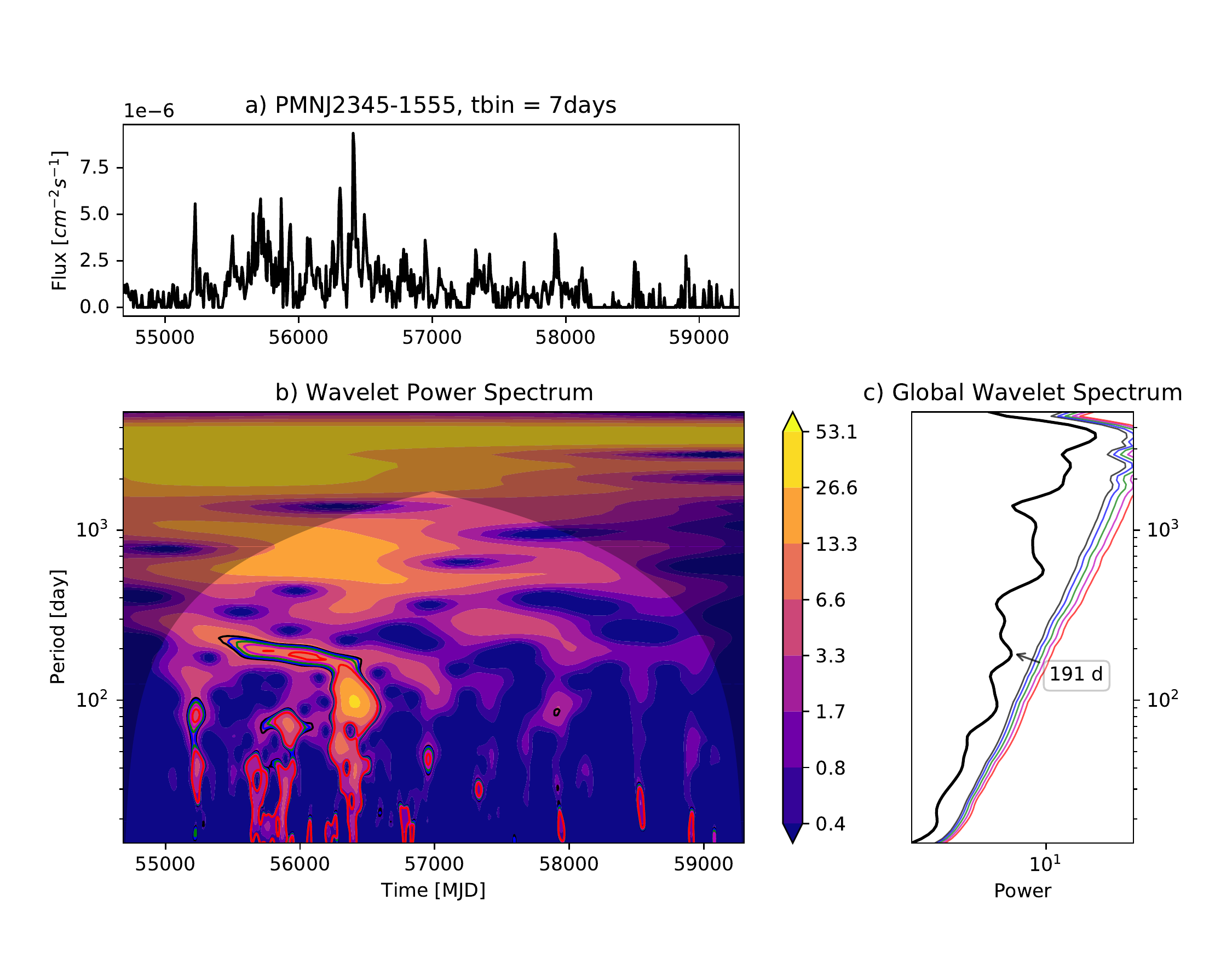}
	\end{subfigure}
	\vskip\baselineskip
	
	\begin{subfigure}[b]{0.48\textwidth}  
		\centering 
		\includegraphics[width=\textwidth]{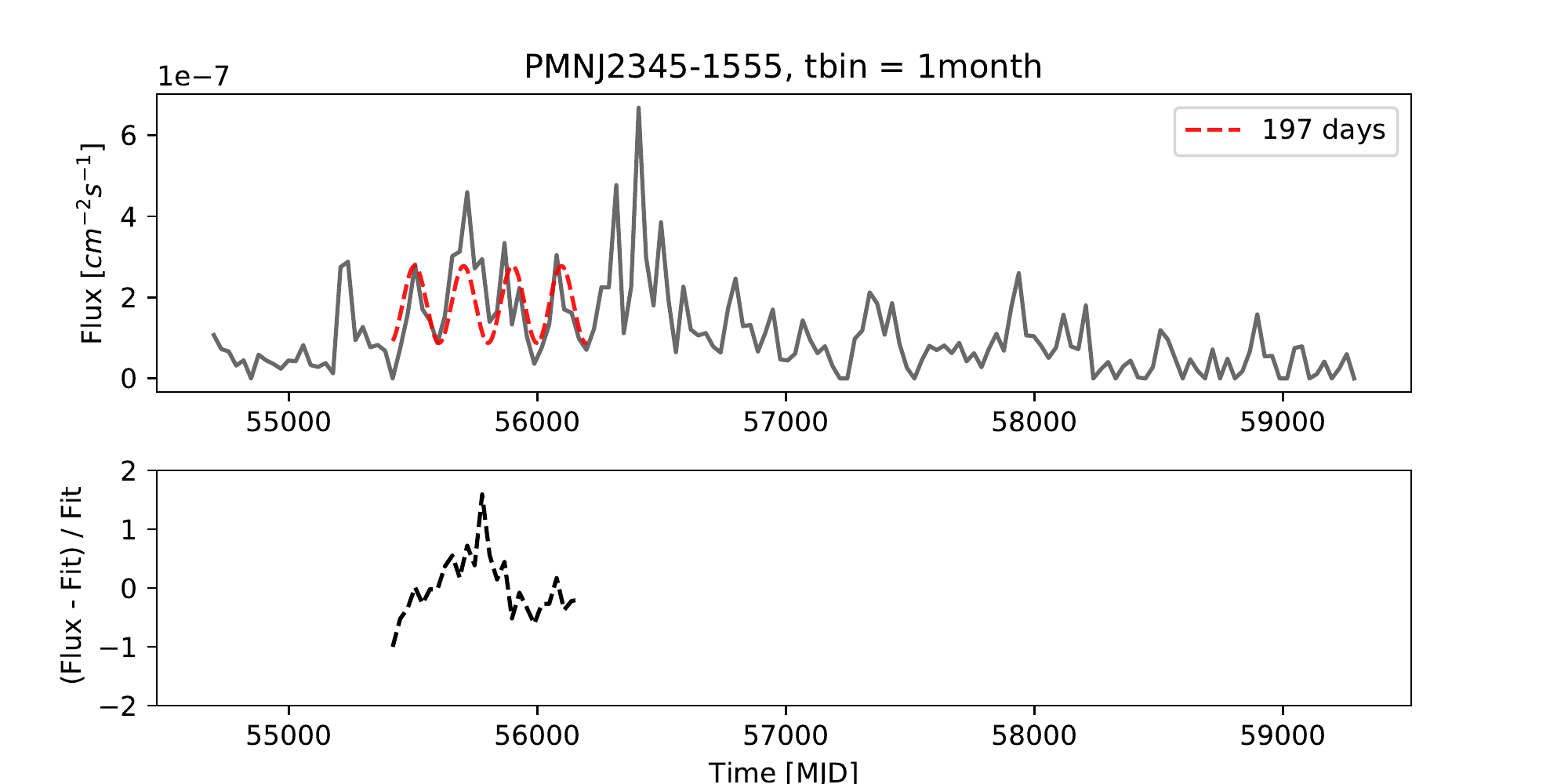}
	\end{subfigure}
	\hfill
	\begin{subfigure}[b]{0.48\textwidth}  
		\centering 
		\includegraphics[width=\textwidth]{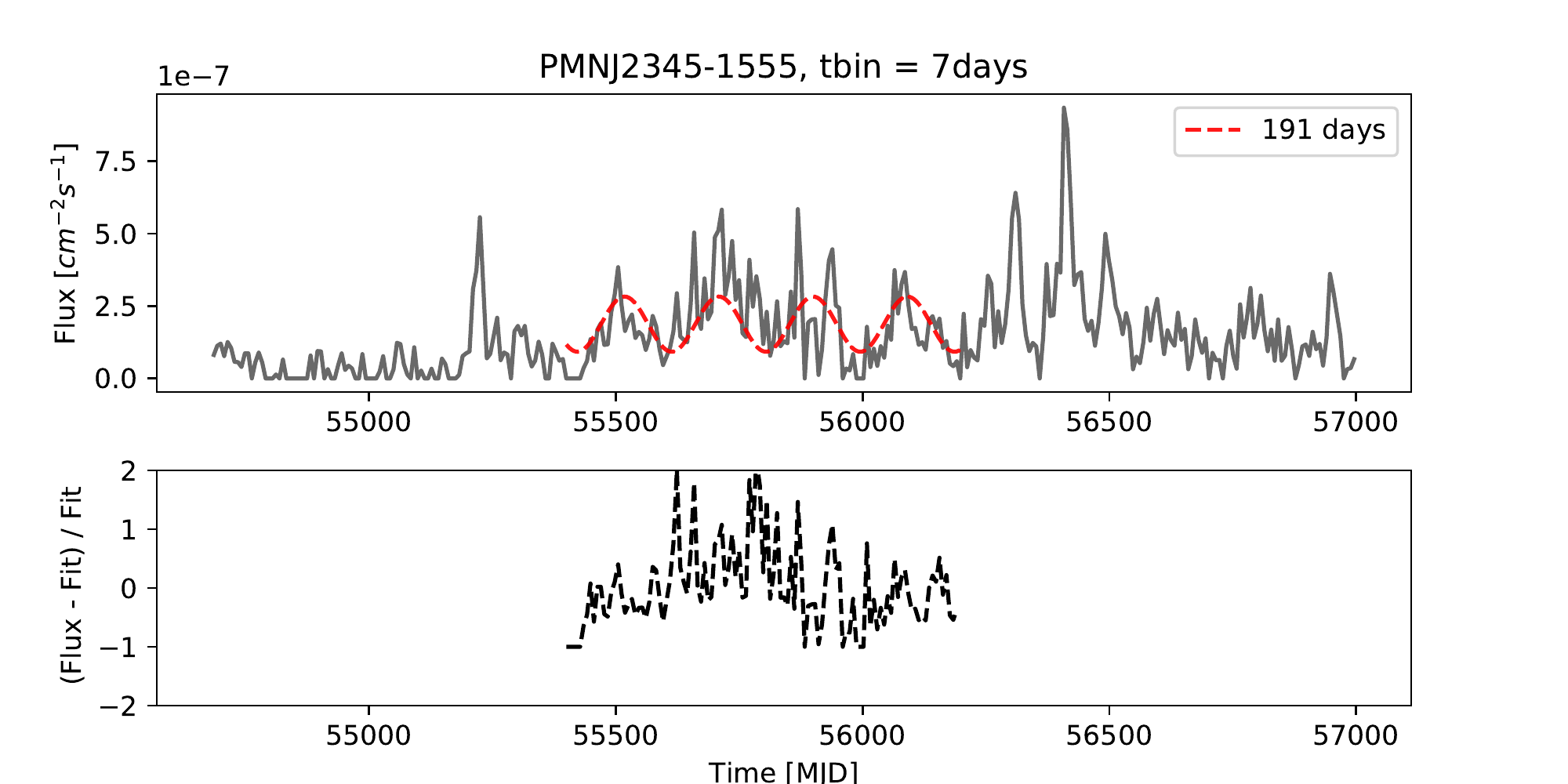}
	\end{subfigure}
	
	\caption{CWT map for monthly binned light curve (left) and weekly binned light curve (right) of PMN~J2345-1555, and the fitted light curves.}
	\label{fig:CWT16}
\end{figure*}


\end{document}